\documentclass[a4paper,11pt,oneside]{thesis}

\begin{document}
\pagestyle{plain}
\pagenumbering{roman}

\begin{titlepage} 
\begin{adjustwidth}{-4em}{-4em}

\begin{center} 
\begin{figure}[tr]
\includegraphics[width=3cm,height=1.0cm]{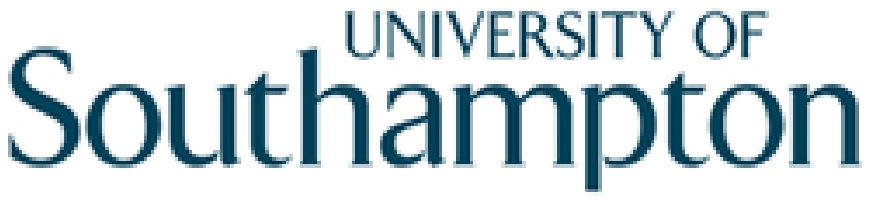} 
\vspace{2.0cm}
\end{figure}

\huge \scshape{\textcolor{Red}{O}bservational St\textcolor{Red}{u}dies of} 
 
\huge \scshape{ -- \textcolor{Red}{H}ighly Evol\textcolor{Red}{v}ed Catacl\textcolor{Red}{y}smic \textcolor{Red}{V}ariables --}
\vspace{0.3cm}

\begin{figure}[h]
\centering
\begin{adjustwidth}{-4em}{-4em}
\includegraphics[width=16.5cm,height=0.35cm]{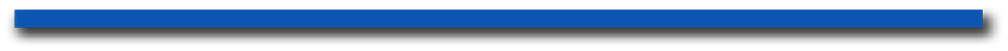} 
\end{adjustwidth}
\end{figure} 
 
\begin{figure}[h]
\begin{adjustwidth}{-2em}{-4em}
\includegraphics[scale=1.0]{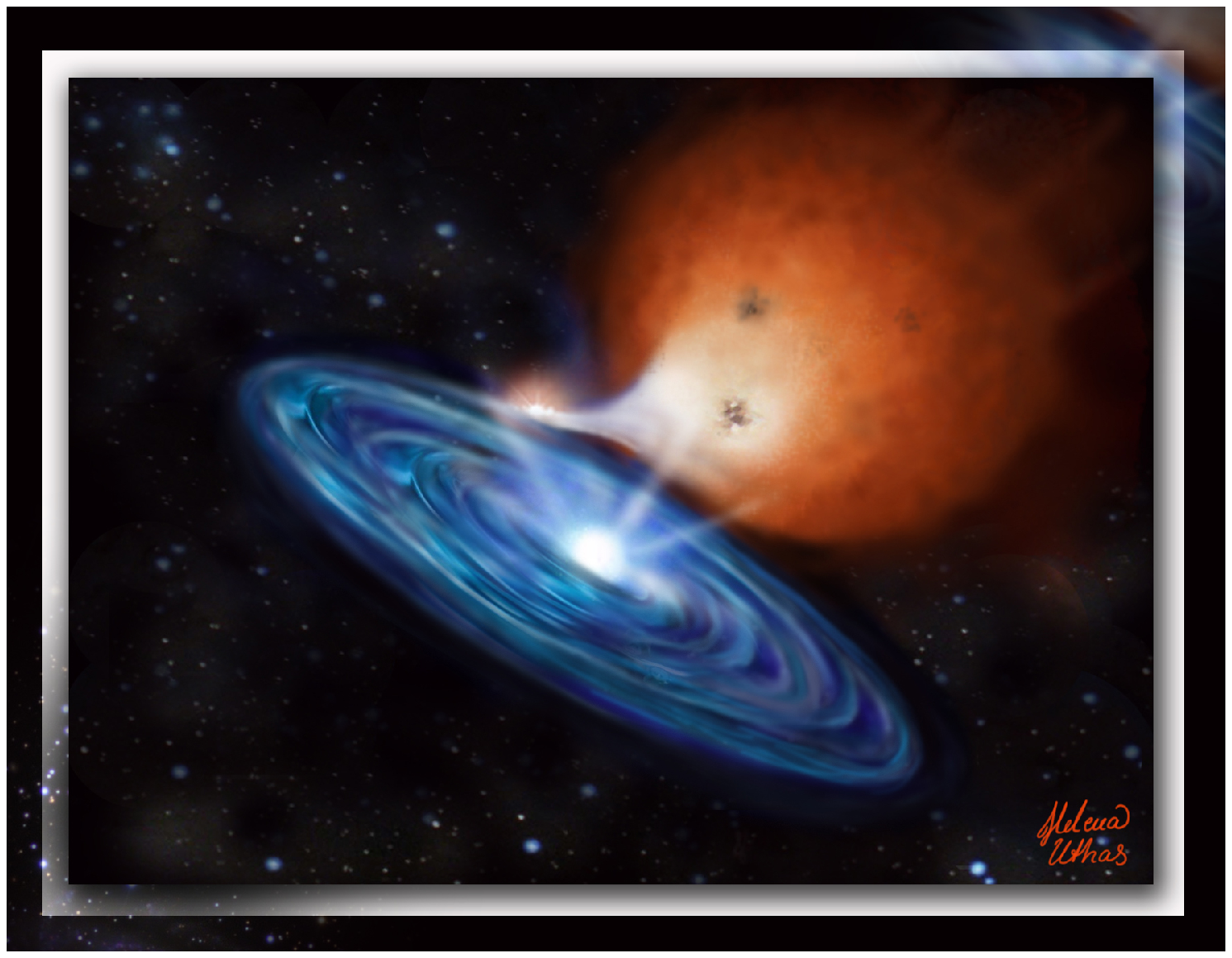} 
\end{adjustwidth}
\end{figure}  
\vspace{0.3cm}
\large \scshape{By \textcolor{Red}{H}elena Ut\textcolor{Red}{h}as}
\\ 
\vspace{0.5cm}
\sc{February 2011}
\\
\normalsize \sc{\textcolor{Red}{Thesis submitted for the degree of Doctor of Philosophy}}
\\
\textsl{School of Physics \& Astronomy -- University of Southampton} 
 
\end{center}
\end{adjustwidth}
\end{titlepage}
\newpage \thispagestyle{empty} \mbox{} 
\newpage \thispagestyle{empty} \mbox{} 


\begin{center}
\sc{University of Southampton}
\\
\sc{Faculty of Engineering, Science \& Mathematics}
\\
\sc{School of Physics \& Astronomy}
\\
\rm \underline{Doctor of Philosophy}
\\
\vspace{0.8cm}
\sc \textbf{Observational Studies of Highly Evolved Cataclysmic Variables}
\\
\sl{by Helena Uthas}
\end{center}

\section*{Abstract}

{\footnotesize Cataclysmic Variables (CV) are binary systems where a main-sequence star transfers mass onto a white dwarf (WD). According to standard evolutionary theory, angular momentum loss drives CVs to initially evolve from longer to shorter orbital periods until a minimum period is reached ($\approx$ 80 minutes). At roughly this stage, the donors becomes degenerate, expand in size, and the systems move towards longer orbital periods. Theory predicts that 70\% of all CVs should have passed their minimum period and have sub-stellar donors, but until recently, no such systems were known. I present one CV showing evidence of harbouring a sub-dwarf donor, SDSS J1507+52. Due to the system's unusually short orbital period of $\approx$ 65 minutes, and very high space velocity, two origins for SDSS J1507+52 have been proposed; either the system was formed from a young WD/brown-dwarf binary, or the system is a halo CV. In order to distinguish between these two theories, I present UV spectroscopy and find a metallicity consistent with halo origin.   

Systems close to the minimum period are expected to be faint and have low accretion rates. Some of these CVs show absorption in their spectra, implying that the underlying WD is exposed. This yields a rare opportunity to study the WD in a CV. I introduce two new systems showing WD signatures in their light curves and spectra, SDSS J1457+51 and BW Sculptoris. Despite the fact that CVs close to the minimum period should be faint, we find systems that are much too bright for their orbital periods. Such a system is T Pyxidis -- a recurrent nova with an unusually high accretion rate and a photometrically determined period < 2 hours. The system is $\sim$ 2 times brighter than any other CV at its period. However, to confirm the status of this unusual star, a more reliable period determination is needed. Here, I present a spectroscopic study of T Pyxidis confirming its evolutionary status as a short-period CV.  

In this thesis, I discuss what implications these systems may have on the current understanding of CV evolution, and the importance of studying individual systems in general.}


\newpage \thispagestyle{empty} \mbox{} 
\newpage \thispagestyle{empty} \mbox{} 
\renewcommand\contentsname{List of Contents}
\tableofcontents 
\newpage

\section*{Declaration of Authorship}
\vspace{0.5cm}
I, Helena Uthas declare that the thesis entitled, \emph{Observational Studies of Highly Evolved Cataclysmic Variables}, and the work presented in the thesis are both my own, and have been generated by me as the result of my own original research. I confirm that:

\begin{itemize}
\item this work was done wholly or mainly while in candidature for a research degree at this University;

\item where any part of this thesis has previously been submitted for a degree or any other qualification at this University or any other institution, this has been clearly stated;

\item where I have consulted the published work of others, this is always clearly attributed;

\item where I have quoted from the work of others, the source is always given. With the exception of such quotations, this thesis is entirely my own work;

\item I have acknowledged all main sources of help;

\item where the thesis is based on work done by myself jointly with others, I have made clear exactly what was done by others and what I have contributed myself;

\item parts of this work have been published as:
\begin{itemize}
\item \textbf{\small Uthas, H., Knigge, C., Steeghs, D., 2010, MNRAS, 409, 237}
\item \textbf{\small Uthas, H., Knigge, C., Long,  K. S., Patterson, J., Thorstensen, J., 2011, to appear in MNRAS}
\item \textbf{\small Uthas, H., Patterson, J., Kemp, J., Knigge, C., Monard, B., Rea, R., Bolt, G., McCormick, J., Christie, G., Retter, A., Liu, A., 2011, to appear in MNRAS}
\end{itemize}
\end{itemize}

\noindent Signed (date):


\newpage \thispagestyle{empty} \mbox{} 
\newpage \thispagestyle{empty} \mbox{}

\section*{Acknowledgements}
 
I spent the two first years of my PhD studies high above the clouds watching the stars, during cold, endless nights at the Nordic Optical Telescope (NOT) on La Palma/Spain. I am most grateful to \textbf{Thomas Augusteijn} (Deputy-Director at the NOT) for giving me the opportunity to work as a support astronomer, and for giving me invaluable experience in the field of observational astronomy.  

When I started my PhD studies, I had no previous experience in programming. I read somewhere that, \emph{programming, and especially debugging, brings out strong emotions}. For me, it certainly was both frustrating and emotional, and I would like to thank both \textbf{Ricardo C\'{a}rdenes} (Systems Specialist at the NOT) and \textbf{Tony Bird} (University of Southampton), for your enormous patience and for spending so much time teaching me how to program in Python, even though you had no obligations to do so.

Another person that deserves to be mentioned is \textbf{Danny Steeghs}. You taught me the technique of Doppler tomography, and were always an email away to answer any questions. Using your impressive teaching skills, you somehow managed to help me visualise the most abstract concepts through simple drawings.

I am also most grateful to \textbf{Joe Patterson}, who without knowing me, took me for a 50-night long observing campaign at the MDM telescope, Kitt Peak Observatory in Arizona, where I had a lot of time on my hands to collect data for my thesis. I will never forget the fascinating discussions we had during late nights at the telescope, and the adventurous 13-hour long car trip through the Mojave desert. 
 
Finally, thanks to \textbf{Christian Knigge} for being such an extraordinary supervisor. I feel privileged that I had the opportunity to work with you, and I am most grateful for the outstanding support I got through out the whole time of my PhD studies, no matter where on earth I was currently working from.


\newpage \thispagestyle{empty} \mbox{} 

\pagestyle{fancy}
\pagenumbering{arabic}
\setcounter{page}{0}


\chapter{Cataclysmic Variables and their Evolution}
\label{chap:icm}

\begin{Huge}\color{Red}{I}\end{Huge}n this Chapter, I will give an overview of cataclysmic variables (CVs) and how we think they evolve. Key physical concepts such as Roche-lobe overflow, loss of angular momentum and nova eruptions will be introduced, as well as characteristic observational features in CVs, in particular, their variability. I will then give an overview of the current status of theoretical and observational understanding of these objects.

\section{CVs as Accreting Systems}

\begin{figure}[t]
\begin{center}
\includegraphics[scale=0.75]{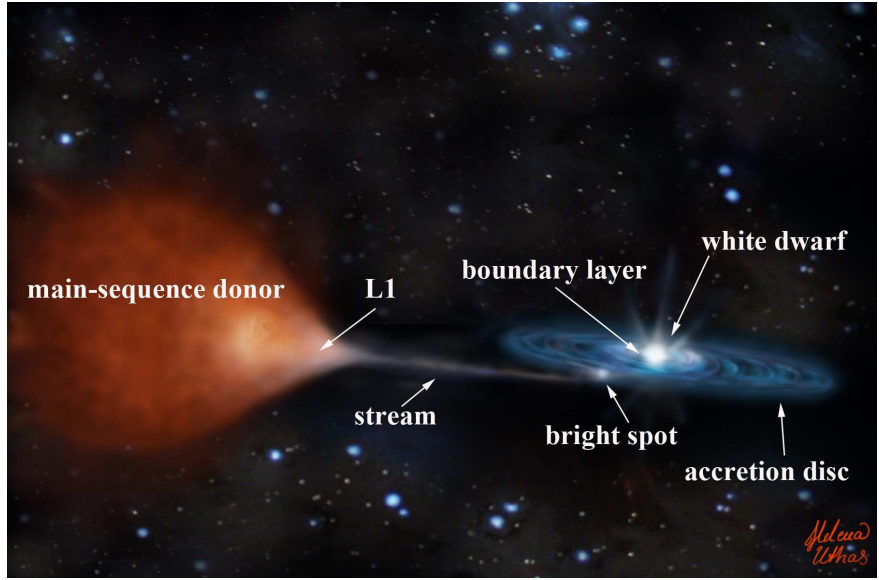}
\caption{\emph{Author's impression of a cataclysmic variable.}}
\label{mypic}
\end{center}
\end{figure}

Cataclysmic variables are interacting binary stars in which a donor star loses mass to a primary white dwarf (WD). The donor is generally near the main sequence (except in rare cases where the donor may already have undergone a significant amount of nuclear evolution). In Figure~\ref{mypic}, I present the standard picture of a CV, showing how matter from the donor is transferred to the WD by means of an accretion disc. In the disc, viscous processes efficiently transport angular momentum outward and matter inward towards the WD. As a result, gravitational potential energy is converted into radiation, and the disc is responsible for a large portion of the detected optical light. The gas emerges from the donor and forms a stream that impacts the disc at a location called the bright spot. Close to the WD, kinetic energy is converted into radiation as the disc material is slowed down before it is accreted by the WD. The boundary layer where this occurs can emit up to half of the total luminosity of the system. In a CV, the two stars are very close together, and the separation is comparable to the diameter of our Sun. This means that CVs have short orbital periods (on the order of hours). As a consequence of this close interaction, the donor is heavily deformed due to tidal and rotational distortion, and it is forced to move synchronously with the orbit. 

\begin{figure}[t]
\begin{center}
\includegraphics[scale=0.4]{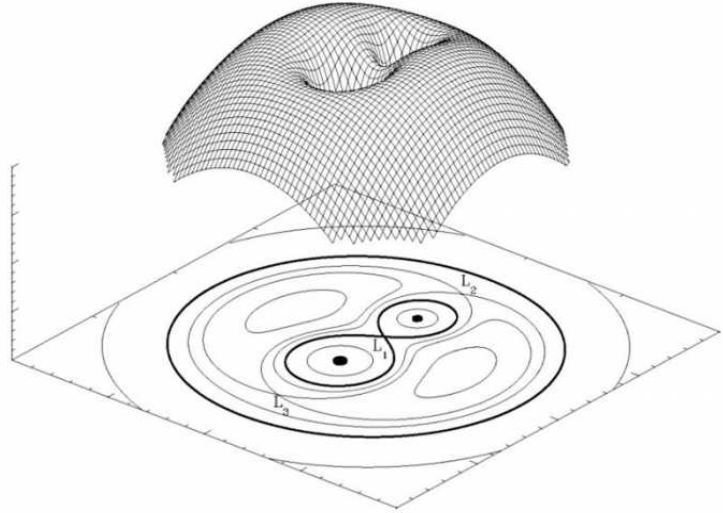}
\caption{\emph{Roche-lobe potential for a binary system with a mass ratio ${q}$ = M$_{2}$/M$_{1}$ = 2. If one of the stars fills its Roche lobe, material will escape through the inner Lagrangian point (here L$_{1}$), and be accreted by the companion. \newline (\small{Source: http://hemel.waarnemen.com/Informatie/Sterren/hoofdstuk6.htmlmtr})}}
\label{roche}
\end{center}
\end{figure}

The interaction between the two stars can be understood by considering the \textbf{Roche-lobe approximation}, in which the full gravitational potential of the binary is approximated as that of two point masses moving in circular orbits. This is expected to be a good approximation, since the donor is forced to rotate synchronously with the orbit, resulting in the removal of any initial eccentricity. The orbital angular momentum of the binary system is given by

\begin{equation}
J_{\text{orb}} = \sqrt{G \frac{M_{1}^{2}\,M_{2}^{2}}{M_{1} + M_{2}} a},
 \label{ang_mom}
\vspace{0.3cm}
\end{equation}

\noindent where $a$ is the binary separation, and $M_{1}$ and $M_{2}$ are the masses of the primary star and donor star, respectively. Figure~\ref{roche} shows a schematic view of the Roche potential for a particular mass ratio, ${q} = M_{2}/M_{1} = 2$. The inner Lagrangian point ($L_{1}$) is the point where the individual Roche lobes of the two stars are in contact. Mass transfer will occur if one of the stars overfills its Roche lobe. This can happen either if the Roche lobe shrinks in size, or if the star expands its radius. Cataclysmic variables are semi-detached systems, which means that the donor star slightly overfills its Roche lobe. The internal gas pressure of the donor then causes it to lose mass to the primary through the inner Lagrangian point.  

Roche-lobe overflow in CVs can push the donor star out of thermal equilibrium (TE). A star is in TE when the energy generated in the core is equal to the amount of energy that is transported to the surface and radiated away. The donor can only stay in thermal equilibrium if the time scale for the mass loss, $\tau_{\dot{M}_{2}} = M_{2}/ \dot{M}_{2}$, is larger than the thermal time scale, also called Kelvin-Helmholtz time scale, $\tau_{\text{th}} = GM_{2}^{2}/L_{2}R_{2}$. This is defined as the time scale on which the donor is able to adjust its structure so as to stay in thermal equilibrium. If the mass-loss rate is very low, $\tau_{\dot{M}_{2}} >> \tau_{th}$, the donor will manage to stay in TE, and its radius will be the same as that of a main-sequence star. If the mass-loss rate is very high, $\tau_{\dot{M}_{2}} << \tau_{\text{th}}$, the process of mass loss will be adiabatic, and the donor will expand its radius (assuming that the donor is a low-mass star with a large convective envelope). As it turns out, for a CV donor, $\tau_{\dot{M}_{2}} \sim \tau_{\text{th}}$, which means that the donor is not \emph{quite} able to adjust its radius fast enough in response to mass-loss. This means that donors in CVs are slightly out of thermal equilibrium, so even though their radii do shrink in response to mass loss, this shrinkage does not quite manage to keep up with the ongoing mass loss. As a result, CV donors become larger than equal-mass main-sequence stars. Observations show that donor stars can be oversized by up to 30\% (\citealt{2005PASP..117.1204P, 2006MNRAS.373..484K}). 

The radius of the donor ($R_{2}$) can be compared to the average radius of the Roche lobe volume, which only depends on the binary separation and the mass ratio ${q} = M_{2}/M_{1}$. In the range $0.1 < q < 0.8$, this volume-averaged radius can be approximated to an accuracy of 2\% (\citealt{1971ARA&A...9..183P}) by:
 
\begin{equation}
\frac{R_{2}}{a} \approx 0.462 \left(\frac{q}{1 + q}\right)^{1/3}.
 \label{paczynski}
 \vspace{0.3cm}
 \end{equation}

\noindent Also, Newton's generalisation of Kepler's third law shows that the orbital period ($P$) depends on the binary separation and the masses of the two components,

\begin{equation}
P = \sqrt{\frac{4\pi^{2} a^{3}}{G(M_{1} + M_{2})}},
 \label{kepler}
 \vspace{0.3cm}
\end{equation}
\newline
\noindent where $G$ is the gravitational constant. Combining Equations~\ref{paczynski} and~\ref{kepler}, yields the period-density relation for Roche-lobe filling stars 

\begin{equation}
\bar{\rho}_{2} = \frac{3 M_{2}}{4 \pi R_{2}^{3}}  \simeq \frac{95.4}{G} \frac{1}{P}.
\label{period_density}
\vspace{0.3cm}
\end{equation}

\noindent If we assume that the donor is a low-mass star near the main sequence, M$_{2}$/M$_{\odot}$ = $f$(R$_{2}$/R$_{\odot}$), where $f \approx$ 1, then the mass and radius of the donor only depend on the orbital period of the system. This results in the mass-radius/period relation 


\begin{equation}
\frac{M_{2}}{M_{\odot}} \approx \frac{R_{2}}{R_{\odot}} \approx 0.1P_{\text{hr}}.
 \label{mass_radius_period}
 \vspace{0.3cm}
 \end{equation}

In a cataclysmic variable, material from the donor emerges from the Lagrangian point at approximately the speed of sound, ${\sim}$ 10 km\,s$^{-1}$, and forms an accretion stream towards the primary white dwarf. Due to the orbital motion of the system, the Lagrangian point moves perpendicular to the line connecting the primary and donor, at a speed of about 100 km\,s$^{-1}$. 

The stream material has too much angular momentum to accrete directly onto the white dwarf. Instead, the material settles down into the orbit with the lowest possible energy for its angular momentum, which is a circular orbit. This is the so-called circularisation radius ($R_{\text{C}}$). The Keplerian velocity at $R_{\text{C}}$ is $v_{K} = \sqrt{GM_{1}/R_{\text{C}}}$. However, the stream material emerging from the $L1$ point has a velocity of $2\pi R_{\text{L1}}/P$. Assuming that the angular momentum of the stream material at the $L1$ radius is conserved at $R_{\text{C}}$, the circularisation radius is expressed as $R_{\text{C}} = 4\pi^{2}R^{4}_{\text{L1}}/GM_{1}P^{2}$. 
Combined with Equation~\ref{kepler}, this yields 

\begin{equation}
R_{\text{C}} = \frac{(1+q)R^{4}_{\text{L1}}}{a^{3}}.
 \label{r_circ}
  \vspace{0.3cm}
 \end{equation}

\noindent Material rotating at $R_{\text{C}}$ loses angular momentum as it spirals down to the WD. Closer to the primary, matter orbits faster than further away from it, so any form of viscosity acting on the gas rotating near $R_{\text{C}}$ will tend to heat it and spread it into a disc. For the total angular momentum to be conserved, a small amount of material carrying angular momentum with it must be transferred back to the outer parts of the disc, causing the disc to extend further out from the WD (\citealt{1981ARA&A..19..137P}). In the outer edge of the disc, angular momentum is fed back to the binary orbit via tidal interactions. However, standard molecular viscosity is too inefficient to be relevant to CVs; the associated time scale for the transfer of material through the disc would be much longer than those observed in CVs.

To further our understanding of this problem,~\cite{1973A&A....24..337S} introduced a so-called \textbf{$\alpha$-disc model} in which turbulence in the gas is assumed to provide the relevant source of viscosity. Their model assumes a geometrically thin disc where the disc radius ($R_{\text{d}}$) is much larger than the disc height ($H$). The kinematic viscosity ($\nu$) at a given radius is taken to be $\nu = \alpha v_{\text{s}}$H, where $v_{\text{s}}$ is the speed of sound, and the parameter $\alpha$ is expected to be less than 1, where 0 corresponds to no accretion.

The specific physical origin of this turbulent viscosity is still under debate, but one of the most likely sources, at lest for highly ionized discs (such as those in nova-like CVs), is the \textbf{magneto-rotational instability} (re-)discovered by~\cite{1991ApJ...376..214B} (earlier discussed by Evgeny Velikhov and Subrahmanyan Chandrasekhar in 1960). According to this, a weakly magnetised disc, orbiting a central compact object will be highly unstable and develop into a turbulent flow, resulting in a redistribution of the angular momentum.

The total luminosity of the disc can be expressed as  

\begin{equation}
L_{\text{disc}} = \frac{1}{2} \frac{G\,M_{1}\,\dot{M}}{R_{1}},
 \label{lum}
  \vspace{0.3cm}
 \end{equation}

\noindent where $M_{1}$ and $R_{1}$ is the mass and radius of the WD. However, this is only half of the total accretion energy ($L_{\text{acc}} = G\,M\,\dot{M}/R$). The other half of the accretion luminosity is released in the so-called \textbf{boundary layer}, which forms the region where the material is transferred from the inner edge of the accretion disc onto the primary WD. This energy release occurs because the white dwarf spins at an equatorial speed of a few hundred km\,s$^{-1}$ (see for instance Chapter~\ref{j1507}) but, just above the WD surface, the Keplerian velocity of the material in the disc is about ten times higher, and this kinetic energy needs to be dissipated. The dissipation is expected to occur within the narrow boundary where material is slowed down and accreted onto the white dwarf (e.g.~\citealt{1981ARA&A..19..137P}).

If the WD has a strong magnetic field, it will partially or completely prevent the formation of an accretion disc, and material will follow the field lines and accrete directly onto the magnetic poles of the WD. In this thesis, I will almost exclusively discuss CVs which are assumed to have dynamically insignificant magnetic fields.

\section{CVs as Variable Stars}

One fundamental observational characteristic of CVs is their variability. CVs vary on many different time scales, from long-term variations of years, down to short-term variability of seconds. When studying these events, we are limited by the fact that we watch the current CV population from a snapshot in time compared to their long-term time scale for evolution. Also, since we are unable to directly resolve the light from the different components within the CV, it is quite a challenging task to distinguish what mechanisms/locations are responsible for causing any particular observed variability.
\newpage 
 
\subsection{Long-Term Variability}

\subsubsection{Classical Nova Eruptions} \label{classnovaeruption}

\begin{wrapfigure}{r}{0.5\textwidth}
  \vspace{-20pt}
  \begin{center}
    \includegraphics[width=0.48\textwidth]{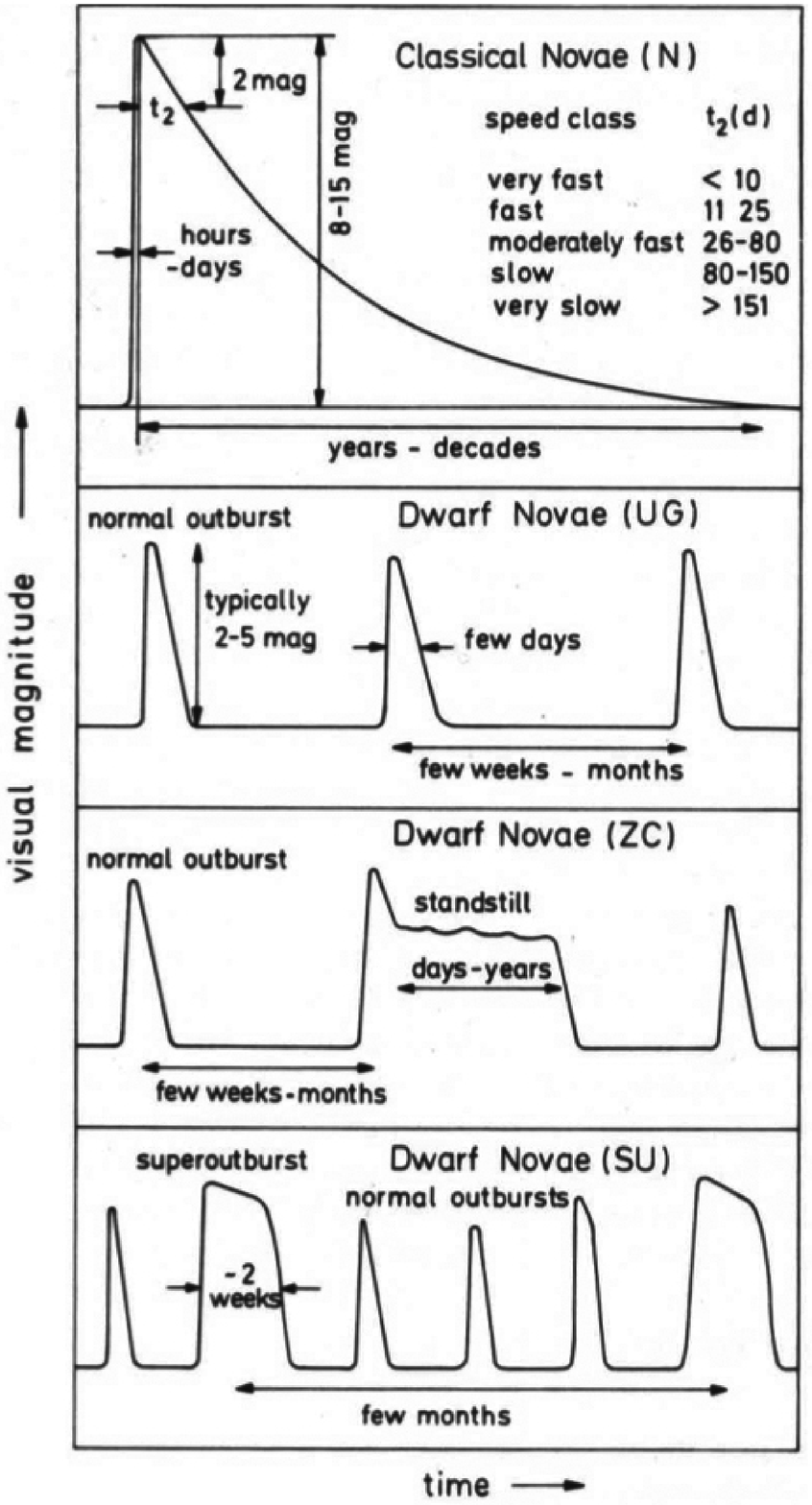}
  \end{center}
  \vspace{-30pt}
  \caption{\emph{Schematic figure showing the eruption behaviour of Classical novae and dwarf novae (from Ritter 1992).}}
  \vspace{-10pt}
\label{novae}
\end{wrapfigure}

As a result of the process of accretion, material from the donor star slowly builds up on the surface of the WD, creating a hydrogen-rich layer that is growing in size. The accreted material is heavily compressed due to the high gravity of the WD, and electrons become degenerate under the intense pressure. As the pressure increases, so does the temperature. However, in a degenerate gas, the pressure does not depend on temperature, and excess heat can not be dissipated by thermal expansion of the gas. The temperature will continue to increase until the hydrogen layer reaches a critical pressure (hence a critical envelope mass), after which the gas ignites and detonates in a nuclear-fusion reaction. This type of runaway thermo-nuclear reaction on the surface of the white dwarf is called a \textbf{classical nova eruption} and is responsible for the most violent events observed in CVs. In the explosion, a large portion of the accreted hydrogen layer is ejected in an expanding shell surrounding the nova. Left in the centre is the WD core, normally composed of carbon and oxygen (more massive WDs show evidence of neon/oxygen cores -- so-called neon novae). After the eruption, the WD will start the slow process of accreting more hydrogen-rich material until enough material is accreted for a new classical nova eruption to take place once more. One of the earliest, but still useful reviews on classical nova eruptions was presented by~\cite{1978ARA&A..16..171G}.

In theory, the expected mass required to ignite the hydrogen layer on the WD surface should be roughly equal to the total mass of the expelled nova shell. A WD of mass 1 M$_{\odot}$ will need to accumulate $\approx 10^{-4}$ M$_{\odot}$, before ignition (\citealt{1982ApJ...257..752F}). Assuming an accretion rate of $\dot{M} \sim 10^{-9}$ M$_{\odot}$ yr$^{-1}$, the time scale between two such eruptions is about $\sim10^{4} -10^{5}$ years. 

CVs that only have one observed eruption are called \textbf{Classical Novae (CNe)}. Observationally, CNe with more massive WDs are brighter during the eruptions, but decline faster, compared to CNe with lower WD masses. They are therefore divided into classes depending on how fast they fade, with the speed class usually being defined by the quantity $t_{2}$, which is defined as the time it takes for the nova to decline 2 magnitudes below maximum brightness. In the fastest novae, $t_{2}$ < 10 days, and they decline with a rate up to 0.2 mag\,d$^{-1}$. A slow nova has $t_{2}$ > 150 days and drops with a rate of 0.008 -- 0.013 mag\,d$^{-1}$ (\citealt{1995CAS....28.....W}). Spectroscopically, due to a redistribution of the flux to shorter wavelengths, novae become bluer as they decline. After the initial decline in the visual wavelength range, the flux in the ultraviolet region rises (see Figure 1 in~\citealt{2010AN....331..160B}). The top panel in Figure~\ref{novae} shows the general behaviour and speed classification of classical nova eruptions.

There is another subclass of CVs that also undergo classical nova outbursts, the \textbf{Recurrent Novae (RNe)}. They are observationally defined as novae for which more than one outburst has been observed, implying a nova recurrence time scale of approximately 10 years -- 100 years. According to standard nova models (e.g. \citealt{2005ApJ...623..398Y}), this can only occur for systems which have a high accretion rate (of order $10^{-8}$ M$_{\odot}$\,y$^{-1}$) and a massive WD (M$_{1}$ > 1 M$_{\odot}$). In RNe, the WD is expected to gain more mass between eruptions than it ejects during them. This could make the already high WD mass exceed the Chandrasekhar limit (the limit for which the degeneracy pressure cannot support the WD anymore, M$_{\text{CH}} \approx 1.4$\,M$_{\odot}$). Therefore, RNe are considered as candidate supernova Type 1a progenitors.

One intriguing possibility in RNe is that the expanding shells of material from subsequent eruptions may interact with each other. For instance, bright knots are observed in the shell surrounding T Pyxidis which are thought to be formed as material from different shells collides and ignite the gas (\citealt{2010ApJ...708..381S}). In Chapter~\ref{tpyx}, I present a spectroscopic analysis of the recurrent nova T Pyxidis.

\subsubsection{Dwarf Nova Eruptions} \label{dne}

\textbf{Dwarf nova eruptions} are less violent than the classical nova eruptions described in the previous section. However, they occur more frequently and are caused by instabilities in the accretion disc. As more material builds up in the disc, the disc heats up and eventually a critical density is reached, causing the stored material to be rapidly transported onto the WD. As a result, a large amount of gravitational potential energy is released.  

\begin{figure}[t]
\begin{center}
\includegraphics[scale=0.4]{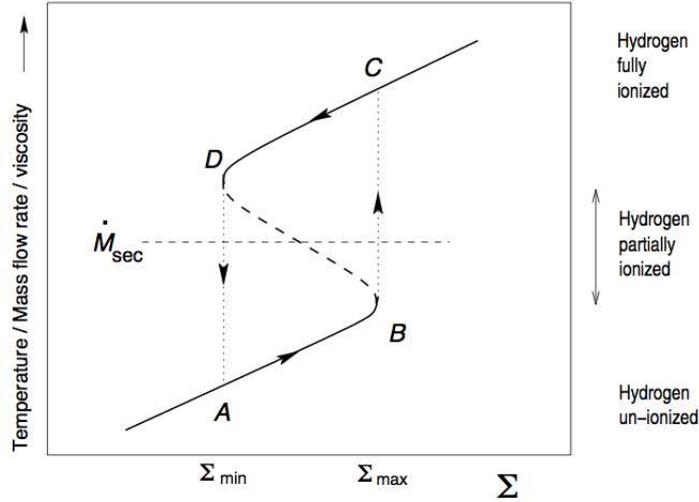}
\caption{\emph{The stability of a disc annulus can be understood from the dependence of viscosity ($\nu$) on the surface density ($\Sigma$), which is represented by the so-called S-curve (Figure from Connon-Smith 2007). A cool, un-ionized disc would be situated on the stable lower branch (A -- B), while a hot, ionized disc would be located at the upper stable branch (C -- D). Systems located between the two stable states (B -- D) have thermally unstable discs and exhibits dwarf nova eruptions. CVs belonging to the class dwarf novae will move between the two states, in the direction indicated by the arrows.}}
\label{scurve}
\end{center}
\end{figure}

Dwarf nova eruptions can be understood in the context of the \textbf{thermal-viscous disc instability model} (DIM) (first proposed by \citealt{1974PASJ...26..429O}, see reviews by \citealt{1996PASP..108...39O, 2001NewAR..45..449L}). Generally, material from the secondary reaches the disc at a rate $\dot{M}_{\text{sec}}$ and will flow through the disc until it is accreted by the primary WD. The rate at which material flows through the disc ($\dot{M}$) is a function of the viscosity ($\nu$). If $\dot{M} < \dot{M}_{\text{sec}}$, material will start to build up in the disc. However, if $\dot{M} > \dot{M}_{\text{sec}}$, gas is removed from the disc. The stability of a disc annulus depends on the viscosity ($\nu$), and the surface density ($\Sigma$), and can be represented by a so called \textbf{S-curve}. Such a curve is shown in Figure~\ref{scurve} (from \citealt{2007astro.ph..1654C}). In a disc annulus, $T$ and $\dot{M}$ increase with $\nu$, which means that the y-axis in Figure~\ref{scurve} can be expressed in any of those units. Each point on the curve indicates the thermal equilibrium for a given disc annulus, i.e the locus of points in the $\Sigma$ versus $\dot{M}$, where viscous heating equals radiative cooling. Since this thermal equilibrium is established on a very short time scale, disc annuli can effectively only exist with parameters that place them on the S-curve. In other words, disc annuli can only move {\em along} the S-curve in response to changes in $\dot{M}_{\text{sec}}$ or temporary fluctuations in the physical parameters. Any disc annulus for which the externally imposed $\dot{M}_{\text{sec}}$ intersects the lower or upper branch of the S-curve is stable.

Systems for which $\dot{M}_{\text{sec}}$ intersects the unstable middle branch display frequent disc instabilities and are called \textbf{dwarf novae (DNe)}. Consider a system starting at point A for which $\dot{M}_{\text{sec}}$ intersects the unstable branch. In such a system, a disc annulus is cold and un-ionized. Here $\dot{M} < \dot{M}_{\text{sec}}$, which means that the disc as a whole accumulates mass. As more mass is being stored, the surface density of the disc annulus increases along with the mass-flow rate, pushing the disc annulus up along the path from A to B. Eventually, the critical surface density $\Sigma_{\text{max}}$ (point B) is reached. Here, $\dot{M}$ is still less than $\dot{M}_{\text{sec}}$, and therefore, $\Sigma$ still increases. At this point, the disc annulus heats up and the gas becomes partially ionised as it moves towards the next stable state, which is on the upper branch (point C). Local heating in a disc annulus can trigger heating in adjacent annuli, and eventually cause the whole disc to heat up globally, resulting in accumulated disc material to be released onto the WD. In point C, the gas in the disc annulus becomes fully ionised. Here, $\dot{M} > \dot{M}_{\text{sec}}$, which results in a cooling as the disc annulus moves down along the stable branch from C down to D. 

Any disc annulus for which $\dot{M}_{\text{sec}}$ intersects the unstable branch between points B and D is unstable, and even if an annulus were to start with $\dot{M} = \dot{M}_{\text{sec}}$ on this branch, any small fluctuation in $\dot{M}_{\text{sec}}$ or $\Sigma$ would be sufficient to push the system away from the intersection point, and towards either point B or D. The critical surface densities $\Sigma_{\text{min}}$ and $\Sigma_{\text{max}}$, represent the value above which no cold equilibrium state exists ($\Sigma_{\text{max}}$), and the value below which no hot equilibrium state exists ($\Sigma_{\text{min}}$). These limits are specific for every system and depend on, for instance, the viscosity parameter and the WD mass (e.g. \citealt{2001NewAR..45..449L}). 

In systems with lower $\dot{M}_{\text{sec}}$, the outbursts are triggered from a location close to the WD and are then spread through the disc (so called inside-out outbursts), while systems with higher $\dot{M}_{\text{sec}}$ have so called outside-in eruptions that start from the outer edge of the disc and move inward (\citealt{2001A&A...370..488O}). 

DNe have typical inter-outburst time scales ranging from weeks to years. The eruptions can last a couple of days up to a couple of weeks, and correspond to increases in the optical luminosity by 3 -- 5 magnitudes. In quiescence, the DNe have low accretion rates. DNe dominate the short-period CV population below the period gap (P $\leappeq$ 2 hours; see Section~\ref{pgap}). They are further divided into subclasses depending on their outburst behaviour. A system that belongs to the \textbf{U Gem} class of stars displays normal dwarf nova eruptions. \textbf{Z Cam} stars have accretion rates high enough to sometimes get caught at a more or less constant brightness level on the upper stable branch (\citealt{1996PASP..108...39O}).  \textbf{SU UMa} stars occasionally show superoutbursts in-between the normal dwarf nova eruptions. These superoutbursts last longer and are thought to be triggered due to a combined tidal-thermal instability, as the disc reaches the 3:1 resonance radius (\citealt{1988MNRAS.232...35W, 1989PASJ...41.1005O}). A particularly interesting and important feature of these superoutbursts, the so-called \textbf{superhumps}, is discussed in more detail in Section~\ref{superhps}. Figure~\ref{novae} illustrates the different outburst behaviours of U Gem, Z Cam and SU UMa stars.  

Whether or not a disc is stable is determined by $\dot{M}_{\text{sec}}$. Systems with a constant high $\dot{M}_{\text{sec}}$ will constantly be on the upper stable branch in the S-curve, thus they do not exhibit dwarf nova eruptions. Such systems are called \textbf{nova-like variables} and include all systems where no eruption has been observed. Nova-like variables have hot WDs, high accretion rates and preferentially populate the orbital-period region between 3 -- 4 hours.
\newpage

\subsection{Variability on Orbital Timescales}

The orbital modulation dominates the light curves of eclipsing CVs and provides the most accurate orbital period measurements. In non-eclipsing systems, the bright spot still causes modulations on the orbital period. The reflection effect caused by irradiation of the donor star by the hot WD and disc, can also produce a signal on the orbital time scale, although such a signal would roughly be in anti-phase with any signal from the bright spot. Also, tidal deformation of the disc can produce an ellipsoidal disc shape, causing superhumps, which are detected on orbital times scales in systems with very low mass ratio.   

A key marker for observationally identifying orbital modulations is that such signals usually are coherent, i.e. stable in frequency and phase. Also, orbital signals are always present in quiescence, but might be less obvious during outbursts as the orbital signal then might be depleted temperately due to strong disc emission.

\subsubsection{Superhumps}  \label{superhps}

Superhumps are high-amplitude signals with periods a few percent ($\sim$ 3\%) longer than the orbital period and were first detected during a dwarf nova outburst of VW Hyi (\citealt{1974A&A....36..369V}). As described above in Section~\ref{dne}, dwarf nova outbursts are caused by thermal instabilities in the disc. In addition, systems with very low mass ratio and low accretion rates show instabilities that are thought to arise from an eccentric instability at the 3:1 resonance in the tidally unstable accretion disc (\citealt{1988MNRAS.232...35W}). 

More specifically, dwarf novae that display superhumps are called SU UMa stars. These systems show both normal dwarf nova outbursts lasting a few days, and superoutbursts lasting more than ten days. Generally, superhumps are only present during the superoutbursts and are, in most cases, not seen during the normal outbursts (see bottom panel in Figure~\ref{novae}). 

\begin{figure}[t]
\begin{center}
\includegraphics[scale=0.42]{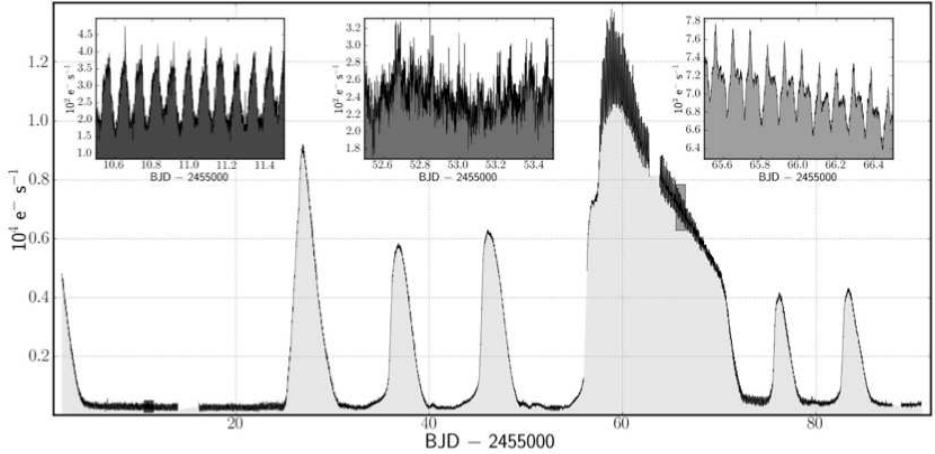}
\caption{\emph{Kepler light curve of V344 Lyrae showing several normal outbursts and one superoutburst (from~\citealt{2010ApJ...717L.113S}).}}
\label{still_suma}
\end{center}
\end{figure}

To explain the existence of superoutbursts,~\cite{1989PASJ...41.1005O} presented the \textbf{thermal-tidal instability model}, which combines the thermal disc instability (DIM) and the tidally driven eccentric instability. In this model, the disc is circular and tidally stable at the early phase of the cycle for the superoutbursts. However, during every normal outburst, not all the accreted material is transferred onto the WD, resulting in an accumulation of mass in the accretion disc. Along with an increased disc mass, the total angular momentum of the disc increases, which results in a gradually increasing disc radius. Eventually, during a normal outburst, the radius of the disc reaches a critical value at the 3:1 resonance with the donor, which causes the disc to become tidally unstable. The normal outburst has turned into a superoutburst. When the disc eventually cools down, the disc becomes circular and tidally stable again (see~\cite{2009AcA....59..121S} for a modification to the theory of superhumps). During the superoutbursts, the ellipsoidal disc precesses in relation to the orbital motion, and the beat frequency is detected as a superhump signal, with a frequency typically a few per cent less than the orbital frequency. The Kepler light curve of V344 Lyrae demonstrates a real example of how a normal outburst turns into a superoutburst (see Figure~\ref{still_suma} obtained from~\citealt{2010ApJ...717L.113S}).

~\cite{2005PASP..117.1204P} have systematically observed more than 200 dwarf novae, and find that all systems with mass ratios $q$ < 0.25 exhibit superhumps, while systems with $q$ > 0.36, do not. Between these two limiting mass ratios, there is a smooth monotonic decline in the fraction of systems showing superhumps. Measurement of the superhump period allows mass ratios as well as the component masses to be estimated (e.g.~\citealt{1998PASP..110.1132P}), since the precession rate, $P_{\text{prec}}^{-1} = P_{\text{orb}}^{-1} - P_{\text{sh}}^{-1}$, is expected to be proportional to the mass of the donor star, and hence also to the mass ratio. Also, the fractional period excess of superhump period $\epsilon$ scales with orbital period, $\epsilon = (P_{\text{sh}} - P_{\text{orb}})/P_{\text{orb}}$ (e.g.~\citealt{2000MNRAS.314L...1M} and~\citealt{2007MNRAS.379..183P}). From the mass-period relation, $\epsilon$ is therefore also a function of $q$. From the $\epsilon$ -- $q$ relation, the donor mass can be estimated by assuming a WD mass; for example,~\cite{2005PASP..117.1204P, 2006MNRAS.373..484K, knigge2011} adopt $M_{WD} = 0.75 M_{\odot}$, based on observational estimates in eclipsing systems.    

In this section, I have only discussed the most common so-called \textbf{positive superhumps}. In addition, there are also \textbf{negative superhumps}, which produce power excess at slightly higher frequency than the orbital frequency. This type of superhump is thought to be associated with a disc tilt that is undergoing retrograde precession (\citealt{1995PASP..107..551H, 2009AcA....59..419S} and \citealt{2010ApJ...722..989M}). For further information about different types of superhumps, see Appendix A1 -- \emph{Hump zoology in cataclysmic variables} of~\cite{2002PASP..114..721P}.

Although superhumps are most commonly observed during superoutburst, similar signals have also been detected for a few systems in quiescence (for instance V344 Lyrae,~\citealt{2010ApJ...717L.113S}). One such system -- BW Sculptoris -- is discussed in Chapter~\ref{j1457_bwscl}.

\subsection{Short-Term Variability}

\subsubsection{Signals On the White Dwarf Spin Period}  \label{dqher}

In magnetic systems, variability on time scales of tens of seconds to tens of minutes is commonly seen and is usually understood as arising from the accreting magnetic poles on the rotating WD. This phenomenon was first detected in the system DQ Herculis (\citealt{1956ApJ...123...68W}) and is sometimes referred to as \textbf{DQ Her modulations}. 


\subsubsection{Quasi-Periodic Oscillations and Dwarf Nova Oscillations} \label{qpo_dno}

Oscillations on similar time scales to the DQ Her modulations have been found in some nova-like variables, as well as in dwarf novae during outburst. However, neither class of systems are thought to harbour strongly magnetic white dwarfs. These oscillations became known as \textbf{dwarf nova oscillations} (DNOs) (reviewed by~\citealt{2004PASP..116..115W, 2008AIPC.1054..101W}). DNOs are further distinguished from DQ Her modulations by exhibiting considerably lower coherence. The so-called quality factor ($Q$) is often used as a measure of the coherence of the signal, where $Q = \left| \frac{dP}{dt}\right| ^{-1}$. DNOs are characterised by quality factors ranging from $10^{3} < Q < 10^{7}$, which can be compared to the factors typical of DQ Her modulations, which are $Q \sim 10^{12}$ (\citealt{2008AIPC.1054..101W}). DNOs typically have periods of 5 s -- 40 s, where the period is somewhat different depending on CV subclass. In some systems, a second DNO is detected at a slightly longer period than the first one. As suggested by their relatively low quality factors, DNO periods are not constant. In fact, they appear to follow a period-luminosity relation, in the sense that the minimum DNO period corresponds to when the system is at a maximum $\dot{M}$ (\citealt{2008AIPC.1054..101W}). Thus, DNO periods are shorter during the rise of a dwarf nova eruption and longer during the declining phase.
 
\textbf{Quasi-periodic oscillations} (QPOs) appear at longer periods than the DNOs and are characterised
by a yet much lower coherence (typically $5 < Q < 20 $;~\citealt{2008AIPC.1054..101W}). QPOs were first discovered as, and named for, a broad excess of power seen during a dwarf nova eruption of RU Pegasi (\citealt{1977ApJ...214..144P}). Since then, they have been commonly found with periods of 10 -- 20 minutes in the high-state light curves of many nova-like variables, as well as in dwarf novae during outburst. 

DNOs and QPOs can exist simultaneously. In particular, in systems showing double DNOs, the beat period between the two DNOs is roughly equal to the QPO period. The ratio $P_{\text{QPO}}/P_{\text{DNO}}$ is constant in systems with high $\dot{M}$, indicating that the two signals are related and that they might be caused by the same (or related) underlying physical process (\citealt{2008AIPC.1054..101W}). However, QPOs can also exist independently of the DNOs and have been detected in dwarf novae during quiescence. 

The origin of DNOs and QPOs is poorly understood, but could be connected to the magnetic field of the WD, even for systems that are considered to be non-magnetic.~\cite{1995ASPC...85..343W} and~\cite{2002MNRAS.335...84W} constructed what they call the \textbf{Low Inertia Magnetic Accretor model} (LIMA), in which the rapid release of material onto the WD during outbursts, causes hot, rapidly spinning equatorial belts to form on the WD surface. Any magnetic field anchored in these belts would create accretion shocks, similar to those seen in intermediate polars. DNOs would occur as the equatorial belts spin up and down in response to the varying accretion rate in the outburst cycle. The connection between DNOs and equatorial belts was first discussed by~\cite{1978necb.conf...89P}. As a consequence of the radiation from these belts, QPOs are caused by reprocessing in a traveling wave at the inner edge of the accretion disc, resulting in a vertical thickening of the disc, that propagate through the disc in prograde direction in respect to the motion of the disc material. QPOs would then occur as this traveling wave obscures and reflects the light from the primary WD (\citealt{2002MNRAS.335...84W}). 

\subsubsection{Non-Radial Pulsations} \label{nrps}

\begin{figure}[t]
\begin{center}
\includegraphics[scale=0.35]{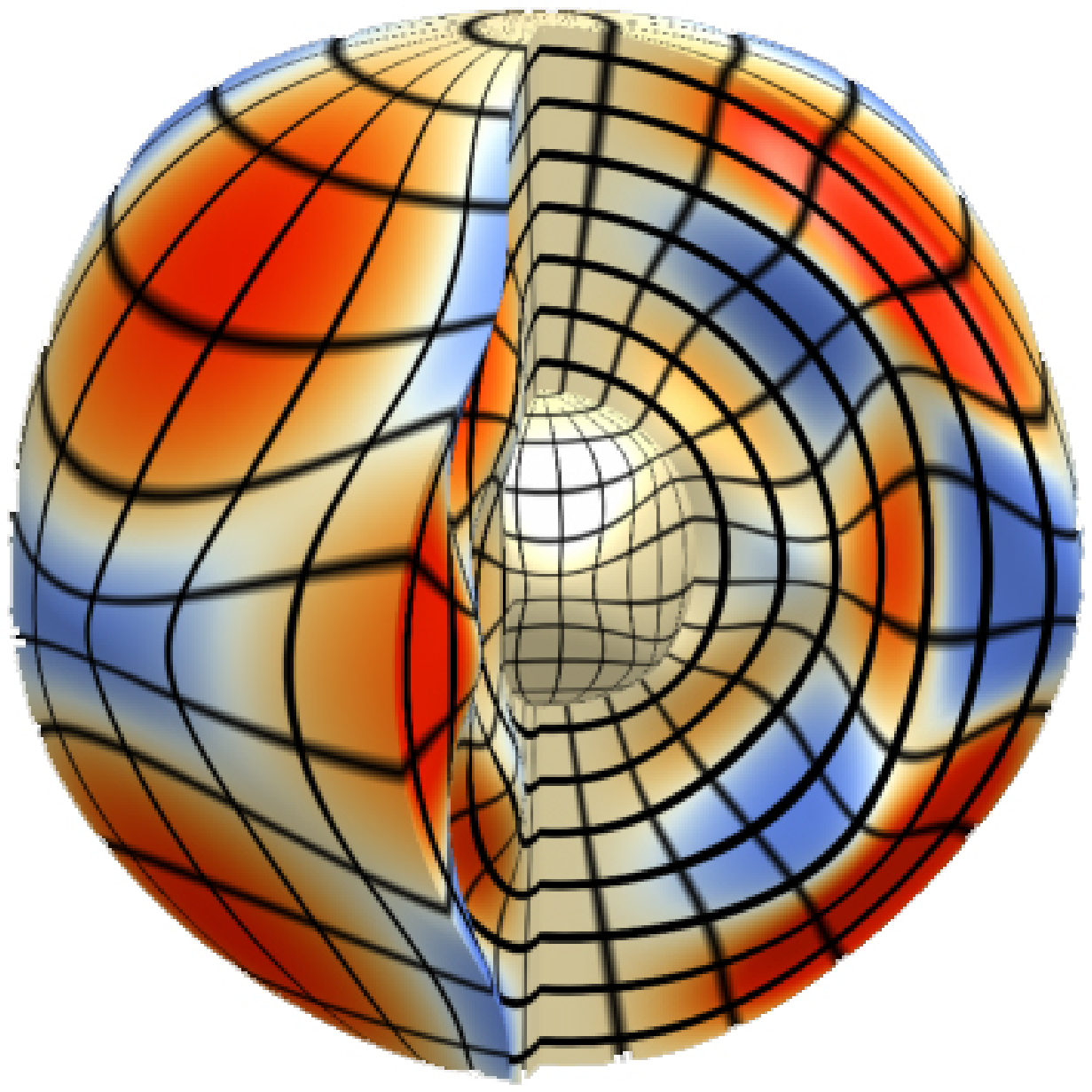}
\caption{\emph{General model showing the deformation of a star caused by g-modes. Here the number of modes is $m$ = 6 in the azimuthal direction, and $l$ = 3 in the longitude direction. \newline (\small{Source: http://www.astro.wisc.edu/$\sim$townsend/resource/news/nrp-surface-cutaway.png})}}
\label{nrp}
\end{center}
\end{figure}

Non-radial pulsations (NRPs) are gravity wave pulsations within the star itself and are commonly found in isolated white dwarfs of DA type, so called ZZ Ceti stars. These stars have hydrogen-rich atmospheres, and pulsations occur as the WD cools and passed through a phase of pulsational instability, detected mainly as g-modes (\citealt{2006AJ....132..831G}). The oscillations themselves are described by three quantum numbers, the radial order $n$, degree $l$, and the azimuthal number $m$. In practice, $n$ specifies the number of nodes between the centre and surface of the star, $m$ describe modes in the azimuth direction, and $l$ in longitude direction. Figure~\ref{nrp} shows a general model of g-modes in a star where $m$ = 6 and $l$ = 3.  

\begin{figure}[t]
\begin{center}
\includegraphics[scale=0.4]{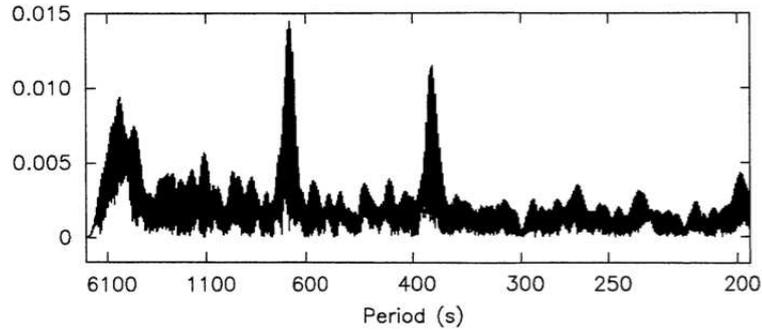}
\caption{\emph{Power spectrum showing the observed NRPs in GW Lib as presented by Warner \& van Zyl  (1998). GW Lib was the first CV to be found to harbour a non-radially pulsating WD.}}
\label{gwlib}
\end{center}
\end{figure}

NRPs are detected with a range of periods from a few seconds to a couple of minutes in isolated WDs. During the last decade, similar signals, generally interpreted as non-radial WD pulsations, have also been detected in faint cataclysmic variables. The first CV proposed to harbour a pulsating white dwarf was GW Librae.~\cite{1998IAUS..185..321W}, found rapid, periodic, and non-commensurate signals in its light curve, suggesting non-radial pulsations of the underlying white dwarf (see Figure~\ref{gwlib}). In most CVs, the accretion energy tends to dominate the luminosity, and the white dwarf itself, shining with $M_{V} \sim$ 10 -- 12, is seldom seen. However, for some of the most intrinsically faint CVs, spectroscopy and time-series photometry does reveal signatures of the underlying white dwarf, such as broad absorption features in the spectrum, sharp eclipses and, sometimes, non-radial pulsations in the light curve. These signals have now been detected in about a dozen CVs, all quiescent systems of low luminosity. In this thesis, these systems are called \textbf{GW Lib stars}, after the first discovery.~\cite{2010ApJ...710...64S} and~\cite{2009JPhCS.172a2069M} present recent reviews of this group of stars. 

\begin{figure}[t]
\begin{center}
\includegraphics[scale=0.38]{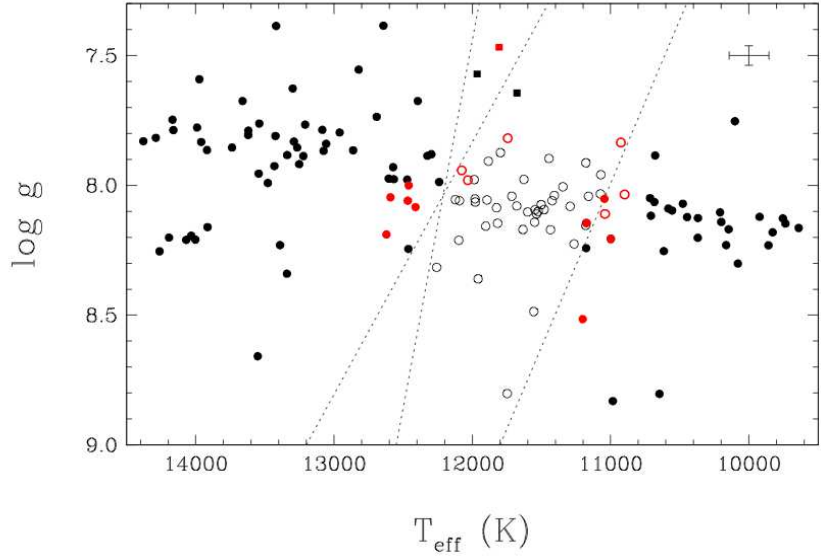}
\caption{\emph{The instability strip for isolated WDs of type DA, where pulsating WDs are marked as black and {\color{Red} red} open circles. Solid dots represent non-pulsating DA stars (figure from~\citealt{2006AJ....132..831G})}}
\label{giannis}
\end{center}
\end{figure}

\begin{figure}
\begin{center}
\includegraphics[scale=0.35]{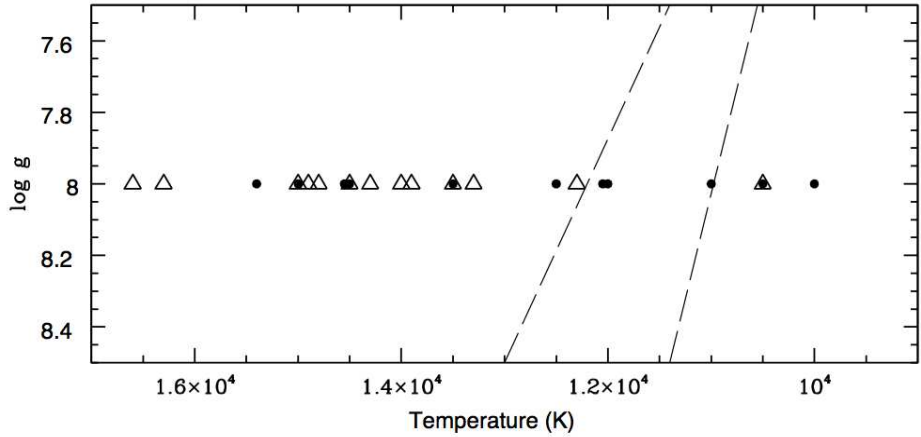}
\caption{\emph{The $T_{\text{eff}}$ -- $\log g$ distribution of pulsating (marked as dots) and non-pulsating (marked as triangles) WDs in CVs (figure from~\citealt{2010ApJ...710...64S}).}}
\label{szody}
\end{center}
\end{figure}

WDs in CVs are different from isolated ones since they undergo accretion, which gives them atmospheres of approximately solar composition. The white dwarfs in CVs are hotter and are also found to be spinning faster compared to isolated ones (\citealt{2009ASPC..404..229S}). In isolated WDs with pure hydrogen atmospheres, pulsations are only observed in stars with temperatures located within a so-called \textbf{instability strip} in the $\log g$ -- $T_{\text{eff}}$ plane, spanning the temperature range $T_{\text{eff}}$ = 10900 K -- 12200 K (see Figure~\ref{giannis} obtained from~\citealt{2006AJ....132..831G}). However, there is no clear instability strip for the GW Lib stars (see Figure~\ref{szody} taken from~\citealt{2010ApJ...710...64S}), and pulsations are found in systems with WD effective temperatures up to at least 15000 K. 

NRPs typically have periods between 80 s -- 1300 s (\citealt{2008AIPC.1054..101W}), with a very low coherence in the signal. Since their periods are of the same order as both DNOs, QPOs and DQ Her modulations, one has be careful when identifying the origin of signals found in a light curve. An important indicator is the coherence of the signal, which is specific for each type of modulation.  

\subsubsection{Flickering}

Low-amplitude stochastic variability on the order of 0.1 magnitudes on time scales of seconds up to minutes is seen in almost all CV light curves. This variability is called \textbf{flickering} and is associated with the process of accretion. Though the physical origin of flickering is not well understood, it may arise preferentially in the bright-spot region, where the accretion stream impacts the disc and/or in the boundary layer in the innermost region of the disc (\citealt{2000A&A...359..998B}). Some faint systems with low accretion rates and faint discs show very little flickering (e.g. SDSS J1457+51 and BW Sculptoris, see Chapter~\ref{j1457_bwscl}).

\section{CVs as Close Binaries: Secular Evolution}

\subsection{Pre-CV Evolution} 

The evolution leading to the formation of a cataclysmic variable starts with a binary system containing two main-sequence stars. The separation between the two stars is about $a \geq 50$ R$_{\odot}$ (e.g.~\citealt{1995CAS....28.....W}), but for the system to become a cataclysmic variable, a substantial amount of angular momentum needs to be removed to decrease the size of the orbit. The more massive of the two stars will be the first one to evolve into a cool red giant and fill its Roche lobe. Due to Roche-lobe overflow, the giant then starts to lose mass to its companion. The centre of mass is therefore moved closer to the companion, resulting in a decreased size of the giant's Roche lobe. However, it cannot decrease its radius to stabilise the mass transfer, since the equilibrium radius of a giant does not depend on its total mass, but only on the mass of its degenerate core (e.g.~\citealt{1995CAS....28.....W}). This will lead to a run-away mass-transfer phase. During this phase, gas is transferred faster than it can be accepted by the companion, and eventually the accreted gas will overflow the Roche lobe of the companion and form a common envelope (CE) enclosing both stars. The binary can stay in this phase for about 10$^{3}$ -- 10$^{4}$ years. As the stars spiral around their centre of mass inside the envelope, they lose orbital energy due to frictional interactions with the envelope material. As a result, the orbital separation shrinks drastically. The CE phase is expected to end when the deposited energy exceeds the envelope's own binding energy (e.g. \citealt{1984ApJ...277..355W}). At this point, the envelope is ejected and forms a planetary nebula. Left in the now close binary system are a primary WD and a main sequence secondary. This second detached phase can last 10$^{7}$ -- 10$^{8}$ years until the loss of angular momentum due to gravitational radiation or magnetic braking (see below) has shrunk the orbit until the point where the secondary fills its Roche lobe. This marks the beginning of a new phase of mass transfer and the start of the system's life as a stable, semi-detached CV.

\subsection{Mechanisms for Angular-Momentum loss} 

In general, the driving mechanism for the evolution of cataclysmic variables is loss of angular momentum through the processes of gravitation wave emission (GR) and/or magnetic braking (MB). 

\subsubsection{Gravitational Radiation}

For systems with periods below $\sim$ 2 hours, the main mechanism operating is usually taken to be gravitational radiation (\citealt{1962ApJ...136..312K, 1981ApJ...248L..27P, 1982ApJ...254..616R}). The rate at which angular momentum is lost depends on the binary separation $a$ and the masses of the two components M$_{1}$ and M$_{2}$, 

\begin{equation}
\frac{\dot{J}}{J} = \frac{32}{5}  \frac{G^{3}}{c^{5}}  \frac{M_{1} M_{2}(M_{1} +M_{2})}{a^{5}}.
 \label{grav_rad}
\vspace{0.3cm}
 \end{equation}

\noindent The corresponding mass-loss rate is $\dot{M} \approx 10^{-11} - 10^{-10}$ M$_{\odot}$\,y$^{-1}$, which is broadly in line with observations (\citealt{1998PASP..110.1132P, 2008MNRAS.388.1582L, 2009ApJ...693.1007T}).

\subsubsection{Magnetic Braking} 

The observationally inferred accretion rates for systems with periods above 3 hours ($\dot{M} \approx 10^{-9} - 10^{-8}$ M$_{\odot}$\,y$^{-1}$ \citealt{1984ApJS...54..443P, 2009ApJ...693.1007T, knigge2011}) are much higher than can be accounted for by GR. Therefore, another, more efficient mechanism for loss of angular momentum is needed to explain the high accretion rates in systems with $P \gtappeq$ 3 hours.~\cite{1981A&A...100L...7V, 1981AcA....31....1P, 1983ApJ...275..713R}, proposed mass transfer driven by a mechanism called \textbf{magnetic braking}. In this scenario the donor has a magnetic field, as well as a weak, ionised wind that travels along the field lines out to the Alfv{\'e}n radius. Up to this radius, the magnetic pressure in the wind material exceeds the gas pressure. The wind is thus forced to co-rotate with the star out to this radius, producing a magnetic braking torque on the secondary star. Due to this, the rotational speed of the secondary is reduced. Also, in a close binary system, tidal effects force the donor to rotate synchronously with the binary orbit. As a result, the magnetic braking torque ultimately removes orbital angular momentum from the system. This leads to a decreased orbital period. As long as the process of magnetic braking continues, the orbit continues to shrink, and the secondary loses mass steadily. Magnetic braking in CVs is now fairly widely accepted as the main mechanism for angular-momentum loss in systems with periods above 3 hours.

\subsection{The Period Gap}  \label{pgap}

There is a dearth of systems with orbital periods between $\approx$ 2 and 3 hours. According to the standard evolutionary model for CVs, this is explained by assuming that CVs evolve through this period range as detached systems. The mass-period relation for main-sequence donor stars (Equation~\ref{mass_radius_period}), shows that for an orbital period of 3 hours, the mass of the donor is ${\sim}$ 0.3 M$_{\odot}$. At this stage, the secondary becomes fully convective. Since the magnetic field in a low-mass main-sequence star is thought to be anchored at the interface between the convective envelope and radiative core, the standard model for CV evolution posits that the field will disappear (completely or partially) as the star becomes fully convective. With a weakened or non-existent magnetic field, the process of magnetic braking is disrupted or considerably reduced, which means that the accretion rate ($\dot{M}$) drops substantially. As a result, the donor adjust its radius closer to the one it would have for thermal equilibrium. The donor will therefore shrink and detach from the Roche lobe, leading to an interruption of mass transfer. The CV will stay in this phase for approximately 10$^{8}$ years (e.g.~\citealt{knigge2011}). This is long enough for the secondary to get back to its equilibrium radius. However, angular-momentum loss from the process of gravitational radiation is still active, shrinking the binary orbit, and so the donor is able to catch up with the Roche lobe radius again at an orbital period of about ${\sim}$ 2 hours. Mass transfer is once again resumed, now driven by gravitational radiation. Empirically, this period gap spans from the range 2.15 ${\leq}$ P$_{orb}$  ${\leq}$ 3.18 hours (\citealt{2006MNRAS.373..484K}). 

The disrupted magnetic-braking theory just outlined has been discussed in more detail and/or used by~\cite{1982ApJ...254..616R, 1983ApJ...275..713R}, ~\cite{1983A&A...124..267S}, ~\cite{1993A&A...271..149K} and ~\cite{2001ApJ...550..897H}. 


\subsection{The Minimum Period}  \label{pmin}

As the donor loses even more mass, the thermal time scale increases more quickly than the mass-loss time scale, so that eventually $\tau_{\dot{M}_{2}} < \tau_{\text{th}}$, and the donor is driven further and further out of thermal equilibrium again. Adopting the power-law mass-radius relation $R_{2} \propto M_{2}^{\alpha}$ (where $\alpha \sim$ 0.8 corresponds to low-mass main sequence stars in thermal equilibrium), and combining this with the period-density relation $P^{-2}\,\propto\,M_{2}R_{2}^{-3}$, gives $P^{2}\,\propto\,M_{2}^{3\alpha -1}$. Differentiating logarithmically yields

 \begin{equation}
\frac{\partial (\ln\,P[t])}{\partial t} = \left(\frac{3\alpha -1}{2}\right)  \frac{\partial (\ln\, M_{2}[t])}{\partial t},
 \label{diff3}
 \vspace{0.3cm}
 \end{equation} 
  
\noindent and we obtain a relation between the orbital period-derivative and the total mass-loss rate from the donor (mass loss both from wind and mass transfer),

\begin{equation}
\frac{\dot{P}}{P} = \frac{3\alpha-1}{2} \frac{\dot{M}_{2}}{M_{2}}.
 \label{p_min}
 \vspace{0.3cm}
 \end{equation}
 
\noindent This shows that the system reaches a period minimum, $\dot{P} = 0$, when it has been pushed so far out of thermal equilibrium that the mass radius index (evaluated along the CV evolution track) becomes $\alpha$ = 1/3 (\citealt{2006MNRAS.373..484K, knigge2011}). The minimum period is theoretically predicted to occur at $P \approx$ 65 min (\citealt{1993A&A...271..149K, 2001ApJ...550..897H}). The latest point at which a system can reach its minimum period is roughly when the donor mass falls below the hydrogen-burning limit and the donor becomes a sub-stellar object. This is because a sub-stellar object responds differently to continued mass-loss and expands or stays constant in size ($\alpha \leappeq 0$). The hydrogen-burning limit for isolated objects is $\approx$ 0.072 M$_{\odot}$  (\citealt{1998A&A...337..403B}). 

~\cite{1993A&A...271..149K} and ~\cite{1997MNRAS.287..929H} theoretically predicted that $\simeq$ 99\% of all CVs should have periods below the period gap and that $\simeq$ 70\% should have passed their minimum period (these systems are often referred to as \textbf{period bouncers}). The reason for these high percentages is the drastically longer evolutionary time scale as systems evolve closer to the minimum period. As a result, a larger portion of systems should be found close to P$_{\text{min}}$. This \emph{pile-up} of short-period systems is often called the \textbf{period spike} and is predicted, for instance, by~\cite{1993A&A...271..149K} and~\cite{1999MNRAS.309.1034K} (a further discussion is presented below in Section~\ref{obs_the}). Observationally, systems close to P$_{\text{min}}$ typically have $ \dot{M} = (3 - 10) \cdot 10^{-11}$ M$_{\odot}$\,y$^{-1}$ (\citealt{1998PASP..110.1132P,  2006Sci...314.1578L}). 

\subsection{Theory versus Observations} \label{obs_the}

Observations of cataclysmic variables are broadly consistent with the standard evolutionary theory described in the previous sections. Yet there are some observational facts than cannot be reproduced by standard theory. The most striking difference is the minimum period. Theoretically, it is predicted to be at $\approx$ 65 minutes (\citealt{1993A&A...271..149K}), but observations indicate a minimum period of P$_{min}$\,=\,82.4 $\pm$ 0.7 minutes (\citealt{2009MNRAS.397.2170G}). If one allows for tidal distortion and an inflated donor radius associated with magnetic activity, the theoretically predicted period becomes somewhat longer, but is still inconsistent with the observed value (\citealt{knigge2011}). Also, according to the theoretical predictions of~\cite{1993A&A...271..149K} and ~\cite{1997MNRAS.287..929H}, only about 1\% of the observed CVs should have periods above the period gap. This is in conflict with observations that show a far larger population of long-period systems (e.g. \citealt{2007MNRAS.374.1495P, 2008MNRAS.385.1485P}).

\begin{figure}[t]
\begin{center}
\includegraphics[scale=0.5]{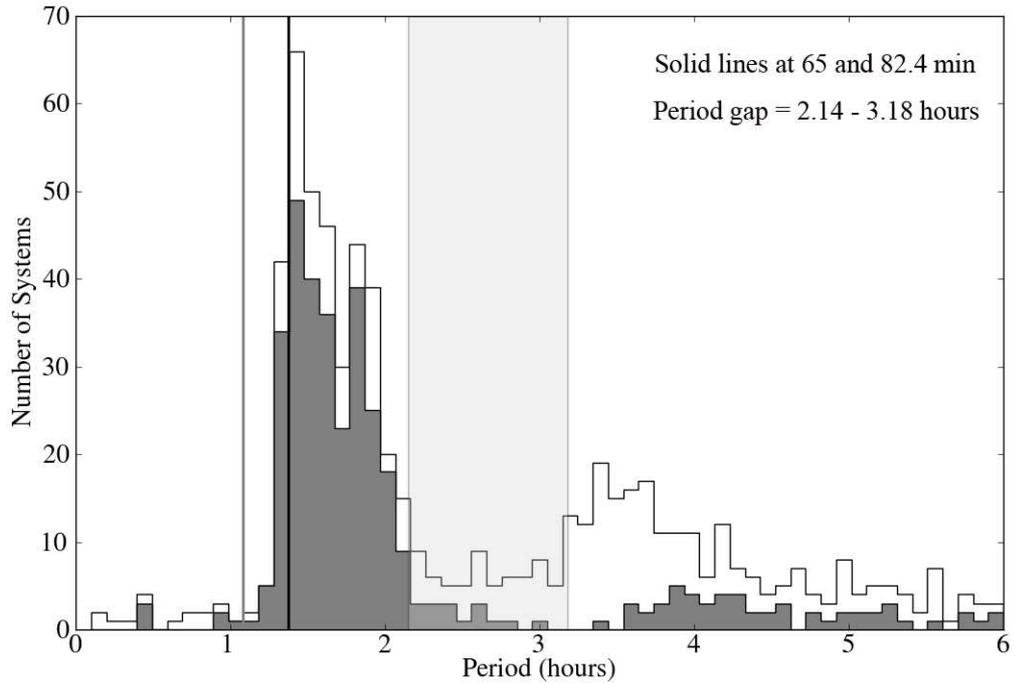}
\caption{\emph{The CV distribution from the Ritter \& Kolb catalogue, version 7.14 (2010). The white region is the distribution of all non-magnetic systems, while the dark grey area only includes dwarf novae. The period gap is marked as a transparent area between 2.14 - 3.18 hours. The two solid lines represent the theoretical minimum period at 65 minutes in grey (Kolb 1993), and the observational one at 82 minutes in black (G\"{a}nsicke et al. 2009).}}
\label{porb_dist}
\end{center}
\end{figure}

The two main tracers of the evolutionary state of a CV are the orbital period and the donor mass. Luckily, it is often a rather simple task to find the orbital period of a system (see Chapter~\ref{methods}). To a first approximation, the orbital period is enough to tell a younger (long-period) system apart from an older (short-period) system. In Figure~\ref{porb_dist}, I present the current period distribution of non-magnetic CVs. Periods were obtained from the catalogue by~\cite{2010yCat....102018R}, version 7.14. The white area shows all non-magnetic systems while the dark grey area only shows the distribution of dwarf novae. An accumulation of systems close to the period minimum is seen. This is the period-spike predicted by theory. Earlier versions of the Ritter \& Kolb catalogue did not show this peak, and~\cite{2009MNRAS.397.2170G} were the first to observationally confirm an accumulation of systems near the period minimum using the Sloan Digital Sky Survey (SDSS) CV sample. The reason why such a pile-up of systems close to the period minimum has not been observed before is simply because these short period systems are difficult to detect due to their low accretion rates and faint luminosities. Therefore short-period systems tend to be under-represented among observed CVs, although more recent and deep surveys (such as the SDSS) are beginning to uncover these systems.

As already noted in Section~\ref{pmin}, standard evolutionary theory predicts that 70\% of all CVs should have passed their minimum period and have sub-stellar donors (\citealt{1993A&A...271..149K, 1997MNRAS.287..929H}). However, few such systems have been successfully confirmed, and, until recently, almost no CVs containing donors with masses below the hydrogen-burning limit had been found. Nevertheless, a few CVs containing sub-stellar donors have now been convincingly identified by~\cite{2006Sci...314.1578L, 2007MNRAS.381..827L, 2008MNRAS.388.1582L}. The system discussed in Chapter~\ref{j1507}, SDSS J1507+52, is one of these CVs, and turns out to be quite unusual in other respects as well. 

\begin{figure}
\begin{center}
\includegraphics[scale=0.45]{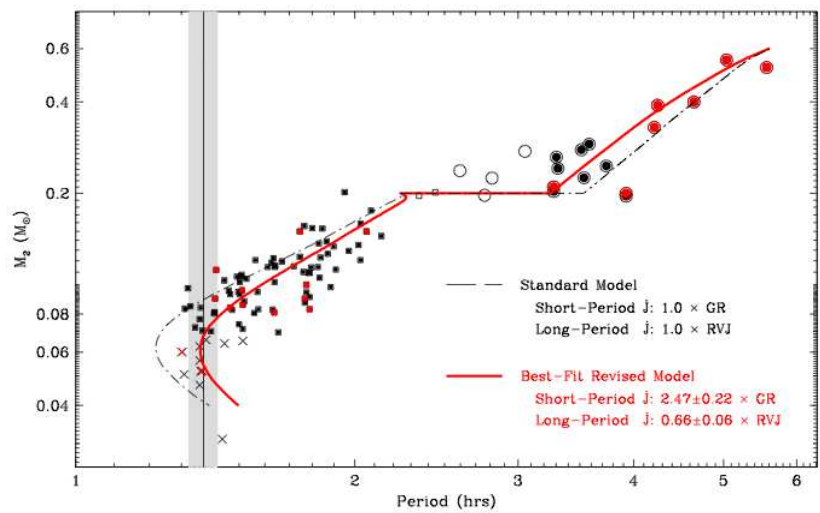}
\caption{\emph{Figure from Knigge, Patterson \& Baraffe (2010) showing the mass-period relation, both from observations and theory. The different colours and shapes correspond to different type of systems and how their masses were estimated. Filled squares correspond to short-period CVs while filled circles are long-period CVs. Black points represent systems where the masses were estimated from superhump-periods while the masses of the {\color{Red} red} points where found from eclipse analysis.}}
\label{knigge_mp}
\end{center}
\end{figure}

In order to distinguish between short-period systems that have not yet reached their minimum period and period bouncers, one has to consider both their orbital periods and donor masses. According to Equation~\ref{mass_radius_period}, there is a relation between the orbital period and the mass of the donor. However, this is only valid for systems close to thermal equilibrium (main-sequence stars), whereas sub-stellar donors will follow quite a different period-mass relation. To account also for sub-stellar donors, the more general power-law mass-period relation can be employed $P^{2}\,\propto\,M_{2}^{3\alpha -1}$ (see Section~\ref{pmin}), where main-sequence donors have $\alpha \sim$ 0.8, and sub-stellar donors have $\alpha \sim$ 0. A plot showing donor masses as a function of their orbital periods should then reveal the two branches of CV evolution, and this mass-period relation for CV donors has been studied in detail by~\cite{2008MNRAS.388.1582L, 2010ApJ...721.1356S, knigge2011, 2011MNRAS.tmp...27P}. In Figure~\ref{knigge_mp}, one such relation is presented (Figure from~\citealt{knigge2011}), where the different colours and shapes of the data points indicate the type of system and how the masses were found. The solid line is the best fit to the data, while the dashed line shows the standard evolutionary model. The flat line between $\approx$ 2.3 h -- 3.2 h shows how the donor mass stays constant through the period gap.

Figure~\ref{knigge_mp} demonstrates the discrepancy between the theoretical and observational minimum period. Many studies have been devoted to solve this problem, as well as to find a solution that would account for the fact that we have not found as many post-bounce systems as predicted by theory. Generally, an enhanced (i.e. stronger than GR) angular momentum loss rate below the gap and/or a larger donor radius is proposed to solve the problem with the minimum period (\citealt{1998PASP..110.1132P, 2000NewAR..44..161M, knigge2011}). For example,~\cite{2008MNRAS.388.1582L} discussed the possibility that stellar activity/star spots may significantly inflate the donor radii in CVs, which would result in a minimum period closer to the observed one. Conversely~\cite{knigge2011}, discussed the implications of residual magnetic braking below the gap. They found that by allowing for a larger donor radius (due to tidal and rotational deformation), increasing the effect of angular-momentum loss below the gap by a factor of about 2.5, and by lowering the effect of angular momentum loss above the gap (by a factor of 1.5), they were able to account for the main differences between theory and observations.  Also, their model yields a minimum period consistent with observations. Since a lower level of angular-momentum loss above the gap would make systems stay above the gap for a longer time, their model are also able to explain why we find more systems above the period gap than predicted by theory. Although Knigge et al.'s (2011) model would require magnetic braking in fully convective donor stars, this is not necessarily in conflict with the disrupted magnetic braking model usually employed to explain the existence of the period gap. This is because a reduction in  $\dot{J}$ by a factor of about 5 is sufficient to ensure that mass transfer will cease at the upper edge of the gap (\citealt{knigge2011}).  
\newpage 


\section{CVs in This Thesis}

Statistical analyses of observed CV samples, such as those presented by~\cite{2008MNRAS.388.1582L, 2009ApJ...693.1007T, 2009MNRAS.397.2170G, 2011MNRAS.tmp...27P} and~\cite{knigge2011}, are important and powerful tools when studying cataclysmic variables. However, to test proposed models quantitatively, it is valuable to also study individual CVs. Systems that appear to be outliers could, in fact, be more important to CV evolution than initially thought. For instance, there could be systems that represent stages of rapid (and potentially destructive) evolution -- as believed is the case for T Pyxidis presented below.
 
In order to establish the evolutionary status of a system, the orbital period, mass ratio and component masses have to be accurately determined. Chapter~\ref{methods} presents an overview of some methods that are commonly used to determine some the most important system parameters.

In Chapters~\ref{j1457_bwscl},~\ref{tpyx} and~\ref{j1507}, I present analyses of four systems close to the minimum orbital period that all are particularly interesting in the context of studying CV evolution. Two of them, SDSS J1457+51 and BW Sculptoris, are consistent with belonging to the ordinary CV population, as described by the standard evolutionary models proposed for cataclysmic variables. The other two systems, T Pyxidis and SDSS J1507+52, are more exotic.

\subsection{Two Ordinary CVs}

\textbf{SDSS J1457+51} and \textbf{BW Sculptoris} are two faint CVs with low accretion rates and thin discs. Both have short orbital periods $\approx$ 80 minutes (Uthas et al. in press), which is right at the mean observed minimum period. Absorption from hydrogen and helium is present in both their spectra, indicating that the underlying WD is exposed (\citealt{2005AJ....129.2386S}).

In Chapter~\ref{j1457_bwscl}, I present the discovery of non-coherent pulsations with main periods of about 10 and 20 minutes in both systems, and discuss the possible origin of these signals. Chapter~\ref{j1457_bwscl} also discusses an interesting signal detected in BW Sculptoris, which would be connected to the unusual phenomenon of quiescent superhumps.

\subsection{Two Peculiar CVs}

The system discussed in Chapter~\ref{tpyx}, \textbf{T Pyxidis}, is a luminous RN that accretes at a much higher rate than is expected for its photometrically estimated orbital period of just under 2 hours (\citealt{1998PASP..110..380P}). T Pyxidis used to have a recurrence time scale between eruptions of about 20 years. However, the last eruption was in 1966, which means that the system has passed its mean recurrence time by more than 20 years.

According to standard evolutionary theory, a system with a period less than 2 hours should be faint and have a low accretion rate. However, T Pyxidis is about two times more luminous than expected for a CV driven by GR at this period. In order to explain its high luminosity, it has been suggested that it could be a wind-driven supersoft source (\citealp{1998PASP..110..380P, 2000A&A...364L..75K}). Theoretically, RNe both have high accretion rates and high WD masses (\citealt{2005ApJ...623..398Y}). This implies that RNe, in general, are candidate progenitors for Supernova type 1a. Also, the current evolutionary timescale of T Pyx is only a few million years. If other CVs would go through similar phases of high accretion rates, this could speed up of their evolution, resulting in a reduction of the number of short-period systems.

Compared to other short-period systems, the high luminosity and accretion rate in T Pyxidis would be highly unusual. However, a photometrically established period of a non-eclipsing system is never as reliable as a spectroscopically determined one. Therefore, to confirm the current status of T Pyxidis as a short-period system, spectroscopic data is highly desirable. In Chapter~\ref{tpyx}, I present a detailed analysis of such data of T Pyxidis, revealing its current evolutionary status.
\newline
\newline
The eclipsing system \textbf{SDSS J1507+52}, which is the subject of Chapter~\ref{j1507}, was first observed by the SDSS and quickly recognised to be odd due of its short orbital period of 67 minutes (\citealp{2005AJ....129.2386S}). Hydrogen absorption originating from the primary WD was detected in its optical spectrum (\citealp{2005AJ....129.2386S}), and together with the short orbital period, this indicates a system with a low-mass secondary, a low accretion rate and a faint disc. 

~\cite{2007MNRAS.381..827L} performed eclipse analysis of SDSS J1507+52 and found a donor mass of $\approx$ 0.05 M$_{\odot}$, which would imply that the secondary is a sub-stellar object. However, if SDSS J1507+52 is placed on the plot presented in Figure~\ref{knigge_mp}, the system does not appear on either the evolutionary branch for period bouncers or the one for main-sequence donors. In addition to the low donor mass,~\cite{2007MNRAS.381..827L} found that its donor radius is much smaller than expected for sub-stellar donors at this mass. They also point out that these unusual donor properties may indicate a young system, in which the secondary has not yet had time to adjust its radius in response to mass loss. They therefore propose that mass transfer started only recently in SDSS J1507+52, i.e. that its progenitor was a detached WD-brown dwarf binary system.

Another study of SDSS J1507+52 was carried out by~\cite{2008PASP..120..510P}, at approximately the same time. They found another peculiarity of the system -- it has a very high space velocity, much like stars in the galactic halo. If SDSS J1507+52 is a halo CV, it should have a low metallicity and therefore also a smaller donor radius and shorter orbital period. Thus~\cite{2008PASP..120..510P} suggest that the system is a member of the Galactic halo population. 

The effective temperature for SDSS J1507+52 was estimated to be $\approx$ 11000 K -- 11500 K by both~\cite{2007MNRAS.381..827L} and~\cite{2008PASP..120..510P}. At this low temperature, strong absorption features due to Fe II and  III are expected in the UV spectrum. A simple test to distinguish between the two proposed theories would thus be to obtain UV spectra to find out if this absorption is present. In Chapter~\ref{j1507}, I present the analysis of such a study of SDSS J1507+52, which settles the question of its origin as either a young WD-brown dwarf system, or a halo CV.


\chapter{Methods and Techniques \\\Large \textsc{- for analysis of CVs -}} \label{methods}
\label{chap:icm}

\begin{Huge}\color{Red}{T}\end{Huge}he first step in confirming the current evolutionary status of a CV is to determine accurately its key system parameters, such as the orbital period and mass-ratio. In this chapter, I will discuss what methods are commonly used to determine some of these key systems parameters. I will also give a brief summary of the observational properties of CVs and of some common techniques used to analyse such observational data. In particular, I will focus on those techniques and methods applied in Chapters~\ref{j1457_bwscl}, ~\ref{tpyx} and~\ref{j1507}.

\section{Observational Properties of CVs}

\begin{figure}[t]
\centering
\includegraphics[scale=0.16]{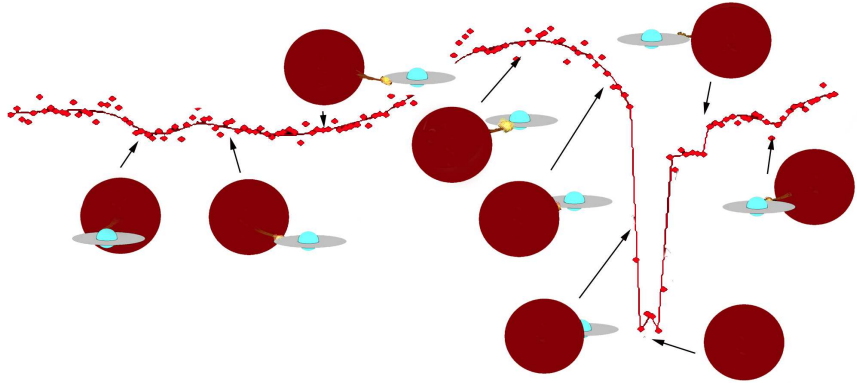}
\caption{\emph{The optical light curve of SDSS J1507+52 obtained at the Nordic Optical Telescope, together with models showing the phase-dependent CV geometry (adapted from figures by Are Vidar Boye Hansen, Olesja Smirnova and Arturs Barzdis).}}
\label{light_fig}
\end{figure}

Since telescopes are not able to separate spatially the light originating from the accretion disc from that coming from the WD or donor, a major part of studying CVs involves the development of methods to isolate the light originating from the different components of the system. Eclipsing systems, in particular, provide us with a unique opportunity for this. Figure~\ref{light_fig} shows the optical light curve of the eclipsing system SDSS J1507+52 (see Chapter~\ref{j1507}), together with a sketch of the phase-dependent CV geometry that is thought to give rise to these light curves\footnote{\emph{Both the light curve (obtained at the Nordic Optical Telescope during 2005) and phase models of SDSS J1507+52 are provided by Are Vidar Boye Hansen, Olesja Smirnova and Arturs Barzdis.}}. The light curve for this particular system shows a large hump during phases where the bright spot is visible. The very steep ingress and egress of the eclipse indicates a faint accretion disc. This all implies that the bright spot and WD dominates the flux distribution in SDSS J1507+52 (a detailed eclipse analysis of this system was presented by~\citealt{2007MNRAS.381..827L}).

In addition to giving the most accurate orbital periods, analyses of eclipsing systems can provide us with a number of other system-parameter estimates. For instance, as shown in Figure~\ref{light_fig}, the total duration of the overall eclipse is proportional to the size of the accretion disc. The size of the WD is found from the duration of the sharp ingress and egress of the eclipse. Also, the depth and duration of the eclipse is a function of the mass ratio and inclination (e.g Equation 2.64 in~\citealt{1995CAS....28.....W}). 

No eclipse analyses have been applicable to the studies I have performed in this thesis, and as such, no detailed deduction of this method is given.

\subsection{Multi-Wavelength Properties}

\begin{figure}[t]
\centering
\includegraphics[scale=0.3]{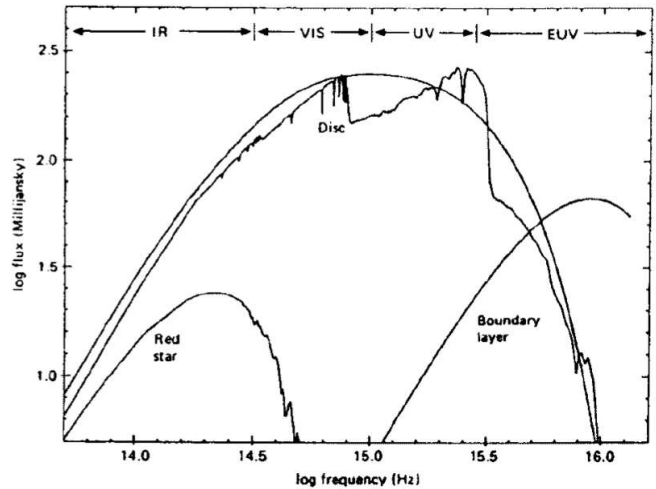}
\caption{\emph{The spectral energy distribution of a CV, comprising a high-state accretion disc (Pringle \& Wade 1985).}}
\label{spectral_cv}
\end{figure}

The various components in a CV have a different relative contribution to the radiation in different wavelength regions. Figure~\ref{spectral_cv} shows an example of the spectral energy distribution (SED) of a CV with a high-state accretion disc (from~\citealt{1985ibs..book.....P}). The following are common to all CVs: 
\newline 

\noindent \textbf{-- Accretion disc}: The flux distribution from the accretion disc spans the whole range from IR to UV. It is usually the dominant feature in the optical wavelength region, making it difficult to observe any of the stellar components directly. The temperature in the disc varies from $\sim$ 100000 K at the inner edge of the disc to $\sim$ 3000 K in the outer edge of the disc. The disc spectrum mostly contains emission lines from low ionisation states, such as H (from Paschen, Balmer and Lyman) and He I. A higher state of ionisation is often detected in systems with high accretion rates. Especially, He II is used as a tracer of high $\dot{M}$ systems. For instance, the system T Pyxidis has an accretion rate of $> 10^{-8}$ M$_{\odot}$ yr$^{-1}$ (\citealp{1998PASP..110..380P};~\citealp{2008A&A...492..787S}), and shows very strong He II lines in its spectrum (see Figure~\ref{fig:spec_mag}).

\begin{figure}[t]
\centering
\includegraphics[width=1.0\textwidth]{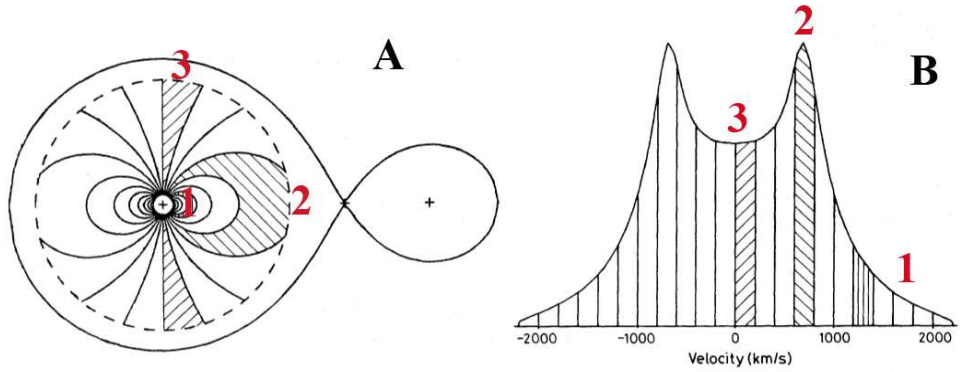}
\caption{\emph{Emission from different regions in a Keplerian disc are marked with its corresponding velocity bin in the line profile (Figure adapted from~\citealt{1986MNRAS.218..761H}, showing a system with $q$ = 0.15).}}
\label{emission}
\end{figure}

Emission lines formed in the accretion disc exhibit double-peaked profiles due to strong rotational Doppler broadening. The splitting of an emission line will be more or less pronounced depending on the orbital inclination and the spectral resolution of the data (quantitatively explored by~\citealt{1981AcA....31..395S}). Furthermore, optically thick disc lines (as considered to be the usual case for disc emission of H and He II), exhibit a deeper central minimum for higher inclinations. For optically thin lines, the central minimum is shallower and has a softer shape. The mechanism of line formation is described in detail by~\citealt{1986MNRAS.218..761H}, and is illustrated in Figure~\ref{emission}. In the left figure (A), the radial velocity contour geometry of a Keplerian disc (seen face-on) is shown, while the figure to the right (B), shows its corresponding emission line profile viewed at an angle $i$. Keplerian rotation causes the lines of constant velocity to form a dipolar pattern aligned with the line of sight. The wings of the line profile are shaped by gas rotating at high radial velocities close to the WD (see position 1 in Figure~\ref{emission}). The radiating surface area $A$ in each velocity bin $dV$ determines the strength of the line at each velocity $V$, and is proportional to $V^{-5}dV$ (since $A \propto RdR$ and $R \propto V^{-2} \Rightarrow dR \propto V^{-3}dV$), which explains the rapid decline in the line wings. The maximum line strength is reached when $V_{\text{D}}= \pm V_{\text{Kep}}(R_{\text{D}}) \sin i$, which corresponds to the outer boundary $R_{\text{D}}$ of the emission line region (position 2). At this point, $A$ starts decreasing, reaching a local minimum at $V = 0$ where the flux arises from gas moving perpendicular to the line of sight (position 3), and a dip in flux, and apparent splitting of the line is seen.

Since the emissivity in the disc is not constant (the inner parts of the disc is much hotter than the outer parts), the line strength does not only depend on the radiating surface area, but also on the disc emissivity.     
 \newline


\noindent \textbf{-- White dwarf}: A WD spectrum displays broad absorption lines (usually of hydrogen), widened by the process of pressure broadening. Average WD temperatures just above the period gap are $\sim$ 26000 K and just below the period gap $\sim$ 18000 K (\citealt{2006ApJ...642.1029U}). These high temperatures imply that the WD is visible in the ultraviolet (UV) spectrum (e.g.~\citealt{2002ApJ...565L..35T}). In short-period system with low $\dot{M}$, observations in the UV can be used to isolate the WD light from the accretion disc and bright spot. 

\begin{figure}[t]
\centering
\includegraphics[width=0.8\textwidth]{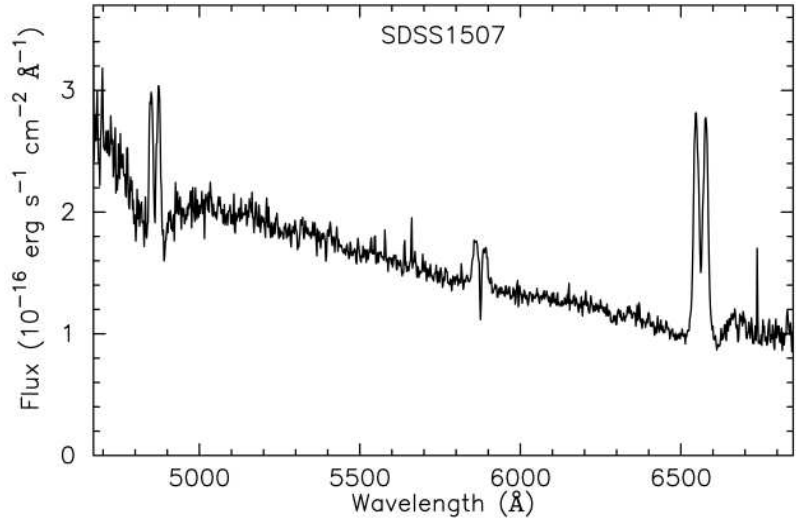}
\caption{\emph{Optical spectrum of SDSS J1507+52 by Patterson, Thorstensen \& Knigge (2008), showing deep double-peaked disc emission lines, superimposed on broad absorption lines originating from the primary WD.}}
\label{spec_1507}
\end{figure}

In the optical, spectral features from the WD are difficult to detect due to the strong emission from the accretion disc. However, there are systems in which absorption features from the WD can be detected also in the optical. These stars typically have low accretion rates and faint discs. Figure~\ref{spec_1507} shows the optical spectrum of the faint system SDSS J1507+52 which shows double-peaked disc emission lines superimposed on absorption from the WD (Chapter~\ref{j1507}). 
\newline

\noindent \textbf{-- Donor star}: The donors in CVs are typically cool, near main-sequence stars of spectral types between G -- M, and temperatures between $\approx$ 3000 K -- 5000 K (\citealt{2006MNRAS.373..484K, knigge2011}). Due to the process of accretion, their radii are larger compared to main-sequence stars of same mass. Also, their spectral types are found to be later than they would have been compared to what a Roche-lobe filling main-sequence star at the same orbital period would have. This is because a larger donor radius results in a lower density, $\rho \propto M_{2}/R_{2}^{3}$ (\citealt{2006MNRAS.373..484K}). 

Infrared (IR) observations are the most promising way to uncover features originating from the cool donor (e.g.~\citealt{2011ApJ...728L..16H}). However, in some CVs, the donors might be detectable also in the optical, for instance, if the hot WD (and/or the inner part of the accretion disc) illuminates the front side of the donor. Such illumination could give rise to a reflection effect, caused by the increase in emission from the irradiation-heated front hemisphere. For instance, this has been detected in systems undergoing nova eruptions.
\newline 

\noindent \textbf{-- Bright spot}: The impact point between the accretion stream and disc is often detected as an variation on the orbital period in the optical light curve. The bright spot is less prominent in the IR, and has not been detected in the UV at all.
\newline

\noindent \textbf{-- Boundary layer}: X-ray observations are sometimes used when studying the process of accretion, and especially for dwarf novae, when studying the extremely hot boundary layer close to the WD. The boundary layer is responsible for a large portion of the total emission from the system, and therefore, studying this region can be considered as highly important. In systems with high $\dot{M}$, the boundary layer is optically thick and will emit in the soft X-ray range, while for systems with lower $\dot{M}$, the boundary layer is optically thin and is observed typically as thermal bremsstrahlung radiation in the hard X-rays (\citealt{2003astro.ph..2351K}).

\section{Analysing Data}

\subsection{Time-Series Analysis} \label{freq}

\begin{figure}
\centering
\includegraphics[scale=0.45]{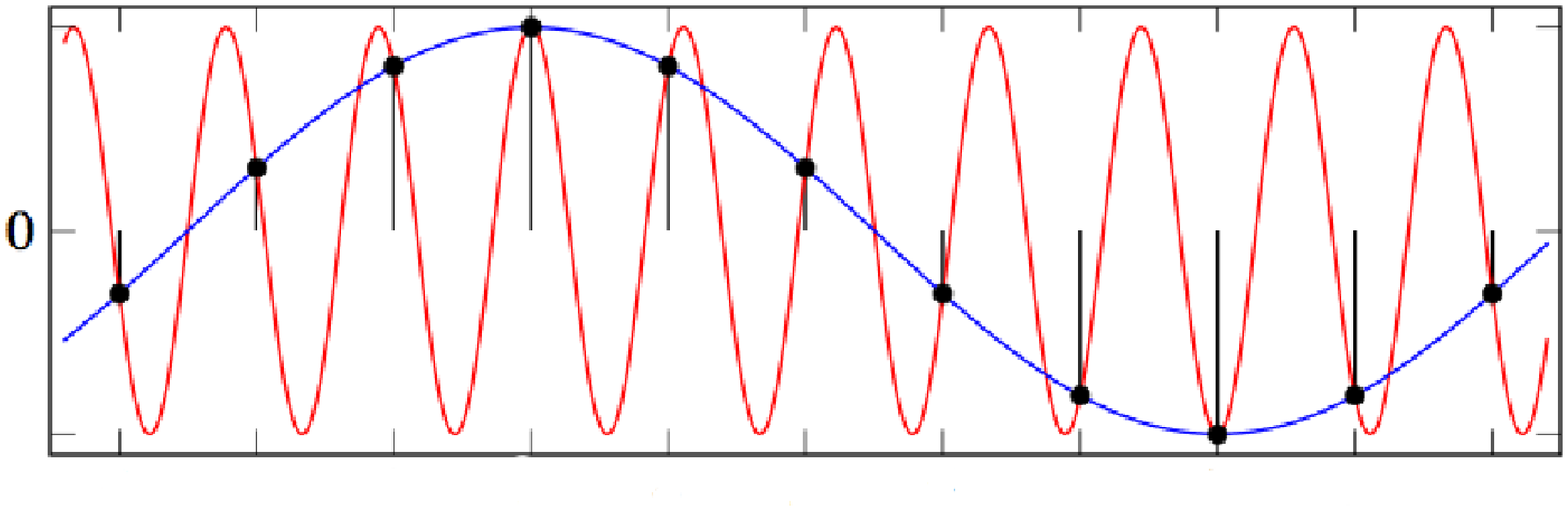}
\caption{\emph{Model showing how the sampling rate (black dots) can produce an alias ({\color{blue} blue}) of the real signal of the data ({\color{Red} red}). \newline \small{Figure from http://sv.wikipedia.org/wiki/Fil:AliasingSines.svg}.}}
\label{sampling}
\end{figure}

A common goal of astronomical time-series analysis is the detection and characterisation of periodic signals in noisy data. Such work is usually carried out by statistical analyses of the frequency content of the data. In this following section, I will provide a brief introduction to this type of analysis. 
\newline
\newline
The lowest frequency that can be detected in a light curve is determined by the length of the data set, while the highest detectable frequency in a light curve is determined by the sampling of the data. In frequency space, the Nyquist sampling theorem states that for evenly spaced data, the sampling frequency $F_{\text{S}}$, should be least twice the highest frequency expected in the data, the so-called \textbf{Nyquist frequency} $F_{\text{N}}$ ($F_{\text{N}} = F_{\text{S}}/2$). A frequency greater than $F_{\text{N}}$ will not appear at the correct frequency. Figure~\ref{sampling} shows an example of this problem, where the sampling rate, marked as black dots, creates a false periodic signal ({\color{blue} blue}) of the original signal in the data (\begin{color}{red}red\end{color}). This problem is called \textbf{aliasing}, and refers to when periodic signals in the source are not reconstructed at the correct frequencies in the time-series analysis. Any data set being sampled will suffer from aliasing. In particular, multi-night observations will, apart from the aliasing associated with the sampling rate of the observations during the night, also show frequencies introduced by the daily sampling. Figure~\ref{fig:s_poworb} shows an example of aliasing present in the power spectrum of SDSS J1457+51. Also, as a consequence of multi-night observations, the data are no longer evenly spaced in time, which means that the Nyquist frequency cannot be specified by a single value. 

To minimise the problem with aliasing, one needs to obtain data sets much longer than the period/periods expected in the data, and if the data are obtained over many nights, as densely and uniformly spaced in time as possible. The Center of Backyard Astrophysics (CBA:~\citealt{1993ApJ...417..298S, patterson_1998}), is a collaboration between small telescopes evenly spaced both in longitude and latitude. This is an efficient way of observing that results in minimal problems with aliasing, since it results in a light curve with fairly regular sampling, free from the problems introduced by daily observations from the same site. Another, more expensive way to reduce aliasing is to use space telescopes that are able to continuously study a source without interruptions. This is done for instance in the Kepler mission (see the observations of V344 Lyrae by~\citealt{2010ApJ...725.1393C}). However, often, the observing circumstances are such that aliasing cannot be completely avoided. In order to manage the problem with aliasing, one could, for instance, perform \textbf{Monte Carlo simulations} or \textbf{bootstrap simulations} on the data. When performing Monte Carlo simulations, for each simulation, the flux values are randomly re-distributed within their errors (see Section~\ref{bot_mont}). In the case of bootstrap simulations, for each simulation, a random subset of the data is selected, altering the sampling pattern and hence also the aliasing pattern. In theory, for a large number of such simulations, the real signal would always be found at the same frequency, while those caused by aliasing would not. As a result, the real signal should statistically be best represented. In practice, one has to be aware that when throwing away data points, one might also throw away points that signify the real period.

The \textbf{Fourier theorem} states that any periodic signal in the light curve can be reproduced by summing a set of sine waves. As a result, the periodicities in a light curve can be found by fitting sine waves of different frequencies to the data, where the amplitude of each sine wave indicates how much signal is present at each frequency. The power of each frequency can be calculated as the square of the amplitude. In a power spectrum, the power is presented as a function of frequency.

All data presented in this thesis are light curves unevenly distributed in time. The most commonly used method to find period signals in such data is to construct a \textbf{Lomb-Scargle periodogram} (\citealt{1976Ap&SS..39..447L, 1982ApJ...263..835S}). However, before such a periodogram can be constructed, the mean value needs to be subtracted from the data, since the sine waves being fitted have a mean of zero. In a Lomb-Scargle periodogram, the power at the frequency $\omega$ of a light curve $y(t_{n}$), is defined as

\begin{equation}
P(\omega) =  \frac{\sigma^{2}}{2} \left(\frac{\left(\sum_{n=1}^{N} y(t_{n}) \cos(\omega(t_{n} - \tau))\right)^2}{\sum_{n=1}^{N} \cos^{2}(\omega(t_{n} - \tau))} + \frac{\left(\sum_{n=1}^{N} y(t_{n}) \sin(\omega(t_{n} - \tau))\right)^2}{\sum_{n=1}^{N} \sin^{2}(\omega(t_{j} - \tau))}   \right),
 \label{scargle}
 \end{equation}

\noindent where $\sigma$ is total variance of the data, and $\tau$ is

\begin{equation}
\tan (2\omega\tau) = \frac{\sum_{n=1}^{N} sin 2\omega t_{n}}{\sum_{n=1}^{N}cos 2\omega t_{n}}.
 \label{scargle_tau}
 \end{equation}
\newline

\noindent In the case where a light curve contains periodic but non-sinusoidal signals, additional higher frequency components are required to reconstruct the data. As a result, harmonics of the fundamental frequency are detected in the power spectrum (e.g. Figure~\ref{fig:s_poworb}). 

The accuracy of the location of a peak in the power spectrum is dependent on the length of the data set, and longer observations results in narrower peaks. While the obvious reason for constructing a power spectrum is to identify frequencies with power excess, the significance of a peak needs to be evaluated. Visually, this is often done by considering signals that have significantly higher power than the highest noise peaks in the power spectrum. However, a signal at low power can still be of interest, for instance, if it is detected in many data sets. A more quantitative approach to find the significance of a peak is to perform a randomisation test, which redistributes the data points in time while maintaining the original sampling pattern. In doing so, any periodic signal present in the data is deliberately destroyed, which allow us to find the probability of detecting excess power by random chance. Confidence limits can then be set on the chance of detecting excess power at a certain frequency. However, this method is not valid if the data are dominated by red noise, i.e. if the data points are correlated. 

\subsection{Trailed Spectrogram}  \label{trails}


\begin{figure}[ht]
\begin{minipage}[b]{0.5\linewidth}
\centering
\includegraphics[scale=0.32]{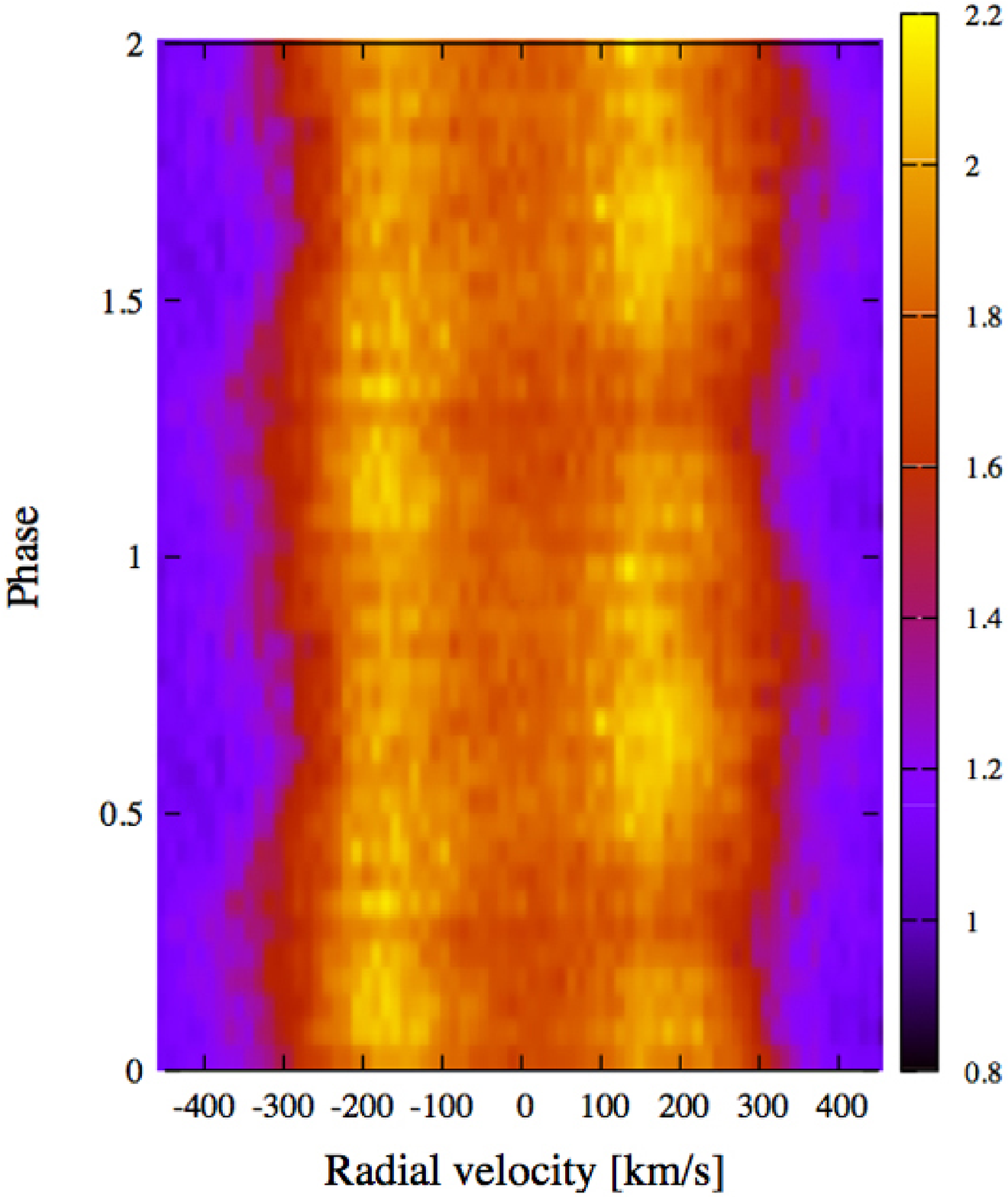}
\end{minipage}
\hspace{0.15cm}
\begin{minipage}[b]{0.5\linewidth}
\centering
\includegraphics[scale=0.66]{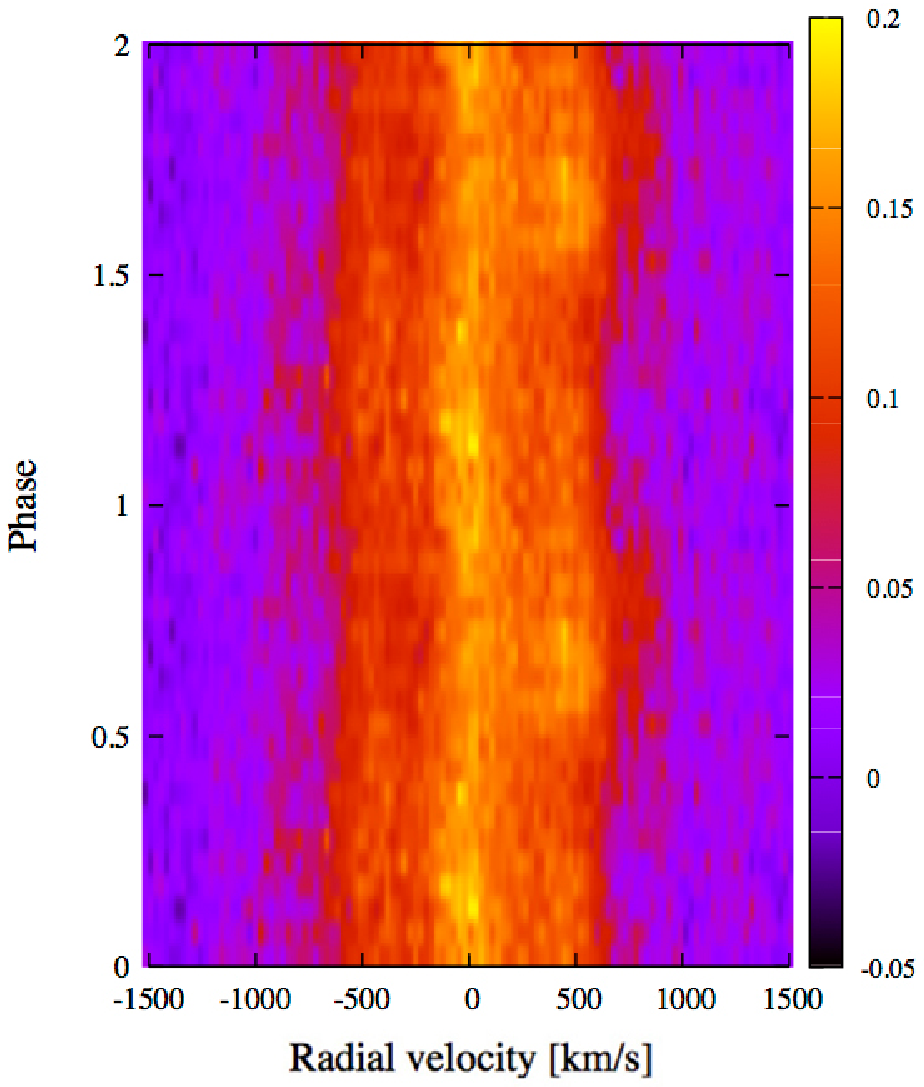}
\end{minipage}
\caption{\emph{Trailed spectrograms of the HeII line at 4686\,\AA\,(left) and the Bowen blend at 4640\,\AA\,\,-- 4650\,\AA\,\,(right) in T Pyxidis. Data are phase binned and plotted over 2 orbital periods.}}
\label{color_trail}
\end{figure}


In Chapter~\ref{tpyx}, I present time-resolved optical spectroscopy of T Pyxidis. Such data can be effectively visualised by plotting the line profiles as a function of time. A trailed spectrogram is a 2-dimensional plot where the region around a chosen line, expressed in velocity units ($V$), is plotted as a function of orbital phase ($\phi$), $F(V, \phi$). The orbital motion of the gas in the binary will, due to Doppler shifts, generate spectral lines that follow the shape of a S-curve. The trailed spectrogram is then the superposition of all such S-curves. 

A trailed spectrogram can be very useful, especially in the early stages of the analysis, since it can give a rough idea of the orbital period and helps visualise the overall orbital behaviour of the lines. Also, if studying disc emission lines, the wings of the line profiles are expected to roughly track the motion of the WD, and can therefore provide an estimate of the radial velocity of the WD ($K_{1}$). Also, the mean offset in velocity units between the measured line centre and the rest wavelength of the line immediately yields the systemic velocity ($\gamma$). It is also useful to compare trailed spectrograms for different lines to each other. If there are phase shifts and/or amplitude differences between them, they most certainly originate from different regions within the CV system (for example, Figure 2 in~\citealt{2002ApJ...568..273S} shows phase shifts between lines from the WD and donor).   

Figure~\ref{color_trail} shows trailed spectrograms constructed from the double-peaked HII line at 4686\,\AA\,\,(left) and the Bowen blend (consisting of NIII and CIII lines) at 4640\,\AA\,\,-- 4650\,\AA\,\,(right) in the spectrum of T Pyxidis. The data were binned in phase and velocity, and the systemic velocity was removed. Here, the trails are shown repeated over 2 orbital cycles for clarity. The colour scale represents normalised fluxes. The relative amplitudes of the two peaks in the HeII line vary over the orbital cycle, with first one peak, and then the other dominating. This could be due to a non-Keplerian velocity distribution of the emission (e.g. a slightly elliptical disc). The trailed spectrograms for both the HeII line and Bowen blend show spectral features, moving together in phase, which indicate that they most likely are formed in the same region.   

 \subsection{Doppler Tomography} \label{doppler}

Doppler tomography is an indirect imaging method developed by~\cite{1988MNRAS.235..269M} to isolate the different components of a CV in velocity space (see reviews by~\citealt{2001LNP...573....1M, 2005Ap&SS.296..403M, 2004AN....325..193M, 2004AN....325..185S}). This method is most commonly used to gain insight into the structure of accretion discs and other accretion flows (e.g.~\citealt{2003MNRAS.344..448S}) and is analogous to the method used in medical X-ray tomography. Doppler tomography uses the fact that the spectral lines follow S-shaped radial velocity curves as a function of orbital phase as described in the previous section (e.g. Figure~\ref{color_trail}). A Doppler tomogram therefore presents the flux distribution from a spectral line in velocity space. The velocity centre in the map corresponds to the centre of mass, which will be close to the WD. Doppler tomogram are often rotated so that $\phi$ = 0 is at the superior conjunction of the WD (where the donor is closest to us). In this orientation, the donor star appears at $\phi$ = 0.5, where orbital phases are defined to increase in the clockwise direction. However, before the data can be put on a 2-dimensional velocity map, the wavelengths and orbital phases needs to be converted into the velocity components V$_{x}$, V$_{y}$. A position within the binary has an associated velocity vector in the co-rotating frame, reflecting the motion of the gas. The observed velocity V($\phi$) is then the radial velocity component of that gas vector at a given binary orientation, also correcting for any systemic component ($\gamma$),

\begin{equation}
V(\phi) = \gamma - V_{x} \cos(2 \pi \phi) + V_{y} \sin(2 \pi \phi).
 \label{radial_vel}
 \vspace{0.3cm}
 \end{equation}

\noindent There are several methods for converting the spectral components into velocity space, i.e. a Doppler tomogram. The simplest method compares predicted S-curves for every possible velocity in the map with the data in the trailed spectrogram, by assigning the integrated flux $D$ along each S-curve in the trailed spectrograms to the corresponding V$_{x}$, V$_{y}$ point in the Doppler map ($M$).~\cite{horne_1992} presented this in the following way

\begin{equation}
M(V_{x}, V_{y}) = \frac{\int D(V(V_{x},V_{y},\phi),\phi)\omega(\phi)d\phi}{\int \omega(\phi)d\phi},
 \label{back_proj}
 \vspace{0.3cm}
 \end{equation}

\noindent where $\omega$ is a weighting function. This method is also referred to as \textbf{back-projection} (as described by~\citealt{2001LNP...573....1M, 2004AN....325..185S}). In this method each spectrum is projected onto the map, rotated with an angle corresponding to the phase of the observation. An extension to the method described above is to apply an iterative approach to construct an optimal map, such as the \textbf{maximum entropy regularisation}, first used by~\cite{1985MNRAS.213..129H} when performing eclipse mapping.  

\begin{figure}[t]
\centering
\includegraphics[scale=0.4]{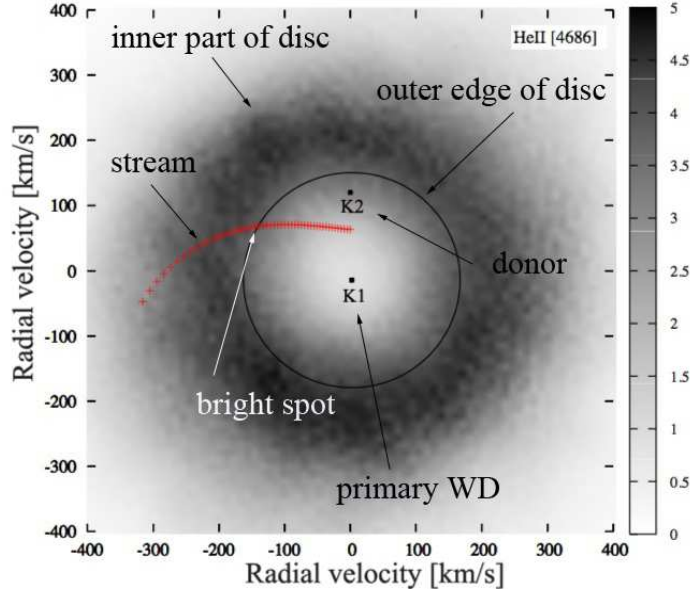}
\caption{\emph{Doppler tomogram of T Pyx constructed from the He II line at 4686\,\AA. The different components in the system are marked.}}
\label{dopp_ill}
\end{figure}
 
\noindent Figure~\ref{dopp_ill} presents a Doppler map constructed from the trail of the HeII line seen to the left in Figure~\ref{color_trail}. The most prominent features, such as the positions of the WD, donor star, bright spot and accretion disc/stream are marked. Since the outer part of the disc has lower velocity than the inner parts, the disc is turned \emph{inside out} in velocity space. The velocity semi-amplitude of the WD ($K_{1}$) and the donor ($K_{2}$) could, in principle, be found directly from the map (however, to find $K_{2}$, one obviously would need to study lines which actually show emission from the donor). Also, the velocity at the outer disc radius, $v_{\text{R}_{\text{disc}}}\!\sin i$ (marked as a black circle) can be measured directly from the Doppler map (see Section~\ref{vodr}).

Doppler tomography is not restricted to the mapping of accretion discs, and has been used, for instance, for studying the accretion flow in magnetic CVs (e.g. ~\citealt{1997A&A...319..894S}). It is possible to convert the velocity map into a spatial map, but this is a poorly constrained problem. In order to make this conversion, one must make assumptions about the distribution of the velocities of the material in the system, for example, it is often assumed that material in the accretion disc moves in a Keplerian fashion. Since important information might be lost in the conversion process, we might as well study the Doppler tomograms as they are, in velocity space.

\section{Estimating System Parameters}

\subsection{Orbital Period} \label{orbital}

In the context of CV evolution, the orbital period ($P$) is the most important system parameter for a CV since it indicates the evolutionary state of the system. There are several methods that can be used to obtain the orbital period, the most reliable one being eclipse timing, though this is obviously only available in systems viewed at sufficiently high inclination (> 75$^\circ$). 

In non-eclipsing systems, orbital modulations from the bright spot are normally detected, and the orbital period can be found from extracting the frequency content of any photometric light curve obtained for a longer time than $P$ (see Section~\ref{freq}). Any signal of orbital nature appearing in a power spectrum should be stable in frequency, and therefore appear as a narrow peak. For systems with low accretion rates, the first harmonic of the orbital frequency is commonly found to have higher power than the orbital frequency itself: these kind of double-humped waves are, for instance, seen in the power spectrum for both of CVs presented in Chapter~\ref{j1457_bwscl}.  

A spectroscopically determined orbital period is generally more reliable than a photometrically determined one. This is due to the fact that one can never be sure that the period detected in the light curve is of orbital nature, or that the correct alias has been identified in the resulting power spectrum. In order to find $P$ spectroscopically, radial velocity curves can be constructed from the time-resolved spectroscopic data. Such a study often involves fitting Gaussian profiles to the Doppler-broadened disc emission lines. The line centre of each spectrum is then retrieved as a function of time, yielding a sinusoidal radial velocity curve with a period equal to the orbital period for the system. Radial-velocity curves can also be constructed by using more advanced techniques, such as the double-Gaussian method described below in Section~\ref{svvsa}, which specifically aims to track the innermost disc regions, and hence the orbital motion of the WD.

When a radial-velocity curve has been constructed, $P$ is found either by fitting a sinusoid to the curve, which has a period correspond to the orbital period, or by constructing a power spectrum of the radial-velocity data. 

Spectroscopically determined orbital periods also suffer from the problem with aliasing in the resulting power spectrum. In Chapter~\ref{tpyx}, I use bootstrap simulations to find the most likely orbital frequency in T Pyxidis (described in Sections~\ref{freq} and~\ref{bot_mont}).

\subsection{Systemic Velocity and Velocity Semi-Amplitude} \label{svvsa}

The systemic velocity ($\gamma$) of a system is the velocity at the centre of mass of the binary. The velocity semi-amplitudes of the WD ($K_{1}$) and the donor ($K_{2}$), are proportional to the masses of the two components ($K_{1}/K_{2} = M_{2}/M_{1} = q$), and can be measured as the amplitudes of their individual radial-velocity curves.

\begin{figure}[t]
\centering
\includegraphics[scale=0.6]{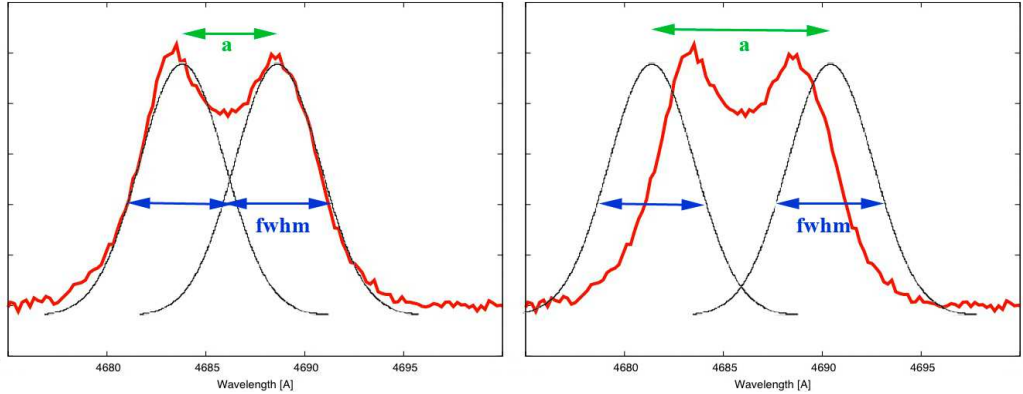}
\caption{\emph{Illustration of the double-Gaussian technique, showing two different separations ($a$) between the two Gaussians (here with a fixed \emph{FWHM}). The {\color{Red} red} line shows the He II line in T Pyxidis.}}
\label{double_gaussian}
\end{figure}

In Chapter~\ref{tpyx}, I estimate $\gamma$ and $K_{1}$ from the radial-velocity curves obtained from disc emission lines in T Pyxidis, by using the \textbf{double-Gaussian technique} (\citealt{1980ApJ...238..946S}), where the velocities are extracted from the wings of the lines. These are thought to be formed in the innermost regions of the accretion disc, closest to the WD (see Figure~\ref{emission}). As a result, one might hope that they will track the orbital motion of the WD. In this method, a chosen double-peaked emission line profile is convolved with two Gaussian profiles, separated with a distance $a$. The weighted flux average in the red part of the line, with the weighting function being the Gaussian profile, is compared to that of the blue part of the line, with the aim to find for which value of the radial velocity, the two weighted averages are equal. Assuming that the double-peaked line profile is symmetric, this radial velocity corresponds to the Doppler shift. This is done for each of the spectra in the time series, for a wide range of central positions of the Gaussian profiles, and values of the \emph{FWHM} of the two Gaussians. In Figure~\ref{double_gaussian}, I illustrate the double-Gaussian technique performed on the He II line in T Pyxidis, where the separation $a$ is changed for a fixed \emph{FWHM}.   

Resulting radial-velocity curves are created for each set of $a$ and \emph{FWHM}, and a sinusoid of the form $V(\phi ,a) = \gamma(a) - K_{1}(a) \sin(\phi)$, where $V$ is the velocity at phase $\phi$, $\gamma$ is the mean velocity, and $K_{1}$ is the amplitude, is fitted (for instance, by nonlinear least-squares fitting) to the data. In order to find what combination of $a$ and \emph{FWHM} that corresponds to the best output radial-velocity curve, a \textbf{diagnostic diagram} can be constructed. In this diagram, $\gamma$, $K_{1}$ and $\phi$ are plotted as a function of $a$ for different values of the \emph{FWHM}. Such a diagram, constructed from radial velocity curves of the HeII line at 4685\,\AA\, in T Pyxidis, is seen in Figure~\ref{fig:dd}. The best set of $a$ and \emph{FWHM} are found from inspecting for which $a$ and \emph{FWHM} each of the variables $\phi$, $\gamma$ \& $K_{1}$ are stable. The corresponding \emph{best} values of $a$ and \emph{FWHM} are then used to construct a final radial velocity curve from which final values of $\gamma$, $K$ and $\phi$ are extracted.


\begin{figure}[t]
\centering
\includegraphics[scale=0.3]{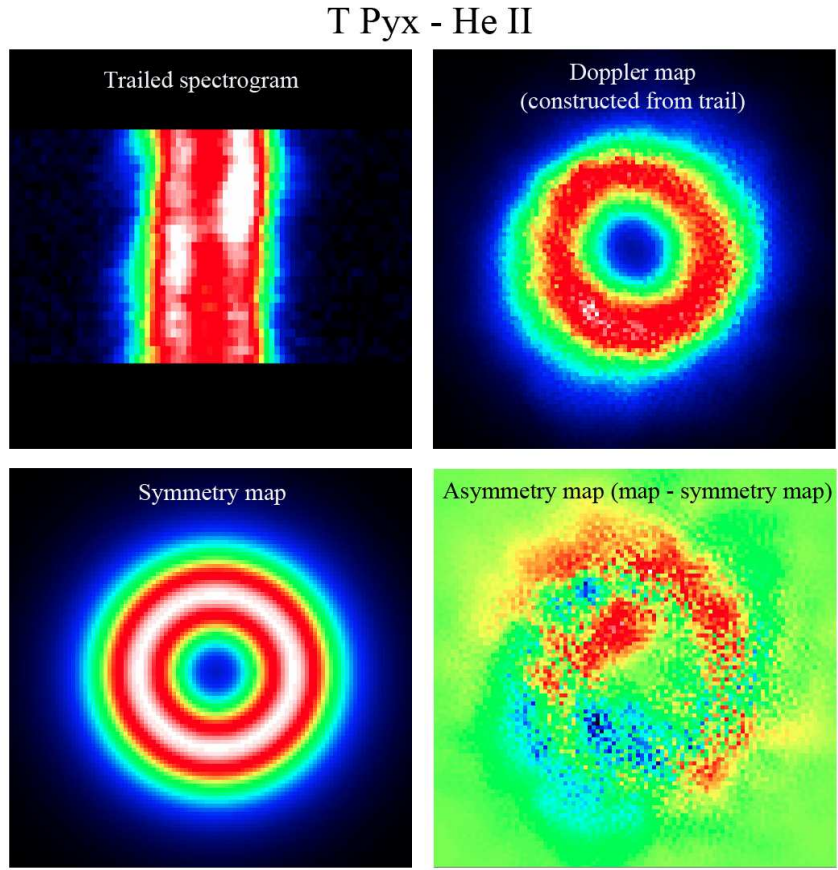}
\caption{\emph{Top left figure shows a trailed spectrogram for the HeII line at 4686\,\,\AA\,in T Pyxidis. Its corresponding Doppler tomogram is seen in the top right figure. Bottom left figure shows the symmetry map. In the bottom right figure, the asymmetry map constructed from subtracting the symmetry map from the original map is plotted.}}
\label{tpyx_4plots}
\end{figure}

There are additional methods to find the $K$ velocities of the components in a CV, for instance, $K_{1}$ can be found by locating the centre of symmetry of the disc emission in a Doppler tomogram. The centre should be located at  $V_{x} = 0$ and $V_{y}= - K_{1}$  (e.g. \citealt{2002ApJ...568..273S}). First, a Doppler tomogram is constructed from a trailed spectrogram. After masking out any regions of strong asymmetric non-disc emission, the centre of symmetry is found by taking a trial point, smoothing the map azimuthally about this point, and then subtracting this smoothed map from the real map. If the trial point is far from the centre of symmetry, the resulting difference map will contain large asymmetric residuals. The optimal centre of symmetry is thus the trial point for which the difference map exhibits the smallest residuals (as measured by the standard deviation of the difference map). 

Figure~\ref{tpyx_4plots} illustrates how $K_{1}$ can be obtained using the method described above for the He II line at 4686\,\AA\,\,in T Pyxidis. The upper left frame shows the trailed spectrogram used to construct the Doppler tomogram presented in the upper right frame. The corresponding symmetry map centred on $V_{x}, V_{y}$, is presented in the lower left figure. The lower right figure show the asymmetry map, constructed by subtracting the symmetry map from the original map. 

In the Doppler tomogram, the donor star is positioned at $V_{x} = 0$ and $V_{y} = + K_{2}$. However, in order to locate the position of $K_{2}$ in maps which are not constructed from donor lines, one needs to know $K_{1}$ and $q$ ($K_{2} = K_{1}/q$).

\subsection{Velocity at the Outer Disc Radius} \label{vodr}

The projected velocity at the outer disc radius ($v_{\text{R}_{\text{disc}}}\!\sin i$) can be found by measuring half of the peak-to-peak ($\Delta\!V_{\text{peak-to-peak}}$) separation in the double-peaked emission lines, and/or by measuring the radius of the inner part of the disc structure in the Doppler tomogram (see Figure~\ref{dopp_ill}).

~\cite{1981AcA....31..395S} investigated the exact relationship between the true $v_{\text{R}_{\text{disc}}}\!\sin i$ and the measured $\Delta\!V_{\text{peak-to-peak}}$, taking the effects of instrumental broadening and different disc emissivity distributions into account. He found that, for a wide range of realistic situations, $v_{\text{R}_{\text{disc}}}\!\sin i  = 0.5\,\Delta\!V_{\text{peak-to-peak}} / u$, where $u=1.05 \,\pm\, 0.05$.

\subsection{Mass-Ratio} \label{mass_ratio}

~\cite{1973MNRAS.162..189W} showed that the radial velocity of the WD ($K_{1}$) divided by a characteristic disc radius ($v_{\text{R}_{\text{C}}} \sin i$), is a function of only the mass-ratio ($q$). Such a relation implies that $q$ can be estimated from time-series spectroscopy, where $K_{1}$ and $v_{\text{R}_{\text{C}}} \sin i$ can be measured as described in Sections~\ref{svvsa} and~\ref{vodr}. In order to derive a relation for $q$,~\cite{1973MNRAS.162..189W} assumed that a specific particle leaving the Lagrangian point $L_{1}$, will conserve its angular momentum in the orbit around the primary WD. Thus, the circularisation radius is used as a characteristic disc radius. From Kepler's third law for circular orbits, a particle in an orbit around the WD has a velocity $v$

\begin{equation}
v= \frac{2 \pi a_{1}}{Pf^{2}(q)} \frac{(1+q)^{2}}{q^{3}},
 \label{warner_1973a}
 \vspace{0.3cm}
 \end{equation}

\noindent where $a_{1}$ is the distance between the centre of the primary and the centre of mass for the binary, and $f(q)$ is the distance between the centre of the primary and the $L_{1}$, measured in units of $a1$. Combining Equations~\ref{warner_1973a} with the expression $K_{1} = 2 \pi a_{1}\sin i /P$,~\cite{1973MNRAS.162..189W} found the following relation

\begin{equation}
\frac{K_{1}}{v_{\text{R}_{\text{C}}} \sin i} = \frac{f^{2}(q)q^{3}}{(1+q)^{2}}.
 \label{warner_1973}
 \vspace{0.3cm}
 \end{equation}

\noindent In Chapter~\ref{tpyx}, I estimate $q$ for T Pyxidis in a similar manner. However, instead of using a characteristic disc radius ($v_{\text{R}_{\text{C}}} \sin i$), I use the velocity at the outer disc radius ($v_{\text{R}_{\text{disc}}}\!\sin i$), which can easily be measured as described in Section~\ref{vodr}, assuming a circular accretion disc. Such an assumption is plausible for any late-type binary system since the tidal circularisation time scale must be very short (\citealp{1989A&A...223..112Z}). Also, in the case of T Pyxidis, I assume that the disc radius extends all way out to the tidal radius.~\cite{1996MNRAS.279..219H} performed empirical measurements of the radii of accretion discs for eclipsing systems, and found that systems with high $\dot{M}$ have larger discs than systems with lower $\dot{M}$. Since T Pyxidis is a high $\dot{M}$ system, a tidal limitation can be expected to be a reasonable approximation, thus, its disc radius is given by the following equation (\citealt{1995CAS....28.....W}),

\begin{equation}
\frac{R_{\text{disc}}}{a} = \frac{0.60}{(1+q)}.
 \label{warner_1995a}
  \vspace{0.3cm}
 \end{equation}
 
\noindent These assumptions, together with Kepler's third law, yield the expected relationship (please see Appendix A for a full derivation of this expression), 


\begin{equation}
\frac{K_{1}}{v_{\text{R}_{\text{disc}}}\!\sin i}= 0.77 \frac{q}{1+q}.
 \label{mod_warner_1973}
  \vspace{0.3cm}
\end{equation} 

\begin{figure}
\centering
\includegraphics[scale=0.38]{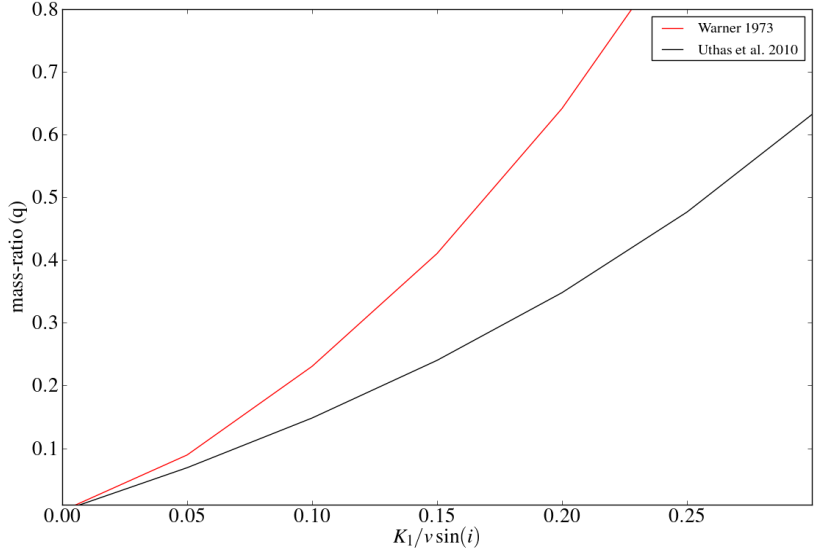}
\caption{\emph{The relation between $q$ and the quote $K_{1}/v_{\text{R}_{\text{disc}}}\!\sin i$ is presented as derived in Equation~\ref{warner_1973} ({\color{Red} red} line) and in Equation~\ref{mod_warner_1973} (black line).}}
\label{war_hel}
\end{figure}

\noindent In Figure~\ref{war_hel}, I compare the ratio $K_{1}/v_{\text{R}_{\text{disc}}}\!\sin i$ as a function of $q$, as derived from the expression in Equation~\ref{warner_1973} from~\cite{1973MNRAS.162..189W} (although this is not specifically the outer disc radius), with the expression I present in Equation~\ref{mod_warner_1973} (\citealt{2010MNRAS.409..237U}). The function $f$ in Equation~\ref{warner_1973} is highly dependent of $q$, and was here derived from tabular values of Roche-lobe models presented by~\cite{1964BAICz..15..165P}. High values of $q$ result in a smaller value of the quote $K_{1}/v_{\text{R}_{\text{disc}}}\!\sin i$ for Equation~\ref{warner_1973} compared to Equation~\ref{mod_warner_1973}. For example, the system T Pyxidis, has a measured value of $K_{1}/v_{\text{R}_{\text{disc}}}\!\sin i = 0.130$, resulting in $q = 0.20$ if obtained from Equation~\ref{mod_warner_1973}, while the same quote would result in a higher $q$ ($\approx 0.32$), if calculated from Equation~\ref{warner_1973}. Since~\cite{1973MNRAS.162..189W} used the fact that the disc radius cannot be smaller than the circularisation radius, this means that Equation~\ref{warner_1973} effectively yields an upper limit on $q$.
 
\subsection{Inclination}

The inclination ($i$) of a CV can be accurately determined for eclipsing systems, where the duration of the eclipse is a function of $i$, $q$ and $P$. In order for an eclipse to occur, $i$ has to be $\gtappeq$ 75\,$^{\circ}$. For non-eclipsing systems, $i \leappeq 75\,^{\circ}$, and can be estimated if $P$, $K_{1}$ and $q$ are known, from $K_{1} = v_{1}\sin i$, where $v_{1}$ is defined as

\begin{equation}
v_{1} = \frac{2 \pi a}{P} \frac{q}{(q+1)},
 \label{v1}
   \vspace{0.3cm}
\end{equation}   

\noindent where the binary separation $a$ can be obtained from Kepler's third law; see Equation~\ref{eq3} in Appendix A. The system inclination can also be estimated using the mass functions, assuming that $P$, $K_{1}$ and either the WD or donor mass is known (Equations~\ref{eq5} and~\ref{eq6} in Appendix A).

\subsection{Temperature and Metallicity of the White Dwarf} \label{temp_met}

As noted above, in most CVs, the primary WD is difficult to observe since it is often obscured by light from the accretion disc and bright spot. Nevertheless, in some systems with orbital periods close to the minimum orbital period, the accretion discs are faint and the WD contributes significantly to the overall light from the system, as evidenced by the presence of its absorption lines in the spectrum. Magnetic CVs also provide a good opportunity to study the primary WD since they do not have pronounced accretion discs. Also, in some CVs, the accretion temporarily switches off from time to time, providing a good opportunity for studying the WD without interference from the accretion disc. Such low states
have been observed, for instance, in VY Scl stars -- a subtype of nova-like variables (e.g.~\citealt{2009ApJ...693.1007T}). Since WDs are hot, they are bright in the UV. Therefore, UV  observations are commonly used to study WDs in CVs. By comparing the UV data spectrum with synthetic model-spectra, a number of WD parameters can be established, such as the effective temperature (T$_{\text{eff}}$), surface gravity ($\log g$), metallicity ($[Fe/H]$) and rotational velocity ($v \sin i$).   

\begin{figure}[t]
\centering
\includegraphics[scale=0.37]{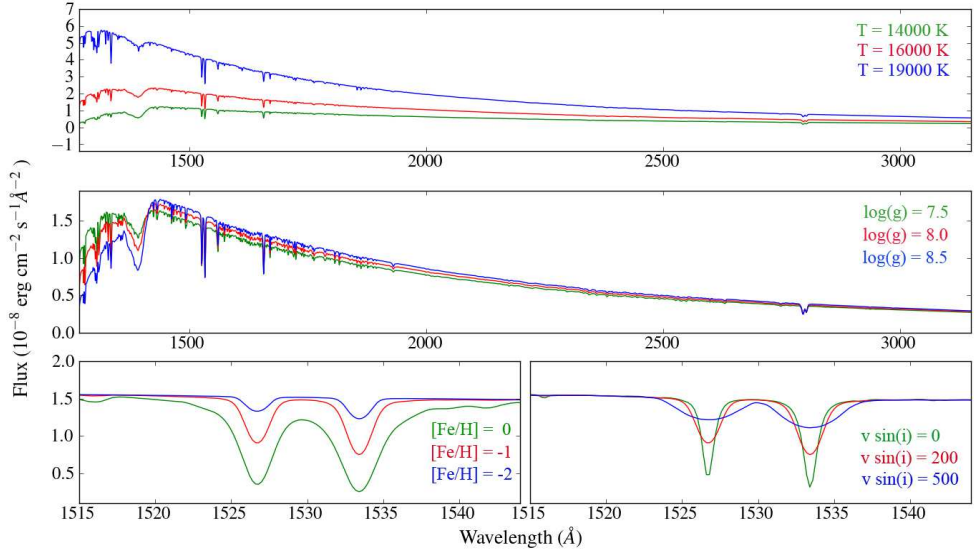}
\caption{\emph{In the top figure, synthetic model-spectra for $T_{\text{eff}}$, 14000 K ({\color{Green} green}), 16000 K ({\color{Red} red}) and 19000 K ({\color{blue} blue}), is shown (other parameters are fixed at $log(g)$ = 8.0, $v\sin i$ = 200 km$\,s^{-1}$ and $[Fe/H]$ = -1). The middle panel show three different values of $log(g)$, 7.5 ({\color{Green} green}), 8.0 ({\color{Red} red}) and 8.5 ({\color{blue} blue}). Here, $T_{\text{eff}}$ = 15000, $v\sin i$ = 200 km$\,s^{-1}$ and $[Fe/H]$ = -1. The bottom left panel show $[Fe/H]$ at, 0 ({\color{Green} green}), -1 ({\color{Red} red}) and -2 ({\color{blue} blue}) ($T_{\text{eff}}$ = 15000, $log(g)$ = 8.0 and $v\sin i$ = 200 km$\,s^{-1}$). The bottom right panel present $v\sin i$ at, 0 km$\,s^{-1}$ ({\color{Green} green}), 200 km$\,s^{-1}$ ({\color{Red} red}) and 500 km$\,s^{-1}$ ({\color{blue} blue}) ($T_{\text{eff}}$ = 15000, $log(g)$ = 8.0 and $[Fe/H]$ = -1). Models are produced by \textsc{Synspec}.}}
\label{model_spec}
\end{figure}

The $T_{\text{eff}}$ and $\log g$ affect the gradient and overall shape of the spectrum and are highly correlated with each other. The two top panels in Figure~\ref{model_spec} show synthetic model-spectra produced by \textsc{Synspec} at three different $T_{\text{eff}}$ (14000 K [\begin{color}{Green}green\end{color}], 16000 K [\begin{color}{red}red\end{color}] and 19000 K [\begin{color}{blue}blue\end{color}]) and at three different values of $\log g$ (7.5 [\begin{color}{Green}green\end{color}], 8.0 [\begin{color}{red}red\end{color}] and 8.5 [\begin{color}{blue}blue\end{color}]). All other parameters were kept at fixed values. As seen from the figures, higher effective temperatures and lower surface gravity produce bluer spectra. The two top panels in Figure~\ref{model_spec} also demonstrate that, at these sorts of temperatures, it is much easier to determine $T_{\text{eff}}$ and $\log g$ in the blue end of the spectrum compared to the red end (i.e. the differences between two $T_{\text{eff}}$ values, or two $\log g$ values, are much larger in the blue end of the spectrum). Also, there is a higher possibility of contamination from the accretion disc and bright spot at the red end of the spectrum. As a result, the far UV-region may be expected to provide more reliable estimates of $T_{\text{eff}}$ and $\log g$.

The parameter $[Fe/H]$ determines the depth of the spectral lines while $v\sin i$ affects both the depth and shape of the lines. As a consequence, these two parameters are also correlated. The bottom left panel in Figure~\ref{model_spec}, shows models at three different values of $[Fe/H]$ ([\begin{color}{Green}green\end{color}], -1 [\begin{color}{red}red\end{color}] and -2 [\begin{color}{blue}blue\end{color}]). The bottom right panel shows models at three different values of $v \sin i$ (0 km$\,s^{-1}$ [\begin{color}{Green}green\end{color}], 200 km$\,s^{-1}$ [\begin{color}{red}red\end{color}] and 500 km$\,s^{-1}$ [\begin{color}{blue}blue\end{color}]). As seen in the two figures, the lines become shallow for lower values of $[Fe/H]$, and wide and shallow for higher values of $v \sin i$. When making a model fit to a data spectrum, even the best fit might not perfectly agree with the data both in overall gradient, width and depth of the lines. Therefore, it might be better to focus on fitting to narrow line regions to be able to establish reasonable values for $[Fe/H]$ and $v \sin i$. Also, to make such a fit reliable, both $T_{\text{eff}}$ and $\log g$ should be kept kept at fixed values, determined from a fit to the overall spectrum (as described above).      

The models presented in Figure~\ref{model_spec} were used to fit the UV spectrum of SDSS J1507+52, as described in Chapter~\ref{j1507}. Also, the correlation between the parameters is explored in Chapter~\ref{j1507} by constructing contour plots (see Figure~\ref{fig:contours}).

\subsection{Distances} \label{dist}

\subsubsection{From Trigonometric Parallax}

There are many ways of estimating the distance to a source, the best known being trigonometric parallax measurements. In this method, a star's annual movement over the sky as the earth moves around the sun, is compared to more distant stars in the background that appear not to move. The distance $(d)$ to the star in parsecs is given by $d = 1/p$, where the parallax $(p)$ is the angle by which the star has moved. However, this method is only possible to use for stars that are not too far away, and the limitation if using a telescope from earth is an angular resolution of $\sim$ 0.01 arc seconds, corresponding to a distance of $\sim$ 100 pc. In order to measure a smaller parallax, space telescopes are needed. For instance, the Hipparcos satellite (an ESA mission in operation during the years 1989 --1993) could measure a parallax as small as 0.0014 arc seconds, and find distances for stars out to $\sim$ 700 pc (an overview of the Hipparcos Mission is given by~\citealt{1997A&A...323L..49P}).

Due to the fact that CVs are faint objects, historically it has been a rather difficult task to find distances towards CVs from trigonometric parallax measurements. Nevertheless, such studies are nowadays quite common. For instance,~\cite{2003AJ....126.3017T} developed a method that uses measured parallaxes together with proper motions and absolute magnitudes to find distances towards CVs (see also \citealt{2004AJ....127..460H, 2008AJ....136.2107T}).

\subsubsection{From Donors}

The distance towards CVs can also be estimated by assuming that the donor star has the same surface brightness that an isolated main-sequence star with the same spectral type would have (for which the distance is known, for instance, obtained from parallax measurements). The flux in a CV is often dominated by the accretion disc, and therefore the surface brightness of the donor star is often measured from obtained observations in the IR wavelength region, where the donor is at its brightest.

~\cite{1981MNRAS.197...31B} developed a method to find distances to CVs from IR observations, by measuring the mass, spectral type or temperature of the donor, and comparing this with main-sequence stars to find absolute magnitudes and hence also distances. This method has been further developed and presented by~\cite{2005PASP..117.1204P, 2006MNRAS.373..484K, knigge2011}. For instance, observations in the K-band yield

\begin{equation}
\log d \gtappeq \frac{K - M_{K,2}(P) + 5}{5},
 \label{dist}
   \vspace{0.3cm}
\end{equation}   

\noindent where $K$ is the apparent magnitude and M$_{K,2}$ is the absolute magnitude, which can be estimated for normal, unevolved CVs just from the orbital period (see Figure 17 in ~\citealt{knigge2011}).

\subsubsection{From Dwarf-Nova Outbursts}

During every dwarf nova eruption, the accretion disc is expected to become optically thick with a surface temperature of about $T$ = 20000 K. Therefore, these eruptions can act as standard candles and be used for distance measurements.~\cite{2011MNRAS.tmp...27P} presented an empirical study of the relation between the absolute magnitudes and orbital periods during dwarf-nova outbursts for 46 short-period CVs. This study was an improvement on the previous relation presented by~\cite{1987MNRAS.227...23W}.~\cite{2011MNRAS.tmp...27P} finds that for erupting dwarf novae, $(M_{V})_{max}$ = 5.70 -- 0.287 P(hr), resulting in a final error on the corresponding distance of about 10\% -- 30\%. In particular, for superoutburst in SU UMa stars,~\cite{2011MNRAS.tmp...27P} finds that $(M_{V})_{max}$ = 4.95 -- 0.199P(hr). SU UMa stars enter long-lasting high-states between the superoutbursts, and therefore these stars are particularly good for distance estimates, where the distance in units of parsec is $\log d = (V_{\text{plat}} - A_{v} - 0.5)/5$, where $A_{v} = 3.1 E(B - V) < 0.2 $ mag. Using this method, the errors on the distances estimates are $\approx 15\%$.

\subsubsection{From WDs}

If the light from an accreting WD can be isolated (from that of the accretion disc and bright spot), we can fit stellar-atmosphere models to the WD spectrum (as described in Section~\ref{temp_met}) to estimate distances by comparing data and model fluxes. The relation between observed fluxes ($F$) and model fluxes (here Eddington fluxes [$H$]) are found from 

\begin{equation}
F =  \frac{4 \pi R_{1}^{2}\,H}{d^{2}},
 \label{f_d}
   \vspace{0.3cm}
\end{equation}   

\noindent where $R_{1}$ is the radius of the WD. Therefore, the distance is a function only of $R_{1}$ and the normalisation factor $(N)$ between Eddington fluxes and observed fluxes 

\begin{equation}
d = \sqrt{\frac{4\,\pi R_{1}^{2}}{N}}.
 \label{d_n}
   \vspace{0.3cm}
\end{equation}   
   
\noindent The radius of the WD is not known from the stellar atmosphere models, but the fits do provide an estimate of the surface gravity, $\log g$. Since WDs follow a mass-radius relation that depends only weakly on temperature, a fixed value of\,\,$\log g$\,\,also fixes $M_{1}$ and $R_{1}$, via

\begin{equation}
g = \frac{G\,M_{1}}{R_{1}^{2}},
 \label{g_mass_radius}
   \vspace{0.3cm}
\end{equation}  

\noindent where $M_{1}$ is the mass of the WD. In Chapter~\ref{j1507}, I use this method to estimate the distance towards SDSS J1507, as well as the mass and radius of the accreting WD in this system. However, these estimates of $d$, $R_{1}$ and $M_{1}$, rely on the fact that the atmospheric parameters found from the model fitting are determined accurately, for instance, that the spectrum is purely WD with no contamination from disc or bright spot. For eclipsing systems, the mass and radius of the WD may be available directly from eclipse modelling, yielding a more accurate estimate of the distance.
\newpage

\subsection{Finding Errors on Parameters} \label{errors}

\subsubsection{Errors on Individual Measurements and their Mean Value}

The data points for any light curve, spectrum or radial velocity curve have formal errors calculated by the programs used to produce them (I have used \textsc{Iraf}\footnote{\textsc{Iraf} is a astronomical data analysis software for photometric and spectroscopic data, developed by the National Optical Astronomy Observatories (NOAO) (http://iraf.noao.edu/).} for producing light curves and \textsc{Molly}\footnote{\textsc{Molly} is a program by Tom Marsh for analysis of time-series spectroscopy \newline (http://deneb.astro.warwick.ac.uk/phsaap/software/molly/html/INDEX.html).} for calculating radial velocity curves). Formal statistical errors on each data point can be estimated by calculating the statistical properties of multiple measurements, where the standard deviation can be expressed as

\begin{equation}
\sigma = \sqrt{\sum^{N}_{i = 1} \frac{(x_{i} - \bar{x})^{2}}{N}},
 \label{sta_dev}
   \vspace{0.3cm}
\end{equation}  

\noindent where $\bar{x}$ is the mean value and $N$ is the number of points. Generally, formal parameter errors can be calculated on the assumption that measurement uncertainties follow the usual normal distribution (Gaussian distribution). Note that at least one of the methods used to find parameter errors in this thesis (bootstrap simulations), do not make this assumption since it does not take the errors on individual data points into account (see Section~\ref{bot_mont}). The normal distribution shows the probability of finding a value within a certain $\sigma$. The 1-$\sigma$ error range from the mean value corresponds to about 68\% of the area of the normal distribution. About 99.9\% of all points are expected to be found within the 3-$\sigma$ error range. The normal distribution is defined by the probability density function $f(x)$
 
\begin{equation}
f(x) = \frac{1}{\sqrt{2 \pi \sigma^{2}}} e^{\frac{-(x-\bar{x})^{2}}{2 \sigma^{2}}}.
 \label{prob}
   \vspace{0.3cm}
\end{equation}  

\noindent The error on the mean value derived from multiple number of measurements is $\sigma/\sqrt{N}$.

\subsubsection{$\chi^{2}$-fitting to data}

When comparing a model to a dataset, the \emph{goodness-of-fit} is used to describe how well the model matches the data, and is usually evaluated on the basis of the $\chi^{2}$ statistic. The $\chi^{2}$ is defined as

\begin{equation}
\chi^{2} = \sum^{N}_{i = 1}\left({\frac{x_{i} - \bar{x}}{\sigma_{i}}}\right)^{2}.
 \label{chi2}
   \vspace{0.3cm}
\end{equation}  

\noindent The 1-$\sigma$ error on any parameter can be estimated by identifying the parameter ranges that correspond to $\chi^{2} = \chi^{2}_{min} \pm$ 1. The mean of a $\chi^{2}$ distribution is its number of degrees of freedom ($\nu$). The reduced $\chi^{2}$ ($\chi^{2}_{\nu}$) is defined as the $\chi^{2}$ divided by $\nu$, and can be expressed as

\begin{equation}
\chi^{2}_{\nu} = \frac{\chi^{2}}{\nu} = \frac{1}{\nu} \sum^{n}_{i = 1}\left({\frac{x_{i} - \bar{x}}{\sigma_{i}}}\right)^{2}.
 \label{chi2_red}
   \vspace{0.3cm}
\end{equation} 

\noindent If the value of $\chi^{2}_{\nu} <$ 1, the model is likely \emph{over fitting} the data, and the initial errors are overestimated. A value of $\chi^{2}_{\nu} >$ 1 indicates a poor fit to the data, and it is likely that the data errors are underestimated, or that the model is inadequate to describe the data. Throughout this thesis, in those cases where $\chi^{2}_{\nu}$ gave a poor fit, I modified the data errors so that $\chi^{2}_{\nu}$ = 1 (thus making the errors more realistic). Such an adjustment of errors is conservative, since it yields larger errors on the parameters. This was done in two different ways, either by adding an intrinsic dispersion term to the errors, so that the total error corresponds to $\sigma_{total} = \sqrt{\sigma^{2} + \sigma_{int}^{2}}$, or by scaling the errors with a multiplicative factor. Flux/magnitude errors on light curves were rescaled by comparison with a multiple sine-wave model, while errors on radial velocities were modified by comparison against a single sinusoid fitted to the radial velocity curves.

\subsubsection{Correlation between Parameters}

There is at least one additional consideration that has to be taken into account when finding errors on parameters. If two parameters are likely to be correlated, the covariance should be included in the total error. The correlation coefficient $\rho$ between the two parameters $A$ and $B$ can be expressed as

\begin{equation}
\rho_{A,B}  = \frac{cov(A,B)}{\sigma_{A} \sigma_{B}} = \frac{E[(A - \bar{x}_{A})(Y - \bar{x}_{B})]}{\sigma_{A} \sigma_{B}},
\label{covar}
\end{equation} 

\noindent where $E$ is the mean value operator. The value of $\rho_{A,B}$ varies between $\pm$ 1, where a value of 0 indicates that $A$ and $B$ are uncorrelated (\citealt{1989sgtu.book.....B}). Instead of calculating the covariance matrix, 2-dimensional contour plots of the two parameters in $\chi^{2}$ space can visually reveal any correlations between parameters (e.g. Figure~\ref{fig:contours}).

\subsubsection{Finding Errors from Data Simulations}  \label{bot_mont}

In this thesis, the final errors on almost all system parameters presented in Chapters \ref{j1457_bwscl}, \ref{tpyx} and~\ref{j1507}, are found from Monte Carlo or bootstrap simulations. When performing Monte Carlo simulations (first proposed by Enrico Fermi in the 1930s), the measured values are randomly redistributed within their 1-$\sigma$ error range. This process is iterated many thousands of times, and for each time, the analysis is repeated yielding a new result. The distribution of all results from all simulations is then used to find the mean value and its corresponding 1-$\sigma$ error, either from fitting a Gaussian to the output distribution or directly from statistical calculations (i.e. Equation~\ref{sta_dev}). When performing bootstrap simulations, mock datasets are created from points that are drawn at random \emph{with replacement} from the original dataset. In the case where the data are time series, the mock data sets do not share the exact same sampling pattern as the original data. In bootstrapping, typical about 2/3 of the original data will appear in each mock data set, and hence, some values will appear twice. For the error analysis presented in Chapter~\ref{j1457_bwscl}, I have performed both Monte Carlo and bootstrap simulations, to ensure they produce similar results. Since the bootstrap method is independent of the errors on the data points, this provides a cross-check that the errors are sensible. For example, if the errors are under estimated, the Monte Carlo method would yield a lower resultant error than the bootstrap method.

\subsubsection{Systematic Errors}

The error ranges for any of the system parameters presented in this thesis were carefully defined by using the methods described in this section. Sometimes, many different analysis methods were used to find one parameter. For instance, $K_{1}$ can be found both from the sinusoidal radial velocity curve and from the true centre of symmetry in the Doppler tomogram. Another example occurs when retrieving results of the same parameter from different time spans or wavelength regions. In case these different methods and/or choice of data do not yield the same result within their statistical formal errors, the range in which the different results are found could be a better measure of how well we are able to determine the parameter. This will give a lower bound of the error, resulting in a \emph{systematic error} rather than a formal, purely statistical one.

\section{Obtaining Data}

Generally, optical photometric observations of CVs can be done from rather small telescopes ($\leappeq$ 1 m), while spectroscopy normally requires slightly larger telescopes in order to achieve an acceptable signal-to-noise ratio ($S/N$). When observing CVs, the main contributing factor to an increased data quality is time. However, there is a trend at the world-leading observatories to close down smaller telescopes in favour of larger ones. A larger photon counting area is often an unnecessary expense of both time and money. However, larger telescopes are sometimes needed, for instance when observing very faint systems or when a high time resolution or spectral resolution is required. 

When planning observations, one has to make sure that the data being obtained match the science goals. For instance, if the goal is to measure the orbital period from phase-resolved spectroscopy, the time resolution has to be small enough compared to the orbital period not to smear the orbital signal. As a consequence, the chosen optical setup, such as the grism and slit, must allow for enough throughput to obtain a reasonable level of $S/N$ at this required time resolution. Normally, calibration frames such as bias frames, dark frames and flat fields, are obtained close in time to the observations. For flux calibration, a flux standard star observed close in time and at the same airmass at the data observations is required, but can only be obtained if the night has photometric conditions. Below follows a brief description of the calibration frames required for standard data reduction. 

\subsection{Dark and Bias Frames} 
 
On a CCD, after each exposure the analog signal containing the number of electrons in each pixel is converted into a digital signal, often expressed in ADU (Analog Digital Units). The conversion from photons to electrons is not perfect, and various sorts of noise are present on the resulting frame. Each pixel charge carries a fixed offset voltage, a \textbf{bias}. There is also noise caused by thermally generated electrons, the \textbf{dark current}. Thus, even if obtaining an exposure of 0 seconds, there is still a signal varying from pixel to pixel. To compensate for these effects, an exposure of 0 seconds (a bias frame) is taken to determine the bias pattern. A dark frame is obtained with the same exposure time as the science frames, but with the shutter closed. 

Proper cooling of the CCD is necessary to reduce the amount of dark current. At larger telescopes the CCD detectors are normally cooled with liquid nitrogen. As a result, the dark current only becomes significant for very long exposure times. If the CCD is not cooled at all, the dark current dominates the noise (together with the sky background), while the noise contribution from the bias might be negligible.

\subsection{Flat Fields} 

Each pixel on the CCD has a slightly different sensitivity to the incoming photons. \textbf{Flat fields} are images obtained to correct for these pixel-to-pixel variations. A twilight flat uses the twilight sky to uniformly illuminate the CCD, while a dome flat uses that from a bright (usually quartz/Halogen) lamp. The pixel sensitivity depends on wavelength, and therefore a set of flat fields must be obtained for each filter that is being used. The exposure time varies with optical setup, but the pixels should be filled to about half of their full well capacity, or about 60\% of the saturation level, which normally ensures a good $S/N$ in each flat without causing saturation in bright areas of the field. Small-scale variations in the frame, such as dust appearing in the optics, can sometimes also be removed by flat fielding, as well as larger-scale variations, such as vignetting. A vignetting effect is produced when the centre of the CCD is brighter than the corners, due to uneven distribution of light. Also the use of filters which have a higher transmission in the centre causes dark corners on the resulting frame. Other effects, for instance fringing, can also be diminished by a flat-field correction. Fringing is seen as an interference pattern in the output image and is thought to be caused by light being reflected as it enters the CCD. Photons of higher energy are absorbed by the silicon in the CCD and have short absorption depth. However, photons with lower energy have a larger absorption depth, and when this depth becomes equal or larger than the thickness of the CCD, the photons go through the CCD and reflects back in the (backside illuminated) CCD, creating interference pattern both in photometry and spectroscopy. Fringing becomes a problem for longer wavelengths and thin CCDs.


\chapter{SDSS J1457+51 and BW Sculptoris  \\\Large \textsc{- Two new accreting and pulsating WDs - }} \label{j1457_bwscl}
\label{chap:icm}

\emph{I would like to start this chapter by acknowledge the contribution from my collaborators. In particular, I want to thank Joseph Patterson for providing me with data for BW Sculptoris, and for being responsible for the analysis and interpretation of this data. I have formalised and integrated his preliminary work with my own analysis of the target SDSS J1457+51. Due to the similarities between these two systems, comparison with the BW Sculptoris data were invaluable and provided many insights during my own analysis of SDSS J1457+51. For completeness, I here include the analysis of BW Sculptoris, even though this is not primarily my own work.
\\\\
The main contents of this chapter will appear in MNRAS as~\cite{uthas_j1457_bwscl}, with co-authors; Joseph Patterson, Jonathan Kemp, Christian Knigge, Berto Monard, Robert Rea, Greg Bolt, Jennie McCormick, Grant Christie, Alon Retter and Alex Liu.}
\newline
\newline

\begin{Huge}\color{Red}{N}\end{Huge}on-radial pulsations (NRPs) are commonly found in isolated white dwarfs of DA type, often called ZZ Ceti stars. These stars have hydrogen-rich atmospheres. Pulsations occur as the WD cools and pass through a phase of pulsational instability, and are detected mainly as g-modes (\citealt{2006AJ....132..831G}). Similar signals have also been detected in faint CVs, which are interpreted as NRPs from the primary WD. All CVs displaying this type of signal also show absorption from the underlying WD in their spectrum (see Section~\ref{nrps}). Here, these systems are called \emph{GW Lib stars}, after the first discovery (\citealt{1998IAUS..185..321W}).
  
In this chapter, I present time-series photometry of two CVs that appear to be certifiable members of the GW Lib class. Both systems have very low accretion luminosity and show signatures from the white dwarf in their optical spectra. Also, both systems show double-humped orbital signals and non-commensurate periodic signals, indicating non-radial pulsations. One is \textbf{SDSS J1457+51}, which has a photometric wave suggesting an orbital period of $77.885 \pm 0.007$ minutes. The other is \textbf{BW Sculptoris}, with $P_{\text{orb}} = 78.22639 \pm 0.00003$ minutes. Both stars show pulsations mainly near 10 and 20 minutes. These rapid signals drift slightly in frequency, and may consist of several, finely spaced components. BW Sculptoris also shows a remarkable photometric variation at 87 minutes, which could be explained as a \textbf{quiescent superhump}, possibly arising from a 2:1 orbital resonance in the accretion disc.

\section{SDSS J1457+51}

SDSS J1457+51 (hereafter J1457) was first identified in the Sloan Digital Sky Survey by~\cite{2005AJ....129.2386S}. They obtained spectroscopy that showed the broad absorption characteristics of a white dwarf, indicating a system of low accretion rate. The source was found to be faint ($g \approx 19.5$). Due to the double-peaked nature of the emission lines, they suggested the system to be of high inclination.

The same year,~\cite{dillon_2008} obtained both spectroscopy and photometry of J1457, but no orbital period could be established. She found some evidence that the system might harbour non-radial pulsations and also identified a periodic signal at $\sim 85$ minutes, which she concluded must be close to the orbital period.

\subsection{Observations, Data Reduction and Analysis} \label{obs}

Time-resolved photometry of J1457 was obtained with the 1.3 m and 2.4 m MDM telescopes at the Kitt Peak observatory, Arizona, during April and May 2010. The star was observed during 14 nights in total, spread over 47 days. With all data coming from the same terrestrial longitude, we were not immune from aliasing problems and therefore strove to obtain the longest possible nightly time series (generally $\sim$ 4 - 7 hours). Weather conditions such as clouds, snow and full moon prevented us from obtaining times series over more than 4 consecutive nights at a time. The time resolution was in the range of 20 s -- 30 s. Table~\ref{tab:obs1} presents a log of the observations, where the column for quality indicate the quality of the night (in terms of weather and contribution from moon light) A clear filter with a blue cutoff was used to minimise differential-extinction effects and allow for a good throughput.

The data reduction was done in real time during the observations, using standard \textsc{Iraf} routines. The data consisted of differential photometry with respect to the field star USNO A2.0:1350-08528847. The search for periodic signals was initially done for single nights separately, and Lomb-Scargle periodograms (\citealt{1982ApJ...263..835S}) were constructed. Formal flux errors were rescaled so that the $\chi^{2}_{\nu} \approx$ 1. This was done by fitting a fake light curve to the original data, composed of multiple sine waves with periods corresponding to the strongest signals found in the single-night power spectrum. Monte Carlo simulations were performed on every peak of interest in the power spectrum to find the period and its error. In this method, the peak errors are found by randomly re-distributing the points in the light curve within their errors, a repeated number of times, and constructing a Lomb-Scargle periodogram each time. The 1-$\sigma$ error is then found by fitting a Gaussian to the output distribution of the peaks found in the periodograms. Data from several nights were then combined to allow the search for signals with lower amplitude, and also to improve the frequency resolution. Monte Carlo simulations as well as bootstrap analysis was performed to find errors on the peaks, and to distinguish between the most likely aliases. The two methods have different advantages and were used to provide a cross-check of the results. Monte Carlo simulations use the whole dataset and are only truly reliable when the errors are well estimated, while bootstrap analysis (formally bootstrap-with-replacement) does not consider the errors on the data points, and so can be more reliable in cases where the errors are not well estimated. Bootstrap analysis can also help to distinguish between the most likely aliases, since the sampling pattern of the data is changed by creating a mock dataset where a subset of the original dataset is randomly created and so with repeated use this will efficiently destroy or weaken the alias pattern. Results from the two methods agreed well with each other.

\begin{table}
\begin{center}
 \begin{tabular}{llllll}
  \hline
     \hline
\textbf{Date} & \textbf{HJD} & \textbf{Telescope} & \textbf{Length} & \textbf{Quality}\\
& (-2455000) &  & (hours) & \\
  \hline
100414 & 301 &  1.3m &  2.30 & medium\\
100416 & 303   & 1.3m & 7.89 &  excellent\\
100417 & 304   & 1.3m & 4.32 &  bad\\
100419 & 306   & 1.3m   & 4.72 & medium\\
100424 & 311    & 1.3m & 5.42 & medium\\
100425 & 312    & 1.3m  & 7.82 & bad\\
100503 & 320  &  1.3m  & 7.92 & good\\
100504 & 321   & 1.3m  & 8.26 & medium\\
100506 & 323   & 1.3m  & 8.40 &  good\\
100507 & 324   & 1.3m  & 3.50 & medium\\
100508 & 325   & 1.3m  & 5.47 & medium\\
100528 & 345   & 2.4m  & 4.97 & bad\\
100529 & 346   & 2.4m  & 8.11 & good\\
\hline
 \hline
\end{tabular}
\vspace{0.5cm}
 \caption{\emph{Summary observing log for SDSS J1457+51. Data was obtained at the MDM observatory during April and May 2010. The column for quality is a measure of how good the night was (in terms of weather and contribution from moon light).}}
\label{tab:obs1} 
\end{center}
\end{table}

\subsection{Light Curve and average Power Spectrum}

\begin{figure}
\centering
\subfigure[\emph{The mean Lomb-Scargle periodogram for J1457, composed of data from the six nights of best quality. The orbital period at 18.48 c\,d$^{-1}$ and its first harmonic are plotted as solid lines.}] 
{
    \label{fig:s_avpow}
    \includegraphics[width=11.75cm]{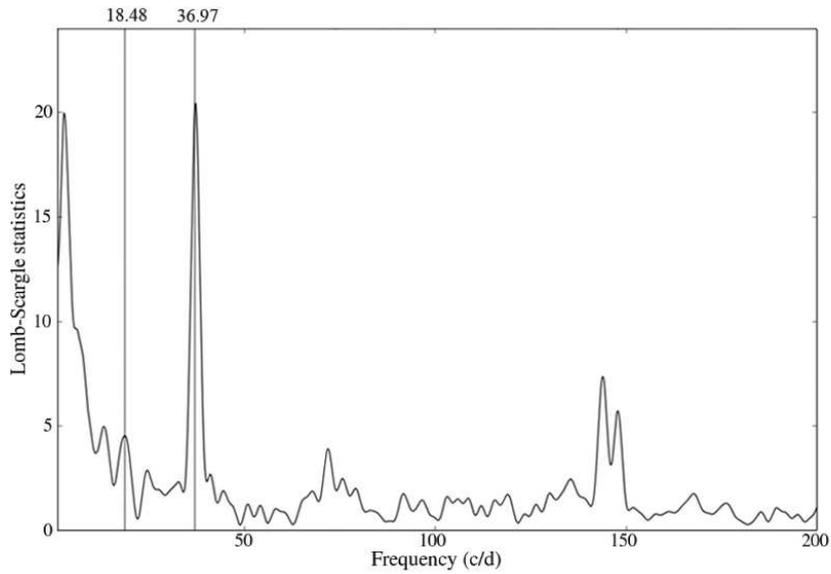}
}
\hspace{0.2cm}
\subfigure[\emph{The smoothed and normalised light curve of J1547 from one single night. A model light-curve constructed from the four strongest periods (including $\omega_{\text{o}}$ and $2\omega_{\text{o}}$) found in Figure~\ref{fig:s_avpow}, is plotted together with the data.}] 
{
    \label{fig:s_light}
    \includegraphics[width=11.12cm]{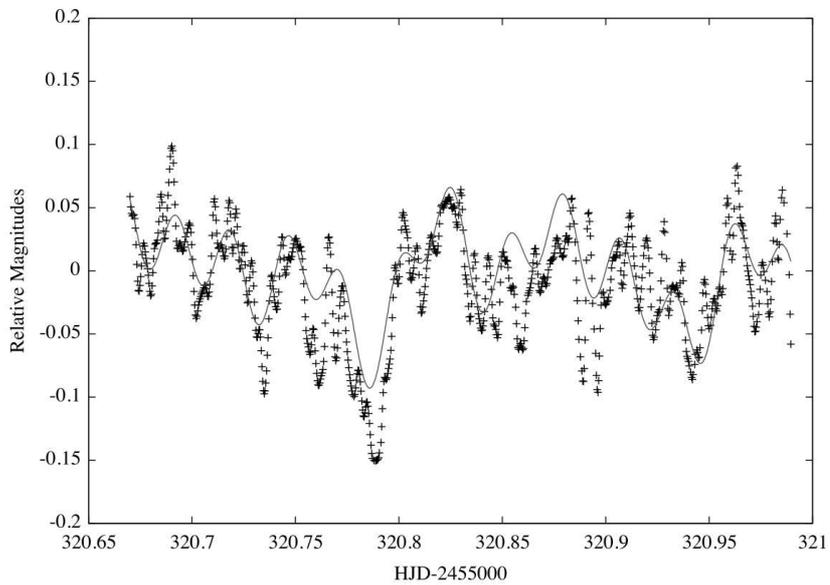}
}
\caption{\emph{Mean power spectrum and single-night light curve of J1457.}}
\label{fig:sub} 
\end{figure}

The mean power spectrum, averaged over the six best nights, is shown in Figure~\ref{fig:s_avpow}. The 18.48 c\,d$^{-1}$ and 36.97 c\,d$^{-1}$ signals are almost certainly the orbital frequency ($\omega_{\text{o}}$) and its first harmonic ($2\omega_{\text{o}}$). This kind of variations at $2\omega_{\text{o}}$ are commonly seen in the orbital light curves of CVs with low accretion rates (for instance in WZ Sge; see~\citealt{2002PASP..114..721P}). The signals found in the range 142 c\,d$^{-1}$ -- 148 c\,d$^{-1}$ are non-commensurate with the orbital frequency. These peaks vary slightly in period and amplitude when present at all in the nightly power spectrum. The complex structure around them indicates either an unresolved fine structure or periods varying from night to night; it is discussed in detail below. Also, during a few nights, peaks were found at 135 c\,d$^{-1}$ and 72 c\,d$^{-1}$, which are also non-commensurate with the orbital frequency. The lower region of the power spectrum shows strong peaks at 4 c\,d$^{-1}$ -- 6 c\,d$^{-1}$, corresponding to the typical length of a nightly observing run. The unit cycles per day (c\,d$^{-1}$), is used throughout this chapter since it is the natural unit for the sampling pattern of the multi-day light curves. Also, it clearly shows the natural daily alias pattern.  

The average power spectrum in Figure~\ref{fig:s_avpow} has low frequency resolution, since each night is less than 8 hours long. The power spectrum of a spliced light curve spanning several nights (the coherent power spectrum) is in principle better, since the resolution is always near 0.1\,T$^{-1}$ c\,d$^{-1}$, where T is the duration in days. It does, however, make the assumption that a candidate periodic signal is constant in period, phase, and amplitude over the duration of the observation. Power spectra will be difficult or impossible to interpret correctly when this assumption is grossly violated.

Figure~\ref{fig:s_light} shows the normalised and smoothed light curve for a sample night. A model light curve constructed from the four strongest peaks found in the power spectrum that night (including both $\omega_{\text{o}}$ and $2\omega_{\text{o}}$), is plotted on top of the smoothed light curve. Peaks found at frequencies higher than 40 c\,d$^{-1}$ are not represented in the model light curve. During one of the observing nights, we obtained multicolour data and found the brightness of of the star to be V = 19.2 $\pm$ 0.2. The mean brightness on each night was constant within the measurement error of 0.03 mag. There was essentially no evidence of flickering, a nearly universal and obvious feature of CV light-curves. This, together with the fact that absorption lines are seen in the optical spectrum implies that the total light in the range 4000\,\AA\, -- \,7000\,\AA\, is dominated by the white dwarf. 

    
\begin{figure}
\centering
\subfigure[\emph{The power spectrum of J1457 from 11 nights, showing the orbital signal $\omega_{\text{o}}$ and its first harmonic, $2\omega_{\text{o}}$. A zoom of the region around $\omega_{\text{o}}$ is plotted on top.}] 
{
    \label{fig:s_poworb}
    \includegraphics[width=8.15cm]{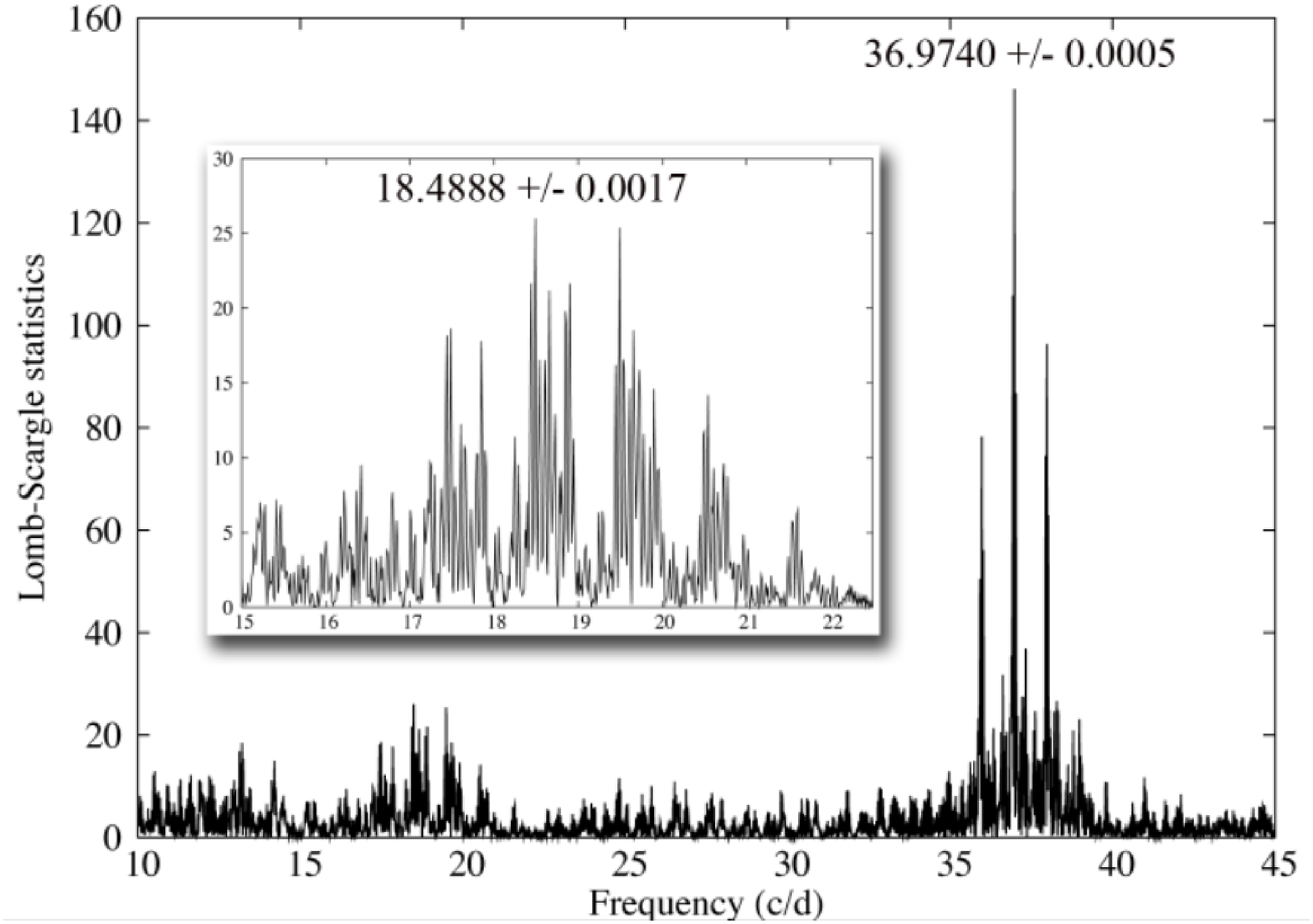}
}
\hspace{0.2cm}
\subfigure[\emph{A model power spectrum of J1457 constructed from two sine waves at $\omega_{\text{o}}$ and $2\omega_{\text{o}}$, using the same sampling pattern as in Figure~\ref{fig:s_poworb}.}] 
{
    \label{fig:s_poworbmod}
    \includegraphics[width=8.15cm]{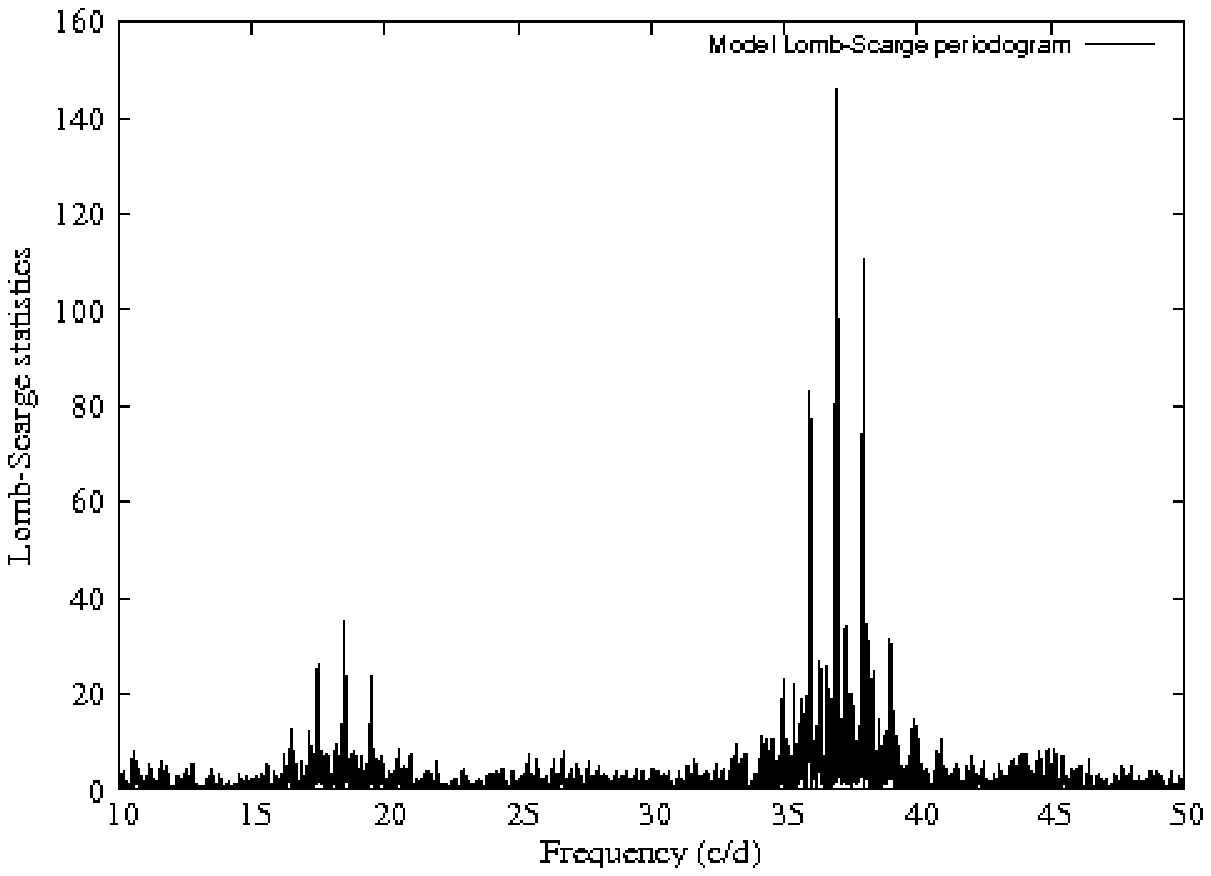}
}
\subfigure[\emph{Light curve of J1457 folded on the $\omega_{\text{o}}$ frequency, showing a double-humped orbital wave.}] 
{
    \label{fig:fold18}
    \includegraphics[width=8.15cm]{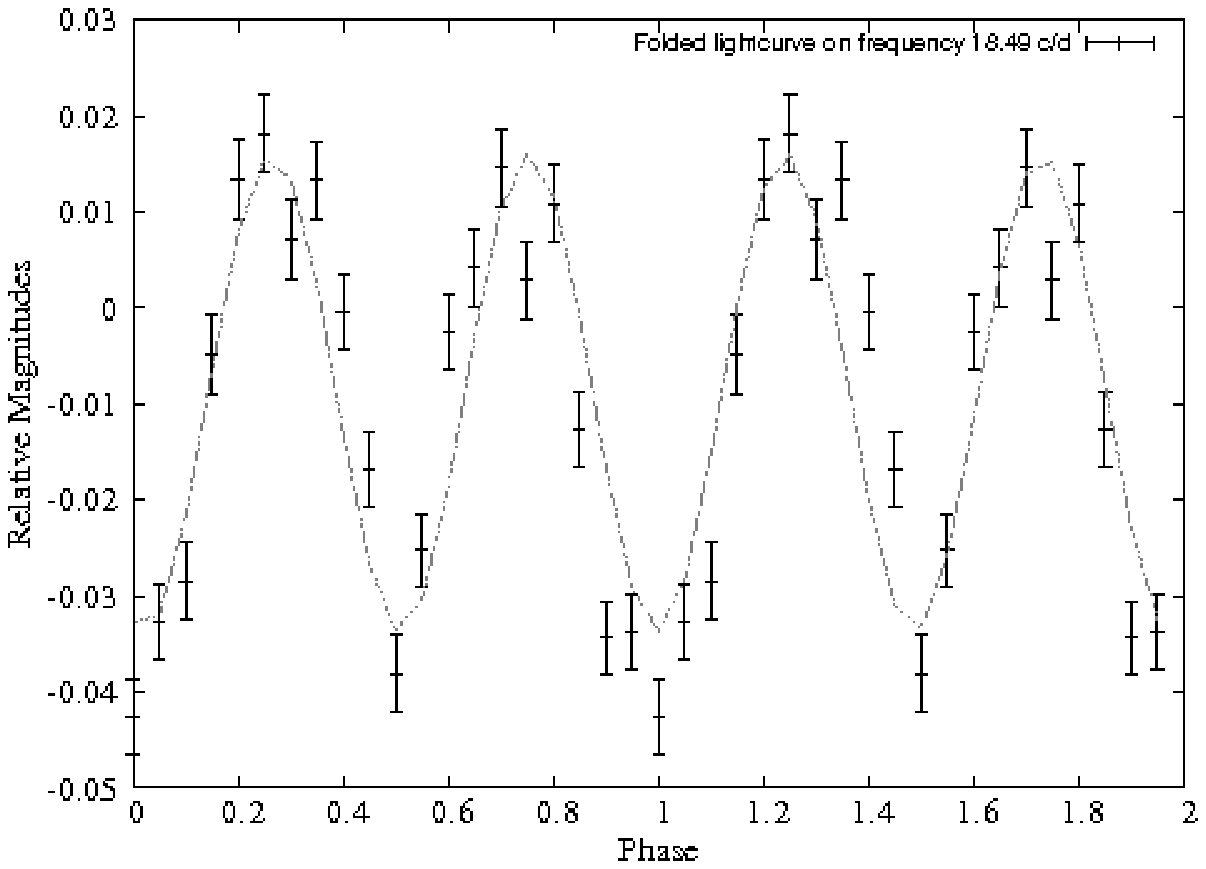}
}
\caption{\emph{The orbital signal in J1457.}}
\label{fig:sub} 
\end{figure}


\subsection{The Orbital Signal}

A dominant, stable feature at $\approx$ 37 c\,d$^{-1}$ is always present in the nightly power spectra. In four of the nights, a weaker, but stable signal is also present at half that frequency,  $\sim$ 18.5 c\,d$^{-1}$. These two signals are interpreted as the orbital frequency and its first harmonic, $\omega_{\text{o}}$ and $2\omega_{\text{o}}$. As mentioned above, double-humped orbital waves are quite common among CVs, especially those of very low luminosity (for instance in WZ Sge and AL Com). A power spectrum composed of 11 nights, spanning 45 days, yields $\omega_{\text{o}} = 18.4888 \pm 0.0017$ c\,d$^{-1}$ and $2\omega_{\text{o}} = 36.9740 \pm 0.0005$ c\,d$^{-1}$. Errors are calculated from bootstrap simulations as described in Section~\ref{obs}.

Figure~\ref{fig:s_poworb} shows the low-frequency portion of the full 11 night power spectrum. A zoom of the region around the orbital frequency is plotted on top. A model power spectrum constructed from two artificial sinusoids at $\omega_{\text{o}}$ and $2\omega_{\text{o}}$, using the exact same sampling as for the original dataset, is shown in Figure~\ref{fig:s_poworbmod}. When comparing model versus data, the surrounding picket-fence pattern is similar in structure and height. This indicates that the orbital signal indeed maintains an essentially constant amplitude and phase. In Figure~\ref{fig:fold18}, data from all 11 nights are folded onto the orbital frequency, showing the double-humped orbital wave at $\omega_{\text{o}}$ and $2\omega_{\text{o}}$.  
 
The lower frequency range of the power spectrum was further investigated to rule out the possibility of signals hiding in the noise (see Section~\ref{superh} for the case of BW Sculptoris). The power spectrum was cleaned from the strongest signals at $\omega_{\text{o}}$, $2\omega_{\text{o}}$ and also from the high-amplitude peaks between 4 c\,d$^{-1}$ -- 6 c\,d$^{-1}$. However, no additional peak was found in this region or in the vicinity of the orbital period.


 \begin{sidewaystable}
\begin{center}
 \begin{tabular}{llllll}
  \hline
  \hline
\textbf{Star} & \textbf{Frequency} (c\,d$^{-1}$) & \textbf{Period} (min) & \textbf{Comments}  \\
 \hline
 J1457 & $18.4888 \pm 0.0017$  & 77.92 & Orbital period ($\omega_{\text{o}}$) \\      
& $36.9740 \pm 0.0005$ & 38.95 & $2\omega_{\text{o}}$ \\
& 71.9 (nightly mean error: 0.5) & 20.0 & NRP, low amplitude, non-stable \\ 
& 135.2/144.3/147.9 (nightly mean error: 0.8) & 10.7/9.9/9.7 & NRP, non-stable \\
\hline
BW Scl & $\sim$ 16.50 $ \pm$ 0.01  & 87.27 & Quiescent superhump, slightly non-stable \\
 & 18.40811 $ \pm$ 0.03  & 78.23 & $\omega_{\text{o}}$  \\ 
& 32.98 $\pm$ 0.01 & 43.66 & Harmonic of quiescent superhump \\
& 36.81622 $ \pm$ 0.03  & 39.11 & $2\omega_{\text{o}}$  \\
& 50.5 $\pm$ 0.5 & 28.5 & period signal, weak  \\
&  69.55 $\pm$ 0.03 & 20.70 & NRP, $\omega_{1}$, non-stable \\
& 103.55 $\pm$ 0.03  & 13.90 & $\omega_{2}$ - $2\omega_{\text{o}}$, non-stable \\
& 140.37 $\pm$ 0.03 & 10.26 & NRP, $\omega_{2}$, non-stable \\
& 121.99 $\pm$ 0.03  & 11.80 & $\omega_{2}$ - $\omega_{\text{o}}$, non-stable   \\
& 153.0 $\pm$ 0.5 & 9.4 & period signal, transient \\
& 307.0 $\pm$ 0.5 & 4.7 & Harmonic of signal at 153 c\,d$^{-1}$, transient \\
  \hline
\hline
\end{tabular}
\vspace{0.5cm}
 \caption{\emph{Summary of frequencies found in SDSS J1457+51 and BW Sculptoris.}}
\label{tab:freq} 
\end{center}
\end{sidewaystable}
 

\subsection{High-Frequency Power Excess}

The complex signals spanning 142 c\,d$^{-1}$ -- 148 c\,d$^{-1}$ move slightly in frequency and are non-commensurate with the orbital frequency. These signals could be interpreted as non-radial pulsations arising from the primary white dwarf. In addition, during five of the observing nights, a broad, low-amplitude peak appeared at 135 c\,d$^{-1}$, along with a signal at 72 c\,d$^{-1}$, neither of which are of orbital origin.  

In order to study these signals in more detail, a coherent power spectrum was constructed by concatenating the data from nights. However, this did not produce clean signals, even though the nightly power-spectrum windows were always clean. The broadness and complexity of the signals indicate a slight shift in amplitude and frequency on the time scale of a few nights and/or an internal fine structure unresolved by our observations. Therefore, analysis was performed on power spectra from separate nights in comparison with the overall mean power spectrum.  


\begin{figure}[t]
\begin{center}
\includegraphics[width=11cm]{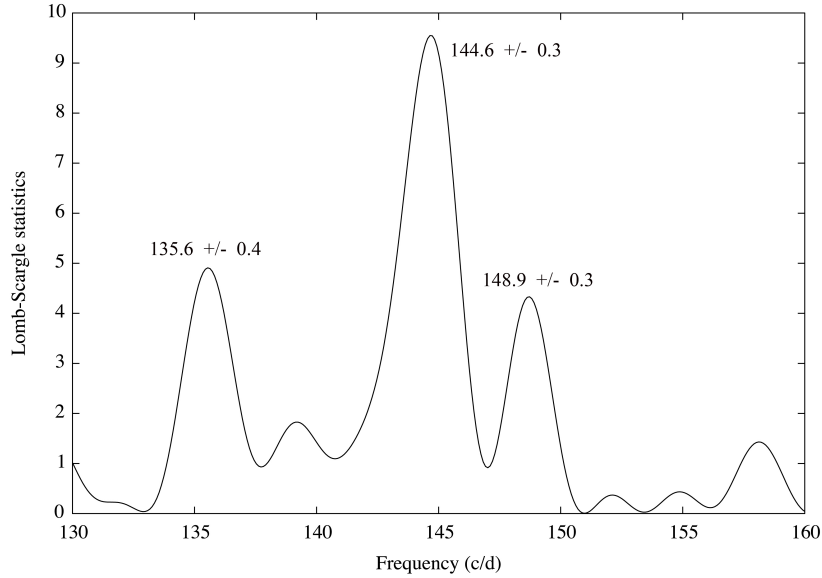}
\caption{\emph{Power spectrum of J1457 from one single night (HJD=2455320). Peaks were seen simultaneously at 135.6 c\,d$^{-1}$, 144.6 c\,d$^{-1}$ and 148.9 c\,d$^{-1}$.}} 
 \label{fig:pow320} 
 \end{center}
\end{figure}


During the first observing night, a peak at 142.6 $\pm$ 1.6 c\,d$^{-1}$ was detected. Over the nights that followed, the signal was found at higher and higher frequencies and two weeks later, a peak was visible, at 147.8 $\pm$ 0.4 c\,d$^{-1}$. When combining data collected over 43 nights, there is a broad power excess around the frequencies, 135.2 c\,d$^{-1}$ (low amplitude), 144.3 c\,d$^{-1}$ and 147.9 c\,d$^{-1}$. This splitting is evident also in the mean spectrum shown in Figure~\ref{fig:s_avpow} and is always seen when combining nights from the start and the end of the campaign. When studying the nightly power spectra in this range, there was (in general) only one peak present at the time. However, during one observing night, three peaks were seen simultaneously at $135.6 \pm 0.4$ c\,d$^{-1}$, $144.6 \pm 0.3$ c\,d$^{-1}$ and $148.9 \pm 0.3$ c\,d$^{-1}$ (Figure~\ref{fig:pow320}). The nightly mean error for any peak appearing at 135 c\,d$^{-1}$ -- 148 c\,d$^{-1}$, is about 0.8 c\,d$^{-1}$. 

    
\begin{figure}
\begin{center}
\includegraphics[width=12cm]{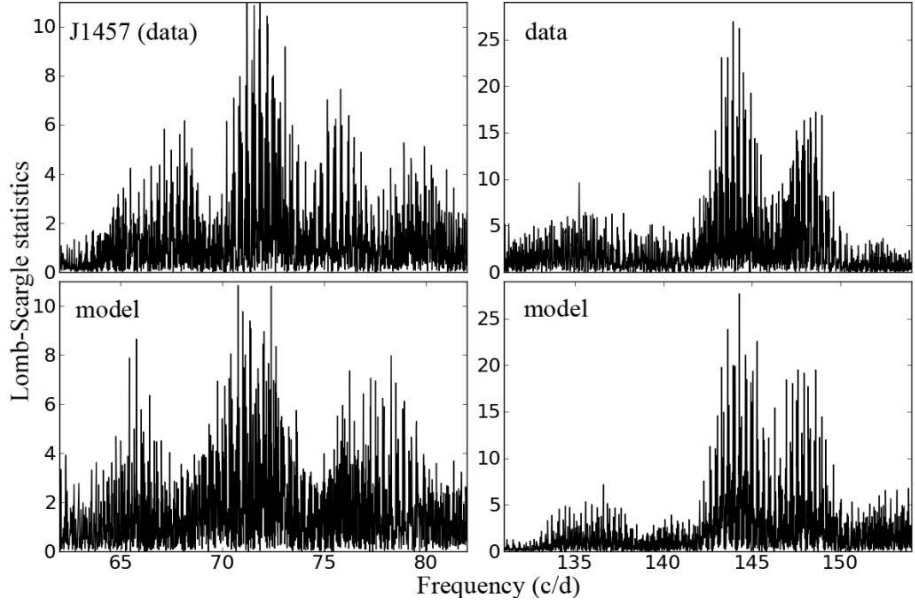}
\caption{\emph{The two top panels show the power spectrum of J1457 from the same set of nights shown in the mean spectrum (see Figure~\ref{fig:s_avpow}). The two bottom frames show models that was constructed from the same sampling pattern as the data. The following frequencies was included in the model; 18.48 c\,d$^{-1}$($\omega_{\text{o}}$), 36.97 c\,d$^{-1}$($2\omega_{\text{o}}$), 71.9 c\,d$^{-1}$, 135.2 c\,d$^{-1}$, 144.3 c\,d$^{-1}$ and 147.9 c\,d$^{-1}$. All main signals along with their line widths seen in the data can be reproduced fairly well by the model, indicating that there is power excess at these frequencies. However, if any of these frequencies are removed from the model, there is no longer a good match between model and data.}} 
 \label{fig:s_pow_best2} 
  \end{center}
\end{figure}


The upper two frames in Figure~\ref{fig:s_pow_best2} show the combined power spectrum of the six nights of best quality. The spectrum spans 43 nights in total, and shows the regions around the signals at 72 c\,d$^{-1}$ and 135 c\,d$^{-1}$ -- 148 c\,d$^{-1}$. The bottom two panels shows model power-spectra constructed from sine waves at 18.48 c\,d$^{-1}$ ($\omega_{\text{o}}$), 36.97 c\,d$^{-1}$ ($2\omega_{\text{o}}$), 71.9 c\,d$^{-1}$, 135.2 c\,d$^{-1}$, 144.3 c\,d$^{-1}$ and 147.9 c\,d$^{-1}$, and the same sampling pattern as the data. The model is able to re-construct the overall appearance and widths of the signals seen in the data reasonably well, indicating that there is power excess at these frequencies. If any of these frequencies are omitted from the model, the data power spectrum cannot be reproduced. The signal at 71.9 c\,d$^{-1}$ is about half that of 144.3 c\,d$^{-1}$, but not exactly, indicating that the signals are not constant in amplitude and phase.

\begin{figure}[t]
\begin{center}
\includegraphics[width=11cm]{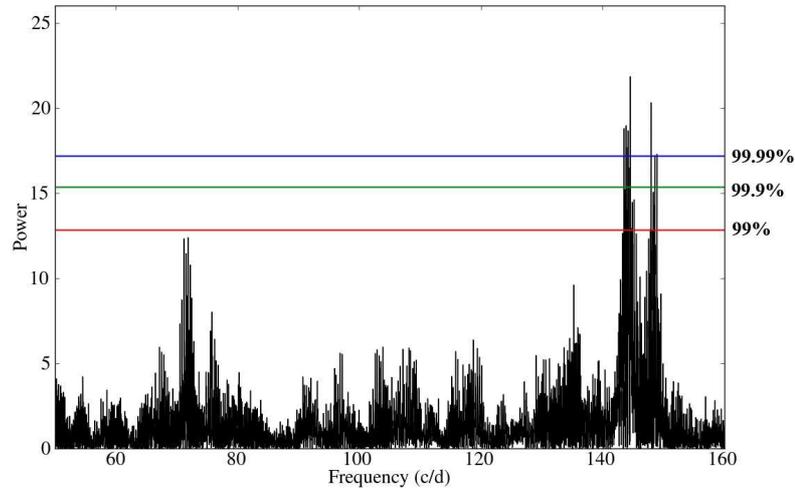}
\caption{\emph{Significance levels ({\color{Red} red} = 99\%, {\color{Green} green} = 99.9\% and {\color{blue} blue} = 99.99\%) in the power spectrum constructed from the combined data of the six nights of best quality. The peak at 72 c\,d$^{-1}$ appears to be below the significance level, but since it is present in many individual data sets, it is still of interest.}} 
 \label{fig:siglim} 
 \end{center}
\end{figure}


Figure~\ref{fig:siglim}, shows a power spectrum of the six nights of best quality, where the significance levels are marked as horizontal lines ({\color{Red} red}= 99\%, \begin{color}{Green}green\end{color} = 99.9\% and \begin{color}{blue}blue\end{color} = 99.99\%). The peak at 72 c\,d$^{-1}$ is below the significance level. However, since it is repeatedly seen in many data sets, and since a signal at this frequency has also been detected in similar targets (see Section~\ref{2009} for BW Sculptoris), it should still be considered as significant.

\begin{figure}
\begin{center}
\includegraphics[width=10cm]{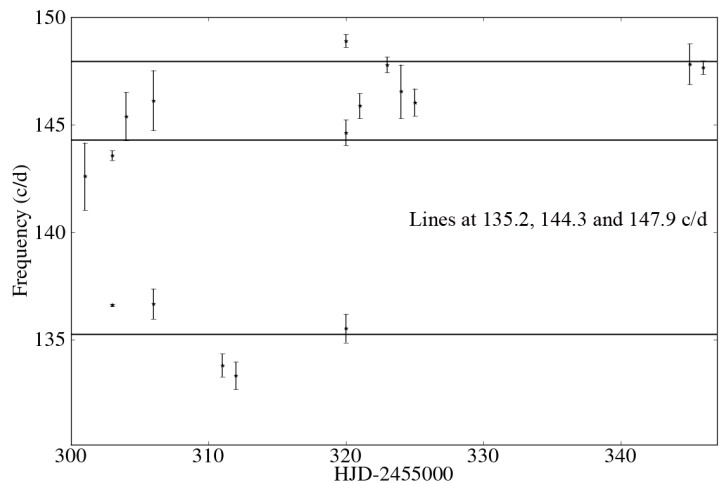}
\caption{\emph{The distribution of frequencies found in J1457, in the range 130 c\,d$^{-1}$ -- 150 c\,d$^{-1}$, plotted as a function of time.}} 
 \label{fig:s_peaks_all3} 
  \end{center}
\end{figure}

In Figure~\ref{fig:s_peaks_all3}, all peak frequencies detected in the range spanning 130 c\,d$^{-1}$ -- 150 c\,d$^{-1}$, are plotted as a function of time. Horizontal lines mark the mean frequencies of the power excess found in the combined power spectrum. The figure shows that the signals do not appear to be intrinsically stable. However, as indicated in Figure~\ref{fig:s_pow_best2}, there could be an underlying stable structure at those frequencies indicated by the solid lines, varying in amplitude, that our sparse data are not able to resolve in detail. 

I conclude that there are persistent periodic signals at the mean frequencies 71.9 c\,d$^{-1}$, 135.2 c\,d$^{-1}$, 144.3 c\,d$^{-1}$ and 147.9 c\,d$^{-1}$, which are non-commensurate with the orbital period. The signals at 72 c\,d$^{-1}$ and 144 c\,d$^{-1}$ are not exactly in a 2:1 ratio (see Section~\ref{2009} for the case of BW Sculptoris). For a complete summary of the frequency analysis, see Table~\ref{tab:freq}.

    
\section{BW Sculptoris}
 
BW Sculptoris (hereafter BW Scl) is a 16th magnitude blue star which was found to
coincide with RXJ2353-0-3852 in the ROSAT bright-source catalogue, and
then identified as a cataclysmic variable by~\cite{1997A&A...318..134A}.~\cite{1997A&A...324L..57A} independently discovered the star as a blue object in the Hamburg/ESO
survey for bright quasi-stellar objects (QSOs). These two studies established the very short
orbital period of 78 minutes. In addition to broad and doubled-peaked H and He
emission lines, BW Scl also shows very broad Balmer and Lyman absorptions,
signifying the presence of a WD of a temperature $\simeq 15000$ K
(\citealt{2005ApJ...629..451G}). If half the visual light comes
from such a white dwarf, then the WD has $V=17.3$ and $M_{V}\sim 12$,
implying a distance of only $\sim 110$ pc. This also agrees with the large
proper motion found in the USNO catalogue (105 ms$\,$yr$^{-1}$; \citealt{2004AJ....127.3060G}). These considerations (a nearby star of very short $P_{\text{orb}}$), and the possibility to study the underlying white dwarf, was the motivation to carry out observations of the star. 

 \begin{table}
 \begin{center}
 \begin{tabular}{llllll}
  \hline
  \hline
\textbf{Year} & \textbf{Spanned}  & \textbf{Observer} & \textbf{Telescope} & \textbf{Nights} \\
& (days) & &  & (hours) \\
 \hline
1999 &  13  &  Kemp & CTIO 91cm & 12/61\\      
2000 & 38 & McCormick & Farm Cove 25cm & 7/35\\
         &      & Rea    &     Nelson 35cm   &  2/9\\
2001  & 77  &    Rea   &           "     &         16/87\\
        &     & Kemp     &   CTIO 91cm       &   14/84\\
           &  & Woudt     &  SAAO 76cm           & 1/4\\
2002   & 0   &   Kemp      &       "               & 1/4\\
2004   & 7    &  Monard   &   Pretoria 25cm  & 7/38\\
2005  & 55    &  Rea       &  Nelson 35cm    & 16/66\\
          &  & Christie   & Auckland 35cm      & 12/48\\
        &  &  Retter/Liu   &                    & 6/32\\
    &   &  Monard     & Pretoria 25cm &  4/22\\
   &   &  Moorhouse   &                    & 3/10\\
2006  & 90   &   Rea        & Nelson 35cm  &  26/112\\
          &  & Monard     & Pretoria 35cm  & 3/21\\
&         &   McCormick  & Farm Cove 25cm & 5/15\\
2007  & 21  &    Rea        & Nelson 35cm   &  7/28\\
          &   & Richards  &  Melbourne 35cm     & 1/3\\
&         & Allen     &  Blenhem 41cm   & 1/3\\
2008  & 71   &   Monard   &   Pretoria 35cm & 22/88\\
&          &  Rea        & Nelson 35cm   & 13/90\\
2009  & 50  &    Rea         &     "             & 15/94\\
          &   & Monard      & Pretoria 35cm & 11/62\\
\hline
  \hline
\end{tabular}
\vspace{0.5cm}
  \caption{\emph{Summary observing log for BW Sculptoris.}}
\label{tab:obs2} 
 \end{center}
\end{table}

\subsection{Observations}
 
In total, BW Scl was observed for about 1000 hours spread over about 200 nights, mainly using the globally distributed telescopes of the Center for Backyard Astrophysics (CBA; \citealt{1993ApJ...417..298S, patterson_1998}). A summary observing log is presented in Table~\ref{tab:obs2}. In order to maximise the signal and optimise the search for periodic features, usually no filter or a very broad filter spanning the range 4000\,\AA\, --\, 7000\,\AA\,\,was used. Occasional runs were obtained in V and I bandpasses to provide a rough calibration, and BW Scl was found to remained within $\sim 0.3 $ mag of $V=16.6$, throughout the campaign. In order to study the periodic behaviour, data was obtained densely distributed in time, preferably with contribution from telescopes widely spaced in longitude (in order to solve problems associated with daily aliases). Most of the analysis was based on long time series from telescopes in New Zealand, South Africa, and Chile, and hence not afflicted by aliasing problems.

\subsection{The 1999 Campaign}

\begin{figure}
 \begin{center}
\includegraphics[width=10cm]{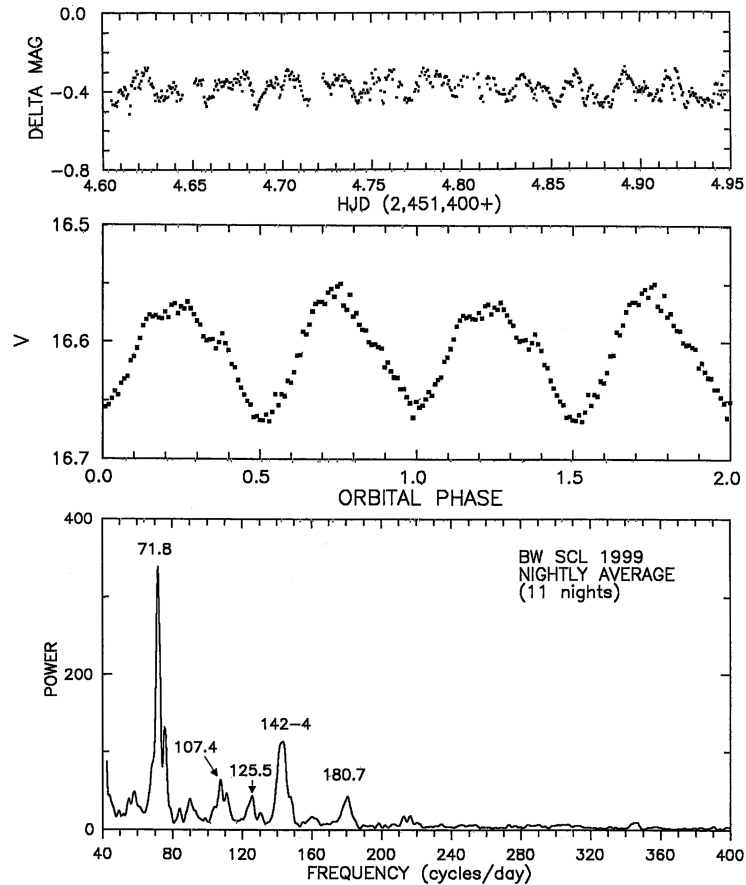}
\caption{\emph{BW Scl in 1999. The upper frame shows an 8-hour light curve, dominated by the orbital wave. The middle frame presents the mean orbital light curve, showing a double-humped waveform. The lowest frame shows the mean nightly power-spectrum, averaged over the 11 best nights. Significant features are labelled with their frequency in c\,d$^{-1}$ (with an average error of $\pm$ 0.7).}} 
 \label{fig:1} 
   \end{center}
\end{figure}

The upper frame of Figure~\ref{fig:1} shows the light curve from one night during 1999. The general appearance is typical of all nights, as well as seen in previously published light curves (\citealt{1997A&A...324L..57A, 1997A&A...318..134A}). As those papers demonstrated, a 39-minute periodicity is always present, suggesting that the fundamental period is actually double this, i.e. 78 minutes (confirmed by spectroscopy). The middle frame shows the 1999 double-humped orbital light curve, similar to that of J1457 (see Figure~\ref{fig:fold18}). 

 The average nightly power spectrum, the incoherent sum of the 11 nights of best quality, is shown in the bottom frame of Figure~\ref{fig:1}. Power excesses near 72 c\,d$^{-1}$ and 143 c\,d$^{-1}$ are evident. In order to study these higher-frequency signals in more detail, adjacent nights were added together, and the coherent power spectrum was constructed. However, even though the 72 c\,d$^{-1}$ and the 143 c\,d$^{-1}$ signals were always present, they were always broad, complex, and slightly variable in frequency (similar to J1457). This is usually a sign that the actual signals violate the assumptions of Fourier analysis; constancy in period, amplitude and phase.


\begin{figure}
 \begin{center}
\includegraphics[width=11cm]{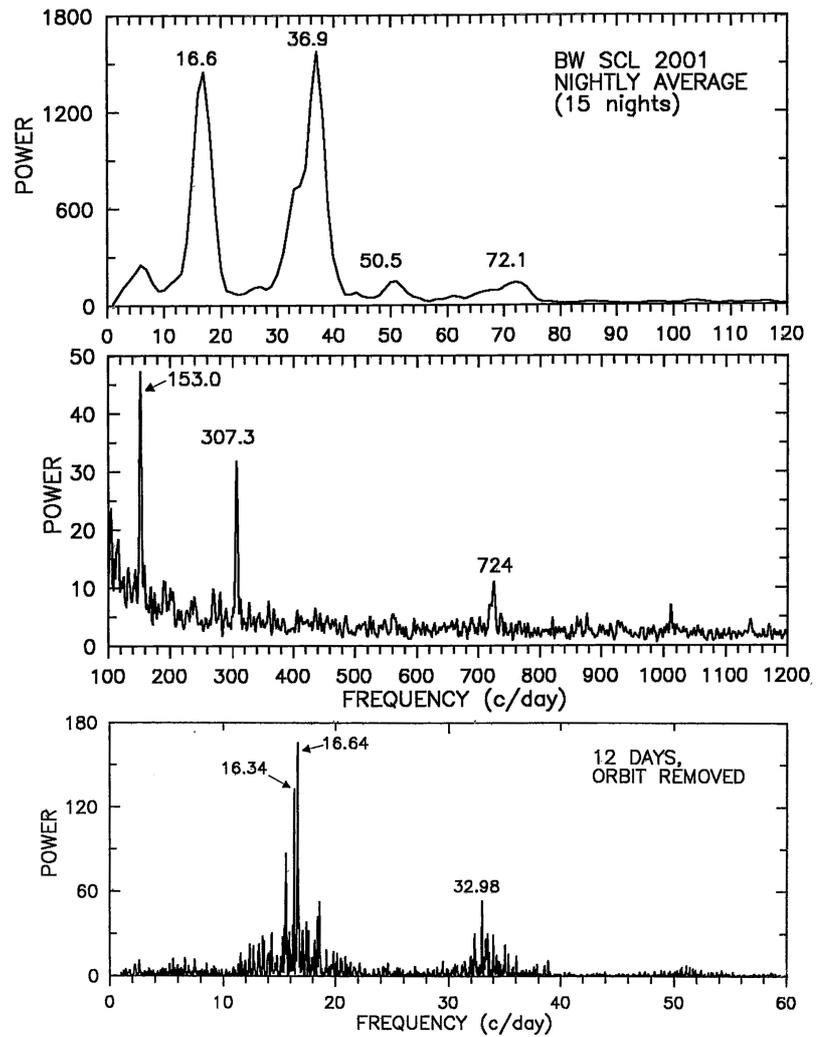}
\caption{\emph{BW Scl in 2001. The upper two frames show the mean nightly power
spectrum, averaged over the 15 best nights. Features of obvious or possible significance are labelled with their frequency in c\,d$^{-1}$ (with an average error of $\pm$ 0.7).}} 
 \label{fig:2} 
  \end{center}
\end{figure}


\subsection{The 2001 Campaign}
 
During 2001, observatories at three longitudes in Chile, New Zealand and South Africa, contributed with good nightly coverage of BW Scl. A very low flickering background is shown in the light curves. Variations as small as $\sim 0.002$ magnitudes could be detected due to the very extensive campaign that was carried out.

The upper frames of Figure~\ref{fig:2} shows the nightly mean power spectrum, with significant signals marked to an accuracy of 0.5 c\,d$^{-1}$. Both the orbital period and the signal at 72 c\,d$^{-1}$ were present along with two other signals at low frequency: a powerful signal at 16.6 c\,d$^{-1}$, and a weak signal at 50.5 c\,d$^{-1}$. At higher frequency, signals are detected at 153 c\,d$^{-1}$, 307 c\,d$^{-1}$, and 724 c\,d$^{-1}$, though the latter is likely to be caused by instrumental effects. Many telescopes have worm gears which turn with a period of exactly 120 sidereal seconds, and this period was reported in research on many types of stars during 1960 -- 1990, i.e. during the photolectric-photometer era. CCDs are much less prone to this error. However, since 724 $\pm$ 2 c\,d$^{-1}$ corresponds to 120 sidereal seconds (to within the measurement error), we interpreted the signal to be caused by this instrumental effect.   
 
The signal at 153 c\,d$^{-1}$ can be interpret as a pulsation frequency, with a significant 
first harmonic. Examination of individual nights showed this signal to be slightly variable, at least in amplitude. This behaviour was seen on 7 of the 15 good-quality nights. Since the orbital frequency is known precisely, and since its photometric signature is powerful and constant, the first harmonic (and also the second harmonic when detected) was subtracted from the central 12-night time-series, prior to analysis. The resultant power spectrum at low frequency is seen in the bottom panel of Figure~\ref{fig:2}. A weak signal appears at the orbital frequency, and stronger signals at 16.34/16.64 c\,d$^{-1}$ and 32.98 c\,d$^{-1}$ (with an accuracy of $\pm\,\,0.01$ c\,d$^{-1}$). It seems likely that these are superhump signals in the quiescent light curve. Specifically, the 16.34/16.64 pair can be interpreted as signifying an underlying frequency of $\sim 16.5$ c\,d$^{-1}$, as such splitting can be produced by amplitude and/or phase changes. The precession frequency $\Omega$ can be expressed as $16.50 = \omega_{\text{o}} - \Omega$, where $\omega_{\text{o}}$ is the orbital frequency (implying that $32.98 = 2(\omega_{\text{o}} - \Omega)$). The precession frequency $\Omega$ is then equal to 1.9 c\,d$^{-1}$.

The primary signal near 16.5 c\,d$^{-1}$ has been seen before. It was the dominant signal reported by~\cite{1997A&A...318..134A}, but was then discounted by~\cite{1997A&A...324L..57A} as probably the result of cycle-count error. The data presented here certainly have no ambiguity in cycle count and reveals this signal very clearly. 


\begin{figure}
\begin{center}
\includegraphics[width=10cm]{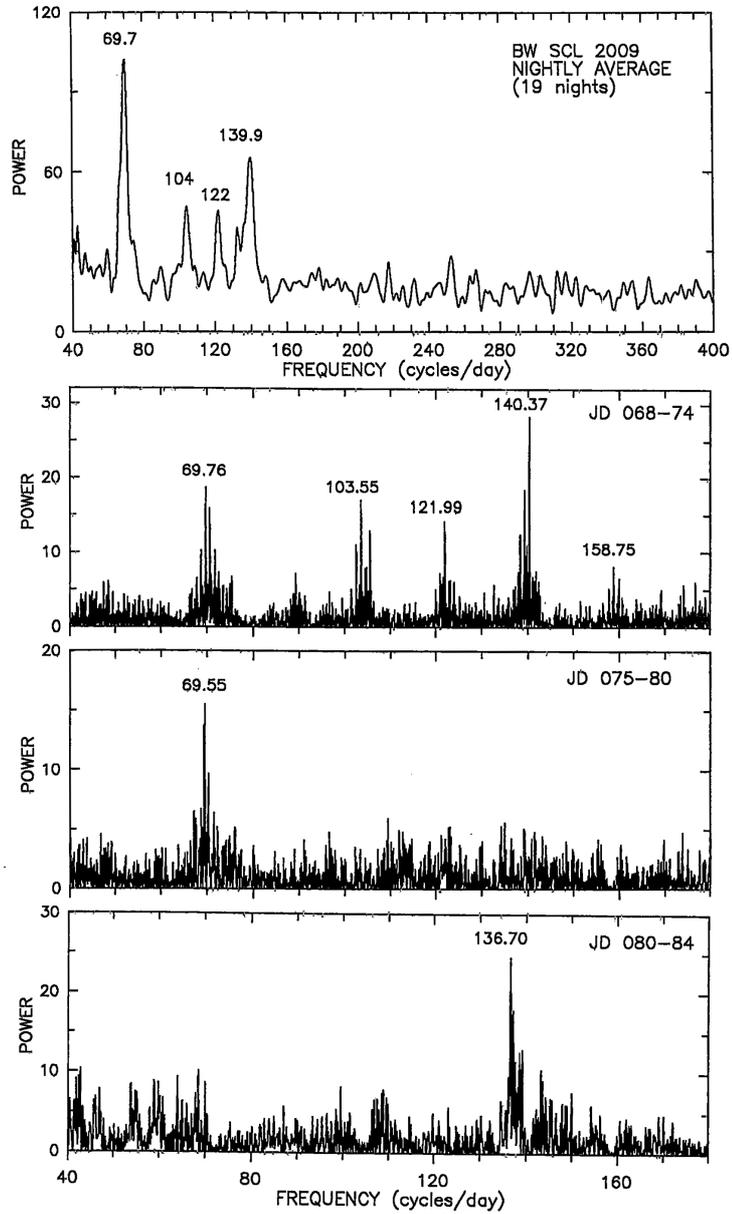}
\caption{\emph{BW Scl in 2009. The top frame shows the mean nightly power spectrum, averaged over the 19 best nights. Features are labelled with their frequency in c\,d$^{-1}$ ($\pm$ 0.6). Other frames show power spectra of $\sim$ 5 day intervals with particularly dense coverage, with significant features labelled ($\pm$ 0.03).}} 
 \label{fig:3} 
   \end{center}
\end{figure}


\subsection{The 2009 Campaign}  \label{2009}

The upper frame of Figure~\ref{fig:3} shows the nightly power spectrum, averaged over the 19 nights of coverage obtained in 2009. Signals near 70 c\,d$^{-1}$ and 140 c\,d$^{-1}$ are present, as well as weaker signals near 104 c\,d$^{-1}$ and 122 c\,d$^{-1}$. In the three lower frames, intervals of particularly dense coverage are shown, each with a frequency resolution of $\pm\,\,0.04$ c\,d$^{-1}$. In these frames, the 70 c\,d$^{-1}$ and 140 c\,d$^{-1}$ signals (both normal features of BW Scl) show their variability in frequency and amplitude (similar to J1457). If we define $\omega_{2}$ = 140 c\,d$^{-1}$, the frequencies at 104 c\,d$^{-1}$ and 122 c\,d$^{-1}$ are displaced by exactly $\omega_{2}$ - $2\omega_{\text{o}}$ and $\omega_{2}$ - $\omega_{\text{o}}$ of the orbital frequency.

\subsection{Superhumps in BW Sculptoris?} \label{superh}

The signal at 87 minutes in BW Scl, displaced by $\sim 11$\% from $P_{\text{orb}}$ seen in the 2001 power spectrum (Figure~\ref{fig:2}), is a transient but repeating feature in the light curve. This feature was always strong when detected, but moved slightly in frequency, even on timescales of a few days. This qualitatively describes a common superhump, which is a well-known feature in CV light curves (see Section~\ref{superhps}). However, there are a few substantial differences: superhumps are found in outburst states, typically at $M_{V} \sim 5$, and not at quiescence when $M_{V} \sim 12$. Also, superhumps occur with fractional period excesses, $(P_{\text{sh}} - P_{\text{orb}})/P_{\text{orb}}$, near 3\%, not 11\%.

Systems with quiescent superhumps are rare, but has been reported before. In AL Com, a nearly identical signal was reported and extensively discussed by~\cite{1996PASP..108..748P}. A similar signal has also been reported in CP Eri (see Appendix A1 of \citealt{2002PASP..114..721P}), V455 And (\citealt{2005A&A...430..629A}), and possibly also in SDSS J1238-03 (\citealt{2010ApJ...711..389A}). Except for CP Eri (which is excluded from consideration because of its helium content), all these stars are all period-bouncer candidates (see Tables 3 and 5 of \citealt{2011MNRAS.tmp...27P}). These candidates were chosen due to their low donor-star masses (or low $q = M_{2}/M_{1}$). A possible account of how stars of very low $q$ might be able to manufacture quiescent superhumps has been given in~\cite{1996PASP..108..748P}, and includes the idea that a low donor mass implies a larger Roche lobe surrounding the disc. Weak tidal torques could then allow the quiescent disc to extend to the 2:1 orbital resonance, where an eccentric instability could drive a fast prograde precession (viz., at $\Omega$ = 1.9 c\,d$^{-1}$), resulting in a superhump with $\omega = \omega_{\text{o}} - \Omega$.

\section{Interpretations as Non-Radial Pulsations}

Both BW Scl and J1457 are quiescent CVs of very low accretion luminosity. Their spectrum show evidence of the primary WD, as do their light curves which contain rapid, non-commensurate signals. This is the general signature of the GW Lib stars, where the periodic signals are believed to represent the non-radial pulsations of the underlying white dwarf. However, no proof of this has ever been found, not for GW Lib or any other of the 10 -- 15 members of the class. In order to explain these signals, two other hypotheses deserve consideration: first, quasi-periodic oscillations (QPOs; see Section~\ref{superhps}) arising from the accretion disc, and second, the spin frequency of a magnetic WD (DQ Her modulations; see Section~\ref{qpo_dno}). QPOs are typically seen as very broad peaks in the power spectrum, $\delta v/v \sim 0.5$ (see for instance Figure 11 and 12 in~\citealt{2002PASP..114.1364P}), while the signals at 70 c\,d$^{-1}$ and 140 c\,d$^{-1}$ ($\omega_{1}$ and $\omega_{1}$), in both BW Scl and J1457 have $\delta v/v \sim 0.01$. Therefore, QPOs are most likely not the origin of these pulsations. A WD spinning with a period of 20 minutes could possibly explain the signals seen at $\sim$ 70 c\,d$^{-1}$ and 140 c\,d$^{-1}$ in both objects (and could possible also explain the orbital sidebands seen in BW Scl). However, such modulations usually result in a much more coherent signal, while those seen in BW Scl and J1457 are broad and shift in frequency and amplitude on a time scale of a few days. Also, in both BW Scl and J1457, $\omega_{2}$ is not exactly $2 \times \omega_{1}$.   

With these considerations taken into account, I conclude that J1457 and BW Scl most likely are new members of the GW Lib class. Common to all known GW Lib stars is that they are low-$\dot{M}$ CVs, showing low-amplitude and non-commensurate periodic signals in their light curves. The signal is roughly constant over a few days, but is slightly shifting in frequency and amplitude on longer timescales. Also, these signals often show strong first harmonics.    
\newpage
\section{Summary}

\begin{enumerate}
\item [1.] Two more CVs with very low accretion rates have shown rapid non- commensurate signals in quiescence, which makes them new members of the GW Lib class. Both J1457 and BW Scl show a complex power spectrum with main signals near 10 and 20 minutes.

\item [2.] The pulsation frequencies in both stars wander by a few percent on a timescale of days. In addition, the power-spectrum constructed from multiple-night light curves, show broad peaks, which might be due to the frequency wandering, or a fine structure not resolved by our data.

\item [3.] BW Scl shows several peaks displaced from the main pulsation frequency by exactly $\omega_{2} \pm \omega_{\text{o}}$ and $\omega_{2} \pm 2\omega_{\text{o}}$, where $\omega_{\text{o}}$ is the orbital frequency $\approx$ 18.4 c\,d$^{-1}$, $2\omega_{\text{o}}$  $\approx$ 36.8 c\,d$^{-1}$ and $\omega_{2} \approx$ 140 c\,d$^{-1}$. Similar behaviour is evident also in other GW Lib stars (SDSS J1507+52, V386 Ser and SDSS J1339+48). The origin of this phenomenon is still unknown.

\item [4.] The orbital light curves of both stars show double-humped waves. From these waves, precise periods are found at $P_{\text{orb}} = 0.05432392(2)$ days for BW Scl, and $0.054087(5)$ days for J1457. Similar systems showing signals non-commensurate with the orbital period, such as SDSS J1339+4847, SDSS J0131-0901 and SDSS J0919+0857, all have orbital periods at $\approx$ 80 minutes. BW Scl sometimes shows a transient wave with a period, $P_{\text{sh}} = 0.060$ days, which is interpreted as a quiescent superhump. It thereby joins a small group of stars who manage a superhump at quiescence, all of which are likely to have very low mass ratios. This might arise from an eccentric instability at the 2:1 resonance in the disc.

\end{enumerate}
   


\chapter{The Recurrent Nova T Pyxidis \\\Large \textsc{- Orbital Period and System Parameters - }} \label{tpyx}
\label{chap:icm}

\emph{All work presented in this chapter was carried out together with Christian Knigge and Danny Steeghs and was published as~\cite{2010MNRAS.409..237U}. In particular, the photometric ephemeris was calculated by Christian Knigge. I also want to thank Joe Patterson for valuable comments and for kindly providing us with the a compilation of recent photometric timings for T Pyx. Also Gijs Roelofs is thanked for obtaining part of the data.} 
\newline
\newline
\begin{Huge}\color{Red}{T}\end{Huge} Pyx is a luminous cataclysmic variable in which the donor star is transferring mass at a very high rate onto its white-dwarf companion, resulting in unusually frequent thermo-nuclear runaways (TNRs) on the surface of the  primary. Between the years 1890 $-$ 1967, T Pyx has undergone five
such nova eruptions, with a recurrence time of about 20 years between the eruptions; it was therefore classified as a member of the recurrent nova subclass. However, the last eruption was in 1966, which means that T Pyx has now passed its mean recurrence time by more than 20 years. The eruptive behaviour of RNe in comparison with classical novae is thought to be due to a high mass-transfer rate in combination with a massive primary white dwarf (see Section~\ref{classnovaeruption}).  

\begin{figure}[t]
\begin{center}
\includegraphics[width=10cm]{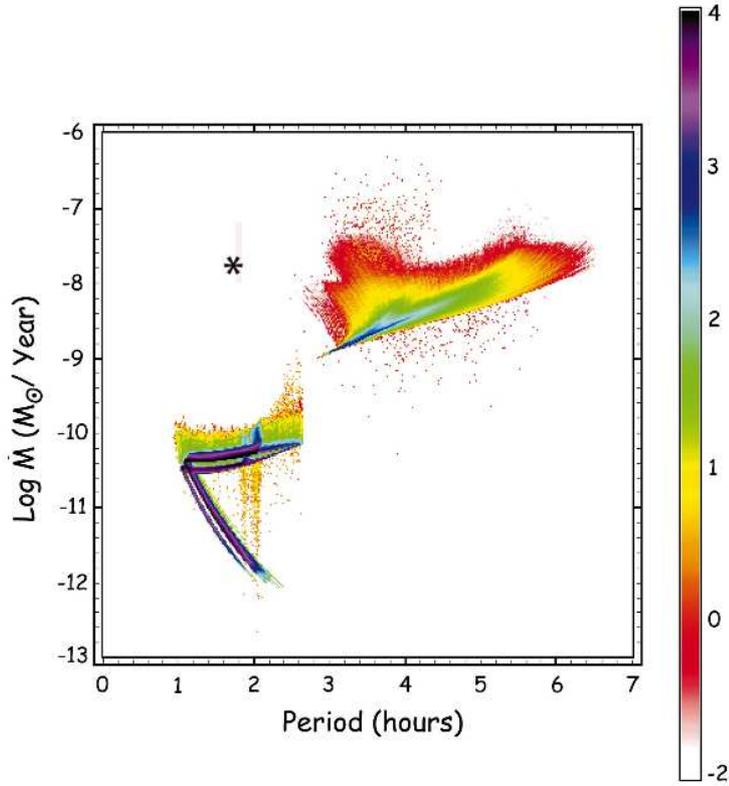}
\caption{\emph{Accretion rates of cataclysmic variables plotted as a function of their orbital periods. The approximate position of T Pyxidis is marked with a black star, showing that the system clearly has a higher accretion rate compared to other systems at the same orbital period (adapted from figure by~\citealt{2001ApJ...550..897H}).}}
\label{howell}
\end{center}
\end{figure}

~\cite{1998PASP..110..380P} carried out an extensive photometric study of
T Pyx and found a stable, periodic signal at $P=1.83$ hours that was
interpreted as the likely orbital period. This would place T Pyx below
the CV period gap and suggests a donor mass around $M_{2} \sim 0.1$
M$_{\odot}$. This is surprising. According to standard evolutionary
models, a CV below the period gap should be faint and have a low
accretion rate driven primarily by gravitational radiation (GR). Yet T Pyx's quiescent luminosity and status as a RN both imply that it has a high accretion rate of $> 10^{-8}$ M$_{\odot}$ yr$^{-1}$ (\citealt{1998PASP..110..380P};~\citealt{2008A&A...492..787S}). This is about two orders of magnitude higher than expected for ordinary CVs at this period (illustrated in Figure~\ref{howell}, adapted from~\citealt{2001ApJ...550..897H}). 

Assuming that the determination of the photometric orbital period is correct, the
existence of T Pyx is interesting for at least two reasons. First, 
unless the system is seen in a transient evolutionary state, its
lifetime would have to be very short, $\tau \sim M_{2}/\dot{M}_{2} \leappeq
10$~Myrs. This would imply the existence of an evolutionary channel
leading to the fast destruction of at least some short-period
CVs. Second, TNRs frequent enough to qualify as RNe are thought to be possible only on high-mass accreting WDs ($M_{1} \gtappeq 1$ M$_{\odot}$). Moreover, RNe are the only class of novae in which the WD is expected to gain more mass
between eruptions than it loses during them. This would make T Pyx a
strong candidate Type Ia supernova progenitor.  

However, the recent study of the system by~\cite{2010ApJ...708..381S} (see also~\citealt{2008A&A...492..787S}) suggests, first, that T Pyx {\em is}, in fact, in a transient evolutionary
state, and, second, that, integrated over many nova eruptions, its WD
does lose more mass than it gains. More specifically,~\cite{2010ApJ...708..381S} suggest that T Pyx was an ordinary cataclysmic variable until it erupted as a nova in 1866. This
eruption triggered a wind-driven supersoft X-ray phase (as 
first suggested by~\citealt{2000A&A...364L..75K}), resulting in an unusually high
luminosity and accretion rate. However, unlike in the original
scenario proposed by~\cite{2000A&A...364L..75K}, the supersoft phase is not
self-sustaining, so that the accretion rate has been
declining ever since the 1866 nova eruption from $\dot{M} \sim 10^{-7}$
M$_{\odot}$\,yr$^{-1}$ to $10^{-8}$\,M$_{\odot}$\,yr$^{-1}$. As a result, T
Pyx has faded by almost 2 magnitudes since the nova eruption
(\citealt{2010ApJ...708..381S}). Based on this, and the fact that T Pyx has already passed its mean recurrence time by more than 20 years,~\cite{2010ApJ...708..381S} argue that T Pyx might no
longer even be a recurrent nova. If these ideas are correct, T Pyx is
not a viable SN Ia progenitor, and its remaining lifetime can be
substantially longer than a few million years. However, if all its
ordinary nova eruptions are followed by relatively long-lived (> 100
yrs) intervals of wind-driven evolution at high $\dot{M}$, its secular
evolution may nevertheless be strongly affected, with significant
implications for CV evolution more generally (see also Section~\ref{SD}).

A key assumption in virtually all of these arguments is that the
photometric period measured by~\cite{1998PASP..110..380P} is, in fact, the orbital period of the system. So far, there has only been one attempt to obtain a spectroscopic period for T Pyx, by~\cite{1990apcb.conf..391V}, who reported a spectroscopic modulation
with $P=3.44$ hours. Such a long orbital period above
the CV period gap would be much more consistent with the high
accretion rate found in T Pyx. In this study, we present the first
definitive spectroscopic determination of the orbital period of T Pyx,
showing that it is, in fact, consistent with Patterson et al.'s photometric
period. We also use our time-resolved spectroscopy to estimate the
main system parameters, such as the velocity semi-amplitude of the
white dwarf ($K_{1}$), the mass-ratio ($q$), the masses ($M_{1}$ and
$M_{2}$) and the orbital inclination ($i$). Finally, we discuss the
implications of our results for the evolution of T Pyx and related
systems.

\section{T Pyx - Observations and Their Reduction}
 
 \subsection{VLT Multi-fibre Spectroscopy}
 
Multi-fibre Spectroscopy of T Pyx was obtained during five nights in
2004 and 2005 with the GIRAFFE/FLAMES instrument mounted on the Unit
Telescope 2 of the VLT at ESO Paranal, Chile. The data were taken in
the integral field unit mode. The total field of view in this mode
is about 11.5" $\times$ 7.3" and thus covers most of T Pyx's 10"
diameter nova shell (\citealt{1982ApJ...261..170W}). We used the fibre system ARGUS,
which consists of 317 fibres distributed across the field, of which 5
point to a calibration unit and 15 points to sky. The
grating order was 4, which gives a resolution of R=12000. The
wavelength range was chosen between 4501\,\AA\,to 5078\,\AA, so that the
emission-line spectrum would include the Bowen blend at 4645\,\AA\,-- 4650
\AA\, and the HeII at 4686 \AA. With this setup, the dispersion is
0.2 \AA/pix, corresponding to $\approx$ 12.5 km\,s$^{-1}$/pix. The full
widths at half-maximum (FWHMs) of a few spectral lines obtained
simultaneously with the science spectra from fibres pointing to the
calibration unit indicate that the spectral resolution is about 0.4
\AA. A log of the observations can be found in Table~\ref{tab:obs_tpyx}.

The initial steps in the data reduction were performed using the ESO
pipeline for GIRAFFE. The pipeline is based on the reduction software
 \textsc{BLDRS} from the Observatory of Geneva. The basic functions of the
pipeline are to provide master calibration data and dispersion
solutions. The pipeline also provides an image of the reconstructed
field of view, which can be used to associate a specific fibre to a
given object. In order to extract the spectrum from the desired fibre,
and to correct for the contribution from the sky background and cosmic
rays, the output from the pipeline was processed further in  \textsc{IRAF}. The
PSF of our target, T Pyx, is spread out over several fibres and 6 single fibre spectra containing significant target flux were extracted and median combined after weighting each spectrum by the mean flux in the region 4660\,\AA\, -- 4710\,\AA\,(covering HeII at 4686 \AA). Cosmic rays were removed, by first binning the spectra over 7 pixels (the cosmic rays have a typical width of 2 $-$ 6 pixels). The smoothed spectra were then subtracted from the corresponding fibre spectra to only leave the residuals and the cosmic rays. Cosmic rays were then removed and the residuals added back to the target spectra. Fibres containing the sky were extracted and combined to create a master sky spectrum. Finally, the master-sky was subtracted from the combined science spectrum.

\begin{table}[t]
\begin{center}
 \begin{tabular}{llll}
\hline
\hline
 \textbf{Date} &  \textbf{Tel/Inst.} &  \textbf{No. of Exp.} &  \textbf{Exp.} (s) \\
 \hline
041210 & VLT/GIRAFFE & 28 & 180\\
041224 & VLT/GIRAFFE & 28 & 180\\
050127 & VLT/GIRAFFE & 29 & 180\\
050128 & VLT/GIRAFFE & 28 & 180\\
050131 & VLT/GIRAFFE & 29 & 180\\
080316 & MAGELLAN/IMACS & 37 & 240\\ 
080317 & MAGELLAN/IMACS & 62 & 240\\ 
080318 & MAGELLAN/IMACS & 62 & 240\\
080319 & MAGELLAN/IMACS & 39 & 240\\
\hline
\hline
\end{tabular}
\vspace{0.5cm}
\caption{\emph{Log of the observations (the date is according to UT time at the start of the night).}}
\label{tab:obs_tpyx}
\end{center}
\end{table}


\subsection{Magellan Long-slit Spectroscopy}

T Pyx was observed again during four nights in March 2008, this time
on the 6.5 meter Baade telescope Magellan I at Las Campanas, Chile
(see Table~\ref{tab:obs_tpyx} for a log of the observations). Long-slit spectroscopy
was obtained with the instrument IMACS using the Gra-1200-17.45 grating
with a 0.9" slit, resulting in a dispersion of 0.386 \AA/pix,
corresponding to $\approx$ 25 km\,s$^{-1}$/pix. We estimate that the spectral
resolution of these observations is about 1.5 \AA, as measured from
the FWHMs of a few spectral lines in the arc-lamp spectra. The overall
spectra span over four CCDs and cover a total wavelength range of 4000\,\AA\,
-- 4800\,\AA. All frames from the four CCDs were treated separately
during both the 2D and 1D reduction steps.
 
The spectra were reduced in \textsc{IRAF} using standard packages. A master
bias produced from combining all bias frames obtained during the four
nights was subtracted from the science frames, and the overscan region
was used to remove the residual bias. A master flat-field, corrected
for illumination effects, was produced. The spectra were then
flat-field corrected and extracted. Line identifications of the ThArNe
lamp spectra obtained during the nights were carried out with the help
of the NOAO Spectral Atlas Central and the NIST Atomic Spectra
Database. The spectra were cosmic-ray rejected using the method
described previously for the VLT data set. A flux calibration was done
using data of the white-dwarf flux-standard star EG274, obtained
during the same nights as the target spectra. Also, extinction data
from the site and flux reference data of EG274 from~\cite{1992PASP..104..533H} were used.

Analysis of the reduced VLT and Magellan data were carried out using the packages \textsc{MOLLY} and  \textsc{DOPPLER}, provided by Tom Marsh.

\section{Analysis of T Pyx}

\subsection{The Overall Spectrum}

\begin{figure}[ht!]
\begin{minipage}[b]{1.0\linewidth}
\centering
\includegraphics[scale=1.0]{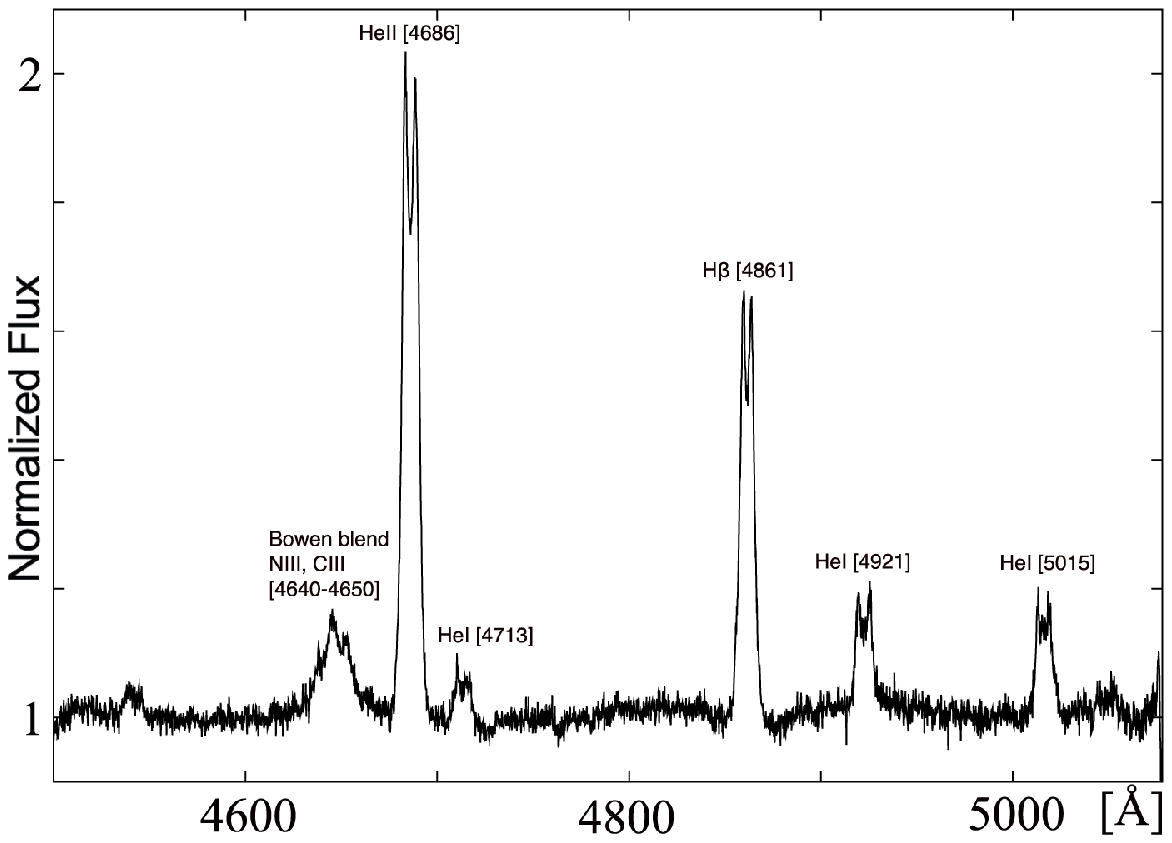}
\caption{\emph{The normalized and averaged spectra constructed from 140 individual exposures obtained at the VLT telescope.}}
\label{fig:spec_vlt}
\end{minipage}
\vspace{1.0cm}
\begin{minipage}[b]{1.0\linewidth}
\centering
\includegraphics[scale=1.0]{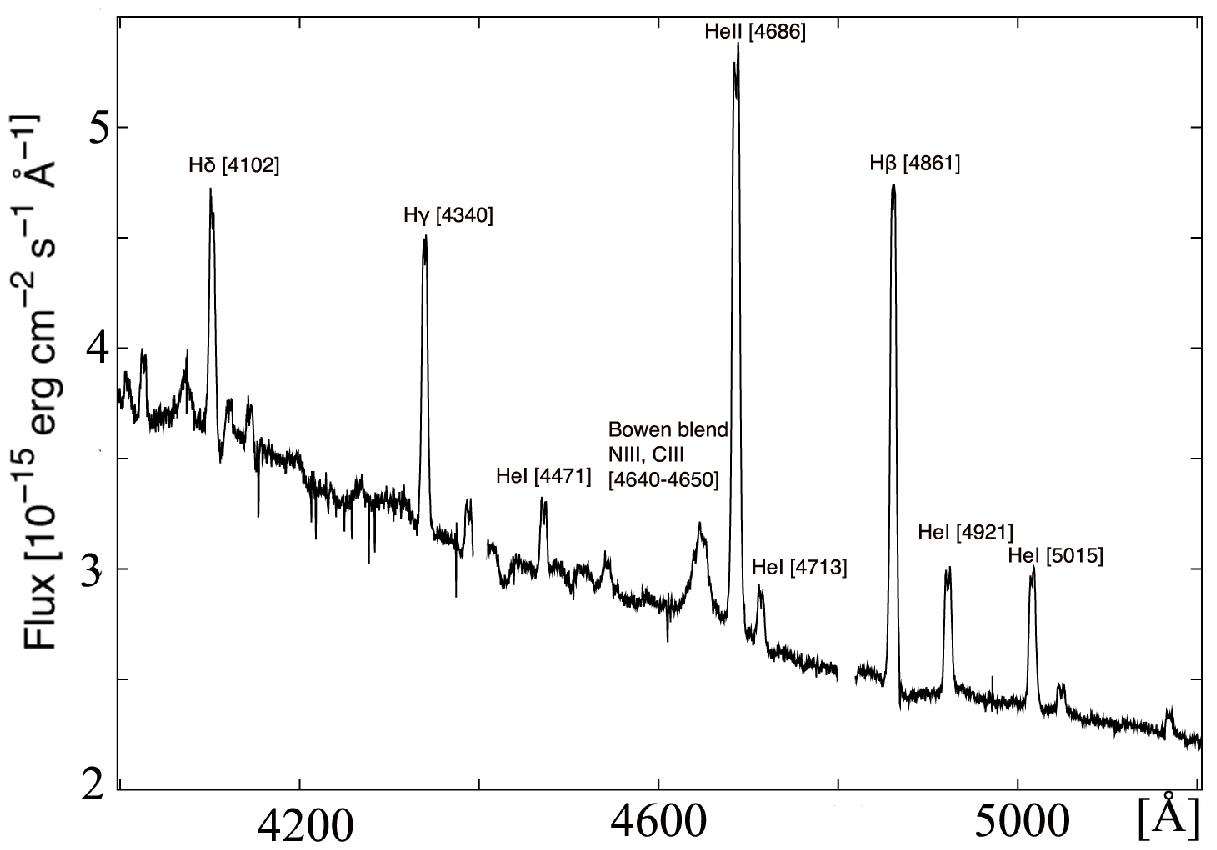}
\caption{\emph{The flux-calibrated average of 200 spectrum observed with the Magellan telescope.}}
\label{fig:spec_mag}
\end{minipage}
\end{figure}

T Pyx has a high-excitation emission-line spectrum that is unusual
when compared to systems with a low $\dot{M}$, but the overall spectrum is similar to that for nova-like variables, supporting the idea that T Pyx has a high accretion rate. Its brightness has been found to be fading for the last century, indicating a decrease in $\dot{M}$,
and its 2009 magnitude in blue was B = 15.7 mag (\citealt{2010ApJ...708..381S}). Double-peaked lines originating from the disc are seen in
the spectrum, and the bright disc outshines any spectral signature of
the primary WD and the donor star. The strongest feature is the
doubled-peaked HeII line at 4686 \AA. Double-peaked lines identified
as HeI at 4713\,\AA, 4921\,\AA\,and 5015\,\AA, as well as
the Balmer lines, are also present (Figure~\ref{fig:spec_vlt} and Figure~\ref{fig:spec_mag}).

The Bowen blend at 4640\,\AA\,-- 4650\,\AA\,is clearly visible
and consists of several lines of NIII and CIII. The Bowen blend is
also seen in other CVs and in low-mass X-ray binaries with
high-accretion rates. In some cases, it has been related to emission
from the donor star (eg.~\citealt{2002ApJ...568..273S}), where it is thought
to be produced by a fluorescence process as the front side of the
donor is strongly irradiated by the hot accretor. The Bowen blend was
clearly visible in the spectrum of T Pyx obtained by~\cite{1998ApJ...498L..61M}, and, based on this, we were hoping to detect narrow components
associated with the donor. However, no narrow donor star features were
found in the blend.  

Data analysis was carried out for all the strongest lines, but our final results are based on analysis of the HeII line at 4686 \AA \, and the HeI line at 4921 \AA.


\subsection{The Orbital Period} 

\subsubsection{The Photometric Ephemeris} \label{PE}

The key goal of our study is to
obtain a definitive determination of T Pyx's orbital period based on 
time-resolved spectroscopy. More specifically, we wish to test if the
photometric modulation reported by~\cite{1998PASP..110..380P} is orbital
in nature. The photometric ephemeris is given by~\cite{1998PASP..110..380P},
\\\\
\textbf{Minimum light = HJD $2446439.428 + 0.0762233E + 3.5 \times 10^{-11}E^{2}$.} 
\\\\
The quadratic term is highly significant and indicates a very high period derivative of $\dot{P}
\simeq 9 \times 10^{-10}$ (this corrects a typo in the original paper). The period is found to be increasing, which is not expected for a system that has not yet passed the minimum period. If the photometric signal is orbital, the current evolutionary time-scale of the system is only $\tau \sim P / \dot{P}
\sim 3 \times 10^{5}$ yrs. Such a short period change time-scale is highly unusual for a CV.  

The Center for Backyard Astronomy (CBA) has continued to gather additional
timings in the period 1996 -- 2009. This new photometry, along 
with an updated ephemeris, will be published in due course (Patterson
et al., in preparation). However, since this new data set is much more
suited to a comparison against our spectroscopic observations (since it 
overlaps in time), Joe Patterson has kindly made this 
latest set of CBA timings available to us. Our own preliminary analysis
of all the existing timings yields a photometric ephemeris of
\\\\
\textbf{Minimum light = HJD(UTC) $2451651.65255(35) + 0.076227249E(16) + 2.546(54) \times 10^{-11}E^{2}$.}
\\\\
The numbers in parenthesis are the uncertainties on the two least significant digits and were estimated via bootstrap simulations. Taking the period change into account, the average period at the time of the optical spectroscopy was $P = 0.076228860 \pm 3.0 \times 10^{-8}$ d ($f = 13.1183912 \pm 5.2 \times 10^{-6}$ c\,d$^{-1}$). 

This new ephemeris confirms that the period continues to increase at a fast
rate.  More specifically, the current estimate of the period
derivative is  $\dot{P} = (6.68 \pm 0.14) \times 10^{-10}$, corresponding to a timescale of
$P/\dot{P} \simeq 3 \times 10^{5}$ yrs.  It is important to take this period
derivative into account when comparing spectroscopic and photometric
periods.

\begin{figure}
\begin{center}
\includegraphics[width=10cm]{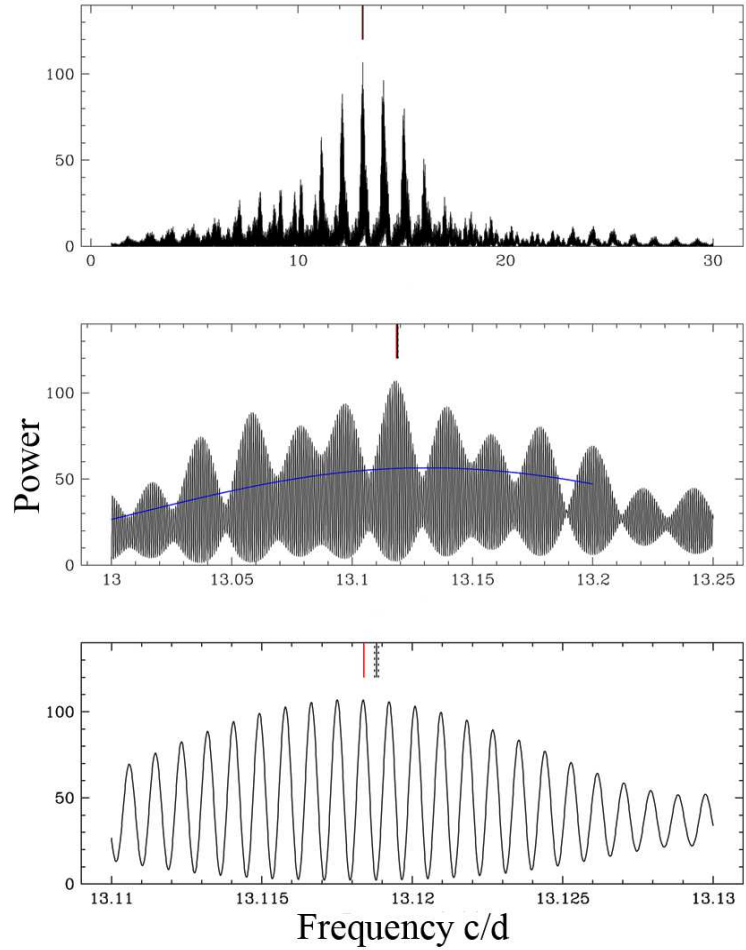}
\caption{\emph{Power spectrum of the HeII radial-velocity data, obtained from the
combined VLT and Magellan data sets. The two bottom panels are zooms of the panel above. The vertical grey line in the bottom panel represents the best period found by Patterson et al. 1998, along with its error. The red line represents the recalculated period using the latest photometric ephemeris (see Section~\ref{PE}) where the period derivative has been taken into account (Patterson et al. in preparation). }}
\label{fig:pow_tpyx}
\end{center}
\end{figure}


\subsubsection{The Spectroscopic Period}

In order to establish the orbital period spectroscopically, we carried
out a radial-velocity study using the double-Gaussian method~\cite{1980ApJ...238..946S}. This method is most effective in the line wings, which are formed in the inner regions of the accretion disc. As discussed further in Section~\ref{RV}, we tried a variety of Gaussian FWHM's, as well as a range of separations between the two
Gaussians.

 The results shown in this section are for a FWHM =  450 km\,s$^{-1}$
and a separation of 300 km\,s$^{-1}$ and are based on the full combined VLT +
Magellan data set, spanning the time period December 2004 to March
2008. The radial-velocity study was carried out for several spectral lines but for the purpose of this section, we will focus exclusively on the radial-velocities derived from the strong HeII line at 4686 \AA, since this line provides the most accurate results. The power spectrum (\citealt{1982ApJ...263..835S}) of this spectroscopic data set is presented in Figure~\ref{fig:pow_tpyx}. The top panel, covering a broad frequency range, confirms that there is a strong signal near the photometrically
estimated 13 c\,d$^{-1}$. No signal is found at $P=3.44$ hours, corresponding to the period found by~\cite{1990apcb.conf..391V}. The middle and bottom panels show close-ups of
narrower frequency ranges around this power spectral peak. Note, in
particular, that the red line in the bottom panel marks the
photometrically predicted frequency for the epoch corresponding to the
mid-point of our spectroscopic data set. For comparison, we also show
the frequency of the linear ephemeris determined by~\cite{1998PASP..110..380P} for just the 1996 -- 1997 timings, along with its error. The spectroscopic and photometric periods appear to be mutually
consistent, {\em but only if we account for the photometry period derivative}. More specifically, the  
$\dot{P}$-corrected photometric prediction lies extremely close to the
peak of the most probable spectroscopic alias. 

This apparent consistency can be checked more quantitatively. The
orbital frequency predicted by the photometry for the average
epoch of our spectroscopy is  $f_{\text{phot}} = 13.1183912 \pm  5.2 \times
10^{-6}$~c\,d$^{-1}$, where the error is once again based on bootstrap
simulations.  Similarly, we have carried out a bootstrap error
analysis for the frequency determined 
from our spectroscopy. Since we are only interested in the agreement
between the photometry and the spectroscopy, we only considered the
spectroscopic alias closest to the photometrically 
determined frequency (this is also the highest alias). This yielded
$f_{\text{spec}} = 13.118368 \pm 1.1 \times 10^{-5}$~c\,d$^{-1}$. The difference
between these determinations formally amounts to just under
2$\sigma$. We consider this to be acceptable agreement. 

The level of agreement between spectroscopic and
photometric periods is important. The spectroscopy on its own is sufficient to establish beyond doubt that T Pyx is a short-period system with $P_{\text{orb}} \simeq 1.8$~hrs. However,
the fact that we can establish reasonable consistency with the
photometric data if (and only if) we account for the photometric
period derivative suggests that (i) we can trust the much higher
precision photometric data to provide us with the most accurate
estimate of the orbital period, and (ii) that the large period
derivative suggested by the photometry is, in fact, correct. 

\subsection{Trailed Spectrograms and Doppler Tomography}

\begin{figure}
\centering
\subfigure[Magellan, HeII (4686\,\AA)] 
{
    \label{fig:sub:b1}
    \includegraphics[width=6cm]{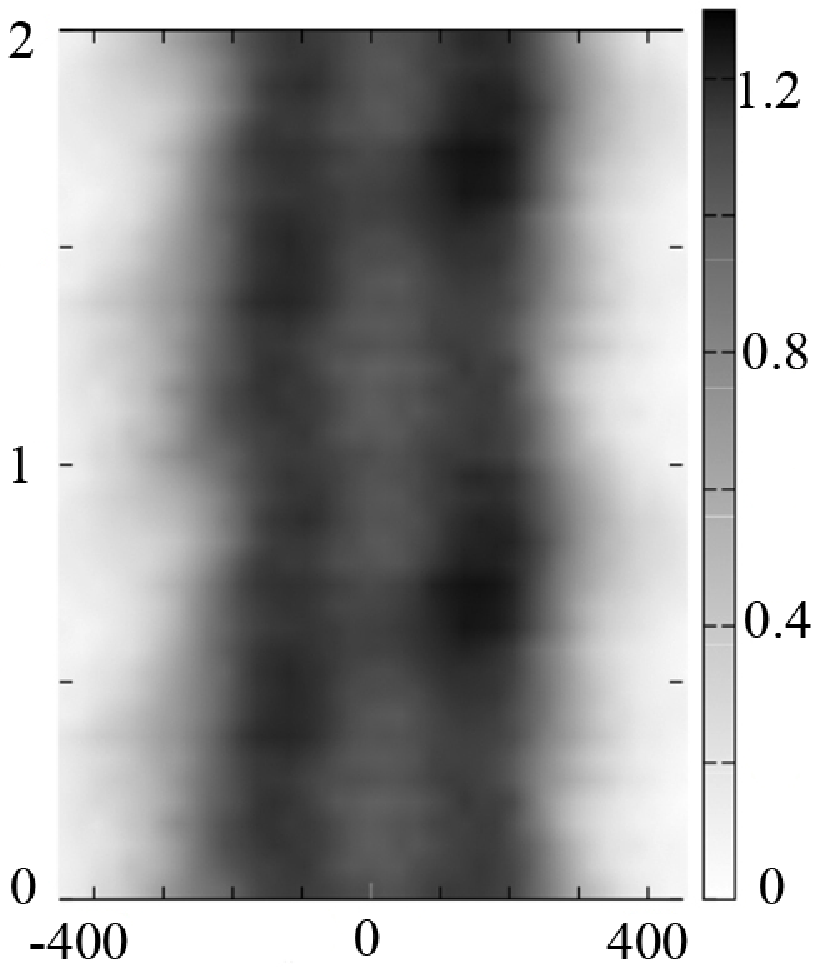}
}
\hspace{0.0cm}
\subfigure[ Magellan, HeI  (4921\,\AA)] 
{
    \label{fig:sub:c1}
    \includegraphics[width=6cm]{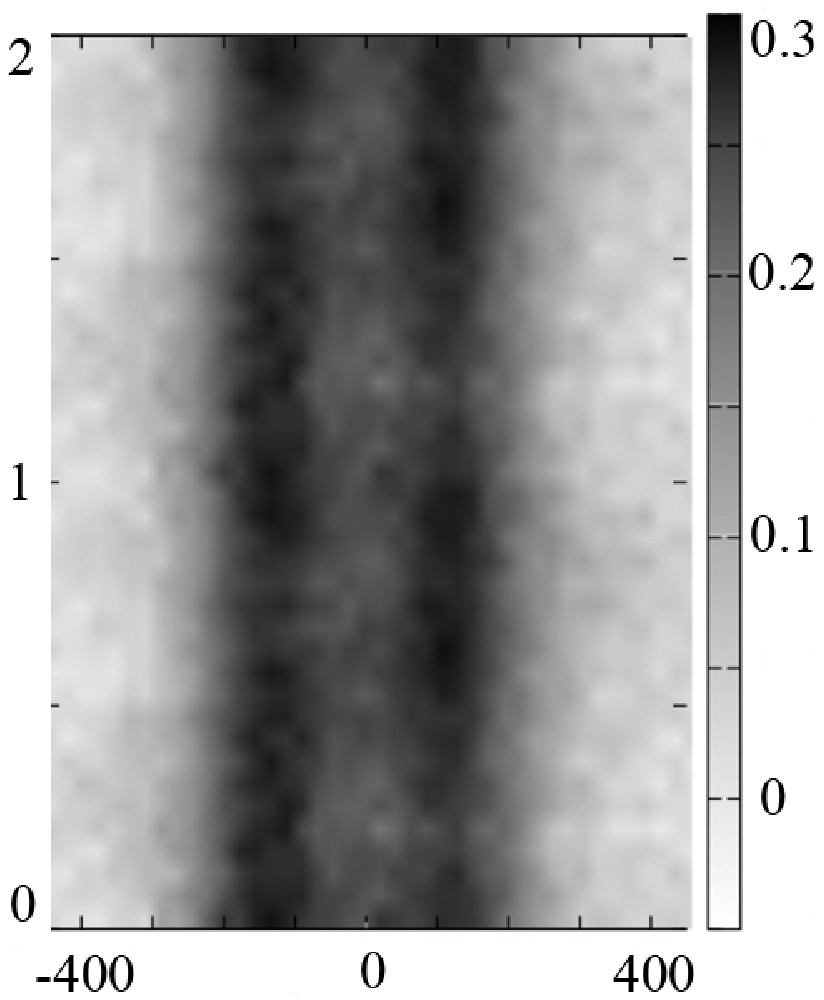}
}
\hspace{0.0cm}
\subfigure[VLT, HeII (4686\,\AA)] 
{
    \label{fig:sub:a1}
    \includegraphics[width=6.6cm]{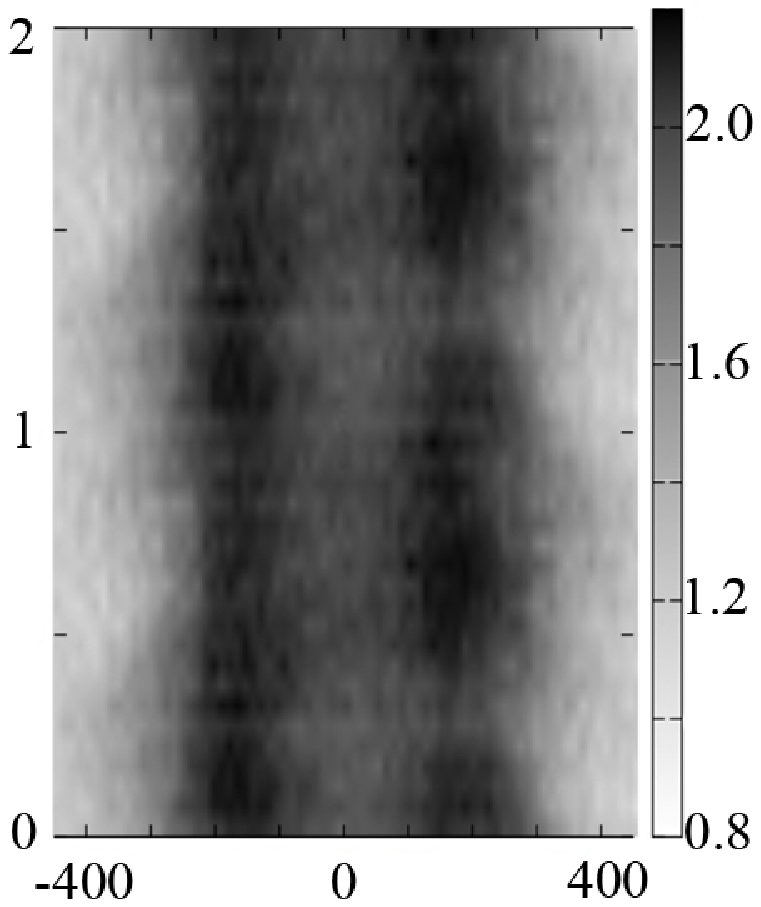}
}
\hspace{0.0cm}
\subfigure[VLT, H$\beta$ (4861\,\AA)] 
{
    \label{fig:sub:d1}
    \includegraphics[width=6cm]{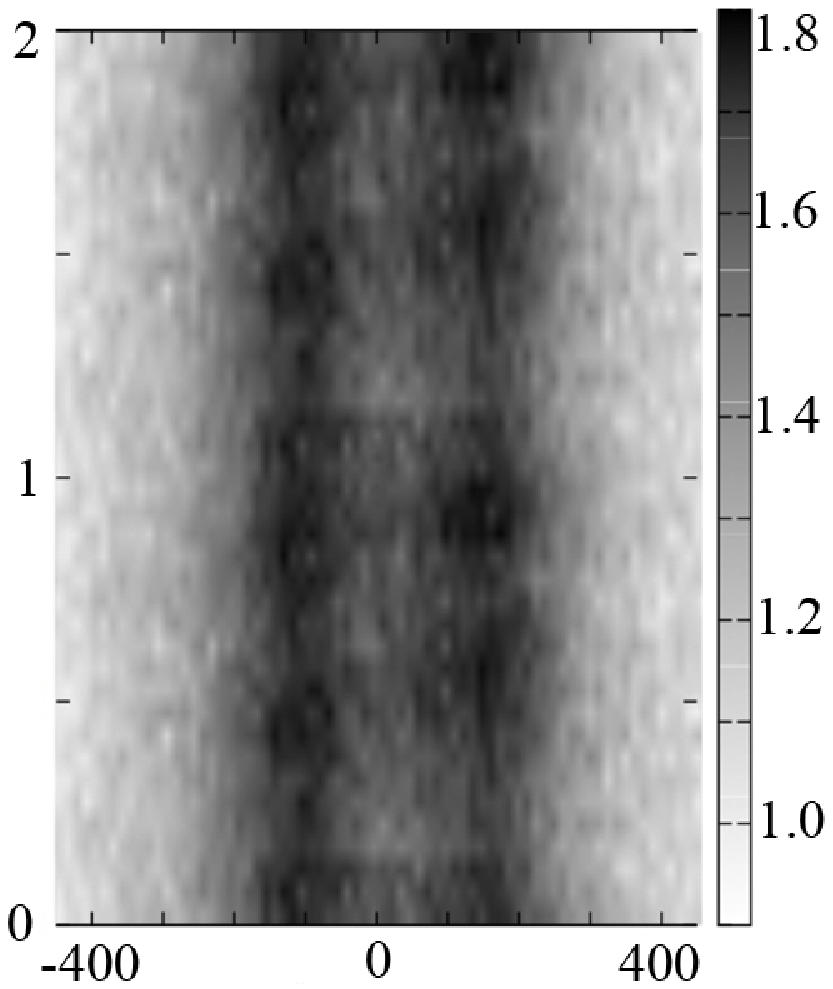}
}
\caption{\emph{Trailed spectrograms, phase binned to match the orbital resolution (37 bins for the VLT data and 27 bins for the Magellan data), are here plotted over 2 periods. Phases are plotted against velocities (in km\,s$^{-1}$).}} 
\label{fig:spec_trail} 
\end{figure}

In order to visualise the orbital behaviour of the spectral line profiles, 
trailed spectrograms were constructed for HeII, HeI, H$\beta$ and the
Bowen blend. These spectrograms, shown in Figure~\ref{fig:spec_trail}, are phase binned to
match the orbital resolution of our data (37 bins for the VLT data and
27 bins for the Magellan data) and are here plotted over 2
periods. All lines show emission features moving from the blue to the
red wing, and localised emission is seen in the red wing at phase, $\phi \approx
0.6$. Similar structures are seen in the spectrograms for both the
HeII and the Bowen blend indicating that they originate in the
same line-forming region, presumably the accretion disc.

\begin{figure}
\centering
\subfigure[Magellan, HeII (4686\,\AA)] 
{
 \label{fig:sub:a2}
\includegraphics[width=5.9cm]{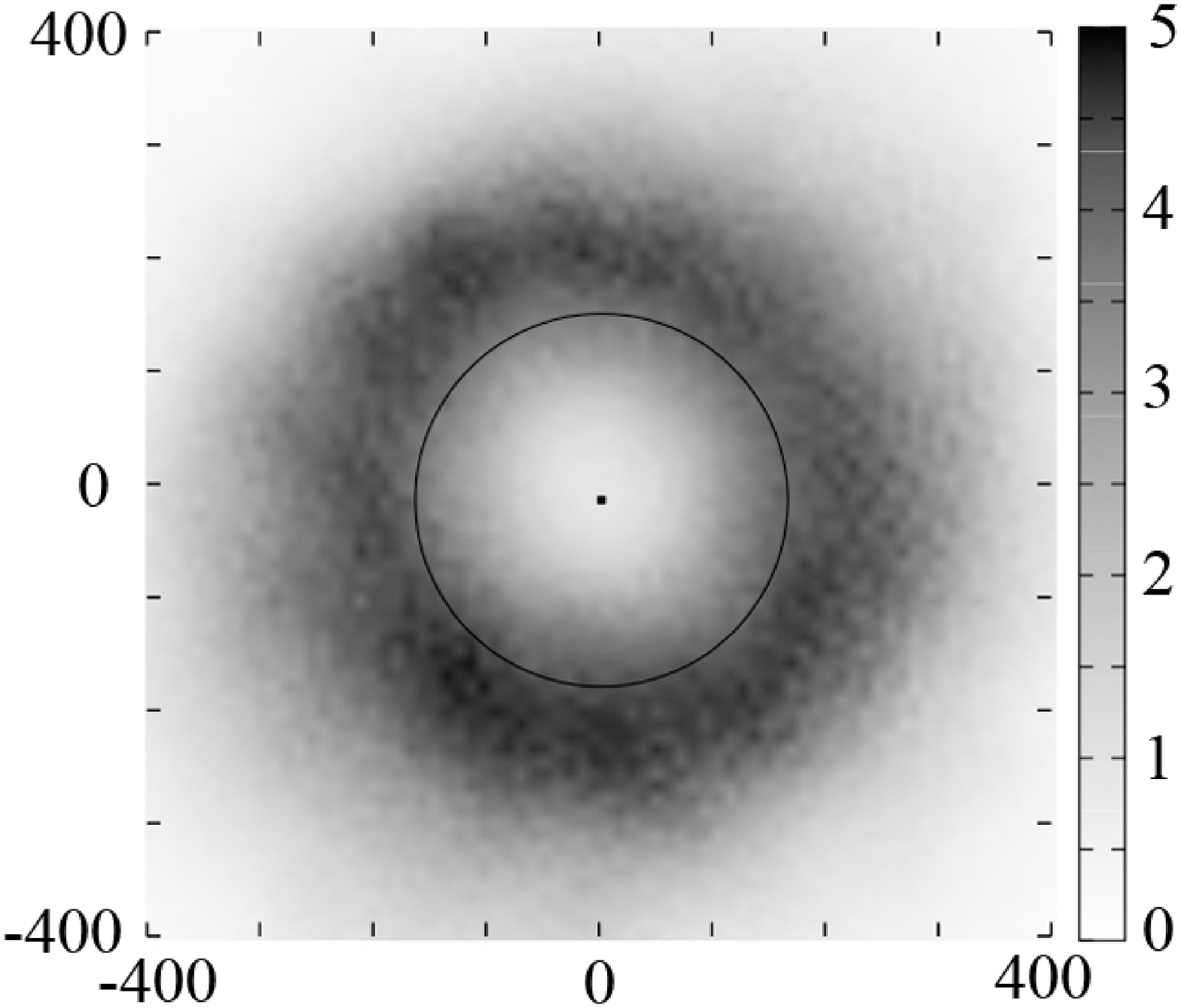}
}
\hspace{0.0cm}
\subfigure[Magellan, Bowen blend \newline (NIII 4641\,\AA\,and CIII 4650\,\AA, 4647\,\AA)] 
{
\label{fig:sub:b2}
\includegraphics[width=5.8cm]{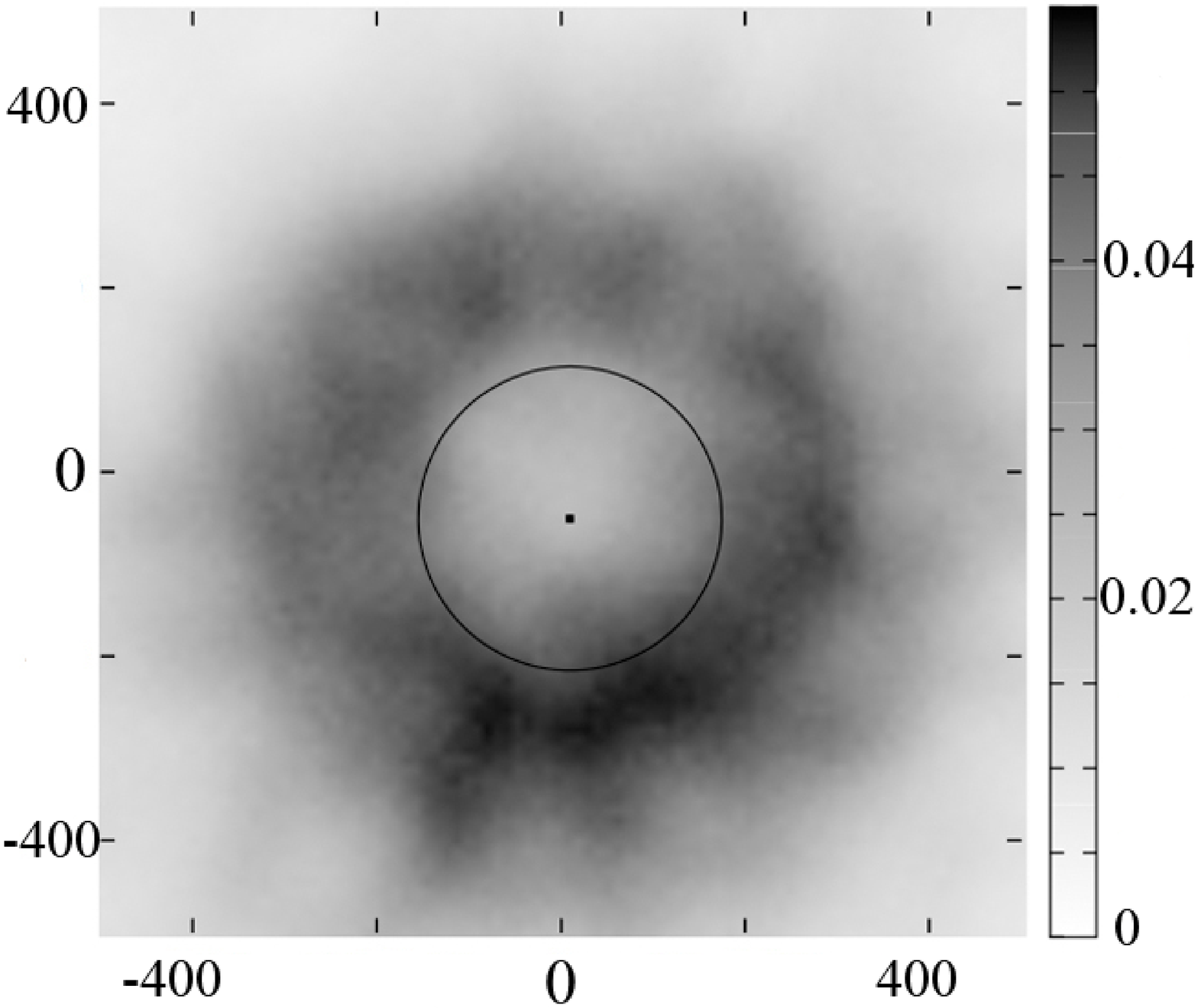}
}
\hspace{0.0cm}
\subfigure[Magellan, HeI (4921\,\AA)] 
{
\label{fig:sub:c2}
\includegraphics[width=5.8cm]{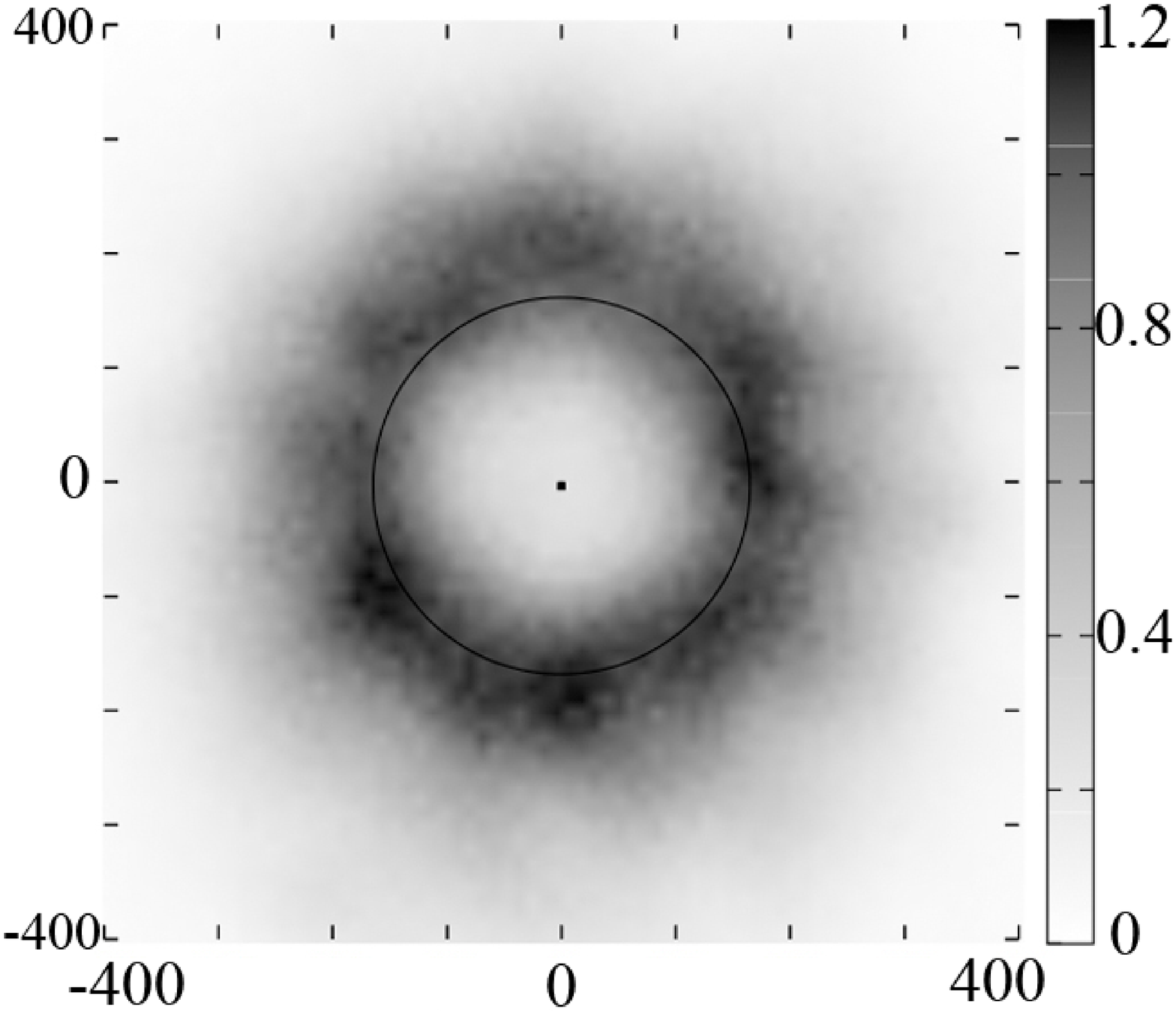}
}
\hspace{0.0cm}
\subfigure[VLT, HeII (4686\,\AA)] 
{
 \label{fig:sub:d2}
 \includegraphics[width=6.0cm]{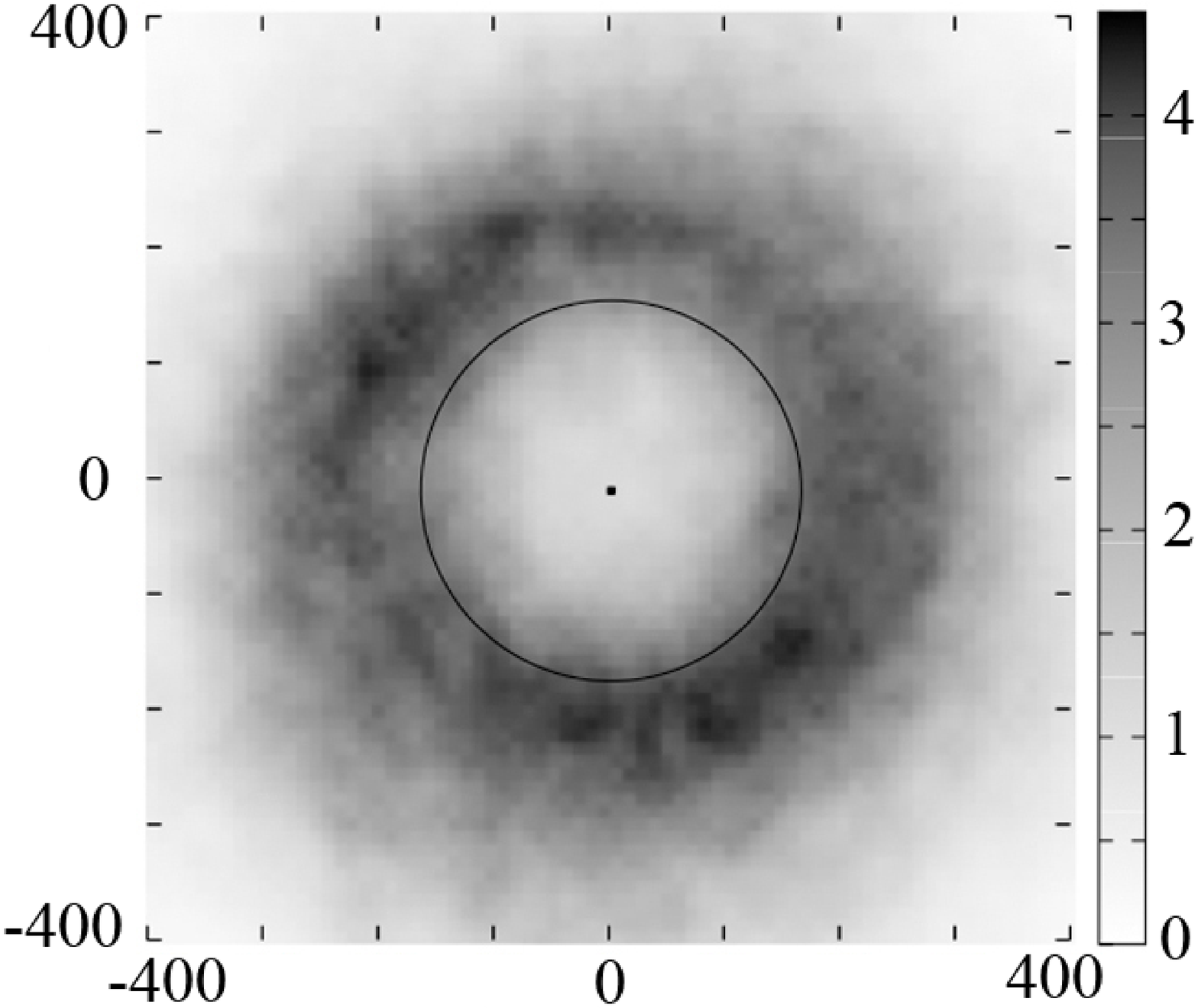}
}
\hspace{0.0cm}
\subfigure[VLT, Bowen blend \newline (NIII 4641\,\AA\,and CIII 4650\,\AA, 4647\,\AA)] 
{
\label{fig:sub:e2}
\includegraphics[width=5.8cm]{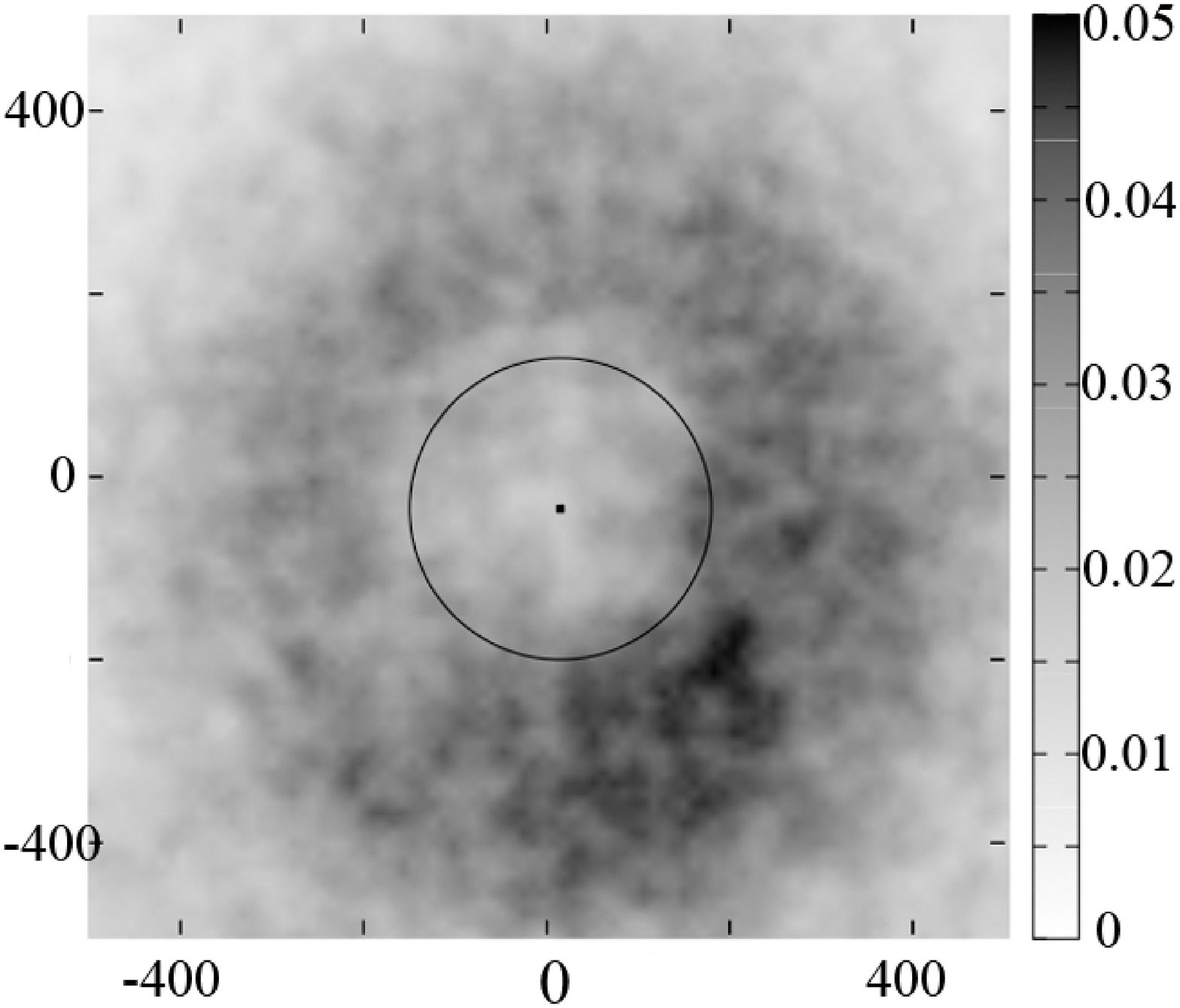}
}
\hspace{0.0cm}
\subfigure[VLT, H$\beta$  (4861\,\AA)] 
{
\label{fig:sub:f2}
\includegraphics[width=5.9cm]{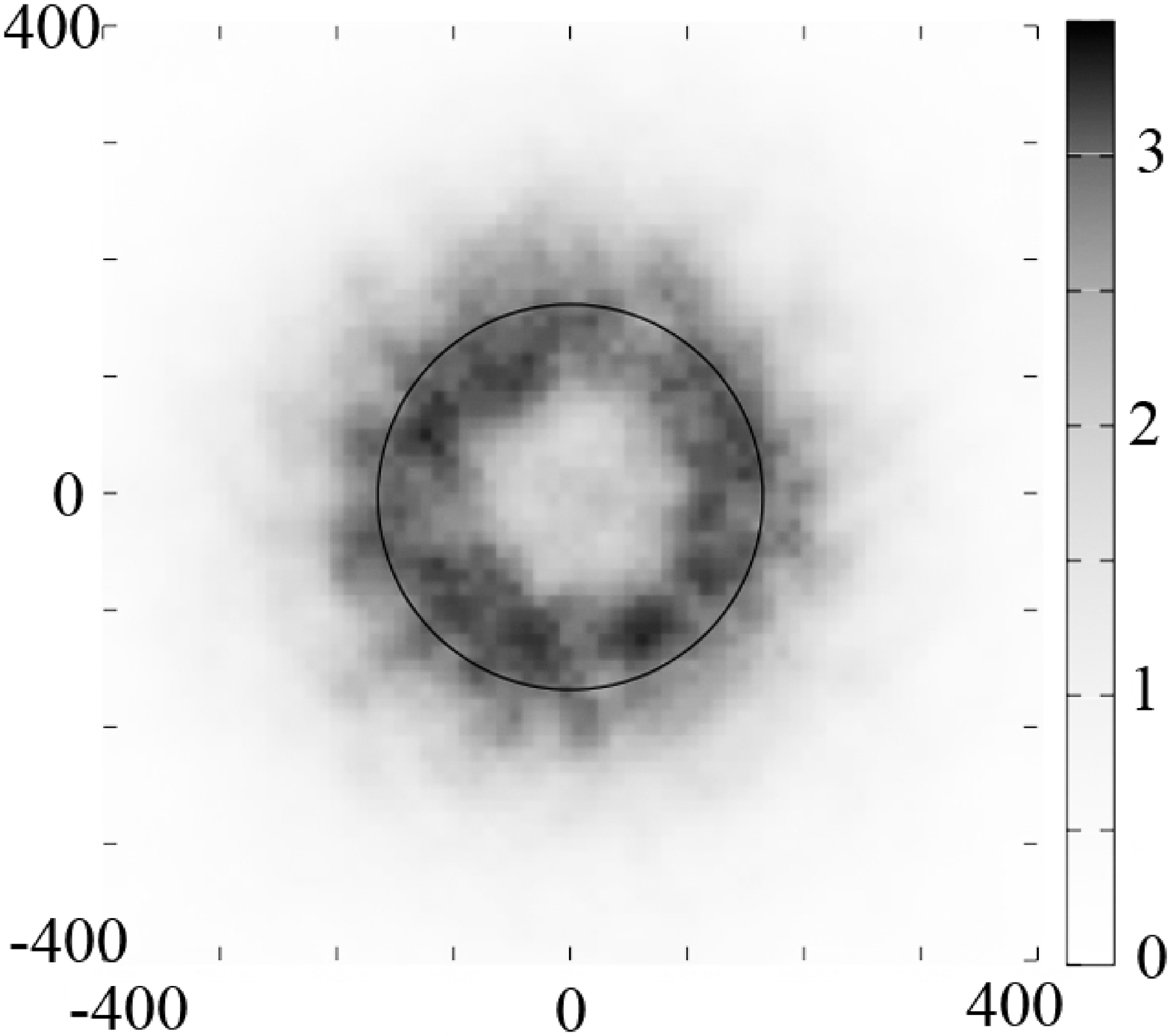}
}
\caption{\emph{Doppler tomograms for the most prominent lines. The center of symmetry in each map is marked as a dot and the outer disc radius is plotted as a circle. Both axis are in km\,s$^{-1}$.}} 
\label{fig:tom} 
\end{figure}

Doppler tomography (\citealt{1988MNRAS.235..269M}) is an indirect imaging
method where the emission-line profiles, depending on phase, are
plotted onto a velocity scale (see Section~\ref{doppler}). The method was used to visualise, in
velocity space, the origin of the emission features found in T Pyx. In
 Figure~\ref{fig:tom}, Doppler tomograms for the most prominent lines are
presented. For the reconstruction of the tomograms, the systemic velocity $\gamma$ was set to zero (see Section~\ref{SV}). All lines show asymmetrically distributed emission from the disc. These asymmetric features are also seen in the trailed spectrograms (Figure~\ref{fig:spec_trail}). No emission can be connected to the bright spot, the accretion stream or the donor star. The map of the Bowen blend was constructed from a composite of three lines, NIII at 4640.64\,\AA\, and CIII at 4650.1\,\AA\, and 4647.4\,\AA\,.

\subsection{A Radial-Velocity Study of T Pyx} \label{RV}  \label{vsa}

\subsubsection{The Systemic Velocity} \label{SV}

We noticed early on in our analysis for both the VLT and Magellan data
sets that the systemic velocity, $\gamma$, appeared to be varying with 
time. In order to confirm the reality of these variations, we
extracted spectra from a few VLT/GIRAFFE/FLAMES fibres that were
centred on knots in the nova shell surrounding T Pyx. The intrinsic
radial velocities of these large scale knots are not expected to change
appreciably over a time-scale of months, so they provide a useful check
on the stability of our wavelength calibration. We found that the
shifts seen in the radial velocity data for the central object are
much larger than those seen for the  knots, suggesting that these shifts cannot be explained by instrumental effects alone. However, since we have only five nights
worth of VLT data  spread over two months, our data set is too sparse
to allow a careful study of trends in the apparent $\gamma$
velocities. For the purpose of the present study, we thus subtracted
the mean nightly $\gamma$ from all radial velocities before carrying
out any analysis. This should minimise the risk of biasing our
results. 

\subsubsection{The Velocity Semi-Amplitude}

\begin{figure}
\begin{center}
\includegraphics[width=10cm]{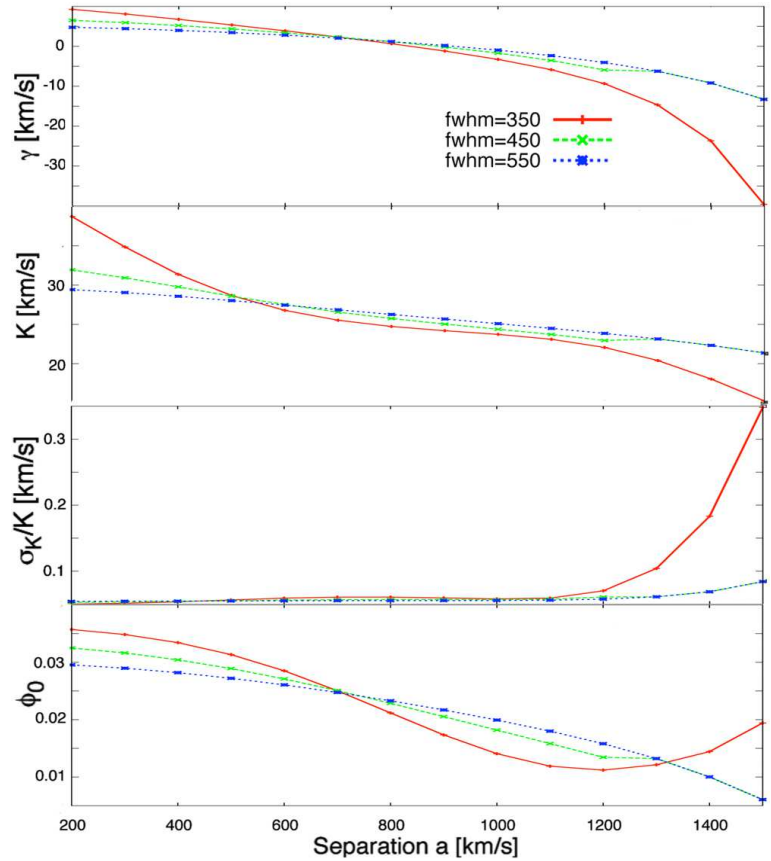}
\caption{\emph{Diagnostic diagram for the HeII line, Magellan. None of the
  parameters ($\gamma$, $K$ and $\phi_{0}$) are converging for any
  combination of the FWHM and separation.}} 
 \label{fig:dd} 
 \end{center}
\end{figure}

In order to establish the velocity semi-amplitude, $K_{1}$, of the WD,
further analysis was pursued on the radial-velocity data obtained from
the double-Gaussian method. A least-squares fit was made to the radial
velocity curves by keeping the orbital period fixed, but allowing the
three parameters, the systemic velocity, $\gamma$, the velocity
semi-amplitude, $K_{1}$, and the phase, $\phi_{0}$, to vary. Phase zero
corresponds to the photometric phase of minimum light. The errors on the input radial velocity data, provided in MOLLY, were used to calculate the $\chi^{2}$ and then
rescaled so that the reduced $\chi^{2}_{\nu} = 1$. The technique was applied to several
strong lines in both the VLT and Magellan data sets, but the radial
velocity data measured from the HeII line at 4686 \AA, and in
particular from the Magellan dataset, gave the most reliable fits. The
results were plotted in a diagnostic diagram (see~\citealt{1983ApJ...267..222S, 1986AJ.....91..940T} and~\citealt{1986MNRAS.219..791H}). However, as can be seen in Figure~\ref{fig:dd}, the key parameters ($\gamma$, $K_{1}$ and $\phi_{0}$) did not converge convincingly for any combination of the FWHM and separation. Normally, it is thought that the bright spot or other asymmetries in the disc tend to be responsible for distorting the diagnostic diagram. However, in T Pyx, the disc appears to be rather symmetric, and no contribution from the bright spot is seen (see Figure~\ref{fig:tom}). We conclude that the radial velocity curves obtained from the double-Gaussian method are, most likely, not tracking the velocity of the primary WD.Thus, $K_{1}$, and $\phi_{0}$ cannot be estimated reliably using this method.


\begin{figure}[t]
\begin{center}
\includegraphics[width=12cm]{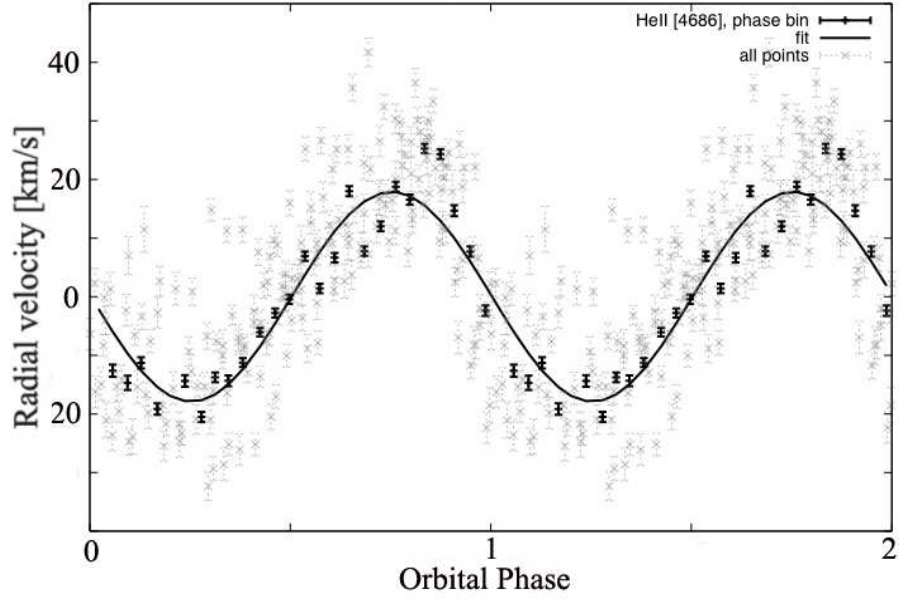}
\caption{\emph{Radial velocity curve of the HeII line, Magellan. Data in black is phase binned while all data are shown in grey. The best sin fit is over-plotted onto the phase-binned data.}} 
\label{fig:rv} 
\end{center}
\end{figure}

\begin{figure}
\begin{center}
\includegraphics[width=14.5cm]{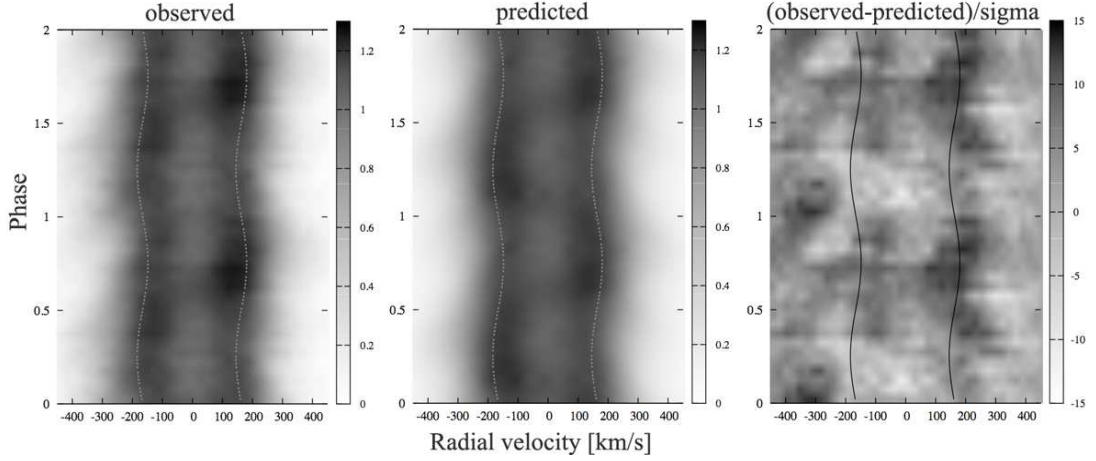}
\caption{\emph{Trailed spectrograms for the HeII line, Magellan.}
\label{fig:trail_heII}}
\end{center}
\end{figure}

\noindent In order to improve on the results obtained from the
double-Gaussian method, we used two other techniques to obtain an
estimate of $K_{1}$. First, we measured the velocity centre of the lines
by fitting Gaussians to the individual spectral lines. Two Gaussians
were fitted to the double-peaked lines, keeping the FWHM and the rest
wavelength of the line fixed, but allowing the peak offsets and the
peak strengths to vary. Radial velocity curves were then plotted for
both blue and red wings. As can be seen from the trailed spectrograms,
Figure~\ref{fig:spec_trail}, the red wing (in particular for the HeII
lines), does not show the same smooth orbital signal as the blue
wing. Consequently, only the data from the blue wing was taken into
account when measuring $K_{1}$. Several lines from both the VLT and
Magellan data sets were analysed. The data were phase binned to the
orbital resolution, velocity-binned in a $\sim$ 100 pixel wide region around
the line, and the nightly mean $\gamma$ was subtracted. HeII at
$\lambda$4686 is the strongest line and was ultimately used to
estimate our best-bet value of the velocity semi-amplitude of the WD:
$K_{1} = 17.9 \pm 1.6$ km\,s$^{-1}$. Figure~\ref{fig:rv} shows the
radial velocities obtained for the HeII line in the Magellan data
set. Data presented in black are phase-binned to the orbital
resolution afforded by our time-resolved spectroscopy, while all
individual data points are shown in grey. Over-plotted onto the radial
velocity curve is the best sinusoidal fit, indicating a phasing close
to zero, $\phi_{0} = -0.03 \pm 0.03$. Figure~\ref{fig:trail_heII}
shows the final trailed spectrograms from the HeII at $\lambda$4686 in
the Magellan data set that was used to determine $K_{1}$.

Second, $K_{1}$ can also be measured from the Doppler tomograms, by locating the centre of symmetry of the disc emission that dominates the maps (this method is described in Section~\ref{svvsa}). This procedure yielded estimates for $K_{1}$ that were consistent with those obtained from the Gaussian fits ($14 < K_{1} < 19$ km\,s$^{-1}$). We thus adopt $K_{1} = 17.9 \pm 1.6$ km$\,s^{-1}$ as our best-bet estimate.

In closing this section, it should be acknowledged that our final
estimate of $K_{1}$ is subject to considerable systematic
uncertainties. In particular, it is difficult to rule out that
whatever is preventing the double-Gaussian method from converging may
also bias the Gaussian fits to the line peaks and the centre of
symmetry of the tomograms. Strictly speaking, our estimate of $K_{1}$
should thus perhaps be viewed as a lower limit. However, we are
reasonably confident that our measurements do trace the orbital motion
of the WD in T Pyx. This is mainly because the presence of persistent
double-peaked lines and ring-like structure in the tomograms implies
that there is a strong accretion disc component to the line
emission. This disc component must ultimately trace the motion of the
WD, and our two preferred methods are designed to exploit this in as
direct a fashion as possible, while simultaneously minimising
contamination from asymmetrically placed emission regions.


\subsubsection{The Velocity at the Outer Disc Radius}

The peak-to-peak separations of the double peaked emission lines can
be used to estimate the projected velocity at the outer disc radius,
$v_{\text{R}_{\text{disc}}} \!\sin i$. We thus fitted Gaussians to the phase-binned, double-peaked H$\beta$, HeI and HeII lines in the Magellan data set. The peak-to-peak separations varies significantly with line, but the mean value is close to that for HeI. We therefore use the HeI peak-to-peak separation as our best-bet estimate
of $v_{\text{R}_{\text{disc}}} \!\sin i$ and adopt half the full range as a rough estimate of the associated error, $\Delta\!V_{\text{peak-to-peak}} = 290\,\pm\,26$ km\,s$^{-1}$. Taking the relation by ~\cite{1981AcA....31..395S} into account (see Section~\ref{vodr}), we find $v_{\text{R}_{\text{disc}}} \!\sin i  = 138 \pm 15$ km\,s$^{-1}$.


\section{Spectroscopic System Parameters}

If we are willing to assume that our estimates of $K_{1}$ and $v_{\text{R}_{\text{disc}}} \!\sin i $ are valid, we can use our data to estimate several other system parameters. 

\subsection{Mass Ratio}

The method we use to estimate the mass ratio, $q$, is similar to that
described in~\cite{1973MNRAS.162..189W}. The basic idea is that, subject to fairly
benign assumption, the ratio $v_{\text{R}_{\text{disc}}} \!\sin i/ K_{1}$ should be a
function of only $q$. More specifically, we assume that the disc is
circular, Keplerian, and, since T Pyx is a high-$\dot{M}$\ system,
that it extends all the way to the tidal radius (Equation~\ref{mod_warner_1973_2}).
It is then possible to show (see Appendix A for a detailed derivation of our modified relation) that

\begin{equation}
\frac{v_{\text{R}_{\text{disc}}} \!\sin i}{K_{1}}= 1.2909 \frac{1+q}{q}.
 \label{mod_warner_1973_2}
\end{equation}

\noindent Based on this, we estimate a mass ratio of $q = 0.20 \pm 0.03$.
Given a mass ratio and disc radius, it is possible to estimate the
expected phasing of the bright spot in the system, if we make the
usual assumption that the bright spot lies at the point where the
ballistic accretion stream from the $L_1$ point meets the disc
edge (see, for example,~\citealt{2008PASP..120..510P, 2000PASP..112.1584P}). Carrying out this calculation for T Pyx suggests that the bright spot should be located at orbital phase $\phi \simeq -0.03$. This is interesting, because it is consistent with the 
negligible phase offset we have found between photometric minimum
light and the red-to-blue crossing of the WD radial velocity
curve. Thus, based on phasing alone, the bright spot could be the
source of the photometric modulation. However, there are also other
possible interpretations for the photometric signature. For example, the
orbital signal could be dominated by a reflection effect, i.e. the
changing projected area of the irradiated face of the donor as it
moves around the orbit. This would explain why inferior conjunction of
the secondary corresponds to {\em minimum} light. On the other hand, 
it raises the question why we do not see the signature of the irradiated
donor in our spectroscopy. As noted above, there is no sign of the
narrow Bowen fluorescence lines that one might expect in the case of
strong donor irradiation. 

\begin{figure}
\begin{center}
\includegraphics[width=9cm]{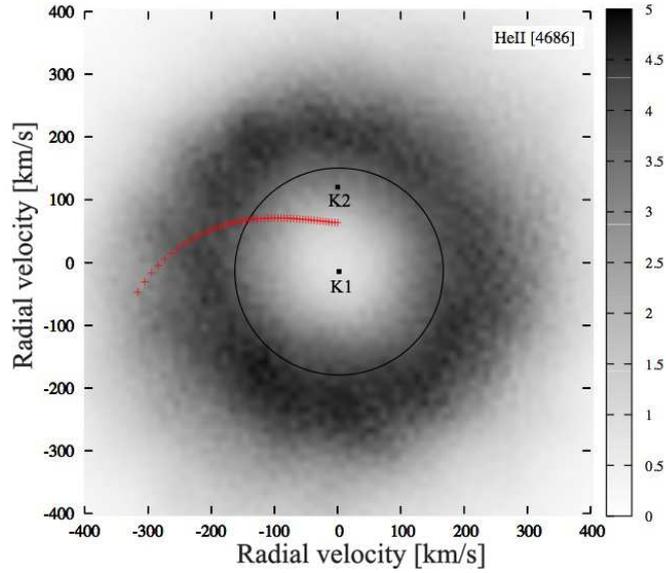}
\caption{\emph{Doppler tomogram for the HeII line, Magellan. The location of the WD, $K_{1}$, the outer disc radius as well as the location of the donor star, $K2$, and the accretion stream is plotted onto the map.}}
\label{fig:dopp_heII} 
\end{center}
\end{figure}

\subsection{Component Masses}

Recurrent novae are expected to harbour a more massive WD than most
ordinary CVs and even than most classical novae. In a classical nova, the mass of the WD is predicted theoretically to be $\sim$ 1 M$_{\odot}$ (\citealt{2005ApJ...623..398Y}), which is also seen observationally. Based on this, both~\cite{2008A&A...492..787S} and ~\cite{2010ApJ...708..381S} agree
on a plausible mass range for the primary WD in T Pyx of 1.25 M$_{\odot}$ -- 1.4 M$_{\odot}$ (using
theoretical models by~\citealt{2005ApJ...623..398Y}). From the theoretical point
of view, a high-mass WD is needed to achieve a sufficiently short
outburst recurrence timescale.

In T Pyx, the bright accretion disc outshines any spectral signature
from the donor star, and we can therefore not constrain the donor mass
spectroscopically. However, we can use the period-density relation for
Roche-lobe filling secondary stars~\cite{1983ASSL..101..239E} to set some
constraints. 

As shown by~\cite{2005PASP..117..427P} and~\cite{2006MNRAS.373..484K}, the donor stars
in ordinary CVs below the period gap are inflated by approximately 10\%
due to mass loss, relative to ordinary main sequence stars of the same
mass. Given T Pyx's peculiar evolutionary state, it is not clear if
this level of inflation is appropriate for T Pyx, and we therefore
adopt a conservative range of 0\% $-$ 20\% inflation. In order to
estimate the mass of the donor star, we thus take the 
theoretical main-sequence mass-radius relation from the 5 
Gyr isochrone of~\cite{1998A&A...337..403B}, adjust the stellar radius to
account for inflation, and then find the secondary mass that yields
the correct density for T Pyx's orbital period. This yields $M_{2} =
0.14 \pm 0.03$ M$_{\odot}$. Note that the inferred donor mass decreases
with increasing levels of radius inflation. The corresponding mass of
the WD is then $M_{1} = 0.7 \pm 0.2\, $M$_{\odot}$. 

Since this estimate of the WD mass is lower than expected for a
recurrent nova, we can also turn the problem around. If the mass of
the WD is $ > 1 $ M$_{\odot}$, as theoretical nova models imply~\cite{2005ApJ...623..398Y},
the mass of the donor becomes  > $0.2 \,$M$_{\odot}$, if our
estimate of the mass ratio is correct. 

All of these estimates should, of course, be taken with a grain of
salt. In particular, they rely on (i) the correctness of our measured
$K_{1}$ and (ii) the assumption that the disc extends all the way to the
tidal radius. Despite these uncertainties, our calculations highlight
an important point: it is highly unlikely that the donor star in T Pyx
is already a brown dwarf. For example, if we retain the assumption of
maximal disc size,  $K_{1} \leq 7$~km\,s$^{-1}$ would be required in order for
$M_{2} \leq 0.07$~M$_{\odot}$ for any $M_{1} > 1~$M$_{\odot}$. We thus rule out
the possibility that T Pyx is a period bouncer in the usual sense,
i.e. that it had already reached the minimum period for ordinary CVs
($\simeq 70$~min theoretically, or $\simeq 80$~min observationally),
in which case its secondary would now be well below the
Hydrogen-burning limit.\footnote{T Pyx, is, of course, a ``period
bouncer'' in the basic sense that its orbital period is currently
increasing.} 

\subsection{The Orbital Inclination}

The orbital inclination for T Pyx is thought to be low due to the
sharp spectral profiles and low radial velocity amplitude.~\cite{1997ApJ...484L..59S} suggested a lower limit of the orbital inclination for T Pyx of
$i \sim 6^{\circ}$ based on the peak-to-peak separation of the
H$\alpha$ line in their spectra.~\cite{1998PASP..110..380P} estimated an
inclination of $i \sim 10^{\circ} - 20^{\circ}$ due to the low amplitude of the
orbital signal, and~\cite{2008A&A...492..787S} estimated $i \sim 20^{\circ} - 30^{\circ}$.  
 
\begin{table}[t]
\begin{center}
 \begin{tabular}{ll}
\hline
\hline
$K_{1} = 17.9 \pm 1.6$ km$\,s^{-1}$ & \emph{velocity semi-amplitude}\\
$v\sin i = 138 \pm 15$ km$\,s^{-1}$ &  \emph{velocity at the outer disc radius}\\
$q = 0.20 \pm 0.03 $ &  \emph{mass-ratio (M$_{2}$/M$_{1}$)}\\
$M_{2} = 0.14 \pm 0.03$ M$_{\odot}$ &  \emph{donor mass} \\
$M_{1} = 0.7 \pm 0.2$ M$_{\odot}$ &  \emph{WD mass}\\
$i = 10 \pm 2$ degrees &  \emph{inclination} \\
\hline
\hline
\end{tabular}
\vspace{0.5cm}
\caption{\emph{The spectroscopic system parameters for T Pyx are presented here, where $K_{1}$ is found from the amplitude of the radial velocity curve of the HII line at 4686\,\AA\,, and $v\sin i$ is estimated from the peak-to-peak separation of the HeI line at 4921\,\AA. The mass-ratio $q$ is calculated from $K_{1}$ and $v\sin i$, and $i$ is calculated from $q$. Component masses were estimated from $q$ and from the mass-radius relation accounted for an inflated donor radius.}}
\label{tab:par_tpyx}
\end{center}
\end{table}


The inclination, $i$, can be estimated from our data via $K_{1} =
v_{1}\sin i$, where $v_{1} = (2 \pi a)/P\,(q/(q+1))$ and $a$ is the distance
between the two components obtained from Kepler's third law. This
provides a constraint on the system inclination of $i = 10 \pm 2$
degrees, for any reasonable combination of the component masses. We
stress that the real uncertainty on the inclination is bound to be
larger, because the formal error does not include systematic
uncertainties associated with, for example, possible bias in our $K_{1}$
measurement and the assumption of tidal truncation in our derivation
of $q$. 
The complete set of spectroscopic system parameters for T Pyx is
summarised in Table~\ref{tab:par_tpyx}.
\newpage

\section{Summary and Discussion}  \label{SD}

The main result of our study is the spectroscopic determination of T
Pyx's orbital period, $P_{\text{orb}} \simeq 1.83$~hours. This confirms that
the system is a CV below the period gap and implies that its current
accretion rate is at least 2 orders of magnitude higher than that of
an ordinary CV at this period. We also find that our spectroscopic
orbital period is consistent with the photometric ephemeris found for
T Pyx (\citealt{1998PASP..110..380P}, an updated version is given in
Section~\ref{PE}). This means not only that photometric timings can be
used as a more precise and convenient tracer of the orbital motion,
but also that the large period derivative required by the photometric
ephemeris marks a genuine change in the orbital period of the system.
In fact, the spectroscopic data are consistent with the photometric
period only if the period derivative is accounted for. The period
derivative obtained here from the latest combined photometric data
($\dot{P} = 6.7 \times 10^{-10}$) is slightly lower than that obtained
by~\cite{1998PASP..110..380P} from data up to 1997 ($\dot{P} \simeq 9 \times
10^{-10}$). A decrease in the rate of period change would be in line with~\cite{2010ApJ...708..381S} scenario
that T Pyx's days as a high-$\dot{M}$ recurrent nova are numbered (at
least until its next ordinary nova eruption). In any case, the average
time-scale for period change found across all the photometric data is about $3 \times 10^5$~yrs.

We have also used our spectroscopic data to obtain estimates of other
key system parameters, most notably the radial velocity semi-amplitude
of the WD, $K_{1} = 17.9 \pm 1.6$ km\,s$^{-1}$, and the mass ratio, $q = 0.20 \pm
0.03$. The latter estimate rests on three key assumptions: first, that our
determination of $K_{1}$ is correct, second, that our estimate
$v(R_{\text{disc}})\sin i$ is correct, and, third, that the accretion disc
around the WD in T Pyx is tidally limited. Taken at face value, this
relatively high value of the mass ratio implies that the donor star in
the system is not a brown dwarf. Thus T Pyx is not a period
bouncer. If we assume that the radius of the secondary is $10\% \pm
10\%$ inflated relative to an ordinary main sequence star of the same
mass, we find that its most likely mass is $M_{2} = 0.14 \pm 0.03$
M$_{\odot}$.

Overall, the physical picture that emerges from our study is
consistent with the scenario proposed by~\cite{2010ApJ...708..381S}. In particular,
they suggest that, prior to the 1866 eruption, T Pyx was an ordinary
CV. That eruption then triggered a  high-$\dot{M}$ wind-driven phase,
as suggested by~\cite{2000A&A...364L..75K} to account for T Pyx's exceptional
luminosity. However, this phase is not quite self-sustaining, so that
T Pyx is now fading and perhaps not even a RN anymore. In line with
this picture, we find that the mass ratio and donor mass we derive are
not abnormally low for a CV at its orbital period. This shows that the
present phase of high-$\dot{M}$ accretion cannot have gone on for too
long already. The hint of a declining period derivative may point in
the same direction, but this needs to be confirmed. 

Does all this mean that the phase of extraordinarily high accretion
rates T Pyx is currently experiencing will have no lasting impact on
its evolution? Not necessarily. In a stationary Roche-lobe-filling
system, the orbital period derivative and total mass-loss rate from
the donor are related via Equation~\ref{p_min}.
In T Pyx, which has an increasing orbital period, we use a mass-radius index $\alpha \simeq -1/3$ (see Section~\ref{pmin}), which is appropriate for adiabatic mass-loss from a fully convective star
(e.g.~\citealt{2000A&A...364L..75K}). We thus expect that $\dot{M_2}/M_2 \simeq
{\dot{P}_{\text{orb}}}/{P_{\text{orb}}}$, which suggests a typical mass-loss rate
from the donor of $\dot{M_2} \sim 5\times10^{-7}$~M$_{\odot}$/yr in its current
state. (The accretion rate onto the WD can be lower than this, since,
in the wind-driven scenario, much of this mass escapes in the form of
an irradiation-driven outflow from the donor.)  If every ordinary nova
eruption in T Pyx is followed by $\sim$ 100 yrs of such
high-$\dot{M}_2$ evolution, the total mass loss from the donor in the
luminous phase is $\sim 5 \times 10^{-5}$ M$_{\odot}$ between any two
such eruptions. This needs to be compared to the mass lost from the
donor during the remaining part of the cycle. If this is driven by
gravitational radiation, the mass loss rate from the donor will be
$\dot{M}_2 \simeq 5 \times 10^{-11}$~M$_{\odot}$/yr~\cite{2000A&A...364L..75K}. The
recurrence time of ordinary nova eruptions for such a system is on the order of $10^5$ yrs~\cite{2005ApJ...623..398Y}, so the total
mass lost from the donor during its normal evolution (outside the
wind-driven phase) is $5 \times 10^{-6}$ M$_{\odot}$. {\em This shows
that the long-term secular evolution may be dominated by its
high-$\dot{M}$ wind-driven phases, even if the duty cycle of these
phases is very low (e.g. 0.1\% for the numbers adopted above).}

It is finally tempting to speculate on the relevance of \emph{T Pyx-like}
evolution for ordinary CVs. At first sight, the numbers above suggest
that the evolution of a CV caught in such a state may be accelerated
by about an order of magnitude. This is interesting, since it could
bear on the long-standing problem that there are fewer short-period
CVs and period bouncers in current samples than theoretically expected
(e.g.~\citealt{1998PASP..110..380P, 2007MNRAS.374.1495P, 2007MNRAS.382.1279P, 2008MNRAS.385.1471P, 2008MNRAS.385.1485P}) Moreover, this channel need not be limited to systems containing high-mass WDs that are capable of
becoming RNe. After all, it is not the recurrent nova outbursts that
are of interest from an evolutionary point of view, but simply the
existence of a prolonged high-$\dot{M}$ phase in the aftermath of nova
eruptions. In CVs containing lower-mass WDs, such a phase may still
occur, although its evolutionary significance could still depend on
the WD mass. For example, the inter-outburst time-scale is longer for
low-mass WDs, and the duration of the high-$\dot{M}$ phase could also
scale with WD mass. One obvious objection to this idea is that, observationally, most nova
eruptions are not followed by centuries- (or even decades-) long
high-$\dot{M}$ phases. However, this need not be a serious issue. Most
observed novae are long-period systems, so if the triggering of a
wind-driven phase requires a fully convective donor, most novae would
not be expected to enter such a phase. It may be relevant in this
context that at least one other short-period nova -- GQ Mus --
exhibited an exceptionally long post-outburst supersoft phase of $\sim
10$~years~\cite{1993Natur.361..331O}.~\cite{2003A&A...405..703G} have also
suggested that the duration of the supersoft phase in novae may scale
inversely with orbital period. 

However, there is another important consequence to the idea that the
evolution of many short-period CVs is accelerated by \emph{T Pyx-like}
high-luminosity phases. The orbital period-derivative is positive in
the high-$\dot{M}$ phase, but negative during the remaining times of
GR-driven evolution. But since $\dot{P}_{\text{orb}}/P_{\text{orb}} \simeq
\dot{M}_2/M_2$ in {\em both} phases, the sign of the secular
(long-term-average) period derivative will generally correspond to the
phase that dominates the secular evolution. So whenever wind-driving
dramatically accelerates the binary evolution, the direction of period
evolution will also be reversed. Is this a problem? Perhaps. The
recent detection of the long-sought {\em period spike} in the
distribution of CV orbital periods~\cite{2009MNRAS.397.2170G} suggests
that there is, in fact, a reasonably well-defined minimum period for
CVs, as has long been predicted by the standard model of CV evolution
(e.g.~\citealt{1993A&A...271..149K}). On the other hand, the location of the observed
spike ($\simeq 83$~mins) is significantly different from the expected
one ($\simeq$ 65 min -- 70~min). Is it possible that the observed period
minimum corresponds to the onset of wind-driving in most CVs? T Pyx,
with $P_{\text{orb}} \simeq 110$~min would then have to be an outlier,
however, perhaps because of an unusually high WD mass.

The idea that T Pyx-like phases may significantly affect the
evolution of many CVs is, of course, highly speculative, and we do 
not mean to endorse it too strongly. However, it highlights the
importance of understanding T Pyx: until we know what
triggered the current high-luminosity state, it will remain difficult 
to assess the broader evolutionary significance of this phase. Note
that the apparent uniqueness of T Pyx is not a strong argument against
such significance. For example, if the duty cycle of high-luminosity
phases is $\sim 0.1$\%, as suggested by the numbers above, we should
not expect to catch many CVs in this state. Thus T Pyx could be the
tip of the proverbial iceberg.


\newpage \thispagestyle{empty} \mbox{} 


\chapter{SDSS J1507+52 \\\Large \textsc{- An eclipsing period bouncer in the Galactic halo - }} \label{j1507}
\label{chap:icm}

\begin{figure}[t]
\begin{center}
\includegraphics[scale=0.8]{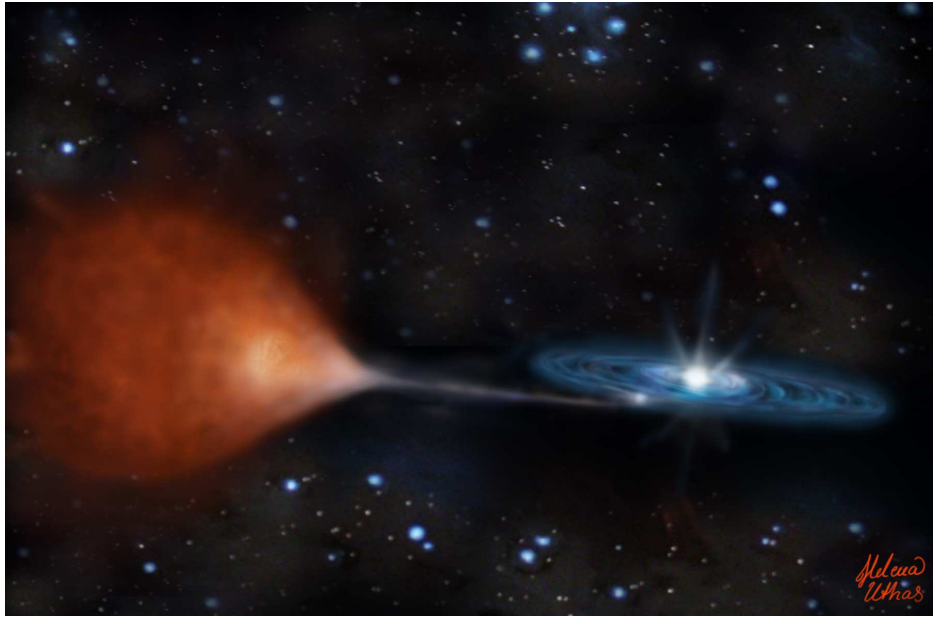}
\caption{\emph{My impression of SDSS J1507+52. The model is scaled from using the estimated binary separation and the radii of the primary and donor as presented by~\cite{2007MNRAS.381..827L}.}}
\label{model_j1507}
\end{center}
\end{figure}

\emph{The main parts of this chapter will appear in MNRAS as~\cite{uthas_j1507}. I would like to thank my collaborators, Christian Knigge, Knox S. Long, Joseph Patterson and John Thorstensen for guidance and valuable comments.} 
\newline
\newline
\begin{Huge}\color{Red}{S}\end{Huge}DSS J1507+52 (hereafter J1507) is an eclipsing cataclysmic variable that consists of a cool, non-radially pulsating white dwarf and an unusually small sub-stellar secondary. The system was first identified from the Sloan Digital Sky Survey (\citealt{2005AJ....129.2386S}) and was quickly recognised to be odd due to its short orbital period of about 67 minutes, which is well below the minimum period characteristic of normal CVs. A few other systems are occasionally found to have periods below the usual minimum period, but, in general, these systems show evidence for abnormally hot and bright donor stars in their optical spectra, suggesting that their secondaries are nuclear-evolved objects. This is not the case in J1507, whose secondary is not visible at all in optical spectroscopy (\citealt{2005AJ....129.2386S};  \citealt{2007MNRAS.381..827L}). Together with its short orbital period, this indicates a system with a faint disc, a low accretion rate and a relatively unevolved, low-mass donor. In Figure~\ref{model_j1507}, I present a scaled model painting of J1507.

\cite{2007MNRAS.381..827L} performed eclipse analysis of J1507 and found a donor mass of 0.056 $\pm$ 0.001 M$_{\odot}$, clearly below the hydrogen-burning limit. However, because of its anomalously short orbital period, the system is not consistent with the standard mass-period relation for CVs (\citealt{2006MNRAS.373..484K}). In line with this,~\cite{2007MNRAS.381..827L} found the donor radius to be smaller than predicted by standard CV evolution sequences. They therefore suggested that the donor might be unusually young, i.e. that J1507 might be a CV that formed recently from a previously detached WD - brown dwarf close binary system. A young brown dwarf has a higher density, and therefore a smaller radius, since the donor has not yet had a chance to expand in response to mass loss. This would explain the short orbital period found in J1507, since according to the period-density relation, a higher density and subsequently a smaller radius, implies a shorter orbital period. The list of photometrically inferred system parameters provided by~\cite{2007MNRAS.381..827L} also includes the effective temperature of the white dwarf, which they estimate to be $T_{\text{eff}}$ = 11000 $\pm$ 500 K. 

Concurrently,~\cite{2008PASP..120..510P} showed that the system has an unusually high space velocity, similar to the velocities of stars in the Galactic halo. A typical star of 1 M$_{\odot}$ in the Galactic disc has a space velocity below 50 km\,s$^{-1}$, while J1507 has a velocity of about 167 km\,s$^{-1}$. If J1507 is a member of the Galactic halo, its donor will be a Population II object with low metallicity. Due to their lower atmospheric opacity, such objects have significantly smaller radii than their solar metallicity Population I counterparts. Membership of the Galactic halo would therefore also account for J1507's small donor radius and short orbital period. However, Population II stars represent only about 0.5\% of the stars in the Solar neighbourhood, which would make J1507 a rare system.~\cite{2008PASP..120..510P} estimated a slightly higher effective temperature of the WD in J1507, $T_{\text{eff}}$ = 11500 $\pm$ 700 K. They also found the system to exhibit multi-periodic variability on time-scales of minutes, which they interpreted as non-radial pulsations originating from the primary white dwarf.

Both hypotheses imply that J1507 is an interesting and important system that can shed light on poorly understood aspects of CV evolution. However, since the models are very different, it is clearly important to determine which -- if either -- of them is correct. As noted above, eclipse-based estimates for the effective temperature of the WD suggested $T_{\text{eff}} \simeq$ 10000 K -- 12000 K. At these temperatures, the ultraviolet (UV) spectrum of a WD accreting metal-rich material would be expected to show strong absorption features due to Fe II and Fe III (c.f. \citealt{1994ApJ...426..294H}). In this chapter, I present far- and near-UV spectroscopy of J1507 that was obtained with the COS and STIS instruments onboard the Hubble Space Telescope, with the aim of measuring the metallicity of the system and establishing its status as either a Pop I (disc) or a Pop II (halo) object.


\begin{figure}[t]
\includegraphics[scale=1.05]{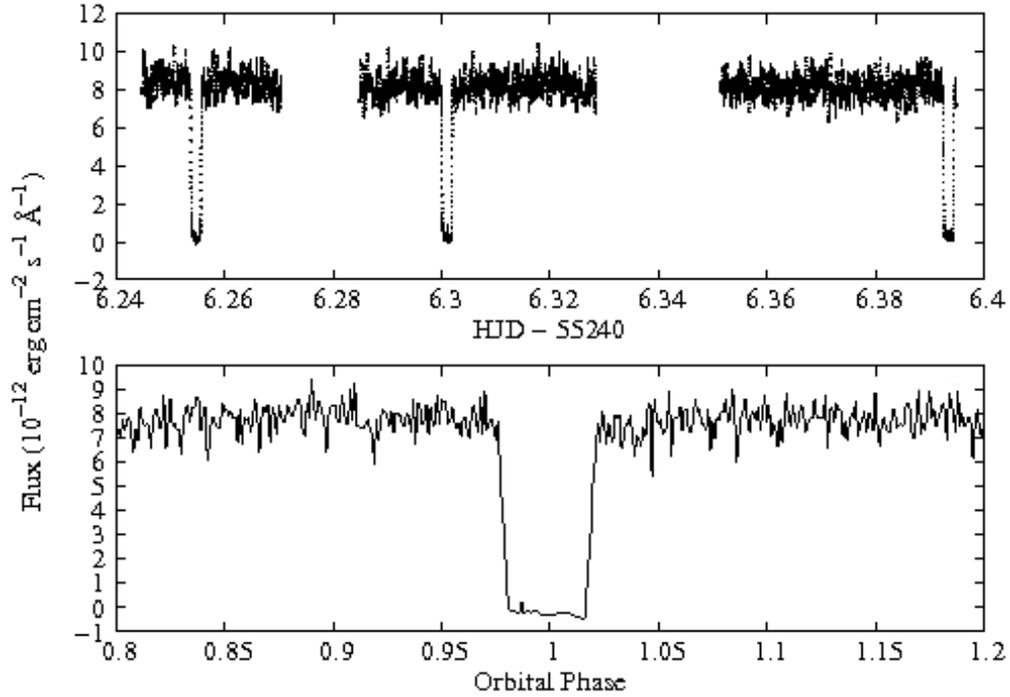}
\caption{\emph{The Gaussian smoothed and flux-calibrated light curve obtained by COS/HST, constructed from the monochromatic flux averaged over the region 1425\,\AA\, -- 1900\,\AA. Bottom panel shows the phase-folded light curve in the vicinity of the eclipse.}} 
\label{fig:light}
\end{figure}


\section{Observations and Reduction} 

Far- and near-UV observations of J1507 were performed in February 2010, using the HST. The observations were obtained in the so-called time-tag mode. In this mode, each photon event is tagged with a corresponding event time, along with their wavelength and spatial position on the detector. This allows us to rebin the data over any time interval and wavelength region. 

Far-UV observations were carried out with the Cosmic Origins Spectrograph (COS) for a total exposure time of about 3 hours, covering 3 eclipses in 3 successive HST orbits. The maximum time resolution of the COS TIME-TAG mode is 32 ms. The G140L grating was used, which gave a spectral range of 1230\,\AA\ -- 2378\,\AA\,and a spectral resolution of 0.5\,\AA.  

Near-UV observations were obtained by the Space Telescope Imaging Spectrograph (STIS) for a total exposure time of about 1.4 hours, at a maximum time resolution 125 $\mu$s. The G230L grating was used, resulting in a spectral range of 1650\,\AA\ -- 3150\,\AA\,and a spectral resolution of 3.16\,\AA. The data set consists of 3 sub-exposures (of length 1600 s, 1918 s and 1708 s). No near-UV data was obtained during the eclipses. Also, there are gaps in the data caused by interruptions coming from Earth occultation. 

All data were reduced and calibrated using the \textsc{Pyraf} package \textsc{Stsdas}, provided by STScI (the Space Telescope Science Institute)\footnote{\textsc{Stsdas} and \textsc{Pyraf} are products of the Space Telescope Science Institute, which is operated by AURA for NASA.}.


\begin{figure}[t]
\includegraphics[scale=0.47]{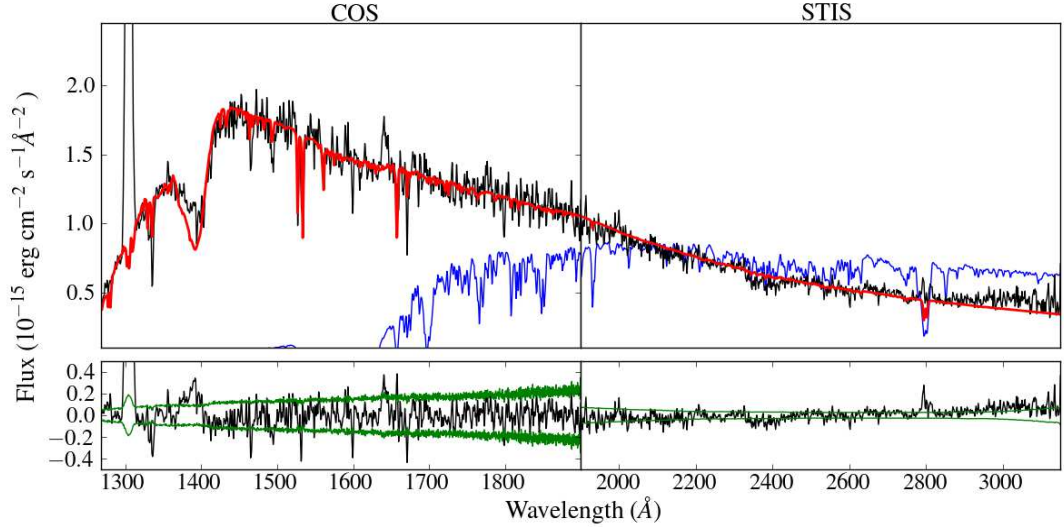}
\caption{\emph{The two top panels show the mean out-of-eclipse spectrum for both the far (left) and near (right) UV regions. A model constructed from parameters obtained from the far UV-region is plotted in red (see Table~\ref{pars_j1507}). A model corresponding to values from Littlefair et al. (2007), $T_{\text{eff}}$\,=\,11000 K and $\log g$\,=\,8.5, is plotted in blue. Underneath each wavelength region is a plot showing the residuals. The dips at (1460, 1530, 1600 and 1670)\,\AA\, coincide with regions of bad pixels. The 1-$\sigma$ flux errors are plotted in {\color{Green} green}.}} 
\label{fig:spec_j1507}
\end{figure}

\section{Analysis and Results}

Figure~\ref{fig:light} shows the Gaussian smoothed and flux-calibrated light curve from the far-UV data. This light curve was constructed by dividing the larger exposures into one-second bins, and by averaging the fluxes over the wavelength range 1425\,\AA\, -- 1900\,\AA. The light curve shows sharp, square-shaped eclipses, close to zero flux at mid-eclipse, and a very flat out-of-eclipse continuum, all strongly suggesting that the UV flux is dominated by the accreting WD. The folded light curve constructed from the combined data for all three HST orbits is presented in the bottom panel. All further analysis was carried out on the unsmoothed data, excluding the eclipses. Ground-based optical observations were performed at the same time as the HST observations, yielding a magnitude for J1507 of $g = 18.44 \pm 0.02$. This is consistent with SDSS, showing that the system was in quiescence during the HST observations. 

The two top panels of Figure~\ref{fig:spec_j1507} show the mean out-of eclipse spectrum (in black), for both the near- and far-UV. The reduction was performed separately for the COS and STIS data, and the mean spectrum for both regions were joined at 1900 \AA\,without overlap for a total range of 1268\,\AA\, -- 3150\,\AA. The COS data were rebinned slightly, to 0.2\,\AA/pix, to increase the S/N and provide sufficient sampling of the G140L grating's spectral resolution. Different dispersions were retained for the the far- and near-UV regions (0.2 \,\AA/pix and 1.58 \,\AA/pix, respectively). The overall UV spectrum for J1507 is relatively sparse in spectral lines. We identify CII at 1335 \,\AA\,, SiII at 1526.7  \,\AA\,, HeII at 1640.5  \,\AA\,, AlII at 1671  \,\AA\, and MgII at 2800 \,\AA. The strong emission feature near 1300 \AA\, is almost certainly due to geocoronal emission in the O I 1304 \AA\, line that could not be cleanly removed by the pipeline background subtraction.

\subsection{Spectral Modelling}

A model grid spanning the four key atmospheric parameters, effective temperature ($T_{\text{eff}}$), surface gravity ($\log g$), metallicity ($[Fe/H]$) and rotational velocity ($v \sin i$), was constructed in order to fit the far- and near-UV spectrum. Atmospheric structures were calculated with \textsc{Tlusty}, with the spectral synthesis being done with \textsc{Synspec}\footnote{\textsc{Tlusty} and \textsc{Synspec} are programs developed for modelling of stellar atmospheres: \newline http://nova.astro.umd.edu/.} (\citealt{1988CoPhC..52..103H}; \citealt{1995ApJ...439..875H}). The grid spanned the region; 12500 K $\leappeq T_{\text{eff}} \leappeq$ 20000 K in steps of 500 K, 7.5  $\leappeq \log g \leappeq 9.5$ in steps of 0.25, -2.0 $\leappeq [Fe/H]  \leappeq$ 0.0 in steps of 0.25, and 0 km\,s$^{-1} \leappeq v \sin i \leappeq$ 1000 km\,s$^{-1}$ in steps of 50 km\,s$^{-1}$. Models at intermediate parameter values were constructed by linear interpolation on this grid. Pixels with non-zero data quality flags set were excluded from our fits, as well as the regions around the spectral lines, OI at 1304\,\AA\, CII at 1335\,\AA\,and the quasi-molecular H feature at $\sim$ 1400\,\AA. 

Before the models could be compared to the observed spectrum, they were convolved with a Gaussian filter to the spectral resolution appropriate to the COS (< 1900\,\AA) and STIS (> 1900\,\AA) data, and linearly interpolated onto the observational wavelength scale. In general, our model fits did represent the main features of the data quite well, but the formal $\chi^{2}$ was somewhat high. In order to obtain more realistic formal parameter errors, we therefore added an intrinsic dispersion term to the flux errors in such a way that $\chi^{2}_{\nu}$ = 1. This dispersion corresponds to about 1\% of the flux at 1900\,\AA. Formal 1-$\sigma$ parameters errors were then estimated in the usual way, by identifying the parameter ranges that correspond to $\chi^{2} = \chi^{2}_{\text{min}} \pm$ 1.

In order to explore the systematic uncertainties affecting our fits, we tried fitting many different wavelength regions, such as the far-UV (COS), near-UV (STIS), regions with a high line concentration and also the whole wavelength range spanning both the far- and near-UV.   
 
\begin{figure}
\includegraphics[scale=0.46]{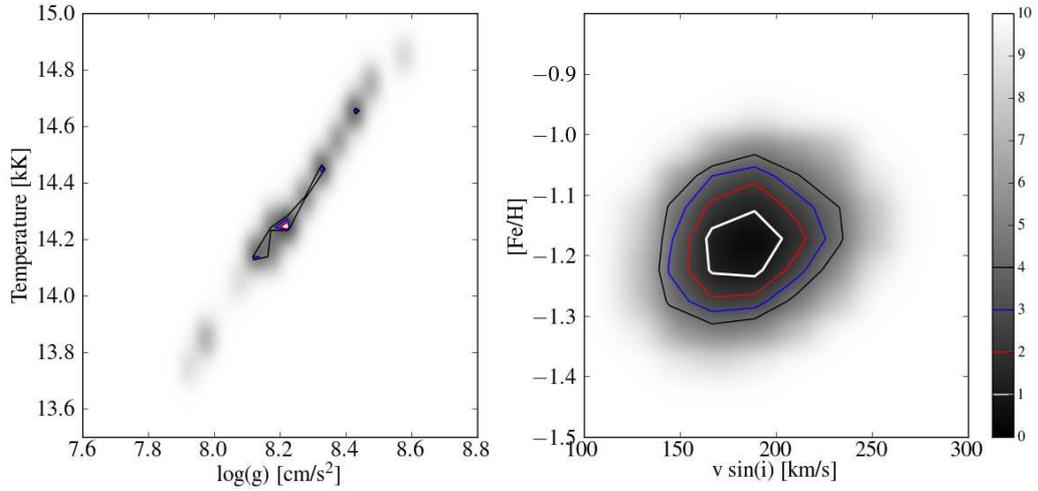}
\caption{\emph{Contour plots in $\chi^{2}$-space showing the correlation between the fit parameters. The left panel shows a strong, almost linear correlation between $T_{\text{eff}}$ and $\log g$. The panel to the right shows that the parameters $[Fe/H]$ and $v \sin i$ are roughly uncorrelated.}} 
\label{fig:contours}
\end{figure}

The correlations between fit parameters were investigated by constructing $\chi^{2}$ contour plots for each possible parameter pair. The left panel in Figure~\ref{fig:contours} show that there is a clear correlation between $T_{\text{eff}}$ and $\log g$. The correlation between $[Fe/H]$ and $v \sin i$ (right panel) is not pronounced.


\subsection{Temperature and Metallicity}

\begin{figure}
\includegraphics[scale=0.46]{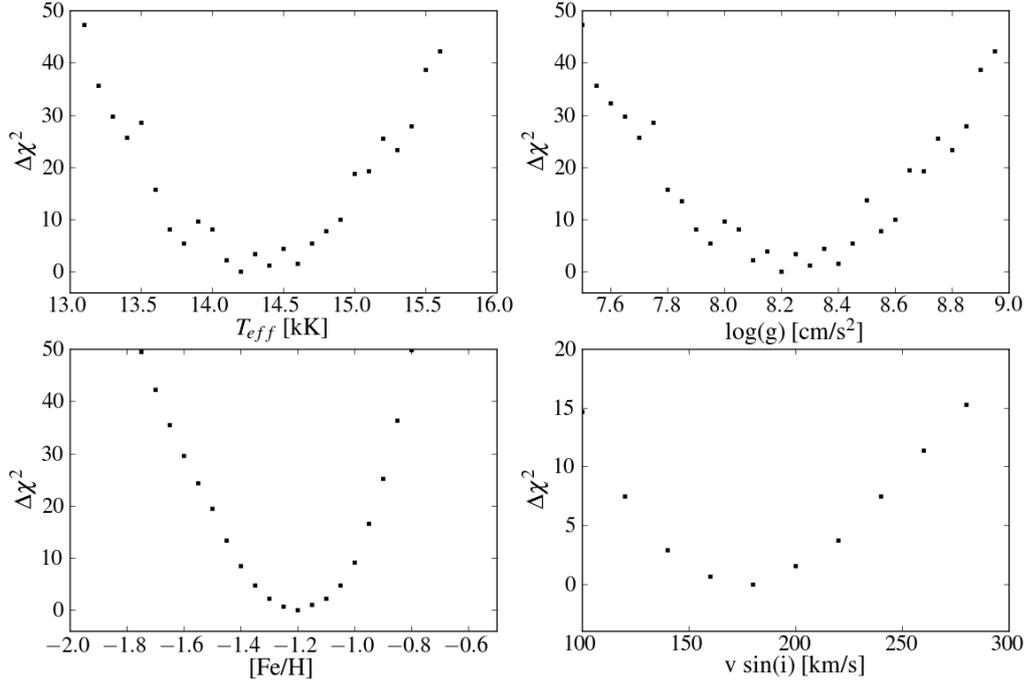}
\caption{\emph{$\chi^{2}$-fitting of the parameters to the COS far-UV spectral region.}} 
\label{fig:chifit}
\end{figure}

A fit to the far UV-region 1268\,\AA\, -- 1900\,\AA\,\,yields $T_{\text{eff}} = 14200 \pm 50$ K, $\log g = 8.2 \pm 0.04, [Fe/H] = -1.2 \pm 0.05 $ and $v \sin i = 180 \pm 20$ km\,s$^{-1}$. These fit parameters provide a good match also for the near-UV spectrum out to 2800\,\AA. We adopt these parameters as our best-fit parameters. The quoted errors are formal errors from the $\chi^{2}$ fitting, shown in Figure~\ref{fig:chifit}. Table~\ref{pars_j1507} presents the best parameters along with their uncertainties. These uncertainties are larger than the formal errors, because they include our best estimates of the systematic uncertainties associated with different choices of fitting windows. However, there are several reasons for adopting the parameters inferred from the fit to the COS far-UV data set as our best-bet estimates. First, this region includes almost all line features, and is also most sensitive to changes in $T_{\text{eff}}$ and $\log g$. Second, the far-UV region is more likely to represent pure light from the WD, while towards redder wavelengths, the spectrum could be affected by the disc and the bright spot. Third, fitting only the COS data removes any possibility that a mis-match in the flux calibration of COS and STIS could affect our results.


The two top panels in Figure~\ref{fig:spec_j1507} show the best-fit model in red, along with the mean out-of-eclipse spectrum for J1507. The corresponding residuals between model and data are plotted underneath each of the COS and STIS spectral regions, together with the 1-$\sigma$ error range marked in \begin{color}{Green}green\end{color}. The dips in the residuals at 1460\,\AA\,, 1530\,\AA\,, 1600\,\AA\, and 1670\,\AA\, are caused by bad pixels in the data, and these regions were excluded from our fits. As a comparison, a model based on the parameters found by~\cite{2007MNRAS.381..827L}, T$_{\text{eff}}$ = 11000 K and $\log g$ = 8.5, is plotted in \begin{color}{blue}blue\end{color}. Clearly, this temperature is far too low to match the data.
 
 The top panel in Figure~\ref{fig:met} shows the mean spectrum together with models at three different values of $[Fe/H]$: 0 (\begin{color}{Green}green\end{color}), -1.2 (\begin{color}{red}red\end{color}) and -2 (\begin{color}{blue}blue\end{color}). It is clear that a model of solar metallicity (\begin{color}{Green}green\end{color} line) does not match the data and overestimates both the number and strength of visible absorption lines in the spectrum. The left panel shows a zoom of the region around the Si II line (the coloured lines correspond to the same values of the metallicity as given for the top panel). To the right is a plot showing the dependence of the model on $v \sin i$, where the coloured lines in \begin{color}{Green}green\end{color}, \begin{color}{red}red\end{color} and \begin{color}{blue}blue\end{color} represent rotational velocities of 0 km\,$s^{-1}$, 180 km\,$s^{-1}$ and 500 km\,$s^{-1}$, respectively. In all panels, the red lines show the best-fit parameters. The metallicity and rotational velocity affect both the line depths and widths. Therefore, as a check, we fixed T$_{\text{eff}}$ and $\log g$ to our best values and made model fits to smaller wavelength regions specifically chosen because they contain strong and/or many spectral lines, with $v \sin i$ and $[Fe/H]$ as free parameters. We find consistent values of $v \sin i$ and $[Fe/H]$, independent of the wavelength region we performed the fits on.   
  
\begin{figure}
\includegraphics[scale=0.46]{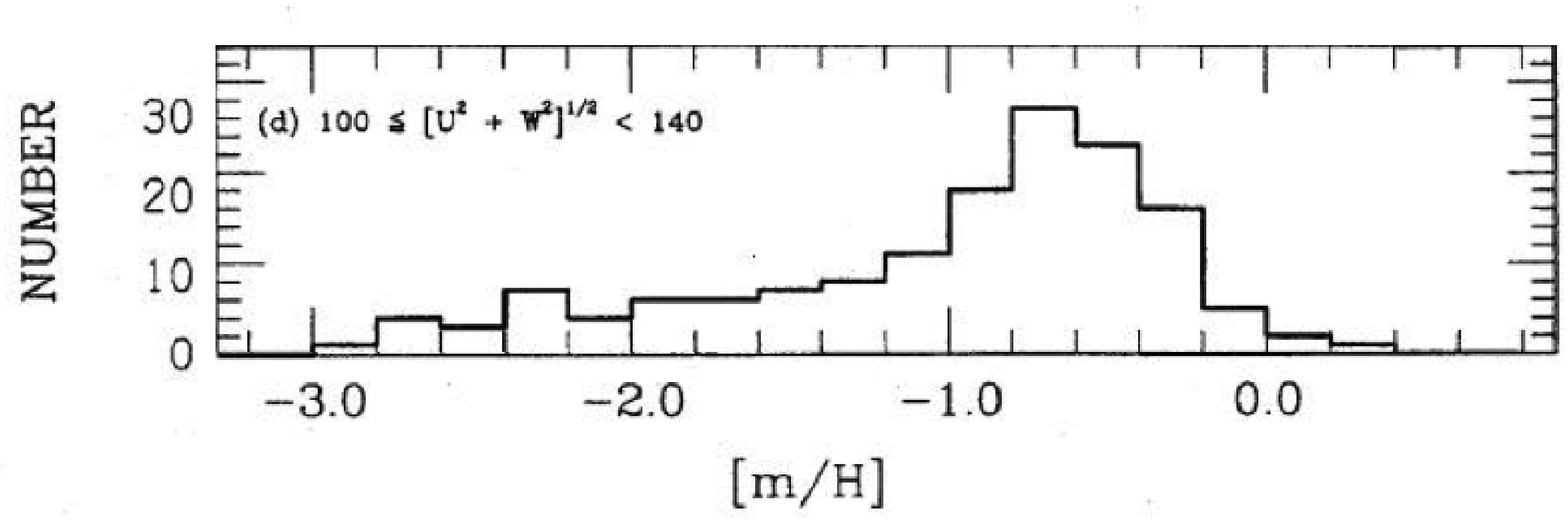}
\vspace{-0.5cm}
\caption{\emph{Figure from~\cite{1996AJ....112..668C}, showing the metallicity distribution for Galactic halo stars.}} 
\label{fig:carney}
\end{figure}
  
The metallicity found for J1507 can be compared with the metallicity distribution for Galactic halo stars at the same space velocity. ~\cite{2008PASP..120..510P} calculated a space velocity for J1507 of 167 km\,s$^{-1}$, which can be broken into the Galactic velocity components U, V and W, where $\sqrt{U^{2} + W^{2}} =139$ km\,s$^{-1}$ for J1507.~\cite{1996AJ....112..668C} presented a study of the metallicity distribution for the Galactic halo stars. They found that for stars with velocities in the range 100 <  $\sqrt{U^{2} + W^{2}}$ < 140, the metallicity distribution peaks around -0.7, but has a long tail towards metallicities as low as -1.5 (see Figure~\ref{fig:carney} from~\citealt{1996AJ....112..668C}). This is consistent with our metallicity estimate of -1.2. We therefore conclude that J1507 is most likely a Galactic halo CV.

\begin{figure}
\includegraphics[scale=0.46]{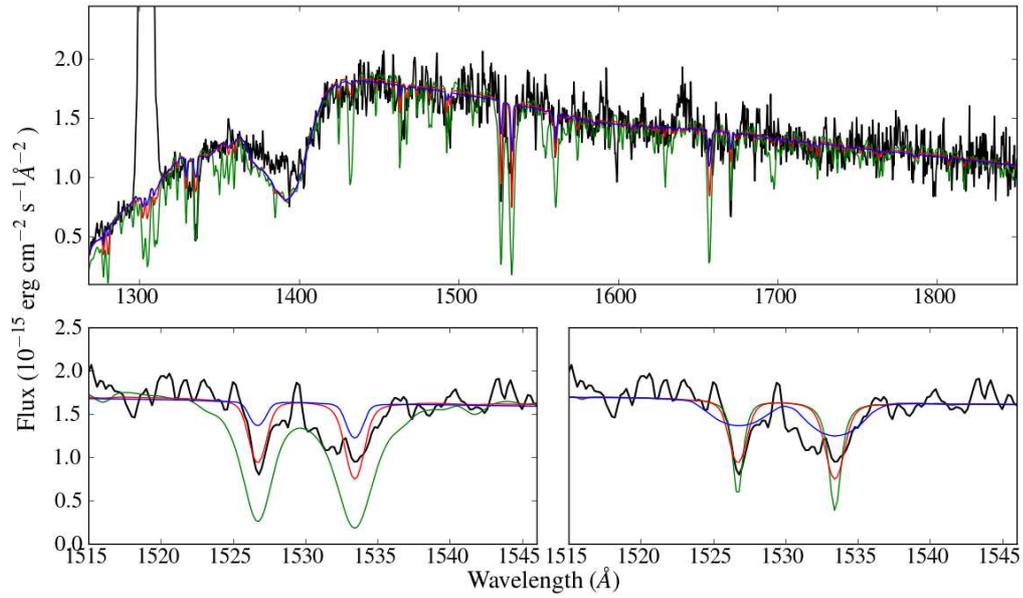}
\caption{\emph{The top panel shows the mean out-of-eclipse spectrum for the far UV region, where models at three different $[Fe/H]$, 0 ({\color{Green} green}), -1.2 ({\color{Red} red}) and -2 ({\color{blue} blue}), are plotted on top. The bottom left panel shows a  zoom of the region around the Si II, where the coloured lines correspond to the same values of the metallicity as in the top panel. In the bottom right panel, the colours correspond to $v \sin i$ at 0 km\,$s^{-1}$ ({\color{Green} green}), 180 km\,$s^{-1}$ ({\color{Red} red}), 500 km\,$s^{-1}$ ({\color{blue} blue}).}} 
\label{fig:met}
\end{figure}

\subsection{Possible Extinction Effects}

Up to now, we have assumed that the observed far-UV spectrum of SDSS J1507 is unaffected by interstellar extinction. Based on the absence of the well-known 2175\,\AA\,absorption feature in the data, we can set E(B-V) $\leq$ 0.05 as a fairly conservative limit on the amount of reddening that may be present. In order to test the effect of extinction at this level on our conclusions, we carried out additional model fits after dereddening the data by E(B-V) = 0.05. We find that none of our inferred parameters would change significantly if extinction at this level were present.



\subsection{Non-Radial Pulsations}

The high effective temperature found for J1507 ($14200\,\pm\,500$ K) means that the system is pushed out of the instability strip for non-accreting WDs ($12000\,\pm\,1200$ K; see Figure 3 of~\citealt{2006AJ....132..831G}). However, observations show that the temperatures found for pulsating WDs in CVs, cannot be defined by this narrow instability strip, and their temperatures are found in a wider range, often towards higher values (\citealt{2010ApJ...710...64S}). 

Time-resolved photometry of J1507 was obtained by~\cite{2006MNRAS.370L..56N} in 2005, but no periodic signals were found. However, also in 2005,~\cite{2008PASP..120..510P} found signals in the light curve for J1507 at 76 c\,d$^{-1}$, 130 c\,d$^{-1}$ and 175 c\,d$^{-1}$ ($\pm\,0.5$ c\,d$^{-1}$), which they identify as non-radial pulsations from the primary WD. These signals were found to have moved to longer periods when measured a year later, and was then found at 75.3 c\,d$^{-1}$, 127.8 c\,d$^{-1}$ and 171.3 c\,d$^{-1}$ ($\pm\,0.6$ c\,d$^{-1}$).
 
\begin{figure}[t]
\begin{center}
\includegraphics[width=0.9\textwidth]{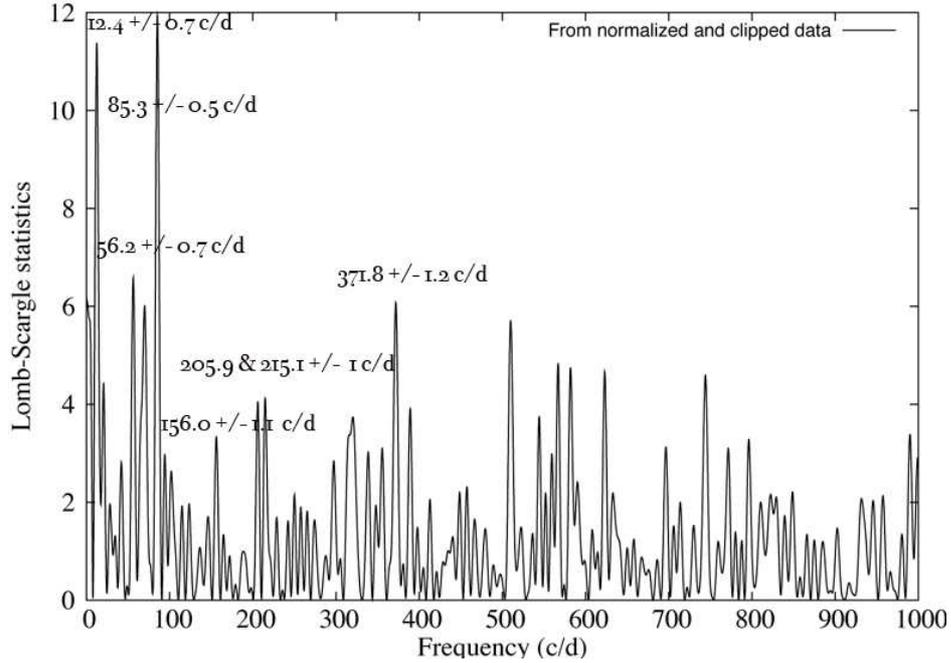}
\end{center}
\vspace{-1pc}
\caption{\emph{Lomb-scargle periodogram constructed from the monocromatic integrated fluxes from the far-UV region from 1268\,\AA\, -- 1900\,\AA\,\,(eclipses were removed).}}
\label{fig:pow}
\end{figure}

I obtained observations of J1507 at the MDM telescopes during one night in May 2010 and recovered two of these signals, now at 90 c\,d$^{-1}$ and 130 c\,d$^{-1}$ ($\pm 1$ c\,d$^{-1}$). The amplitudes of these pulsations are expected to vary with wavelength, and in the COS far-UV data, a strong peak is, in fact, found $85.3\,\pm\,0.5$ d$^{-1}$ (See Figure~\ref{fig:pow}). All eclipses were removed from both the optical and UV light curves before constructing the Lomb-Scargle periodograms. A complete search for the UV counterpart of the optical WD pulsations will be presented in Uthas et al. (in preparation). However, it is already clear that both the new optical and the UV data support the classification of J1507 as a GW Lib pulsator -- the first such object in the halo and thus with a low-metallicity WD envelope.

\subsection{Distance Estimates}

The theoretical relationship between observed flux ($F$) and the Eddington flux ($H$) provided by the \textsc{Synspec} models can be used to estimate the distance ($d$) towards J1507. More specifically, $F =  4 \pi R_{\text{wd}}^{2}H/\,d^{2}$, where $R_{\text{wd}}$ is the WD radius. This means that the normalisation factor needed to optimally match a model spectrum to the data is a direct measure of $R_{\text{wd}}^{2}/d^{2}$. WDs also follow the well-known mass-radius relation (with only a weak temperature sensitivity), so the surface gravity of the model, $g = G\,M_{\text{wd}}/R_{\text{wd}}^{2}$, uniquely fixes $R_{\text{wd}}$. Thus for a WD model with given $T_{\text{eff}}$ and $\log g$, the normalisation factor required to fit the data is a unique measure of distance. 

In practice, the distance was estimated by first fitting a linear function to the relationship between $M_{\text{wd}}$ and $\log g$ in Pierre Bergeron's WD cooling models\footnote{Cooling models by Pierre Bergeron are found at: \newline http://www.astro.umontreal.ca/$\sim$bergeron/CoolingModels/}, at $T_{\text{eff}}$ = 14500 K. Next, this function was used to estimate $M_{\text{wd}}$, and hence $R_{\text{wd}}$, for a given $\log g$. The resulting estimates of the WD mass and radius are found to be; $M_{\text{wd}} = 0.75 \pm 0.15$ M$_{\odot}$ and $R_{\text{wd}} = 0.011 \pm 0.002$ R$_{\odot}$. These values are not very precise because the surface gravity is only poorly constrained by our spectroscopy. In any case, our numbers are consistent with both ~\cite{2007MNRAS.381..827L} and ~\cite{2008PASP..120..510P}. The normalisation constant of the model combined with $R_{\text{wd}}$ then finally yield the corresponding distance estimate. The distance towards J1507 is thus found to be $d = 250 \pm 50$ pc (the effect of reddening is allowed for in the quoted error). This is consistent with~\cite{2008PASP..120..510P} but larger than the distance given by~\cite{2007MNRAS.381..827L}, since the $\log g$ presented here ($\log g$ = 8.2) is lower than the $\log g$ implied by their estimates of M$_{1}$ and T$_{\text{eff}}$ ($\log g$ = 8.5).  


\begin{table}
  \begin{center}
 \begin{tabular}{ll}
 \hline
 \hline
$T_{\text{eff}}$: & 14200 $\pm$ 500 (50) K  \\
$\log g$: & 8.2 $\pm$ 0.3  (0.04)\\
$[Fe/H]$: & -1.2 $\pm$ 0.2 (0.05)   \\
$v \sin i$: & 180 $\pm$  20 (20) km\,s$^{-1}$\\
\hline
\hline
\end{tabular}
\vspace{0.5cm}
  \caption{\emph{Best-fit parameters for J1507 obtained from model fitting to the far-UV region. The errors are defined as the whole range for which good fits are found, irrespective of wavelength region. Formal 1-$\sigma$ errors are given in parenthesis.}}
  \label{pars_j1507} 
 \end{center}
\end{table}


\section{Discussion and Summary}

In this chapter, I have presented HST observations in the UV spectral-range of the cataclysmic variable J1507 that was obtained with the aim of measuring its metallicity to determine whether or not the system is a member of the Galactic halo. 

By comparing the observed spectrum to synthetic spectra described by the four parameters, $T_{\text{eff}}$, $\log g$, $[Fe/H]$ and $v \sin i$, a best fit is found at an effective temperature of $14200 \pm 500$ K. This value of $T_{\text{eff}}$ is based on the assumption that the WD is the only component contributing to the UV flux in J1507. If a hot, optically thick boundary layer would be present, it would bias the estimate of $T_{\text{eff}}$. However, the single-temperature WD models presented here seem to provide a good fit to the data, and the boundary layer in a low $\dot{M}$-system (such as J1507) is expected to be optically thin and thus not contribute significantly to the UV flux.

The best model-fit give a $T_{\text{eff}}$ that is much higher than previous estimates, which implies that the accretion rate is also higher than previously thought. More specifically, since $\dot{M} \propto T^{4}$, the 30\% increase in the estimated $T_{\text{eff}}$ corresponds to an increase in $\dot{M}$ by almost a factor of 3 (\citealt{2003ApJ...596L.227T}). This may have significant implications for the evolution of this system, including the question of whether gravitational radiation alone is sufficient to drive this accretion rate (see \citealt{2009ApJ...693.1007T}; \citealt{knigge2011}).

At the high $T_{\text{eff}}$ presented above, Fe II and III are no longer dominant contributors to the atmospheric UV opacity, making it more difficult than expected to measure the metallicity of the system. Nevertheless, model fits to the data clearly favour a significantly sub-solar metallicity, $[Fe/H] = -1.2 \pm 0.2$, comparable to the typical metallicity found for halo stars at the same high space velocity as J1507. This implies that J1507 is most likely is a halo CV.

This result raises the question if there are any other confirmed Galactic halo CVs. Observationally, they are expected to be difficult to identify, since they are faint and typically will be located at large distances.~\cite{1997A&A...320..136S} performed theoretical population synthesis on Population II CVs, and found that halo CVs would have higher accretion rates and therefore shorter evolutionary timescales compared to CVs with a solar chemical composition. As a result, they would pass through the period gap more quickly and also have a shorter minimum period. This results in an even larger percentage of short-period systems for halo CVs than is expected for the Galactic disc population. Also, due to the lower metallicity, dwarf novae would be expected to exhibit higher outburst amplitudes (\citealt{1988ApJ...333..227C}). So far, only a few surveys have directly aimed to identify halo CVs (\citealt{1990ApJ...356..623H, 1990MNRAS.245..385M, 1992ApJS...78..537S}).~\cite{1990ApJ...356..623H} collected a sample of candidate halo CVs that consisted of 84 high-Galactic latitude objects. They were chosen on the basis that they had latitudes > 30$^{\circ}$ and were very faint, indicating large distances.~\cite{1990ApJ...356..623H} found a slightly larger fraction of short-period systems among their candidates. They also find higher outburst amplitudes. All this might indicate that some of these candidates are in fact halo CVs, although later surveys using infrared photometry data, e.g.~\cite{1995ApJ...439..337H}, show that the method used by~\cite{1990ApJ...356..623H} overestimates the distances, which would imply that their CV sample is probably situated closer to the Galactic plane than they estimated. In order to confirm a membership in the Galactic halo, one needs to measure the space velocities and/or metallicities of any candidates. In the absence of such measurements for any other candidate halo systems, J1507 is the most likely halo CV known to date.  

\cite{2008PASP..120..510P} found periodic signals in J1507, which they identified as non-radial pulsations originating from the primary WD. Non-accreting CVs show pulsations within the instability strip, 10900 K -- 12200 K (\citealt{2006AJ....132..831G}). The effective temperature found for J1507 is well above the instability strip for non-accreting pulsating WDs. However, it is already becoming clear that non-radially pulsating WDs in CVs are found across a wider temperature range, often towards higher temperatures (\citealt{2010ApJ...710...64S}). The fact that non-radial pulsations are present in J1507 is certainly interesting and potentially important, since a higher metallicity in the outer envelope of the accretors might be able to push the instability strip toward higher temperatures (\citealt{2006ApJ...643L.119A}). In this context, it is interesting to note that J1507 has a very low metallicity, but is nevertheless managing to pulsate at an effective temperature of above 14000 K.


\newpage \thispagestyle{empty} \mbox{} 


\chapter{Summary \& Discussion}
\label{chap:icm}

\begin{Huge}\color{Red}{I}\end{Huge} have presented analysis of four cataclysmic variables with orbital periods close to the minimum period for CVs ($\approx$ 83 minutes:~\citealt{2009MNRAS.397.2170G}). All four systems are interesting not just in their own right, but also from the point of view of CV evolution. In order to understand the importance of these systems, and to highlight what impact they may have on our current understanding of CV evolution, I have also given an overview of the theoretical and observational state-of-the-art in CV research in general. In this chapter, I will provide a brief summary and discussion of the main results. The discussion will, in particular, focus on how the study of individual systems -- especially ones that are seemingly rare or unusual -- can provide unique and vital insights that cannot easily be gained by other means. In addition, I will also briefly point out some promising directions for future work suggested by the research described in this thesis.

\section{Summary}

\subsection{SDSS J1457+51 and BW Sculptoris}

In Chapter~\ref{j1457_bwscl}, I introduced two systems that appear to follow the standard evolutionary track for CVs; SDSS J1457+51 and BW Sculptoris. Both systems are faint and show prominent white-dwarf components in their spectra (\citealt{2005AJ....129.2386S}), indicating low accretion rates. I showed that both stars exhibit non-coherent pulsations in their light curves and are likely new members of the GW Lib class of variable stars, in which the periodic and non-commensurate signals are widely believed to arise from non-radial pulsations in the underlying white dwarf. The power spectra of both stars show complex signals with primary periods near 10 and 20 minutes. These signals change in frequency by a few percent on a timescale of weeks or less and probably contain an internal fine structure unresolved by the observations. Double-humped waves were also found, marking the underlying orbital periods, near 78 minutes for both stars. This is right at the observed minimum orbital period for CVs. Their short orbital periods together with their low accretion rates indicate that these stars are likely to be very old CVs. In addition, BW Sculptoris showed a transient, but powerful signal with a period near 87 minutes -- a quiescent superhump. The $\sim$ 11\% excess over the orbital period is difficult to understand and may arise from an eccentric instability near the 2:1 resonance in the accretion disk.

\subsection{T Pyxidis}

T Pyxidis is a luminous recurrent nova that accretes at a much higher rate than is expected for its photometrically determined orbital period of about 1.8 hours. In Chapter~\ref{tpyx}, I presented the first spectroscopic orbital period, $P =1.8295$ hours, based on time-resolved optical spectroscopy obtained at the VLT and the Magellan telescopes, confirming T Pyxidis as a short orbital system. An upper limit of the velocity semi-amplitude of the white dwarf,  $K1 = 17.9 \pm 1.6 \,$km$\,$s$^{-1}$ was derived, as well as an estimate of the mass ratio, $q = 0.20 \pm 0.03$. If the mass of the donor star is estimated using the period-density relation and theoretical main-sequence mass-radius relation for a slightly inflated donor star, one obtains $M_{2} = 0.14 \pm 0.03 \,$ M$_{\odot}$. This implies a mass of the primary white dwarf of $M_{1} = 0.7 \pm 0.2 \,$ M$_{\odot}$. If the white-dwarf mass is $>$ 1 M$_{\odot}$, as classical nova models imply (\citealt{2005ApJ...623..398Y}), the donor mass must be even higher. Therefore, T Pyx has, most likely, not yet evolved beyond the period minimum for cataclysmic variables. I constrained the system inclination to be $i \approx 10$ degrees, confirming the expectation that T Pyx is a low-inclination system. 

\subsection{SDSS J1507+52}

SDSS J1507+52 is an eclipsing cataclysmic variable consisting of a cool, non-radially pulsating white dwarf and an unusually small sub-stellar secondary. The system has a high space velocity and a very short orbital period of about 67 minutes (\citealp{2005AJ....129.2386S}), well below the usual minimum period for CVs. In order to explain the existence of this peculiar system, two theories have been proposed. One suggests that SDSS J1507+52 was formed from a young, detached white-dwarf/brown-dwarf binary (\citealt{2007MNRAS.381..827L}). The other theory proposes that the system is a member of the Galactic halo population (\citealt{2008PASP..120..510P}). 

In Chapter~\ref{j1507}, I presented ultraviolet spectroscopy of SDSS J1507+52 obtained with the Hubble Space Telescope with the aim of distinguishing between these two theories. I found that the UV flux of the system is dominated by its accreting white-dwarf. Fits to model stellar atmospheres yielded physical parameter estimates of $T_{\text{eff}} = 14200\,\pm\,500$ K, $\log g = 8.2 \pm\,0.3$\,cm\,s$^{-2}$, $v \sin i = 180\,\pm\,20$ km\,s$^{-1}$ and $[Fe/H] = -1.2 \pm 0.2$. These fits suggest a distance towards SDSS J1507+52 of $d = 250\,\pm$\,50 pc. The quoted uncertainties include systematic errors associated with the adopted fitting windows and interstellar reddening.

The strongly sub-solar metallicity found for SDSS J1507+52 is consistent with that of halo stars at the same space velocity. I therefore concluded that it is likely that SDSS J1507+52 is a member of the Galactic halo. In addition, I found periodic signals in the UV spectrum of SDSS J1507+52 with frequencies close to those already identified by~\cite{2008PASP..120..510P} in the optical. These signals could be interpreted as non-radial pulsations originating from the primary white dwarf. 

\section{Discussion}

\subsection{Importance of Individual CVs} \label{indi}

With a growing population of newly discovered CVs, it has become the norm to focus on statistical analyses of large CV samples, in efforts to understand their evolution as binary stars (e.g.~\citealt{2008MNRAS.388.1582L, 2009ApJ...693.1007T, 2009MNRAS.397.2170G, 2011MNRAS.tmp...27P, knigge2011}). Such studies are important and yield a fundamental understanding of the common evolutionary behaviour for CVs. However, to quantitatively test the proposed theoretical models, careful studies of well-chosen individual systems, in addition to statistical studies, can also provide us with important and unique insights. Thus statistical studies and studies of individual systems are complementary.

In particular, the study of systems passing through key evolutionary phases, such as the minimum period, is of great importance for the general understanding of CV evolution. All four systems presented in this thesis have orbital periods close to the minimum period, indicating that they are in exactly this transitional phase. Also, studies of CVs that are able to provide us with precise system parameters, such as the well-defined properties found from eclipsing systems, are highly valuable and can provide more stringent tests of theory (for instance SDSS J1507+52).

Another example of individual CVs that deserve detailed study are systems displaying non-radial pulsations (NRPs) from the underlying WD (such as SDSS J1457+51 and BW Sculptoris). The study of NRPs in GW Lib stars may provide important information on how the process of accretion is affecting the evolution of the WDs, compared to isolated ones. Although our current theoretical understanding of the mechanism for exciting non-radial pulsations in CVs is limited, modelling of these signals might ultimately offer the most precise ways of constraining key parameters (see Section~\ref{future}). 

More generally, the study of apparently unusual individual systems will always remain important, since all observational samples are afflicted by selection biases (e.g.~\citealt{2007MNRAS.382.1279P, 2008MNRAS.385.1471P, 2008MNRAS.385.1485P}). Given that even the best available samples are usually imperfect tracers of the underlying population, unusual systems that do not appear to follow the common behaviour (outliers) could be more important than their apparent rarity would suggest.

One example is T Pyxidis. With its unusually high luminosity and high accretion rate compared to other CVs at its period, the system appears to be a clear outlier. However, epochs of enhanced mass transfer (like its present state) may accelerate or even dominate the overall evolution of the system, even if they are relatively short-lived. If most or all systems were to go through similar phases, this could be considered as an evolutionary short-cut that may be relevant to the evolution of cataclysmic variables more generally. Note that even if all CVs did go through such phases, and even if these phases were important to their overall evolution, due to the short evolutionary time scales associated with these phases, we might only ever see a handful of systems like T Pyx observationally. In this context, a considerable fraction of CVs might at some point in their evolution appear as outliers for some short-lived part of their lives. As a consequence, the systems we see as outliers today may very well nevertheless represent a common and important phase of CV evolution.
 
SDSS J1507+52 is another example of a system that clearly appears to be an outlier. Only if one assumes that this system belongs to the Galactic halo population can its main physical properties be explained. Compared to CVs found in the Galactic disc, halo CVs are expected to have higher accretion rates, resulting in a shorter evolutionary time scale, and therefore also a shorter minimum period (\citealt{1997A&A...320..136S}). SDSS J1507+52 is actually interesting in two respects: first, it belongs to the rare sample of CVs that show evidence of  harbouring sub-dwarf donors, and second, it is probably the first confirmed CV in the Galactic halo. This provides a unique opportunity to study the process of almost pure hydrogen accretion in a CV, and also give us the possibility to study NRPs from the primary WD without the complication of metallicity.

\subsection{Future Work} \label{future}

Non-coherent signals in the light curves of the GW Lib stars, are primarily thought to arise from NRPs in the underlying WD, and have been detected for only a decennium. However, to exclude other possible origins of these signals, these group of CVs need to be studied in more detail. Future theoretical modelling of NRPs, predicts more precise constrains of system parameters in CVs. For instance, the WD mass and WD spin periods could be measured in a more precise manner, as well as parameters that we are unable to measure today, such as the accreted envelope mass, which would provide us with a direct estimate of the time since the last nova eruption (\citealt{2006ApJ...643L.119A}).  

In particular, the study of NRPs in SDSS J1507+52 is interesting since it provides an unique opportunity to study the pulsation pattern of an accreting WD in a low-metallicity system. Pulsating WDs in CVs are found in a wider temperature range than isolated WDs displaying NRPs (see Figure~\ref{giannis} and~\ref{szody}), which previously has been suggested could be a result of the higher metallicity contents in the outer envelope of accreting WDs (\citealt{2006ApJ...643L.119A}). However, SDSS J1507+52 is a low-metallicity system, but its WD is still found to be pulsating at an effective temperature of > 14000 K. This implies that we might have to consider alternative explanations of the existents of pulsating and accreting WDs at temperatures above the instability strip. A detailed study of how the pulsations in SDSS J1507+52 vary with wavelength (in the UV region) will be presented by Uthas et al. (in preparation).
\newline
\newline
As mentioned above in Section~\ref{indi} (see also Section~\ref{SD}), phases of enhanced mass-transfer (such as seen in T Pyxidis) may represent an evolutionary stage that is not yet included in any evolutionary model. Therefore, theoretical modelling of how phases of enhanced mass-transfer would affect the overall evolution of CVs, would be highly interesting and potentially important. For instance, population synthesis could indicate what fraction of systems we expect to undergo such phases. Also, the effect of irradiation of the donor star by the WD and/or accretion disc, is not well understood and needs to be studied in more detail. Such irradiation effects might be the cause of these prolonged phases of enhanced mass-transfer in the aftermath of eruptions, and may not only be important for systems with high-mass WDs, such as RNe, but also in systems with lower mass WDs (see Section~\ref{SD}).

In addition to the spectroscopic data of T Pyxidis presented in Chapter~\ref{tpyx}, phase-resolved spectroscopy of the knots in the shell surrounding the system, was also obtained with the VLT in 2004/2005.~\cite{2010ApJ...708..381S} presented a detailed photometric study of the nova shell in T Pyxidis, and concluded that the system most likely, was an ordinary CV before the event of the classical nova eruption in 1866, and that this event triggered the recurrent nova phase. I intend to examine the spectral composition of the nova shell in T Pyxidis, with the aim to study how these eruptions have affected the system. A spectroscopic study of the nova shell in T Pyxidis was also performed by~\cite{1982ApJ...261..170W}, However my initial analysis of the VLT spectroscopic data does not reveal the same spectral composition in the shell as presented by~\cite{1982ApJ...261..170W}.


\pagestyle{plain} 


\appendix

\chapter{Formula Derivations}

\section{Mass function}
 
\begin{Huge}\color{Red}{I}\end{Huge}n this section, I will present brief derivations of the mass functions for both the donor star and the primary WD. We start by expressing the distance ($a_{1}$) between the WD and the centre of mass of the binary system

\begin{equation}
a_{1} = \frac{a\,M_{2}}{M_{1} + M_{2}} = a \frac{q}{1+q},
\label{eq1}
\end{equation}

\noindent where $q = M_{2}/M_{1}$ and $a$ is the binary separation. Assuming circular orbits, the true orbital velocity of the primary $a_{1}$ is

\begin{equation}
v_{1} = \frac{2\pi a_{1}}{P} = \frac{2\pi a}{P} \frac{q}{1+q},
\label{eq2}
\end{equation}

\noindent where $P$ is the orbital period. Newton's generalisation of Kepler's third law yields

\begin{equation}
P^{2} = \frac{4 \pi^{2} a^{3}}{G (M_{1} + M_{2})}.
\label{eq3}
\end{equation}

\noindent If we combine Equations~\ref{eq2} and~\ref{eq3}, and assume that the observed radial velocity of the WD is $K_{1}= v_{1}\sin i, $ we can eliminate $a$ to obtain

\begin{equation}
\frac{K_{1}^{3}P}{2 \pi G} = \sin^{3}\!i \frac{M_{2}^{3}}{(M_{1} + M_{2})^{2}}.
\label{eq4}
\end{equation}

\noindent We can now express the mass function of the donor as

\begin{equation}
\frac{K_{1}^{3}P}{2 \pi G} = \sin^{3}\!i\,M_{2} \left(\frac{q}{(1+ q)}\right)^{2},
\label{eq5}
\end{equation} 

\noindent which provides a lower limit on the donor mass, given only the orbital period of the system and an estimate of the radial velocity of the WD  ($K_{1}$). If we are able to measure the radial velocity of the donor star ($K_{2}$), a lower limit for the WD mass can be estimated. The mass function of the WD is

\begin{equation}
\frac{K_{2}^{3}P}{2 \pi G} =  \sin^{3}\!i\,\,M_{1} \left(\frac{1}{(1 + q)}\right)^{2}  \,\,\,\,\,\,  \text{where}\,\,q = \frac{M_{2}}{M_{1}} = \frac{K_{1}}{K_{2}}.
\label{eq6}
\end{equation} 

\section{Mass Ratio}

In this section, I present a derivation of a relationship that permits a spectroscopic estimate of the mass ratio $q$. In high $\dot{M}$ systems (such as T Pyxidis presented in Chapter~\ref{tpyx}), we can estimate the disc radius $R_{\text{disc}}$ by assuming that the disc extends all the way out to the tidal radius (\citealt{1995CAS....28.....W})
\begin{equation}
R_{\text{disc}} \approx  R_{\text{max}} = \frac{0.60\,\,a}{(1+ q)}. 
\label{eq7}
\end{equation}

\noindent~\cite{1981AcA....31..395S} investigated the relationship between the true projected velocity at the outer disc radius ($v_{\text{R}_{\text{disc}}} \!\sin i$) and the measured spectral peak-to-peak separation ($\Delta\!V_{\text{peak-to-peak}}$), and found that $v_{\text{R}_{\text{disc}}} \!\sin i = \Delta\!V_{\text{peak-to-peak}} / u$, where $u=1.05 \,\pm\, 0.05$. The velocity at the outer disc radius, can then be estimated by 

\begin{equation}
\Delta\!V_{\text{peak-to-peak}} = 1.05 \sin i \sqrt{\frac{G M_{1}}{R_{\text{disc}}}} \pm 0.05.
\label{eq8}
\end{equation} 

\noindent If we rewrite Equation~\ref{eq7} by inserting $a$ from Equation~\ref{eq3}, we find, after
some algebra,

\begin{equation}
R_{\text{disc}} = 0.6 (G M_{1})^{1/3} \left(\frac{P}{2 \pi(1 + q)} \right)^{2/3}.
\label{eq9}
\end{equation}

\noindent Now, combining Equation~\ref{eq8} and ~\ref{eq9} yields

\begin{equation}
\Delta\!V_{\text{peak-to-peak}} = (1.05 \frac{1}{\sqrt{0.60}}) \sin i\,\,M_{1}^{1/3} (1+ q)^{1/3} \left(\frac{G M \pi}{P} \right) ^{1/3} \pm 0.05.
\label{eq10}
\end{equation}

\noindent Finally, dividing Equation~\ref{eq10} by the expression for $K_{1}$ from Equation~\ref{eq4}, the ratio $\Delta\!V_{\text{peak-to-peak}}/K_{1}$ can be expressed as a function of only $q$. This expression can be used to estimate $q$, given only observational estimates of $K_{1}$ and of the peak-to-peak separation of disc-formed emission line, and we have

\begin{equation}
\frac{\Delta\!V_{\text{peak-to-peak}}}{K_{1}} = 1.355 \left(\frac{1 + q}{q} \right) \pm 0.05.
\label{eq11}
\end{equation}

\chapter{SPERIOD  \\\Large \textsc{- a Program for Period Analysis - }} 

\begin{Huge}\color{Red}{F}\end{Huge}or quick and easy analysis of the period contents in a light curve, I have developed routines in the Python programming language\footnote{http://www.python.org/}. In these routines, the light curve is normalised and smoothed (I will explain the advantages of this below), frequencies with power excess are automatically identified (from the unsmoothed light curve), and the errors on these frequencies are calculated. Also, the window function is plotted for any desired frequency, and the light curve is folded on any period. There are two main programs, \textsc{\textbf{Speriod}} and \textsc{\textbf{Speriod-All}}, which are used for analysis of light curves from single nights and light curves from multiple nights, respectively. For each of the two programs, there is a suite of routines, which the user can use interactively (by answering simple questions that appear on the screen). All output data files, log files and figures are saved in folders created by the program. I used these programs for the analysis of SDSS J1457+51 and SDSS J1507+52 (presented in Section~\ref{j1457_bwscl} and~\ref{j1507}). 

Below follows a description of the subroutines in \textsc{\textbf{Speriod}} and \textsc{\textbf{Speriod-All}}. Both programs require that \textsc{python} is installed together with the python packages \textsc{scipy} and \textsc{numpy}.

\section{Analysis of Single-Night Light Curves}
 
\subsection{Part 1}
 
\begin{figure}
\begin{center}
\subfigure[Raw light curve.] 
{
    \label{sub1}
    \includegraphics[width=6.5cm]{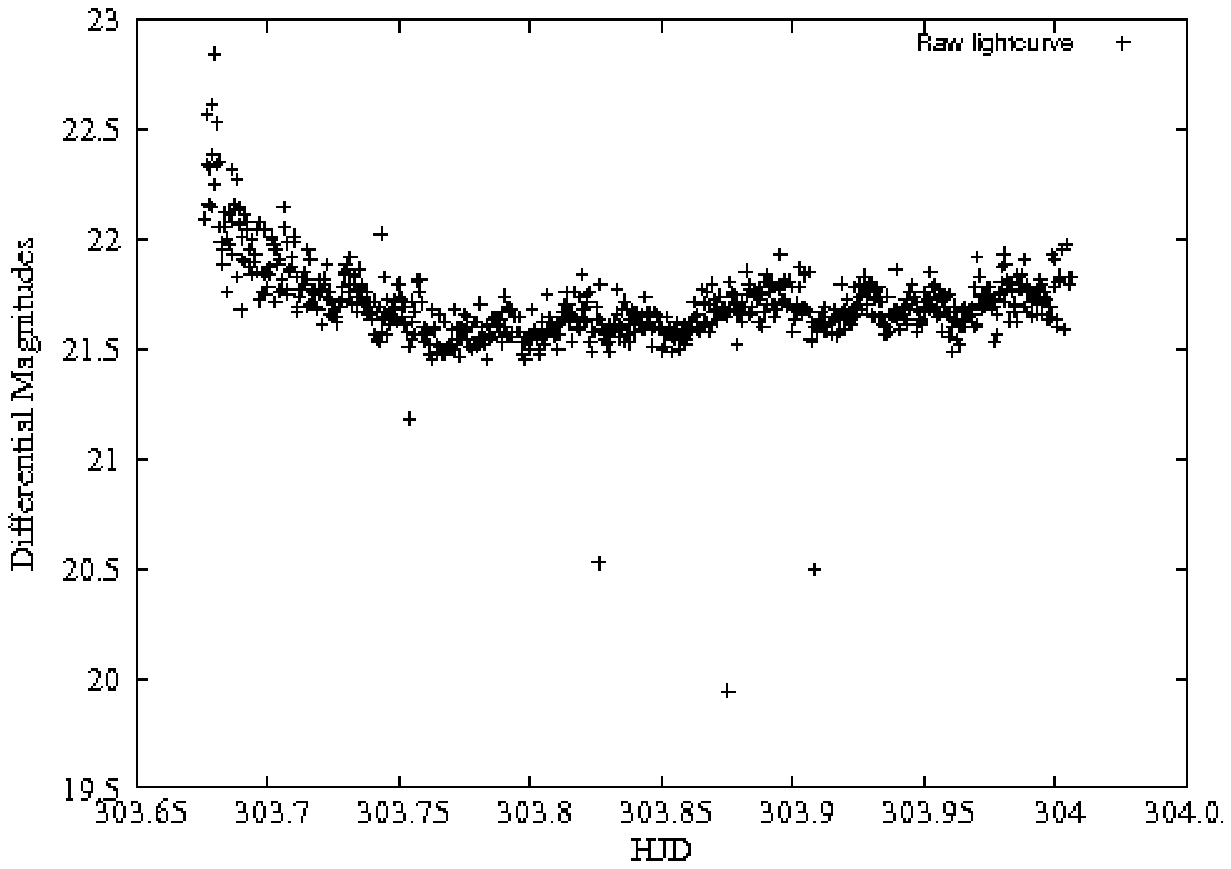}
}
\hspace{0.0cm}
\subfigure[Normalised light curve (mean is removed).] 
{
    \label{sub2}
    \includegraphics[width=6.5cm]{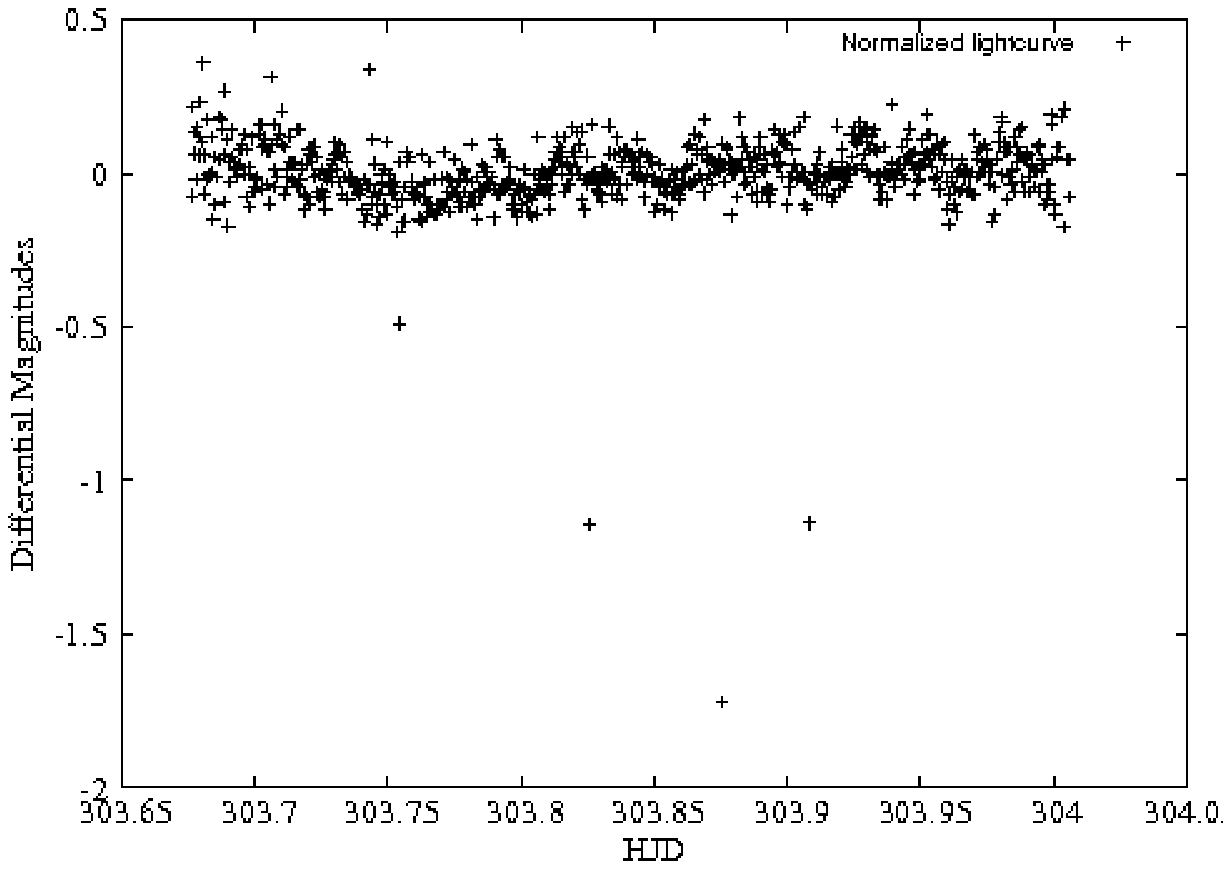}
}
\hspace{0.0cm}
\subfigure[Outliers are removed.] 
{
    \label{sub3}
    \includegraphics[width=6.5cm]{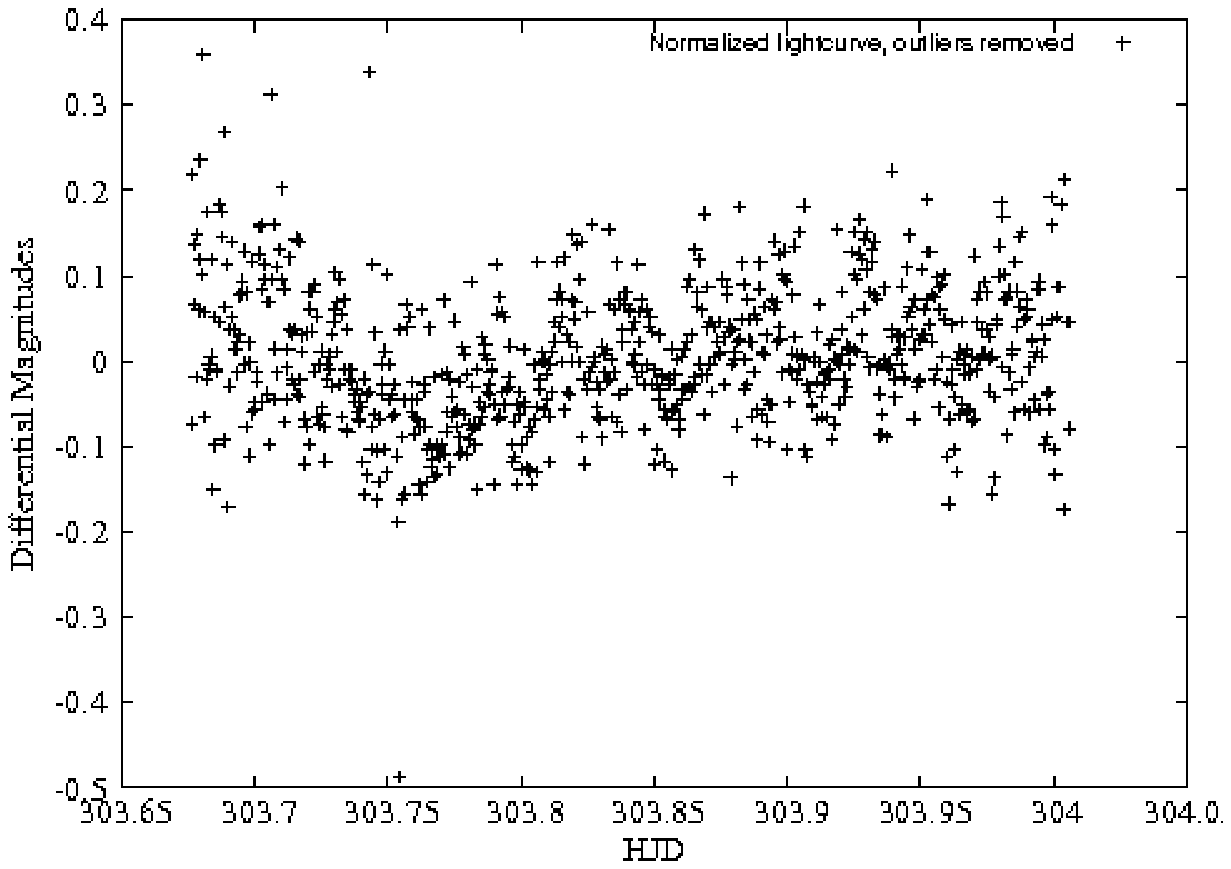}
}
\hspace{0.0cm}
\subfigure[Gaussian smoothed light curve.] 
{
    \label{sub4}
    \includegraphics[width=6.5cm]{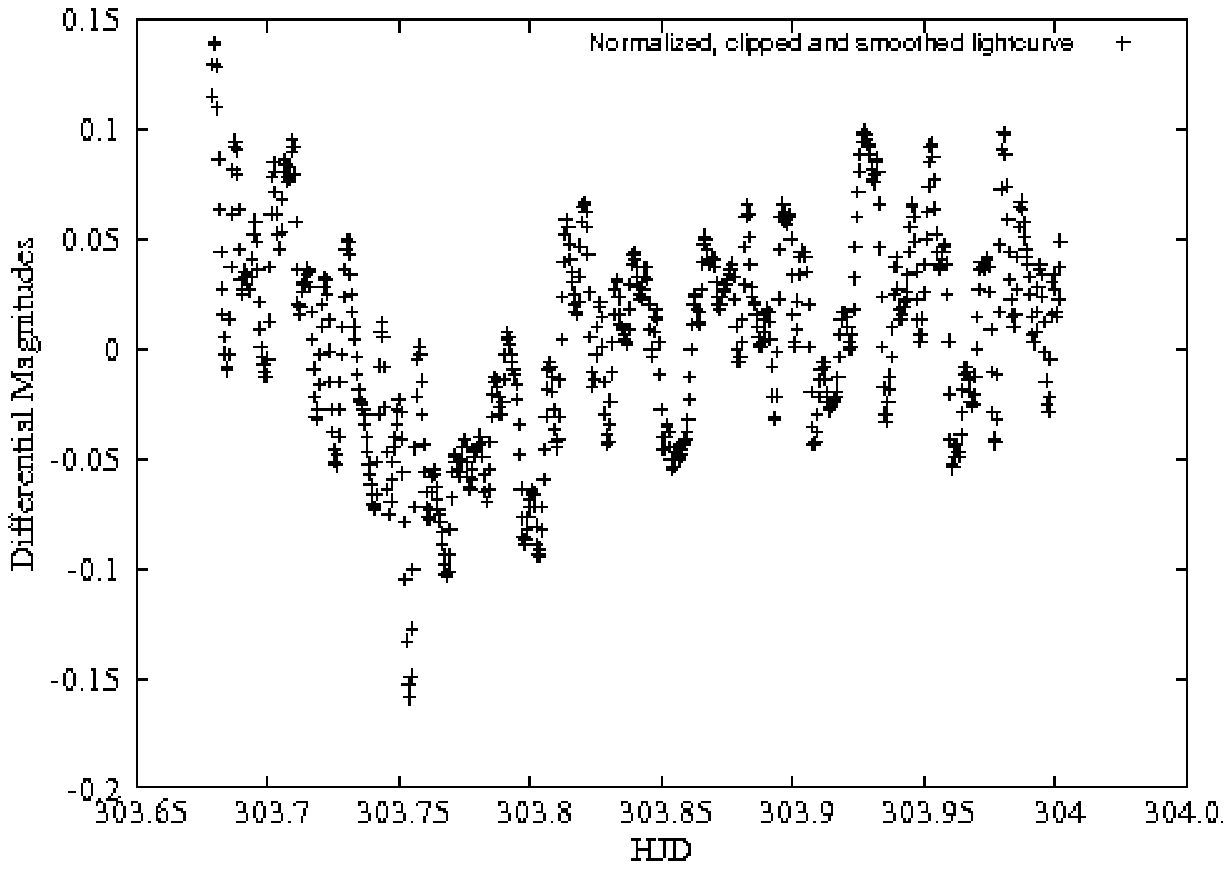}
}
\subfigure[Rebinned light curve.] 
{
    \label{sub5}
    \includegraphics[width=6.5cm]{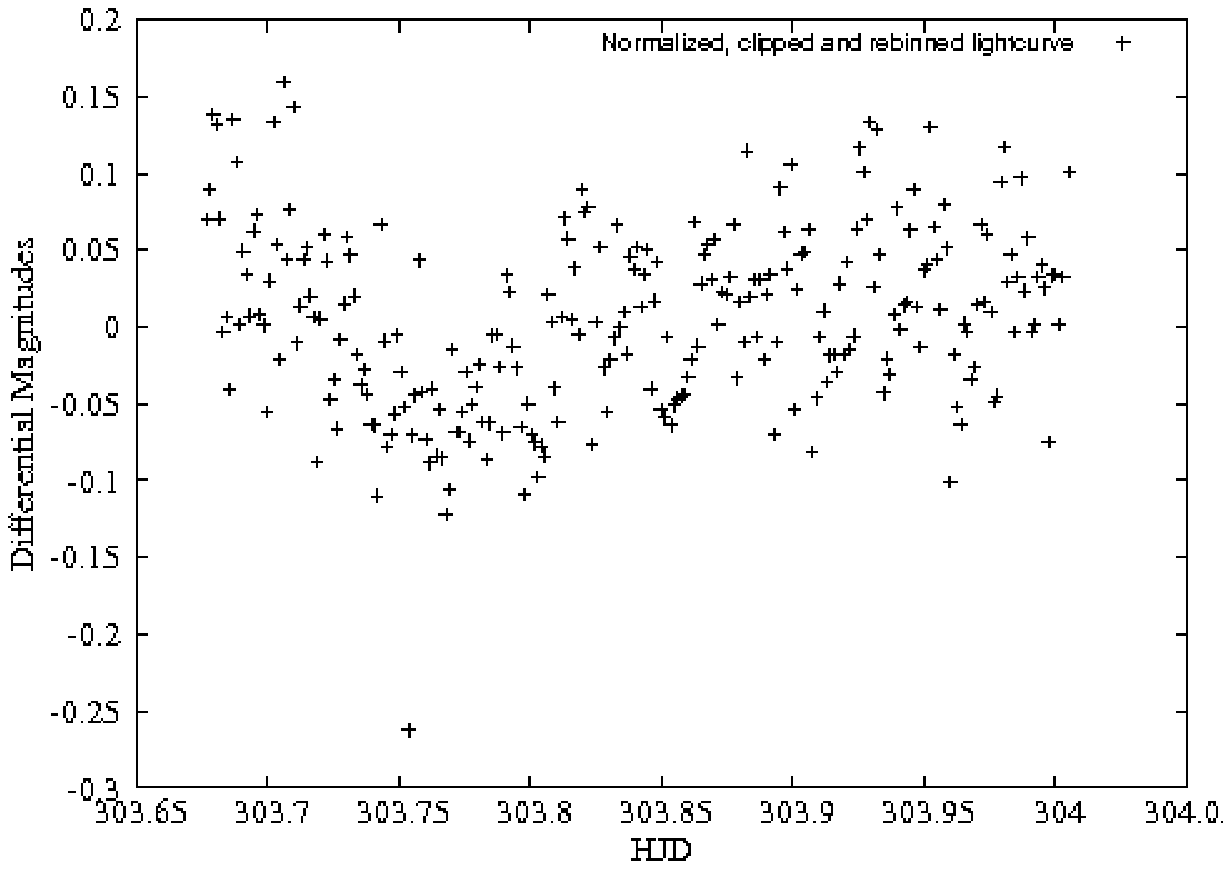}
}
\hspace{0.0cm}
\subfigure[Lomb-Scargle periodogram.] 
{
    \label{sub6}
    \includegraphics[width=6.5cm]{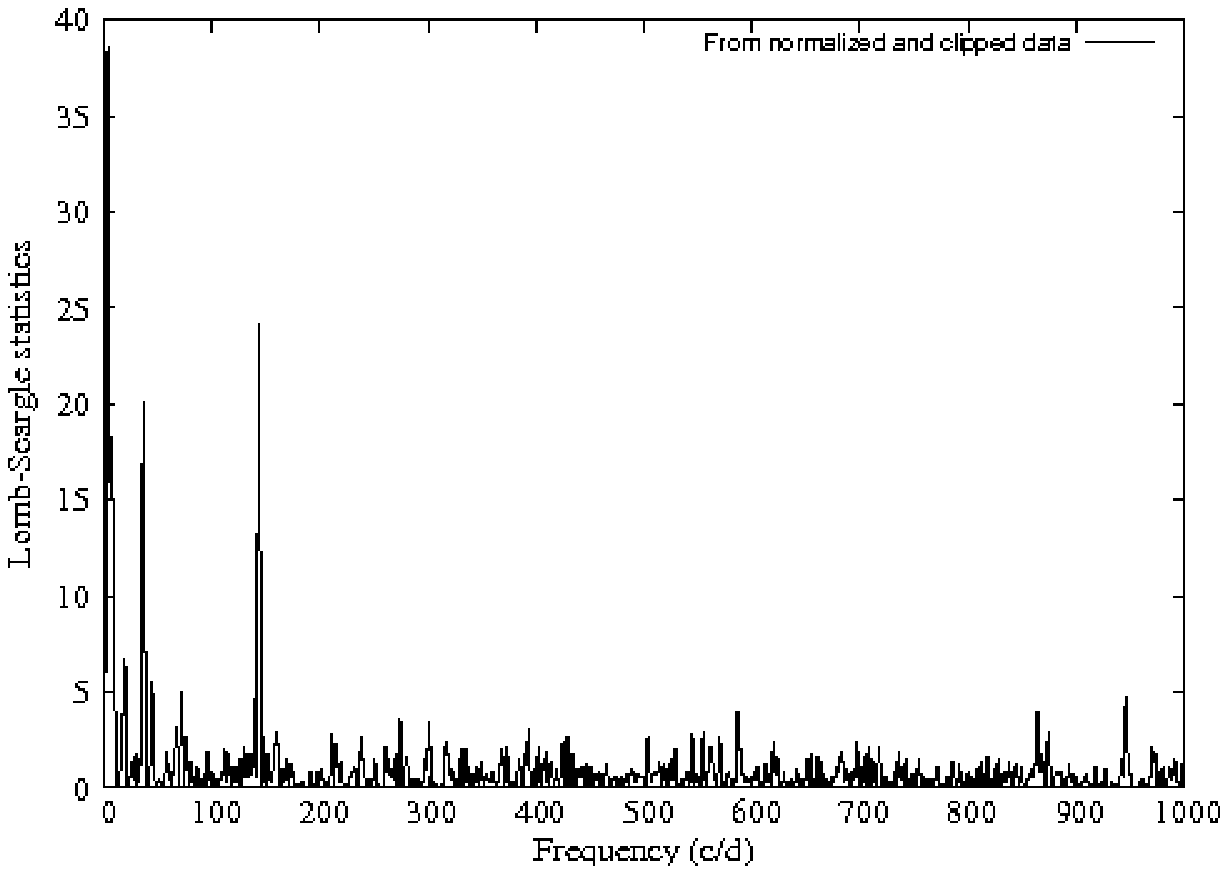}
}
\caption{\emph{The plots above (showing data from a single night of observations of SDSS J1457+51) are produced in the initial steps of the analysis program, \textsc{Speriod}. The normalised light curve where outliers have been removed, Figure~\ref{sub3}, is used for producing the Lomb-Scargle periodogram in Figure~\ref{sub6}.}} 
\end{center}
\label{fig:spec} 
\end{figure}

The input file is a light curve with 3 columns; time, magnitudes or fluxes and their errors. In the first part of the program, all columns in the input file are read and the mean magnitude or flux is calculated and subtracted from the light curve to construct a normalised light curve (Figure~\ref{sub1} and~\ref{sub2}). All examples in this section present data from one observing night of the target SDSS J1457+51.

\subsection{Part 2}
 
In part 2, the user is asked to specify the upper and lower limit for removing outliers. The normalised light curve is plotted to screen (Figure~\ref{sub3}). Next, a Gaussian smoothing is performed (Figure~\ref{sub4}), by passing a window along the light curve and assigning the weighted mean (according to a Gaussian distribution) to the bin in the centre of the window. Equal numbers of points are lost at the start and the end of the data set. As demonstrated in Figure~\ref{sub4}, a smoothing of the light curve can help to reveal periodic signals with periods long compared to the smoothing window but short compared to the total length of the data set. However, this smoothed light curve is only for visually inspecting the data, and the unsmoothed, normalised light curve is used in the following steps. In addition to smoothing the data, a rebinned light curve (Figure~\ref{sub5}) is calculated by taking the mean of every consecutive, independent set of 3 data points. This effectively changes the sampling rate of the data. 
 
\subsection{Part 3}

In part 3, a Lomb-Scargle periodogram (\citealt{1976Ap&SS..39..447L, 1982ApJ...263..835S}) is constructed from the normalised, unsmoothed light curve, with outliers removed. The program will ask the user to specify the frequency range and the oversampling to be used. A larger oversampling causes the program to search for power excess in finer frequency steps. The output power spectrum is printed to screen (Figure~\ref{sub6}).

\begin{figure}
\begin{center}
\includegraphics[scale=0.42]{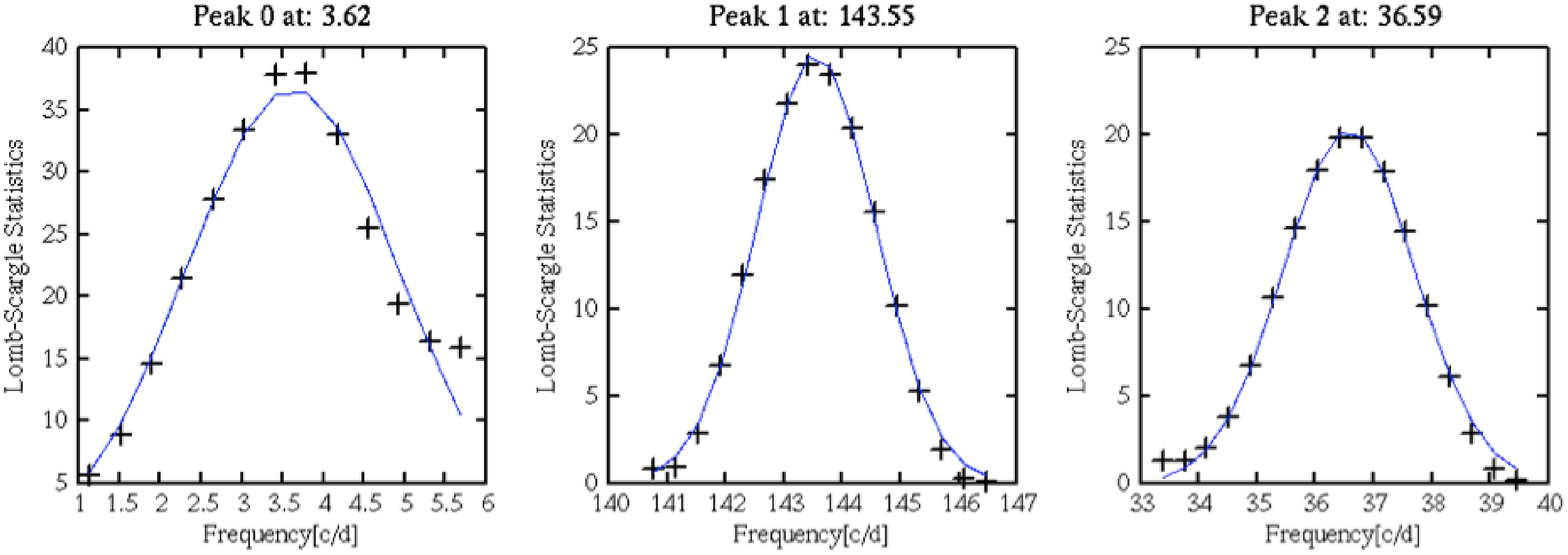}
\caption{\emph{Example of an output from \textsc{Speriod}, showing the signals of highest amplitude found in the power spectrum in Figure~\ref{sub6}.}}
\label{sub7}
\end{center}
\end{figure}

\subsection{Part 4}

Next, frequencies which have excess power (hereafter peaks) are automatically found by looping over all values in the Lomb-Scargle data file, to find the highest amplitudes. The user is asked to specify a lower search limit in units of power:
\\\\
\textbf{\tt Please specify lower search limit for finding peaks in the power spectrum:}
\\  
\textbf{>> 15}
\\\\
All peaks above this value will be registered. A Gaussian is then fitted to each peak. However, this is only to construct a list containing all rough peak frequencies of interest. This list is later used to limit the frequency range when performing Monte Carlo simulations to find the exact frequencies of the peaks and their corresponding errors. Both the peaks and their Gaussian fits are plotted to screen together with the preliminary frequencies (see Figure~\ref{sub7}). 

\subsection{Part 5}

\begin{figure}
\begin{center}
\includegraphics[scale=0.8]{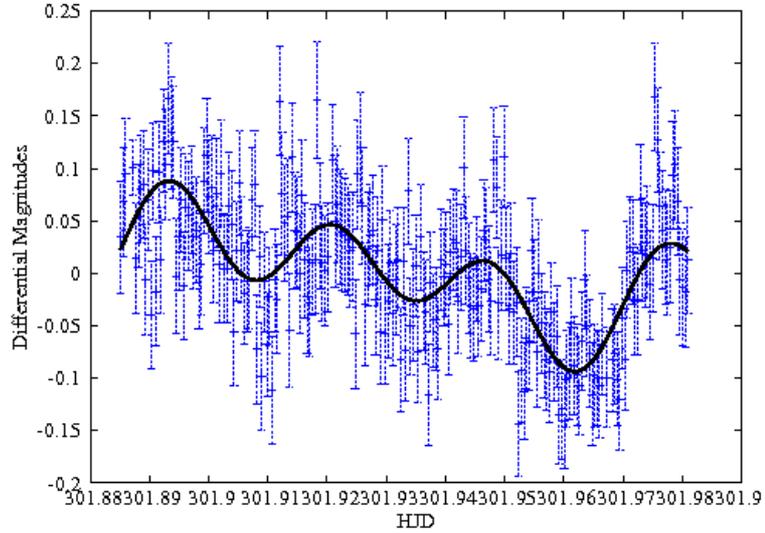}
\caption{\emph{The data light-curve with a fake light curve constructed from the highest frequencies found in the Lomb-Scargle periodogram, plotted on top. The least-square method is used to rescale the errors on the data light-curve.}}
\label{sub8}
\end{center}
\end{figure}

In part 5, the errors on the original magnitude and flux values are rescaled. This is done by fitting multiple sine waves to the normalised and unsmoothed light curve with outliers removed, and scaling the errors until  the reduced $\chi^{2} \sim$ 1 (the fit is a composite of the three highest frequencies found in Part 4). This process is illustrated in Figure~\ref{sub8}. This plot is also saved as an output file.
\\\\
\noindent \textbf{\tt Please wait - rescaling the errors...}
\\
\noindent \textbf{\tt The reduced chi2 is 0.989. Errors are rescaled to 87.0\% of original errors.}
\\\\
\noindent The now normalised, unsmoothed light curve, with outliers removed and with rescaled errors, is saved in an output file to be used in further analysis. 

\subsection{Part 6}

\begin{figure}
\begin{center}
\includegraphics[scale=0.43]{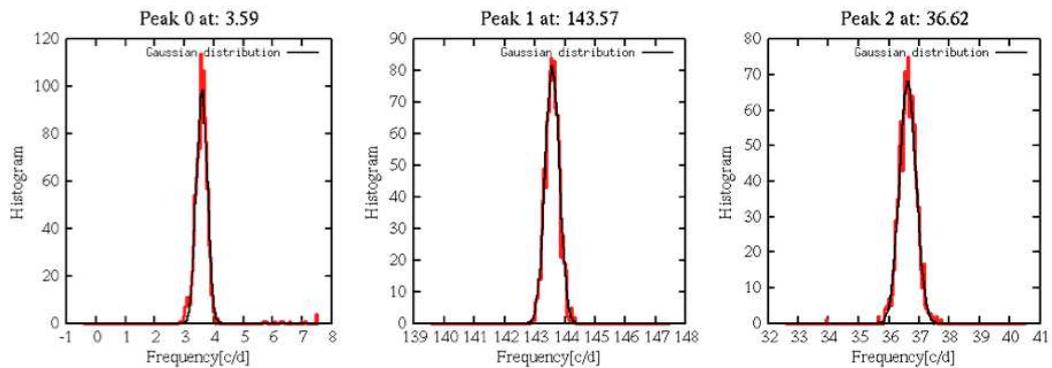}
\caption{\emph{Monte Carlo simulations of the peaks found in Figure~\ref{sub7}.}}
\label{sub9}
\end{center}
\end{figure}

Next, Monte Carlo simulations are performed with the aim of finding the exact peak frequencies and their errors. In this method, the peak errors are found by randomly re-distributing the points in the light curve within their errors a repeated number of times, and constructing a Lomb-Scargle periodogram each time. The 1-$\sigma$ error is then obtained from the output distribution of the peaks found in the periodograms, both statistically, and by fitting a Gaussian. To speed up the program, simulations are only performed in a narrow range around each peak. The frequencies, amplitudes and errors for every peak are printed to screen and the Gaussian fits are plotted (Figure~\ref{sub9}).

\subsection{Part 7}

Next, the user will be asked whether or not a spectral cleaning of the Lomb-Scargle periodogram should be performed:  
\\\\
\textbf{\tt  Do you want to do a spectral cleaning? Answer yes or no. (This will take less than a minute) }
\\
\textbf{>> yes}
\\
\textbf{\tt  How far down in the power noise do you want to continue cleaning? (Give a lower power limit) }
\\
\textbf{>> 3}
\\
\textbf{\tt  Cleaning in progress...}
\\
\textbf{\tt  Cleaning 10\% from peak with frequency 7.209 c/d}
\\
\textbf{\tt ...}
\newline

\begin{figure}
\subfigure[] 
{
    \label{sub10a}
    \includegraphics[width=6.5cm]{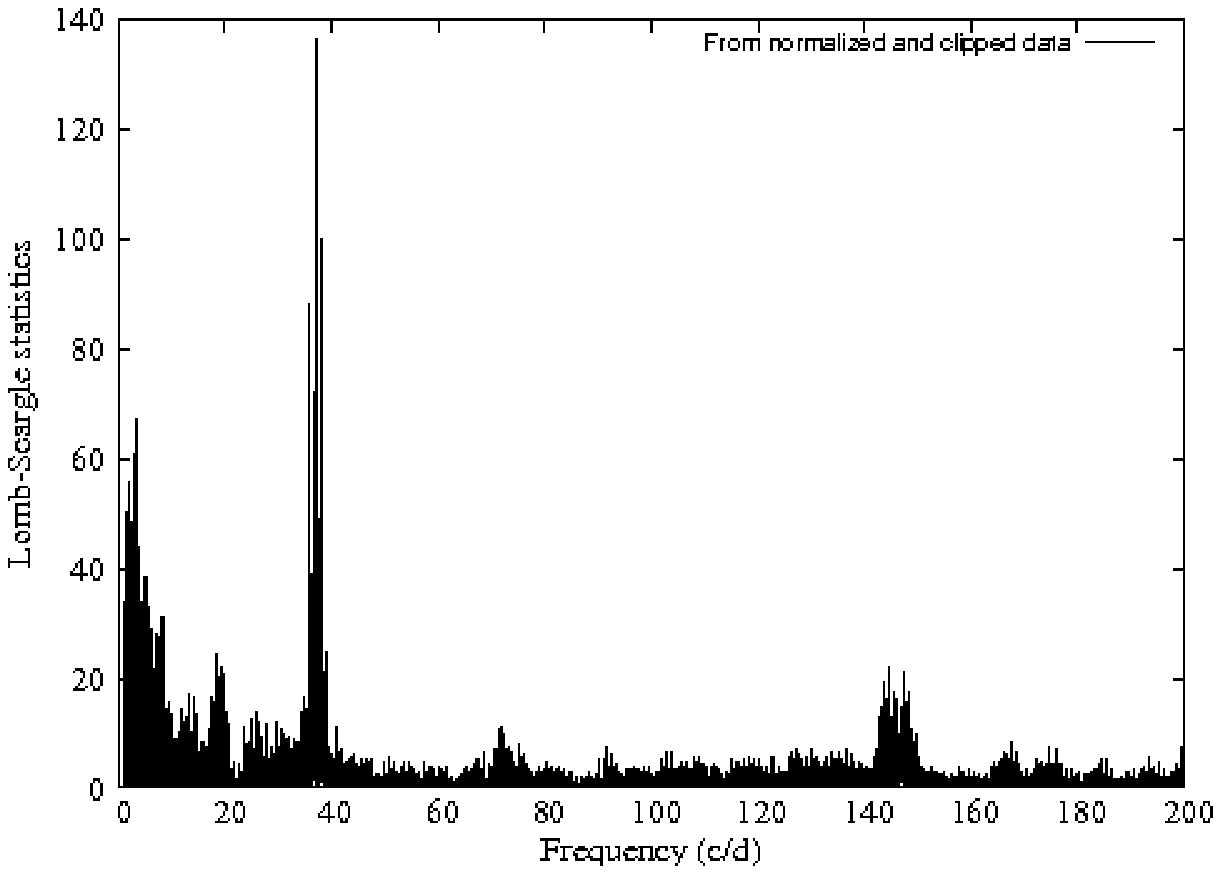}
}
\hspace{0.0cm}
\subfigure[] 
{
    \label{sub10b}
    \includegraphics[width=6.5cm]{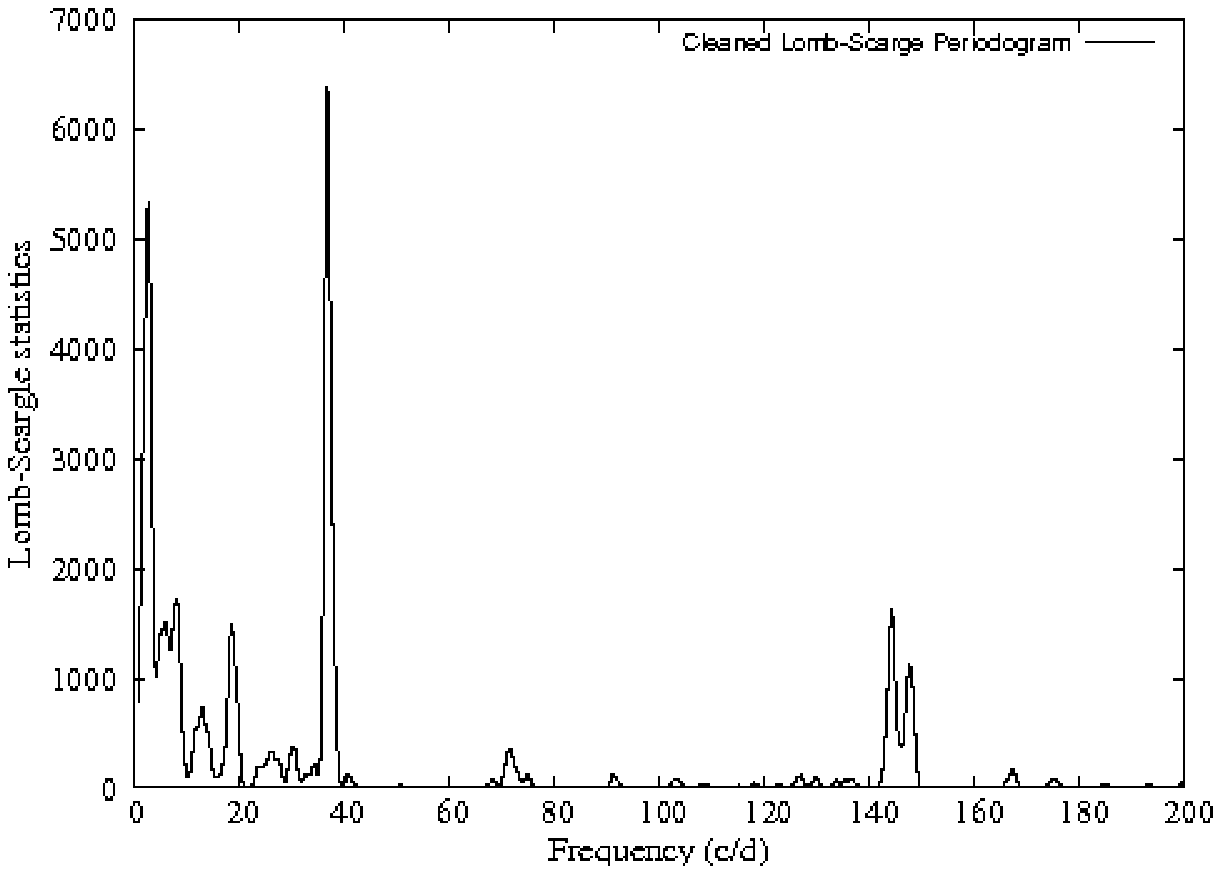}
}
\caption{\emph{Figure~\ref{sub10a} shows the Lomb-Scargle periodogram of SDSS1457+51, constructed from multiple nights. Figure~\ref{sub10b} show the cleaned periodogram of the same data set.}} 
\label{sub10} 
\end{figure}

\noindent If the user answers \emph{yes}, the Lomb-Scargle diagram is imported and the maximum frequency is found. A fake light curve, a sine wave at this frequency, is produced and its corresponding window function is plotted (as a Lomb-Scargle periodogram). The peak amplitude of the window function is scaled so that it matches the original peak in the Lomb-Scargle periodogram. Next, the window function multiplied with a gain factor (10\% of the peak amplitude) is subtracted from the original Lomb-Scargle periodogram. The same cleaning process is then repeated again, first finding the highest peak, now from the partially cleaned Lomb-Scargle periodogram, then a fake light curve is constructed together with its corresponding power spectrum, and so it continues. This will continue until the peaks are below the user-specified lower level. Left are now only the residuals with all the peaks removed. Finally, the peaks are added back to the residual periodogram, constructing the cleaned periodogram. The shape of the peaks in the cleaned periodogram corresponds to a virtual peak of the same width as the most narrow peak found in the original periodogram, and with an amplitude matching the total calculated area of what has been removed. It is important to realise that the cleaned spectrum should not be used to finding frequencies, errors or amplitudes, and is only meant as a tool for better visual inspection of the frequency contents of the data. A cleaned power-spectrum can still be of much use, especially when inspecting data from multiple nights, where the frequency content is more complex. Figure~\ref{sub10} shows the multiple-night power spectrum of SDSS1457+51, before and after cleaning.

\subsection{Part 8}

In part 8, the user is asked whether or not the window function should be calculated. If the user answers \emph{yes}, a list of all frequencies found in previous steps appears on screen. After choosing a peak from the list, a fake light curve at the corresponding frequency and with the same sampling is produced, and a Lomb-Scargle periodogram is constructed.

\subsection{Part 9}

\begin{figure}
\begin{center}
\includegraphics[scale=0.7]{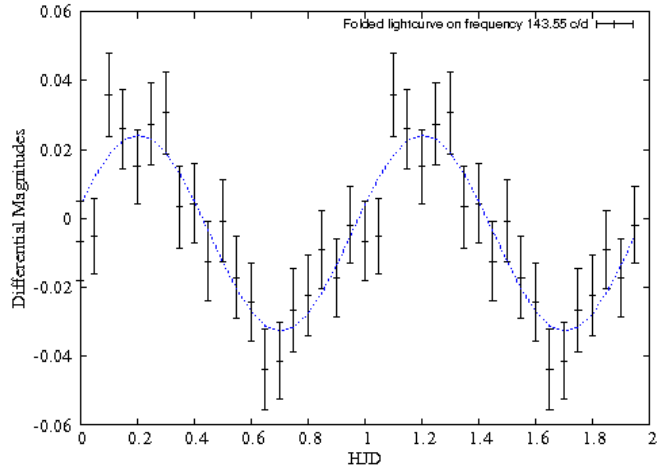}
\caption{\emph{The folded light curve of SDSS 1457+51 from one single night. It is folded on the frequency 143.546 (c\,d$^{-1}$).}}
\label{sub11}
\end{center}
\end{figure}

Finally, the user is asked if folded light-curves should be produced. A list of all frequencies found in previous steps is plotted to screen, and the user is asked to choose which one to fold and plot. 
\\\\
\textbf{\tt Do you want to fold the light curve? (answer yes or no)}
\\
\textbf{>> yes}
\\
\textbf{\tt Choose peak from the list below:}
\\
\textbf{\tt Peak 0 has a frequency of 3.596 (c/d) and a period of 0.278 (d)}
\\
\textbf{\tt Peak 1 has a frequency of 143.546 (c/d) and a period of 0.007 (d)}
\\
\textbf{\tt Peak 2 has a frequency of 36.586 (c/d) and a period of 0.027 (d)}
\\
\textbf{\tt Which peak number do you want to fold? Type q when you want to quit. }
\\
\textbf{>> 1}
\newline
\noindent Folded light curves can be produced at several frequencies and the loop stops when the user types q. Figure~\ref{sub11} shows the folded light curve with a fitted sine wave plotted on top, according to the example above.

\section{Analysis of Multiple Light Curves}

When running the program \textsc{Speriod} for combined light curves from several nights, I quickly realised that the program needed to be adapted to handle the longer data sets and account for the problems associated with aliases. I therefore developed \textsc{Speriod-All} for the analysis of more complex light curves. The input data file should have 3 columns; time, magnitudes or fluxes and errors. In addition to using Monte Carlo simulations to find frequencies with power excess and their associated errors, bootstrap simulations are also performed as a comparison. Bootstrap simulations select a random subset of the light curve containing approximately 2/3 of the the data values, and are useful for dealing with multiple-day datasets as the random re-sampling helps with reduction of aliases caused by the day-to-day data gaps.

\subsection{Part 1}
 
\begin{figure}[h!]
\begin{center}
\subfigure[] 
{
    \label{sub12a}
    \includegraphics[width=6.5cm]{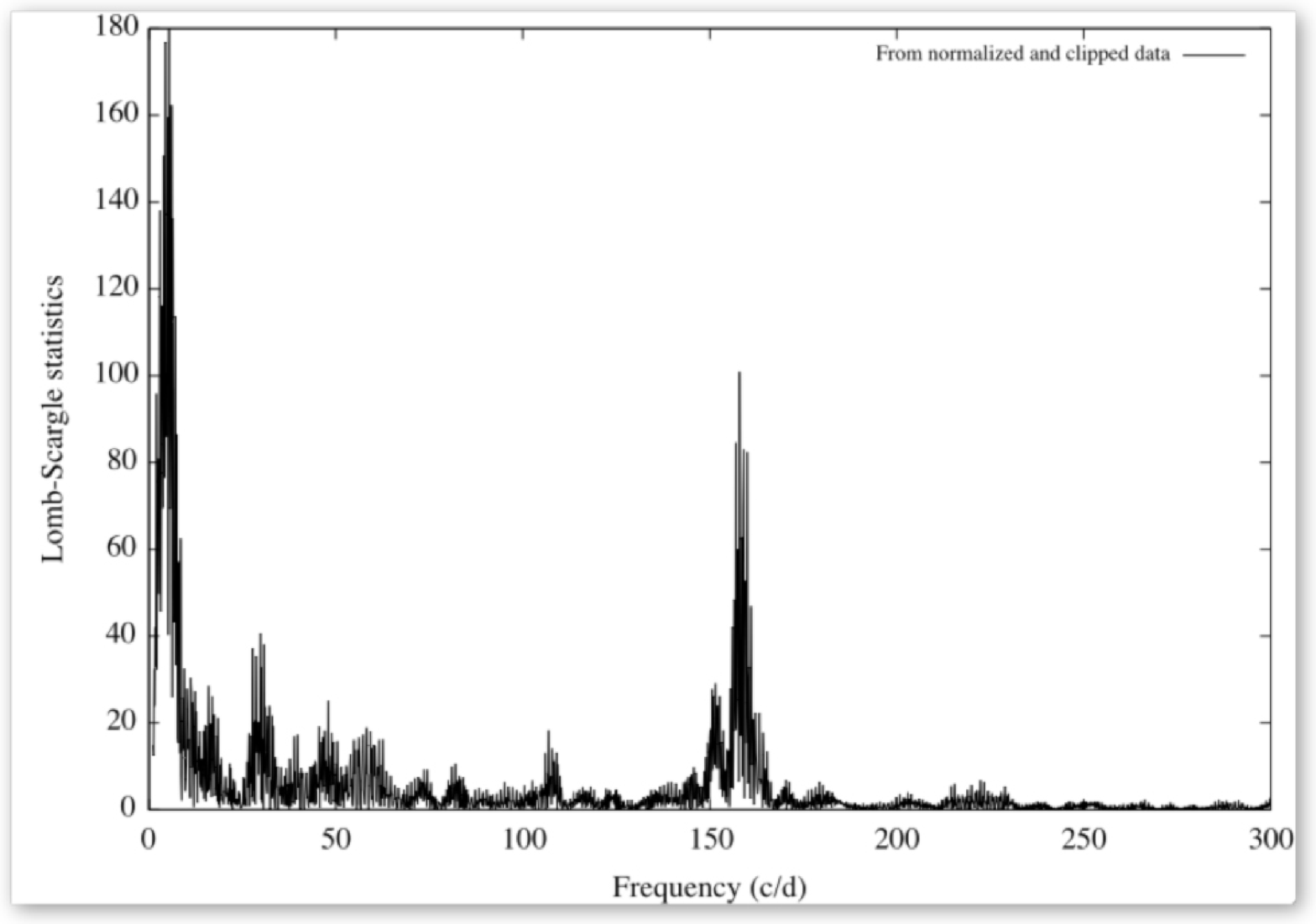}
}
\hspace{0.0cm}
\subfigure[] 
{
    \label{sub12b}
    \includegraphics[width=6.5cm]{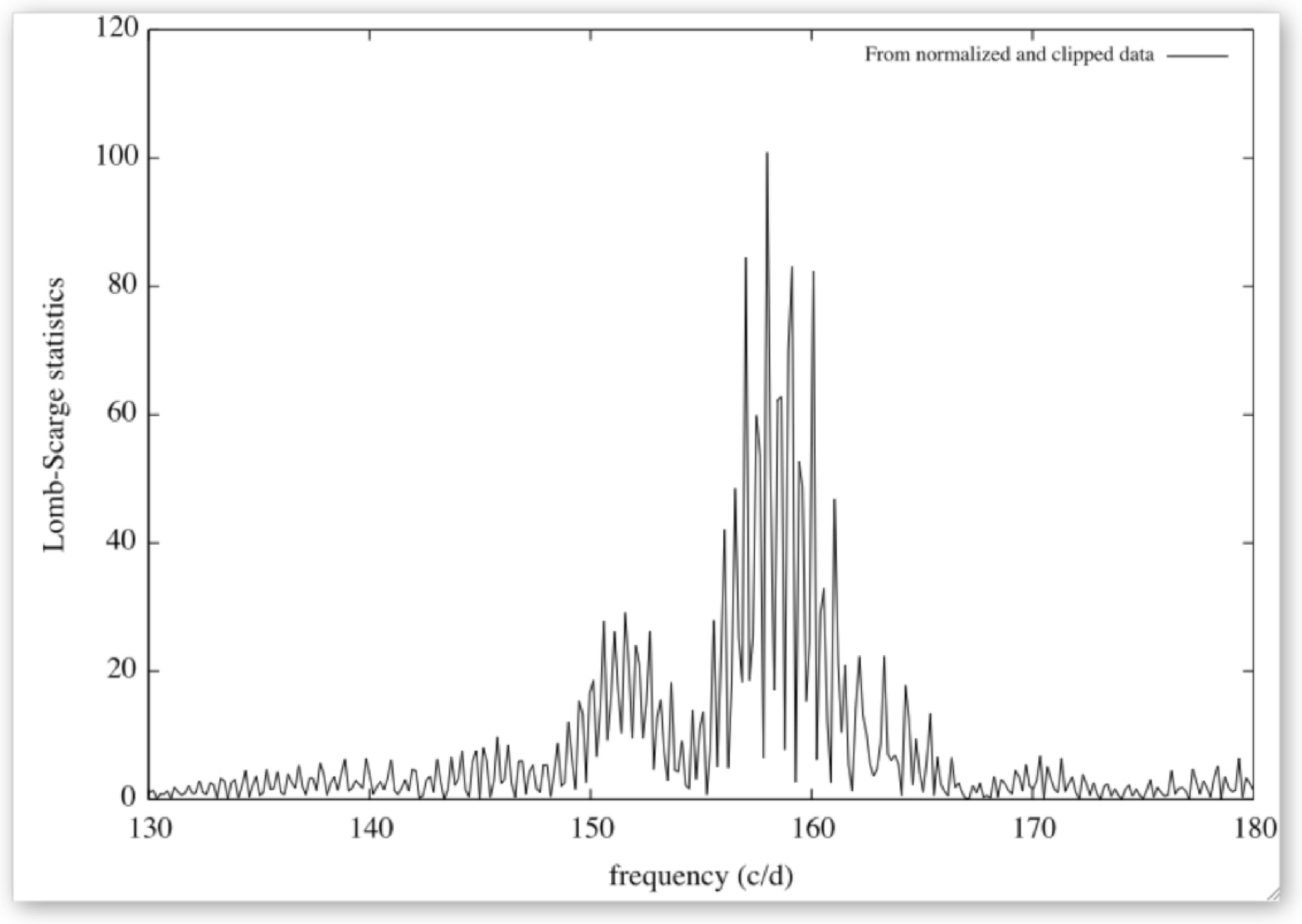}
}
\caption{\emph{Figure~\ref{sub12a} shows the power spectrum for multiple nights, and Figure~\ref{sub12b}, shows a zoom of the same power spectrum in the range between 130 c\,$d^{-1}$ -- 180 c\,$d^{-1}$ (according to the example in the text).}} 
\end{center}
\label{sub12} 
\end{figure}

In part 1, a Lomb-Scargle periodogram is constructed from the input data file and the user is asked to choose which of the frequency regions should be investigated further. The user can zoom in and out in the Lomb-Scargle periodogram, and roughly define the left and right edges of every peak (Figure~\ref{sub12}).
 \\\\
 \textbf{\tt I got 4 options for you: }
\\
 \textbf{\tt  1) If you want to zoom in on a peak, type 1 }
 \\
 \textbf{\tt  2) If you want to zoom out to original size, type 2  }
 \\
 \textbf{\tt  3) If you want to enter a range for the bootstrap analysis, type 3 }
 \\
 \textbf{\tt  4) If you are done, type 4 }
 \\
 \textbf{>> 1}
 \\
 \textbf{>> 130,180}
 
\subsection{Part 2}
 
\begin{figure}
\begin{center}
\includegraphics[scale=0.4]{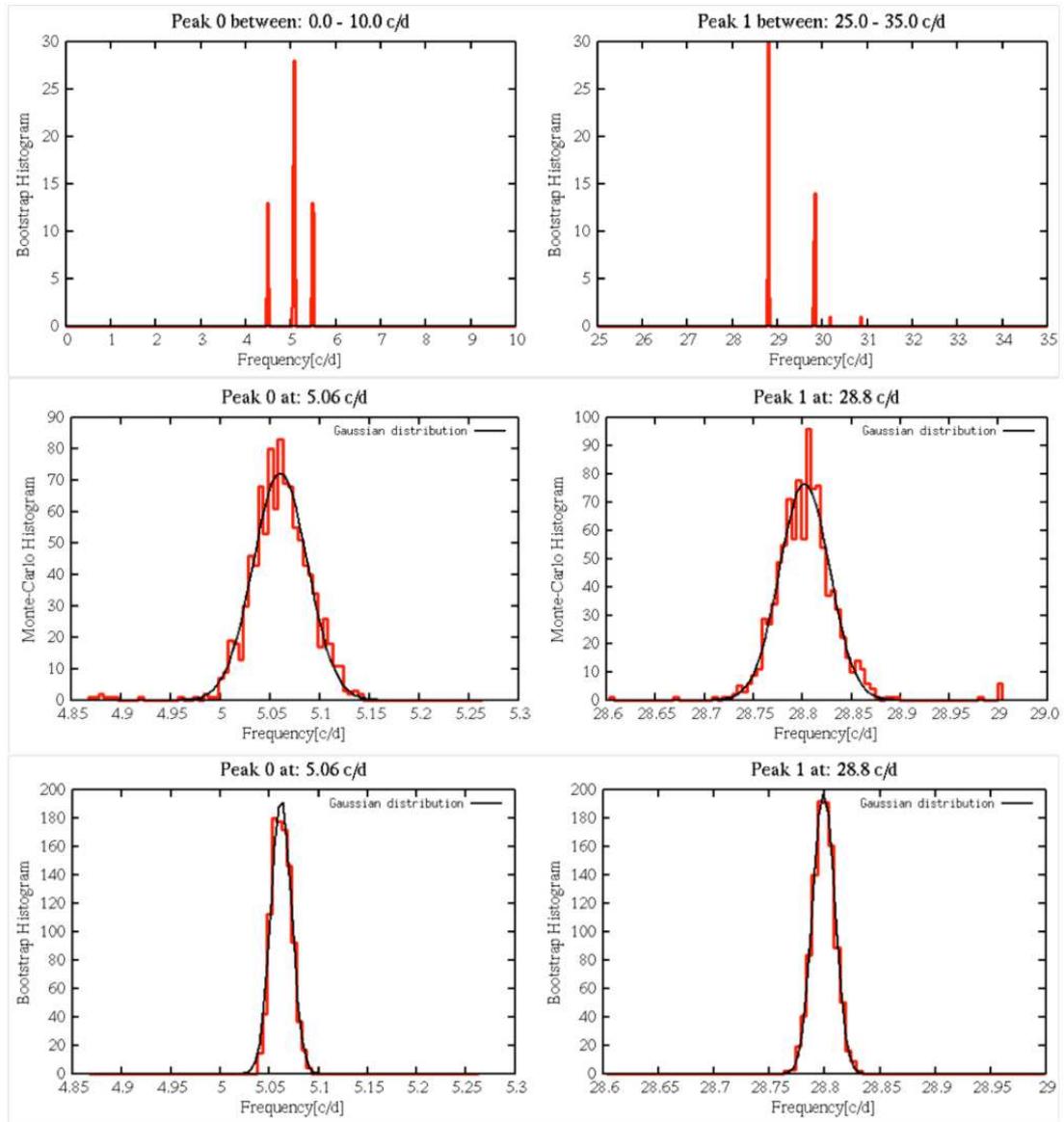}
\caption{\emph{The upper row shows a histogram of two user defined ranges where simulations will be performed. In the middle row, the highest alias found in the Monte Carlo simulations is plotted for each of the two user defined ranges. The bottom row shows the highest alias found in the bootstrap simulations.}}
\label{sub13}
\end{center}
\end{figure}

Next, the user will be asked if Monte Carlo and bootstrap simulations should be performed. The oversampling is set to 64. Three plots are produced: 1) The initial histogram of all peaks plotted over each user-defined frequency range. 2) The highest alias found by the Monte Carlo simulations. 3)The highest alias found by the Bootstrap simulations (see Figure~\ref{sub13}). Errors on the peaks are found from the output histograms of the peak frequencies found by the simulations. 

\subsection{Part 3}

In part 3, a spectral cleaning is performed in the same way as described in the previous section (Figure~\ref{sub10}). 

\subsection{Part 4}

\begin{figure}
\begin{center}
\includegraphics[scale=0.4]{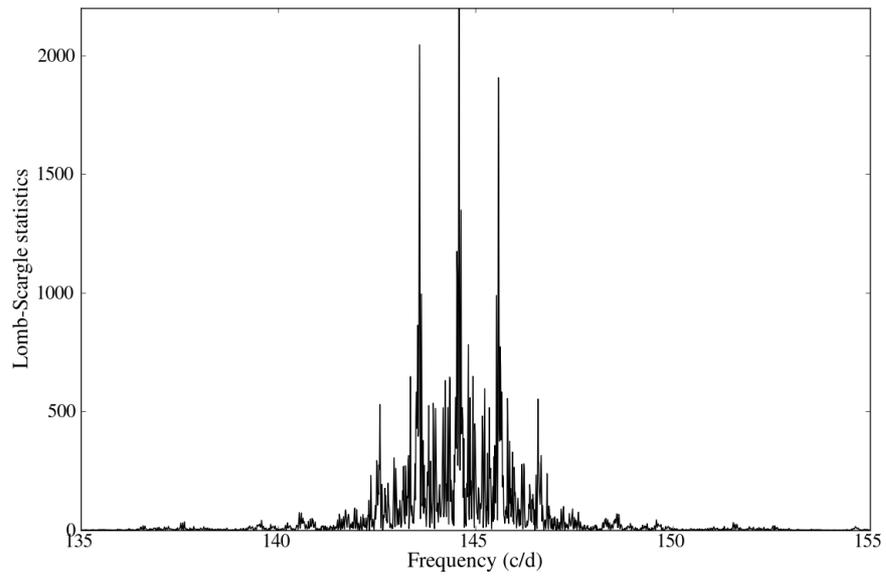}
\caption{\emph{Window function at frequency $\approx$ 145 c\,$d^{-1}$.}}
\label{sub14}
\end{center}
\end{figure}

Next, the spectral window functions are calculated as described in the previous section (Figure~\ref{sub14}).

\subsection{Part 5}

The folding of the light curve is also performed in the same manner as described in the previous section. Finally, all results are printed to screen and to an output log file:
 \newpage
 \textbf{\tt OUTPUT - SPERIOD-ALL}
 \\
  \textbf{**************************************************}
  \\
 \textbf{\tt  Input file: mags.txt has 1876 data points}
 \\
\textbf{\tt Peaks are chosen at:}
 \\
\textbf{\tt Peak no. 0 is in range  0.0 - 10.0  c/d.}
 \\
\textbf{\tt Peak no. 1 is in range  25.0 - 35.0  c/d.}
 \\\\
\textbf{\tt Bootstrap analysis gives...}
 \\
\textbf{\tt Statistics:}
 \\
 \textbf{\tt Period:  5.06267578  +/-  0.00951901 c/d (1 sigma)}
 \\
\textbf{\tt Period:  28.79924526  +/-  0.01006367 c/d (1 sigma)}
 \\
\textbf{\tt Gaussianfit:}
 \\
\textbf{\tt Period:  5.06195725  +/-  0.01053709 c/d (1 sigma)}
   \\
\textbf{\tt Period:  28.79913552  +/-  0.01005182 c/d (1 sigma)}
 \\\\
\textbf{\tt Monte Carlo Simulations gives...}
 \\
\textbf{\tt Statistics:}
 \\
 \textbf{\tt Period:  5.06213765  +/-  0.03155680 c/d (1 sigma)}
 \\
\textbf{\tt Period:  28.80118285  +/-  0.03214551 c/d (1 sigma)}
  \\
\textbf{\tt Gaussianfit:}
 \\
\textbf{\tt Period:  5.05995243  +/-  0.02726467 c/d (1 sigma)}
 \\
\textbf{\tt Period:  28.80196939  +/-  0.02545825 c/d (1 sigma)}
 \\\\
\textbf{\tt Good Bye!}

\newpage \thispagestyle{empty} \mbox{} 
\pagestyle{plain} 
\renewcommand\bibname{\Huge \sc{References}}
\bibliography{ref} 
 
\end{document}